\documentclass[twoside,twocolumn,9pt]{article}
\usepackage{extsizes}
\usepackage[super,sort&compress,comma]{natbib} 
\usepackage[version=3]{mhchem}
\usepackage[left=1.5cm, right=1.5cm, top=1.785cm, bottom=2.0cm]{geometry}
\usepackage{balance}
\usepackage{mathptmx}
\usepackage{sectsty}
\usepackage{graphicx} 
\usepackage{lastpage}
\usepackage[format=plain,justification=justified,singlelinecheck=false,font={stretch=1.125,small,sf},labelfont=bf,labelsep=space]{caption}
\usepackage{float}
\usepackage{fancyhdr}
\usepackage{fnpos}
\usepackage[english]{babel}
\addto{\captionsenglish}{%
  
}
\usepackage{array}
\usepackage{droidsans}
\usepackage{charter}
\usepackage[T1]{fontenc}
\usepackage[usenames,dvipsnames]{xcolor}
\usepackage{setspace}
\usepackage[compact]{titlesec}
\usepackage{hyperref}
\usepackage{units}
\usepackage{placeins} %% to get the float-barrier option --> figures don't float further than barrier
\usepackage{amsmath,amsthm,amssymb,amsfonts}
\usepackage{xcolor}
\usepackage{hhline} 
\usepackage{multirow}
\usepackage{xr} %to quote from the SI
\usepackage{cleveref}
\usepackage{siunitx}
\usepackage{enumitem}
\usepackage{mathtools}
\usepackage{subfig}
\usepackage{siunitx}
\usepackage{xargs}                      % Use more than one optional parameter in a new commands
\usepackage{tikz}   % program images
\usepackage{xfrac}  % nicely display fractions in text
\newcommand{\angstrom}{\mbox{\normalfont\AA}}
\usepackage[colorinlistoftodos,prependcaption,textsize=tiny]{todonotes}
\newcommandx{\unsure}[2][1=]{\todo[linecolor=red,backgroundcolor=red!25,bordercolor=red,#1]{#2}}
\newcommandx{\change}[2][1=]{\todo[linecolor=blue,backgroundcolor=blue!25,bordercolor=blue,#1]{#2}}
\newcommandx{\info}[2][1=]{\todo[linecolor=green,backgroundcolor=green!25,bordercolor=green,#1]{#2}}
\newcommandx{\improvement}[2][1=]{\todo[linecolor=Plum,backgroundcolor=Plum!25,bordercolor=Plum,#1]{#2}}
\newcommand{\defeq}{\stackrel{\text{def}}{=}}

\usepackage{stfloats} % to put 2-column figures at the bottom of the page
\usepackage{chemscheme}
\crefformat{figure}{Figure S#2#1#3}

\newcommand{\comment}[1]{}

\usepackage{epstopdf}%This line makes .eps figures into .pdf - please comment out if not required.

\definecolor{cream}{RGB}{222,217,201}

\begin{document}

\pagestyle{fancy}
\thispagestyle{plain}
\fancypagestyle{plain}{
%%%HEADER%%%
\renewcommand{\headrulewidth}{0pt}
}
%%%END OF HEADER%%%

%%%PAGE SETUP - Please do not change any commands within this section%%%
\makeFNbottom
\makeatletter
\renewcommand\LARGE{\@setfontsize\LARGE{15pt}{17}}
\renewcommand\Large{\@setfontsize\Large{12pt}{14}}
\renewcommand\large{\@setfontsize\large{10pt}{12}}
\renewcommand\footnotesize{\@setfontsize\footnotesize{7pt}{10}}
\makeatother

\renewcommand{\thefootnote}{\fnsymbol{footnote}}
\renewcommand\footnoterule{\vspace*{1pt}% 
\color{cream}\hrule width 3.5in height 0.4pt \color{black}\vspace*{5pt}} 
\setcounter{secnumdepth}{5}

\makeatletter 
\renewcommand\@biblabel[1]{#1}            
\renewcommand\@makefntext[1]% 
{\noindent\makebox[0pt][r]{\@thefnmark\,}#1}
\makeatother 
\renewcommand{\figurename}{\small{Fig.}~}
\sectionfont{\sffamily\Large}
\subsectionfont{\normalsize}
\subsubsectionfont{\bf}
\setstretch{1.125} %In particular, please do not alter this line.
\setlength{\skip\footins}{0.8cm}
\setlength{\footnotesep}{0.25cm}
\setlength{\jot}{10pt}
\titlespacing*{\section}{0pt}{4pt}{4pt}
\titlespacing*{\subsection}{0pt}{15pt}{1pt}
%%%END OF PAGE SETUP%%%

%%%FOOTER%%%
\fancyfoot{}
\fancyfoot[LO,RE]{\vspace{-7.1pt}\includegraphics[height=9pt]{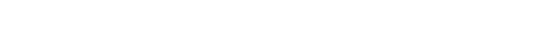}}
\fancyfoot[CO]{\vspace{-7.1pt}\hspace{11.9cm}\includegraphics{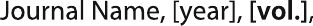}}
\fancyfoot[CE]{\vspace{-7.2pt}\hspace{-13.2cm}\includegraphics{head_foot/RF}}
\fancyfoot[RO]{\footnotesize{\sffamily{1--\pageref{LastPage} ~\textbar  \hspace{2pt}\thepage}}}
\fancyfoot[LE]{\footnotesize{\sffamily{\thepage~\textbar\hspace{4.65cm} 1--\pageref{LastPage}}}}
\fancyhead{}
\renewcommand{\headrulewidth}{0pt} 
\renewcommand{\footrulewidth}{0pt}
\setlength{\arrayrulewidth}{1pt}
\setlength{\columnsep}{6.5mm}
\setlength\bibsep{1pt}
%%%END OF FOOTER%%%

%%%FIGURE SETUP - please do not change any commands within this section%%%
\makeatletter 
\newlength{\figrulesep} 
\setlength{\figrulesep}{0.5\textfloatsep} 

\newcommand{\topfigrule}{\vspace*{-1pt}% 
\noindent{\color{cream}\rule[-\figrulesep]{\columnwidth}{1.5pt}} }

\newcommand{\botfigrule}{\vspace*{-2pt}% 
\noindent{\color{cream}\rule[\figrulesep]{\columnwidth}{1.5pt}} }

\newcommand{\dblfigrule}{\vspace*{-1pt}% 
\noindent{\color{cream}\rule[-\figrulesep]{\textwidth}{1.5pt}} }

\makeatother
%%%END OF FIGURE SETUP%%%

%%%TITLE, AUTHORS AND ABSTRACT%%%
\twocolumn[
  \begin{@twocolumnfalse}
{\includegraphics[height=30pt]{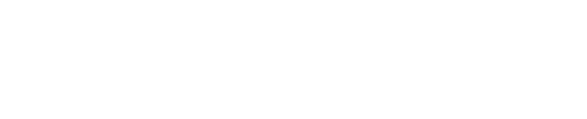}\hfill\raisebox{0pt}[0pt][0pt]{\includegraphics[height=55pt]{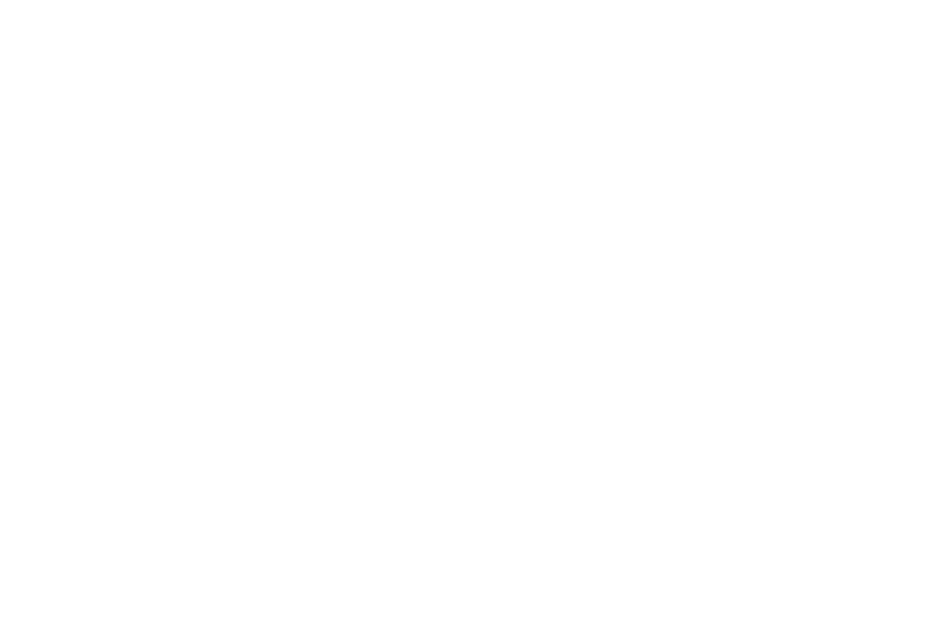}}\\[1ex]
\includegraphics[width=18.5cm]{head_foot/LF}}\par
\vspace{1em}
\sffamily
\begin{tabular}{m{4.5cm} p{13.5cm} }

\includegraphics{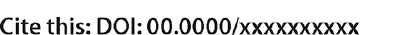} & \noindent\LARGE{\textbf{Controlling $\text{Li}^+$ transport in ionic liquid electrolytes through salt content and anion asymmetry: A mechanistic understanding gained from molecular dynamics simulations$^\dag$}} \\%Article title goes here instead of the text "This is the title"
\vspace{0.3cm} & \vspace{0.3cm} \\

 & \noindent\large{Alina Wettstein,\textit{$^{a}$} Diddo Diddens,\textit{$^{b}$} and Andreas Heuer\textit{$^{a,b}$}} \\%Author names go here instead of "Full name", etc.

\includegraphics{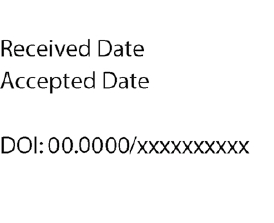} & \noindent\normalsize{In this work, we report the results from molecular dynamics simulations of lithium salt-ionic liquid electrolytes (ILEs) based either on the symmetric  bis[(trifluoromethyl)sulfonyl]imide ($\text{TFSI}^-$) anion or its asymmetric analog 2,2,2-(trifluoromethyl)sulfonyl-N-cyanoamide ($\text{TFSAM}^-$). Relating lithium's coordination environment to anion mean residence times and diffusion constants confirms the remarkable transport behaviour of the $\text{TFSAM}^-$-based ILEs that has been observed in recent experiments: For increased salt doping, the lithium ions must compete for the more attractive cyano over oxygen coordination and a fragmented landscape of solvation geometries emerges, in which lithium appears to be less strongly bound.
We present a novel, yet statistically straightforward methodology to quantify the extent to which lithium and its solvation shell are dynamically coupled. 
By means of a Lithium Coupling Factor (LCF) we demonstrate that the shell anions do not constitute a stable lithium vehicle, which suggests for this electrolyte material the commonly termed "vehicular" lithium transport mechanism could be more aptly pictured as a correlated, flow-like motion of lithium and its neighbourhood. 
Our analysis elucidates two separate causes why lithium and shell dynamics progressively decouple with higher salt content: On the one hand, an increased sharing of anions between lithium limits the achievable LCF of individual lithium-anion pairs. On the other hand, weaker binding configurations naturally entail a lower dynamic stability of the lithium-anion complex, which is particularly relevant for the $\text{TFSAM}^-$-containing ILEs.

} \\%The abstract goes here instead of the text "The abstract should be..."

\end{tabular}

 \end{@twocolumnfalse} \vspace{0.6cm}

  ]
%%%END OF TITLE, AUTHORS AND ABSTRACT%%%

%%%FONT SETUP - please do not change any commands within this section
\renewcommand*\rmdefault{bch}\normalfont\upshape
\rmfamily
\section*{}
\vspace{-1cm}

%%%FOOTNOTES%%%

\footnotetext{\textit{$^{a}$~Institut f\"ur physikalische Chemie, Westf\"alische Wilhelms-Universit\"at M\"unster, Corrensstra{\ss}e 28/30, D-48149 M\"unster, Germany. Fax:+49 (0)251 83 29159; Tel:+49 (0)251 83 29177; E-mail: andheuer@uni-muenster.de}}

\footnotetext{\textit{$^{b}$~Institut f\"ur Energie- und Klimaforschung, Ionics in Energy Storage, Helmholtz Institut M\"unster, Forschungszentrum J\"ulich, Corrensstra{\ss}e 46, 48149 M\"unster, Germany.}}

%Please use \dag to cite the ESI in the main text of the article.
%If you article does not have ESI please remove the the \dag symbol from the title and the footnotetext below.
\footnotetext{\dag~Electronic Supplementary Information (ESI) available}
%additional addresses can be cited as above using the lower-case letters, c, d, e... If all authors are from the same address, no letter is required

%%%END OF FOOTNOTES%%%

%%%%%%%%%%%%%%%%%%%%%%%%%%%%%%%%%%%%%%%%%%%%%%%%%%%%%%%%%%%%%%%%%%%%%
%%%%%%%%%%%%%%%%%%%%%%%%%%%%%%%%%%%%%%%%%%%%%%%%%%%%%%%%%%%%%%%%%%%%%
%%%%%%%%%%%%%%%%%%%%%%%%%%%%%%%%%%%%%%%%%%%%%%%%%%%%%%%%%%%%%%%%%%%%%
%%%%%%%%%%%%%%%%%%%%%%%%%%%%%%%%%%%%%%%%%%%%%%%%%%%%%%%%%%%%%%%%%%%%%
%%%%%%%%%%%%%%%%%%%%%%%%%%%%%%%%%%%%%%%%%%%%%%%%%%%%%%%%%%%%%%%%%%%%%
%%%%%%%%%%%%%%%%%%%%%%%%%%%%%%%%%%%%%%%%%%%%%%%%%%%%%%%%%%%%%%%%%%%%%
%%%%%%%%%%%%%%%%%%%%%%%%%%%%%%%%%%%%%%%%%%%%%%%%%%%%%%%%%%%%%%%%%%%%%
%%%%%%%%%%%%%%%%%%%%%%%%%%%%%%%%%%%%%%%%%%%%%%%%%%%%%%%%%%%%%%%%%%%%%
%%%MAIN TEXT%%%%
%%%%%%%%%%%%%%%%%%%%%%%%%%%%%%%%%%%%%%%%%%%%%%%%%%%%%%%%%%%%%%%%%%%%%
%%%%%%%%%%%%%%%%%%%%%%%%%%%%%%%%%%%%%%%%%%%%%%%%%%%%%%%%%%%%%%%%%%%%%
%%%%%%%%%%%%%%%%%%%%%%%%%%%%%%%%%%%%%%%%%%%%%%%%%%%%%%%%%%%%%%%%%%%%%
%%%%%%%%%%%%%%%%%%%%%%%%%%%%%%%%%%%%%%%%%%%%%%%%%%%%%%%%%%%%%%%%%%%%%
%%%%%%%%%%%%%%%%%%%%%%%%%%%%%%%%%%%%%%%%%%%%%%%%%%%%%%%%%%%%%%%%%%%%%
%%%%%%%%%%%%%%%%%%%%%%%%%%%%%%%%%%%%%%%%%%%%%%%%%%%%%%%%%%%%%%%%%%%%%
%%%%%%%%%%%%%%%%%%%%%%%%%%%%%%%%%%%%%%%%%%%%%%%%%%%%%%%%%%%%%%%%%%%%%
%%%%%%%%%%%%%%%%%%%%%%%%%%%%%%%%%%%%%%%%%%%%%%%%%%%%%%%%%%%%%%%%%%%%%

%% INTRODUCTION::::::::::::::::::::::::::::::::::::::::::::::::::::::
%%%%%%%%%%%%%%%%%%%%%%%%%%%%%%%%%%%%%%%%%%%%%%%%%%%%%%%%%%%%%%%%%%%%%
\section{Introduction} 
Ionic liquid (IL) based electrolytes have emerged as a promising option to replace the state-of-the-art liquid electrolytes relying on organic solvents \cite{eftekhari2016different,ghandi2014review,galinski2006ionic}.
Offering a safer energy storage device due to increasing thermal and electrochemical stability as well as a marginal volatility and risk of flammability \cite{armand2011ionic,macfarlane2014energy,elia2016exceptional,yoon2013fast,wilken2015ionic,lewandowski2009ionic}, ILs are considered as a potentially tailorable material that allows a functional electrolyte design \cite{rogers2003ionic}. However, these key benefits trade-off against a rather poor charge transport performance that manifests in a low conductivity and transference number \cite{zhou2011phase,galinski2006ionic}. 
One strategy to prevent such ionic systems from crystallizing at ambient temperatures lies in the implementation of asymmetric anions \cite{brinkkotter2017influence,giffin2017decoupling,reber2020impact}. 
This approach has been pursued experimentally, for example, on the basis of the well-known $\text{TFSI}^-$ (bis[(trifluoromethyl)sulfonyl]imide) anion via replacement of the trifluoromethylsulfonyl group by a cyano amide moiety, yielding the TFSI-DCA (dicyanamide) hybrid $\text{TFSAM}^-$ anion (2,2,2-(trifluoromethyl)sulfonyl-N-cyanoamide) \cite{shaplov2015new}. The asymmetric substitution is found to improve indeed the liquid range with respect to temperature of the binary IL as well as its conductivity \cite{hoffknecht2017tfsam}.
Considering the strong Coulombic interactions between the exclusively ionic constituents in IL/lithium salt electrolytes an increasing lithium salt content is generally associated with an increasing viscosity and thus a considerable reduction of the Li-ion mobility. 
Surprisingly, superconcentrated "IL-in-salt" electrolytes \cite{marczewski2014ionic}, for which the lithium salt content outweighs that of the IL, have shown even enhanced physicochemical properties as well as improved transport kinetics in terms of the lithium transference number \cite{wilken2015ionic,marczewski2014ionic,yoon2013fast,girard2015electrochemical}. 
A very recent experimental study by N\"urnberg et al. \cite{nuernberg2020} combined these two concepts of anion asymmetry and high lithium doping for $\text{LiTFSAM}_\text{x}/\text{Pyr}_{14}\text{TFSAM}_{1-\text{x}}$ mixtures and checked it against the symmetric anion analogue $\text{TFSI}^-$. Based on an increasing ratio of the diffusion coefficients $\text{D}_{\text{Li}^+}/\text{D}_{\text{anion}^-}$ and Li-ion mobility overtaking that of the anion at elevated salt content, they concluded that the net mode of transport changes from vehicular to structural Li-ion diffusion. 
The decoupling of $\text{Li}^+$ from anion dynamics was found to be significantly more pronounced for the asymmetric $\text{TFSAM}^-$ containing electrolytes and related to dramatic changes in the $\text{Li}^+$ coordination environment. \newline
Inspired by the results by N\"urnberg et al., which paint an intricate picture of the transforming $\text{Li}^+$ solvation sphere and the resulting changes in transport mechanics, we investigate these systems, \textit{i.e.}, $\text{LiTFSI}_\text{x}/\text{Pyr}_{14}\text{TFSI}_{1-\text{x}}$ and $\text{LiTFSAM}_\text{x}/\text{Pyr}_{14}\text{TFSAM}_{1-\text{x}}$ mixtures for x\,=\,0.0-0.7, by means of all-atomistic molecular dynamics (MD) simulations. \newline 
First, we study how the Li-ion coordination numbers are affected by salt content as well as choice of the asymmetric anion. Second, we analyse the transport dynamics via the standard parameters of diffusion coefficients and residence times between $\text{Li}^+$ and anion. In accordance with the experimental results, we observe concomitant with profound changes of the solvation environment a particular enhancement of the Li-ion mobility in the $\text{TFSAM}^-$-based mixtures. 
The focus of this manuscript is a thorough characterisation of the underlying transport mechanics via the Lithium Coupling Factor (LCF) $\lambda$ that measures the dynamic collectivity of $\text{Li}^+$ and its solvating anions. 
Our findings strongly suggest that the generally proposed "vehicular" concept is a deceptive description of the joint motion of a $\text{Li}^+$-shell-complex and urge to be mindful of the fluid nature of ionic liquid electrolytes. 
Based on the observable $\lambda$, we conclude that no fundamental change of transport mechanism occurs with increasing salt concentrations but rather that incorporation of anions in more than one $\text{Li}^+$ solvation environment lowers the dynamic coupling of individual $\text{Li}^+$-anion pairs. Comparison of the LCFs with $\text{TFSI}^-$ to $\text{TFSAM}^-$ shows the power of a tailored anion design where disparate binding sites effectuate low dynamic stabilities of Li-ion and its neighbourhood.\newline  
\comment{\info[inline]{The relatively improved $\text{Li}^+$ mobility, respectively $\text{D}_{\text{Li}^+}/\text{D}_{\text{anion}^-}$ increasing as a function of salt content, emerges naturally in our modeling framework. --> add that part? }} 

\section{Simulation details} 
The MD simulations were conducted with the software package GROMACS (version 2018.8) \cite{VanDerSpoel2005,Pall2015,Abraham2015,Berendsen1995}. All atomic interactions were modelled according to the widely recognised OPLS-AA-derived CL\&P force field, which is developed and maintained by Canongia Lopes and P\'adua particularly for the study of ionic liquids \cite{gouveia2017ionic,CanongiaLopes2012,JoseN.CanongiaLopes2004,Lopes2004,Shimizu2010}. To account for polarization effects in a mean-field sense, all partial charges in the system were scaled down by a factor 0.8 according to prevalent practice \cite{self2019transport,molinari2019general,molinari2020chelation,thum2020solvate,huang2018solvation}. Motivated by the experimental study of N\"urnberg et al., our study of $\text{LiTFSI}_\text{x}/\text{Pyr}_{14}\text{TFSI}_{1-x}$ and $\text{LiTFSAM}_\text{x}/\text{Pyr}_{14}\text{TFSAM}_{1-\text{x}}$ electrolytes covers the lithium salt fractions $x\,=\,[\text{Li}^+]/[\text{anion}^-]\,=\,0.0-0.7$. Since a change of the $\text{Li}^+$ transport mechanism is speculated to occur at $x\approx 0.5-0.65$, this transition regime is sampled in more detail with $x$\,=\, 0.525, 0.5, 0.575 and 0.6. All systems contain in total 1000 ion pairs, \textit{i.e.}, 1000 anions and the according number of $\text{Li}^+$ and $\text{Pyr}_{14}^+$ cations to meet the respective salt concentration. The production runs, which were used for data acquisition, are each 400\,ns long and are carried out in the NPT-ensemble using a Nos\'{e}-Hoover thermostat\cite{Nose1983,Nose1984,Hoover1985}  to maintain the temperature at 400\,K and a Parrinello-Rahman barostat \cite{Parrinello1981} to couple the system to atmospheric pressure.
The details on the system generation, equilibration procedure and overall simulation protocol are provided in the Supplementary Information section A$^\dag$.
%%%%%%%%%%%%%%%%%%%%%%%%%%%%%%%%%%%%%%%%%%%%%%%%%%%%%%%%%%%%%%%%%%%%%
%%%%%%%%%%%%%%%%%%%%%%%%%%%%%%%%%%%%%%%%%%%%%%%%%%%%%%%%%%%%%%%%%%%%%
%%%%%%%%%%%%%%%%%%%%%%%%%%%%%%%%%%%%%%%%%%%%%%%%%%%%%%%%%%%%%%%%%%%%%
%%%  Lithium solvation cage
%%%%%%%%%%%%%%%%%%%%%%%%%%%%%%%%%%%%%%%%%%%%%%%%%%%%%%%%%%%%%%%%%%%%%
%%%%%%%%%%%%%%%%%%%%%%%%%%%%%%%%%%%%%%%%%%%%%%%%%%%%%%%%%%%%%%%%%%%%%
%%%%%%%%%%%%%%%%%%%%%%%%%%%%%%%%%%%%%%%%%%%%%%%%%%%%%%%%%%%%%%%%%%%%%
%%%%%%%%%%%%%%%%%%%%%%%%%%%%%%%%%%%%%%%%%%%%%%%%%%%%%%%%%%%%%%%%%%%%%
%%%  RDF
%%%%%%%%%%%%%%%%%%%%%%%%%%%%%%%%%%%%%%%%%%%%%%%%%%%%%%%%%%%%%%%%%%%%%
\section{Lithium coordination environment} 
\subsection{Radial distribution functions} 
\begin{figure}
  \centering
  \subfloat{\includegraphics[width=0.4\textwidth]{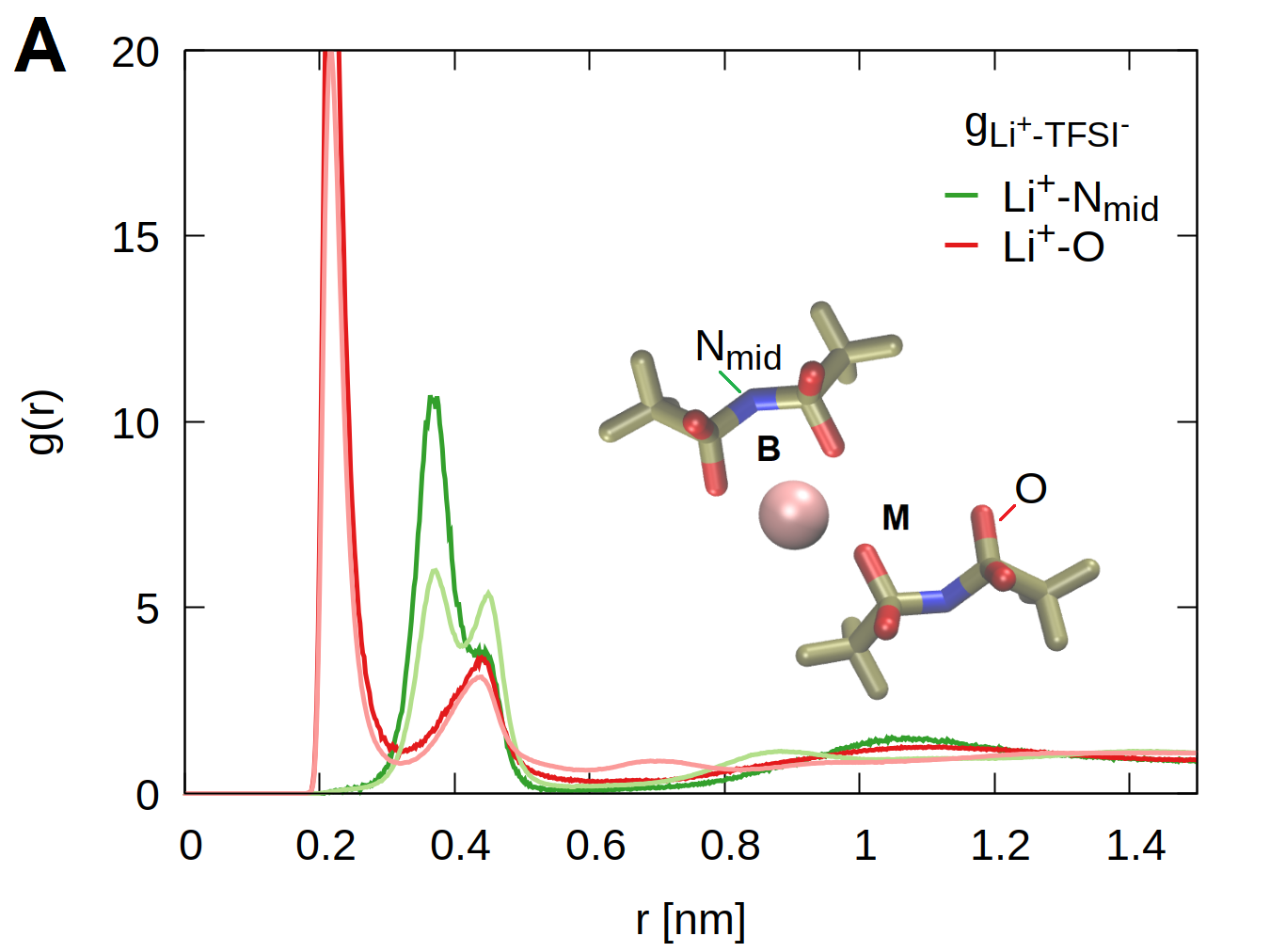}\label{fig:li_coordination_environment_tfsi}}
  \hfill
  \subfloat{\includegraphics[width=0.4\textwidth]{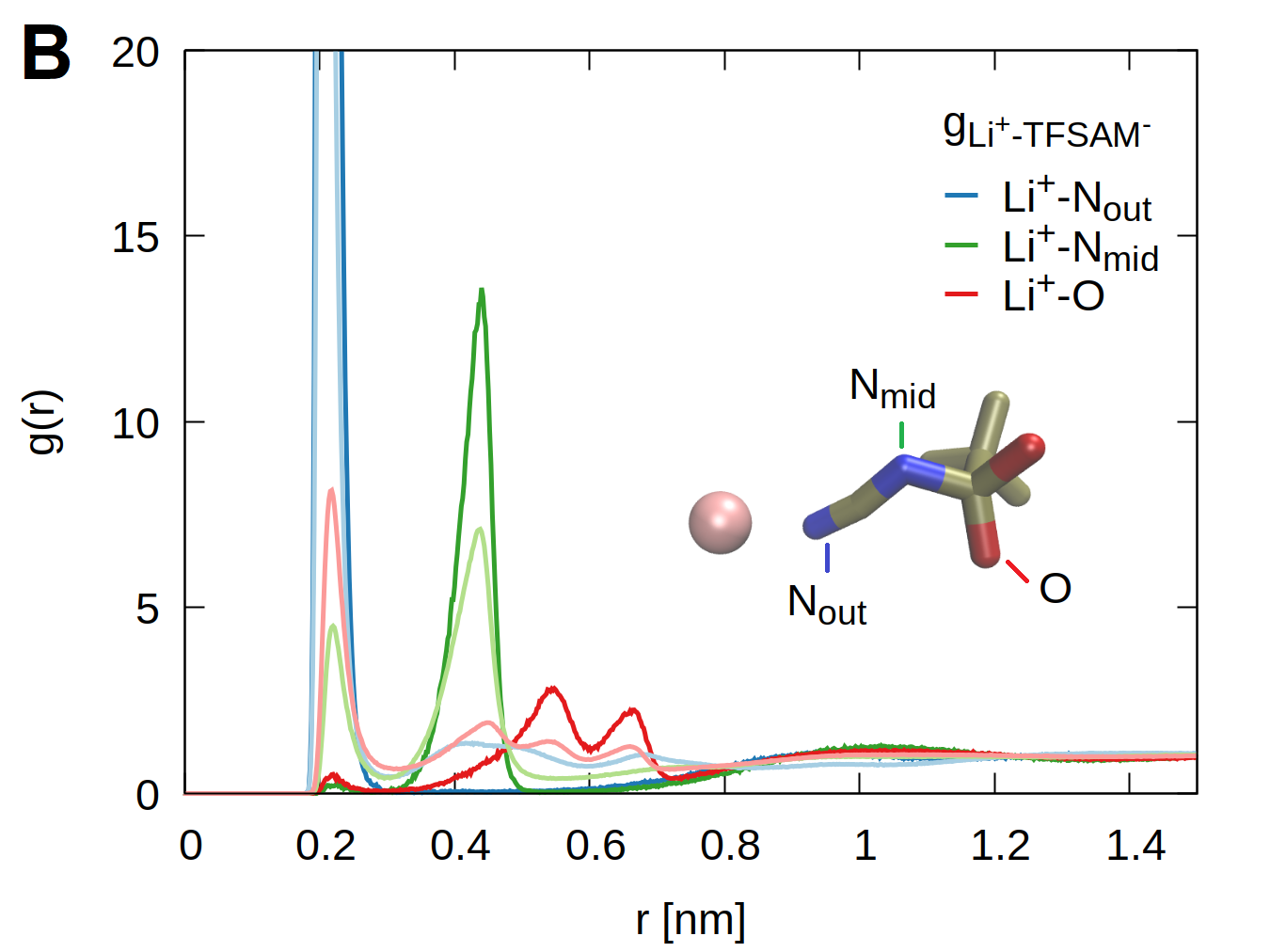}\label{fig:li_coordination_environment_tfsam}}
  \caption{Radial distribution functions $\text{g}_{\text{Li}^+-X}$(r) between Li-ions and nitrogen / oxygen binding sites provided by $\text{TFSI}^-$ (top) and $\text{TFSAM}^-$ (bottom) for lithium salt fractions x\,=\,0.05 (deep color) and x\,=\,0.5 (light color). The schematic snapshots illustrate the possible monodentate (M) and bidentate (B) coordination of $\text{TFSI}^-$ as well as the preferred monodentate $\text{TFSAM}^-$ coordination.}
  \label{fig:li_coordination_environment_tfsi_tfsam}
\end{figure}

We study the lithium solvation structures via radial distribution functions $\text{g}_{\text{Li}^+\text{-X}}$ (RDF) between the Li-ions and the distinct atomic species by means of which the anions form the coordination bonds.
Figure \ref{fig:li_coordination_environment_tfsi_tfsam}A depicts the arrangement of $\text{O}_{\text{TFSI}^-}$ (red) and $\text{N}_{\text{mid,TFSI}^-}$ (green) atoms around $\text{Li}^+$ exemplary for low and high lithium salt contents. 
\begin{figure*}
	\centering
	\includegraphics[width=0.8\textwidth]{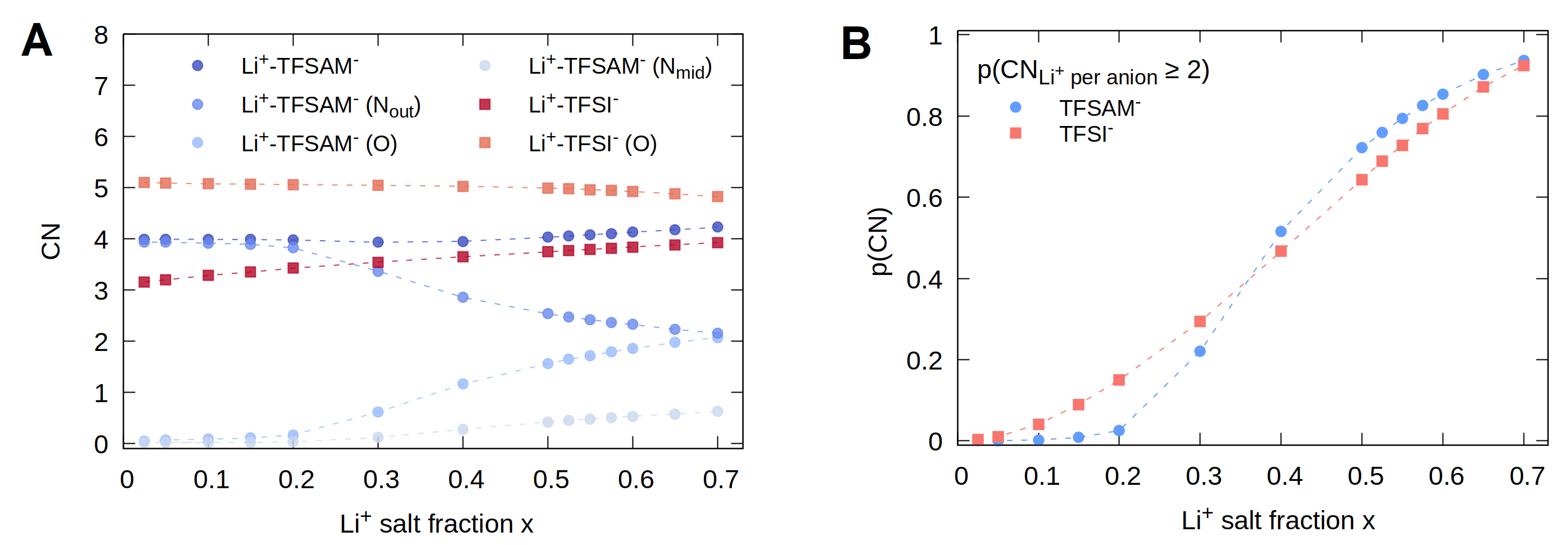}
	\caption{Left: Coordination numbers CN of Li-ions via $\text{TFSI}^-$/$\text{TFSAM}^-$, including resolution regarding the particular atomic binding sites provided by the respective anion. Right: Probability distribution p($\text{CN}_{\text{Li}^+ \text{per anion}} \geq 2$) of at least two $\text{Li}^+$ neighbours to a single $\text{TFSI}^-$/$\text{TFSAM}^-$ as a function of lithium salt content x.} 
	\label{fig:li_coordination_numbers_tfsi_tfsam_distribution}
\end{figure*}
In agreement with other experimental and theoretical studies, we find that $\text{O}_{\text{TFSI}^-}$  accounts for the primary coordination environment with the nearest peak at 2.1 \angstrom\,. 
The double peak structure of $g_{\text{Li}^+-\text{N}_{\text{mid,TFSI}^-}}$ between 3 and 5 \angstrom\, indicates two types of $\text{Li}^+-\text{TFSI}^-$ coordination geometries. 
The nearest peak can be attributed to a bidentate binding, whereas the second peak corresponds to a monodentate contact via one $\text{O}_{\text{TFSI}^-}$ only \cite{borodin2018insights,li2015effect,reber2020impact,nuernberg2020,kubisiak2019molecular,li2012li+}. 
Our findings are in excellent agreement with simulation studies \cite{li2015effect} on $\text{LiTFSI}_\text{x}/\text{Pyr}_{13}\text{TFSI}_{1-\text{x}}$ (x\,=\,0.16) mixtures employing the most recent parametrization of the APPLE\&P polarizable force field as well as a combined experimental and MD investigation \cite{monteiro2008transport} of closely related electrolytes $\text{LiTFSI}\text{x}/\text{BMMI}\,\text{TFSI}_{1-\text{x}}$ (x\,=\,0.24 and 0.38). 
We conclude from the changing relative peak heights of $\text{g}_{\text{Li}^+-\text{N}_{\text{mid,TFSI}^-}}$ that an increasing lithium salt concentration significantly alters the populations of bidentate and monodentate $\text{TFSI}^-$ orientation in favor of the latter\cite{li2012li+,haskins2014computational,chen2018ion,monteiro2008transport}. \newline
The $\text{TFSAM}^-$ anion can coordinate to $\text{Li}^+$ additionally via the cyano-group, \textit{i.e.}, $\text{N}_{\text{out,TFSAM}^-}$ (blue). Figure \ref{fig:li_coordination_environment_tfsi_tfsam}B demonstrates that at low salt concentration $\text{Li}^+$ binds almost exclusively via $\text{N}_{\text{out,TFSAM}^-}$ and vanishingly few direct $\text{O}_{\text{TFSAM}^-}$ contacts are made. From an electrostatic and steric point of view, $\text{N}_{\text{out,TFSAM}^-}$ constitutes the more attractive binding site because it carries the most negative partial charge 
($\text{q}_{\text{N}_{\text{out,TFSAM}^-}} = -0.76\,e, \text{q}_{\text{N}_{\text{mid,TFSAM}^-}} = -0.71\,e ,\text{q}_{\text{O}_{\text{TFSAM}^-}} = -0.53\,e, \text{q}_{\text{F}_{\text{TFSAM}^-}} = -0.16\,e$) \cite{gouveia2017ionic} and is easily accessible for coordination. 
This bias in favor of $\text{Li}^+$-cyano coordination has been confirmed recently by Raman measurements \cite{nuernberg2020} as well as observed in a Raman/MD study of a lithium salt-IL mixture containing both DCA and $\text{TFSI}^-$ anions\cite{huang2018solvation}.
For increasing salt content, we find a dramatic distortion of the initially uniform $\text{Li}^+-\text{N}_{\text{out,TFSAM}^-}$ environment when $\text{TFSAM}^-$ additionally engages in a direct $\text{O}_{\text{TFSAM}^-}$ and $\text{N}_{\text{mid,TFSAM}^-}$ binding to $\text{Li}^+$.

%%%%%%%%%%%%%%%%%%%%%%%%%%%%%%%%%%%%%%%%%%%%%%%%%%%%%%%%%%%%%%%%%%%%%
%%%  Coordination numbers
%%%%%%%%%%%%%%%%%%%%%%%%%%%%%%%%%%%%%%%%%%%%%%%%%%%%%%%%%%%%%%%%%%%%%
\subsection{Coordination numbers} 
\begin{figure}[H]
  
  \subfloat{\includegraphics[width=0.25\textwidth]{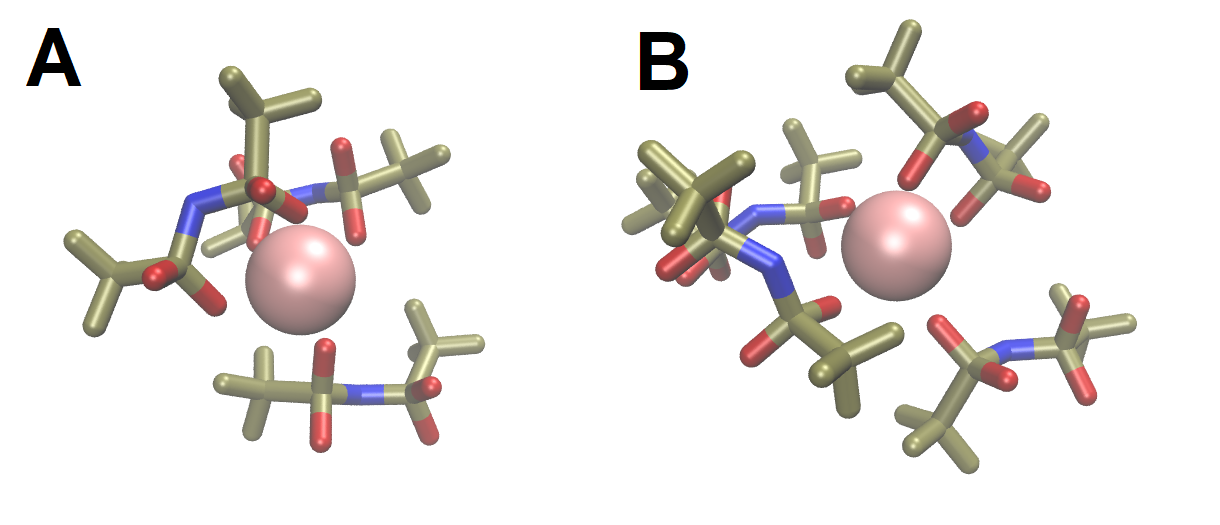}}
  \hfill \\
  \subfloat{\includegraphics[width=0.5\textwidth]{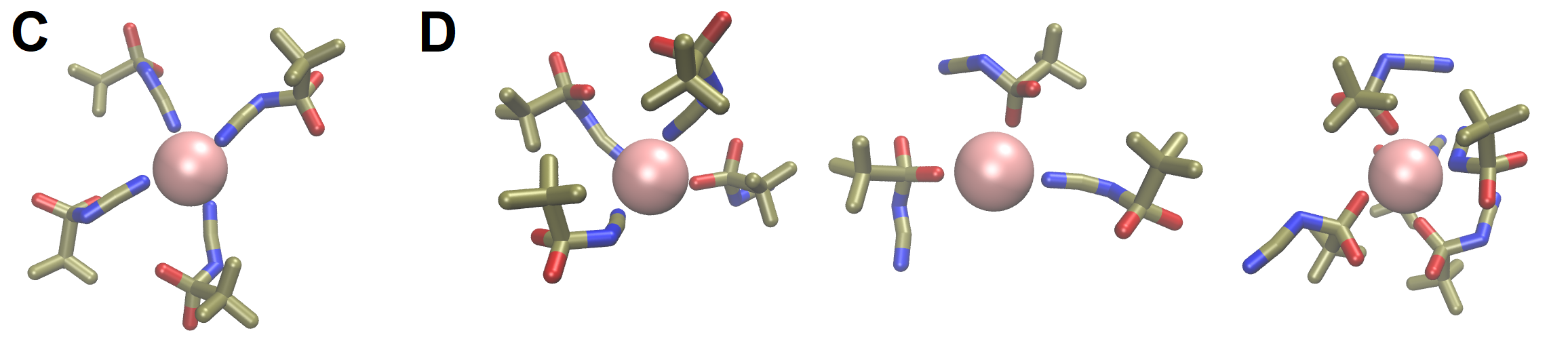}}
	\caption{Examples for Li-ion (pink) solvation shells provided by $\text{TFSI}^-$ (top) and $\text{TFSAM}^-$ (bottom) anions for lithium salt fractions of x\,=\,0.1 (A,C) or x\,=\,0.5 (B,D). Oxygen atoms are displayed in red and nitrogen atoms in blue.} 
	\label{fig:snapshot_coordination_environment_tfsi_tfsam}
\end{figure}

Since $\text{g}_{\text{Li}^+-X}$(r) measures the probability to find the atomic species $X$ within a distance r away from $\text{Li}^+$, the first minimum position can be employed as a structural criterion for present $\text{Li}^+-X$ binding. Coordination numbers $\text{CN}_{\text{Li}^+-X}$ are determined accordingly from integrating the respective RDF up to this cutoff distance $R$
\begin{equation}
\text{CN}_{\text{Li}^+-X} = \rho_{X,\infty} \int_{0}^{R} 4\pi r'^2\,\, g_{\text{Li}^+-X}(r') dr'.
\end{equation}
By explicit counting of the binding anions, which individually may provide $\text{Li}^+$ coordination via multiple $X$, we further obtain a distribution of the molecular $\text{CN}_{\text{Li}^+-\text{anion}}$.
Figure \ref{fig:li_coordination_numbers_tfsi_tfsam_distribution}A depicts $\text{CN}_{\text{Li}^+-X}$ for both electrolyte compositions as a function of salt concentration. \newline
For the $\text{TFSI}^-$ containing mixtures, we observe that the absolute number of $\text{O}_{\text{TFSI}^-}$ contributing to the primary $\text{Li}^+$ environment is barely sensitive to salt content. However, as already deduced from the inverting coordination geometry, the actual number of $\text{TFSI}^-$ anions affording such a solvation shell increases with increasing salt content. While five $\text{O}_{\text{TFSI}^-}$ are initially supplied by on average three $\text{TFSI}^-$ anions, the absolute $\text{CN}_{\text{Li}^+-\text{TFSI}^-}$ accumulates to four at the highest concentration. Figure \ref{fig:snapshot_coordination_environment_tfsi_tfsam} (top) shows snapshots of the $\text{Li}^+$ coordination complexes depicting the coexistence of monodentate and bidentate binding geometries.
Analysis of the distributions of $\text{Li}^+-\text{TFSI}^--$coordination numbers that make up for the average $\text{CN}_{\text{Li}^+-\text{TFSI}^-}$(x) gives evidence of a non-uniform coordination landscape with contributions from $\text{Li}^+\,(\text{TFSI}^-)_3$,  $\text{Li}^+\,(\text{TFSI}^-)_4$ and even $\text{Li}^+\,(\text{TFSI}^-)_5$ complexation at elevated salt content as shown in Figure S3$^\dag$.\newline

\noindent For the low concentration $\text{TFSAM}^-$ based electrolytes, $\text{Li}^+$ is exclusively solvated by four $\text{N}_{\text{out,TFSAM}^-}$ which are naturally supplied by four anions as shown in Figure \ref{fig:li_coordination_numbers_tfsi_tfsam_distribution}A and illustrated by a snapshot in Figure \ref{fig:snapshot_coordination_environment_tfsi_tfsam}C.
In accordance with recent Raman measurements by N\"urnberg et al. \cite{nuernberg2020} our simulations indicate no primary coordination via $\text{O}_{\text{TFSAM}^-}$ for sufficient $\text{N}_{\text{out,TFSAM}^-}$ supply. However, when the ratio of $\text{n}_{\text{Li}^+}:\text{n}_{\text{TFSAM}^-}$ exceeds 1:4, $\text{N}_{\text{out,TFSAM}^-}$ can no longer provide solely for a saturated solvation shell. We even identify direct binding to $\text{N}_{\text{mid,TFSAM}^-}$, which is yet a rare coordination motif. 
Overall, the $\text{Li}^+$ environment becomes increasingly heterogeneous both in terms of the atomic species mediating the binding and the net number of $\text{TFSAM}^-$ molecules constituting the primary shell as shown in Figure S3$^\dag$. A random selection of $\text{Li}^+$ environments is displayed in Figure \ref{fig:snapshot_coordination_environment_tfsi_tfsam}D.\newline
As a final characterisation of the $\text{Li}^+$ shell structure, we adopt the anion's perspective and measure the distribution of $\text{Li}^+$ neighbours around a single anion. Figure \ref{fig:li_coordination_numbers_tfsi_tfsam_distribution}B depicts that $\text{Li}^+$ maintains a self-contained $\text{TFSAM}^-$ solvation environment as long as the preferred $\text{N}_{\text{out,TFSAM}^-}$ binding sites are not saturated. In the low concentrated $\text{TFSI}^-$-based mixtures, however, a steadily growing amount of $\text{TFSI}^-$ bridges at least two Li-ions even though an excess of lithium-free anions is available for coordination. This agrees well with findings from previous MD studies \cite{haskins2014computational,lesch2014combined} where lithium aggregates occurred at salt fractions x$\,\geq\,$0.1. In the regime of high concentrations, we observe that the inclination of $\text{TFSAM}^-$ to lithium clustering rapidly overtakes that of $\text{TFSI}^-$.

%%%%%%%%%%%%%%%%%%%%%%%%%%%%%%%%%%%%%%%%%%%%%%%%%%%%%%%%%%%%%%%%%%%%%
%%%%%%%%%%%%%%%%%%%%%%%%%%%%%%%%%%%%%%%%%%%%%%%%%%%%%%%%%%%%%%%%%%%%%
%%%%%%%%%%%%%%%%%%%%%%%%%%%%%%%%%%%%%%%%%%%%%%%%%%%%%%%%%%%%%%%%%%%%%
%%%  Dynamics
%%%%%%%%%%%%%%%%%%%%%%%%%%%%%%%%%%%%%%%%%%%%%%%%%%%%%%%%%%%%%%%%%%%%%
%%%%%%%%%%%%%%%%%%%%%%%%%%%%%%%%%%%%%%%%%%%%%%%%%%%%%%%%%%%%%%%%%%%%%
%%%%%%%%%%%%%%%%%%%%%%%%%%%%%%%%%%%%%%%%%%%%%%%%%%%%%%%%%%%%%%%%%%%%%

\section{Ion transport properties}

%%%%%%%%%%%%%%%%%%%%%%%%%%%%%%%%%%%%%%%%%%%%%%%%%%%%%%%%%%%%%%%%%%%%%
%%%  self diffusion constants
%%%%%%%%%%%%%%%%%%%%%%%%%%%%%%%%%%%%%%%%%%%%%%%%%%%%%%%%%%%%%%%%%%%%%

\subsection{Diffusion constants}
\begin{figure}[ht!]
  \centering
  \subfloat{\includegraphics[width=0.4\textwidth]{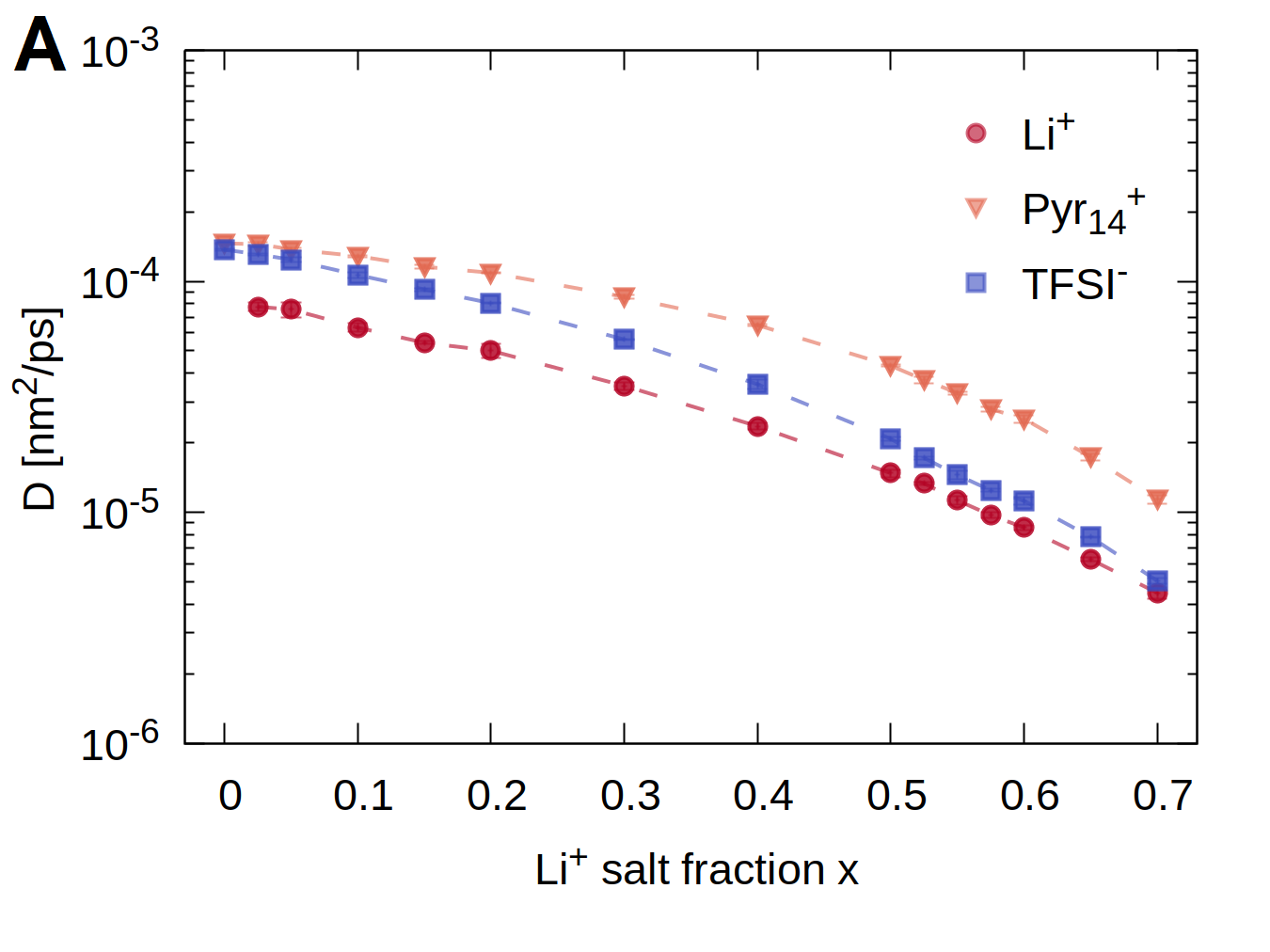}}
  \hfill
  \subfloat{\includegraphics[width=0.4\textwidth]{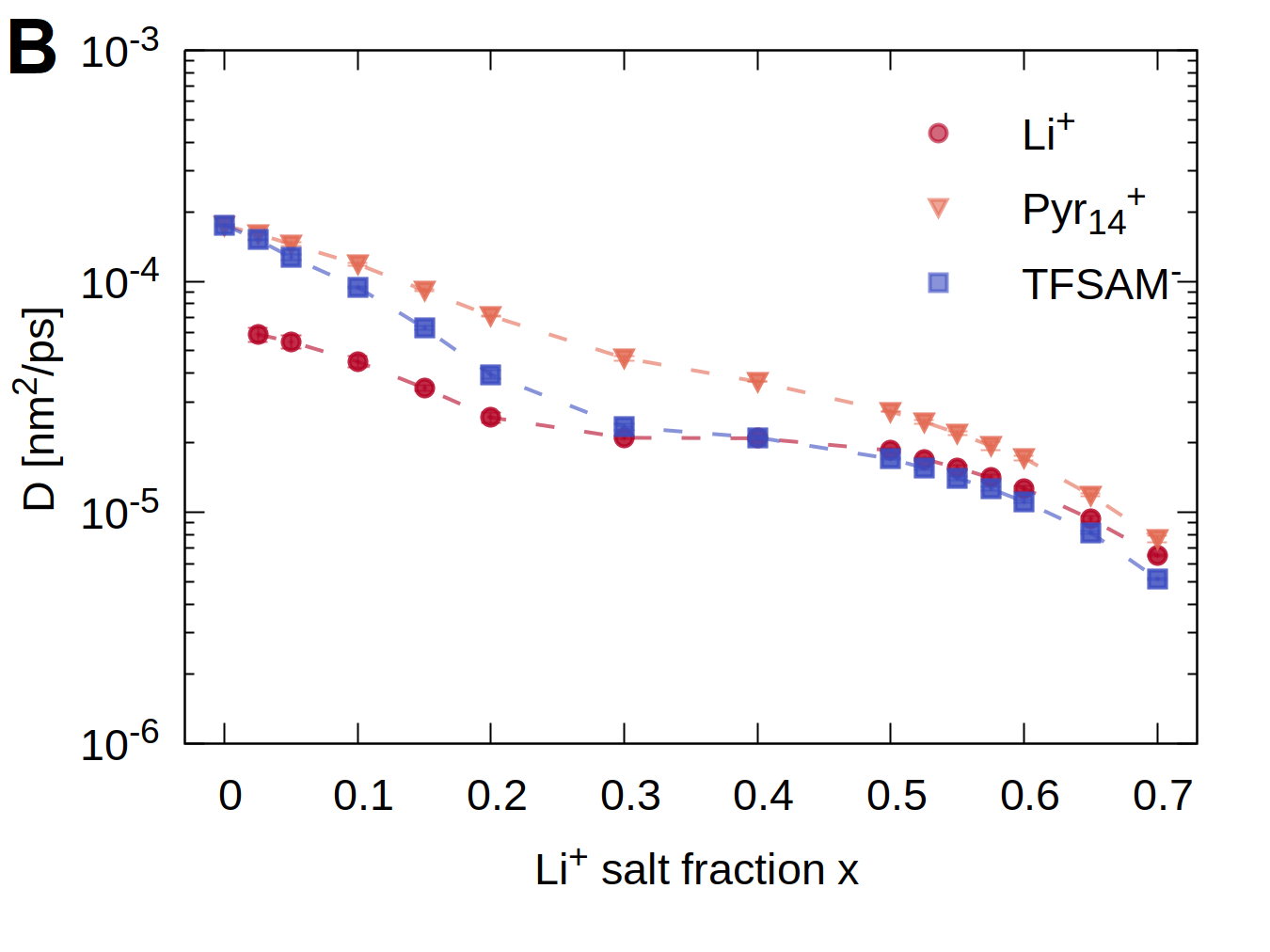}}
    \hfill
  \subfloat{\includegraphics[width=0.4\textwidth]{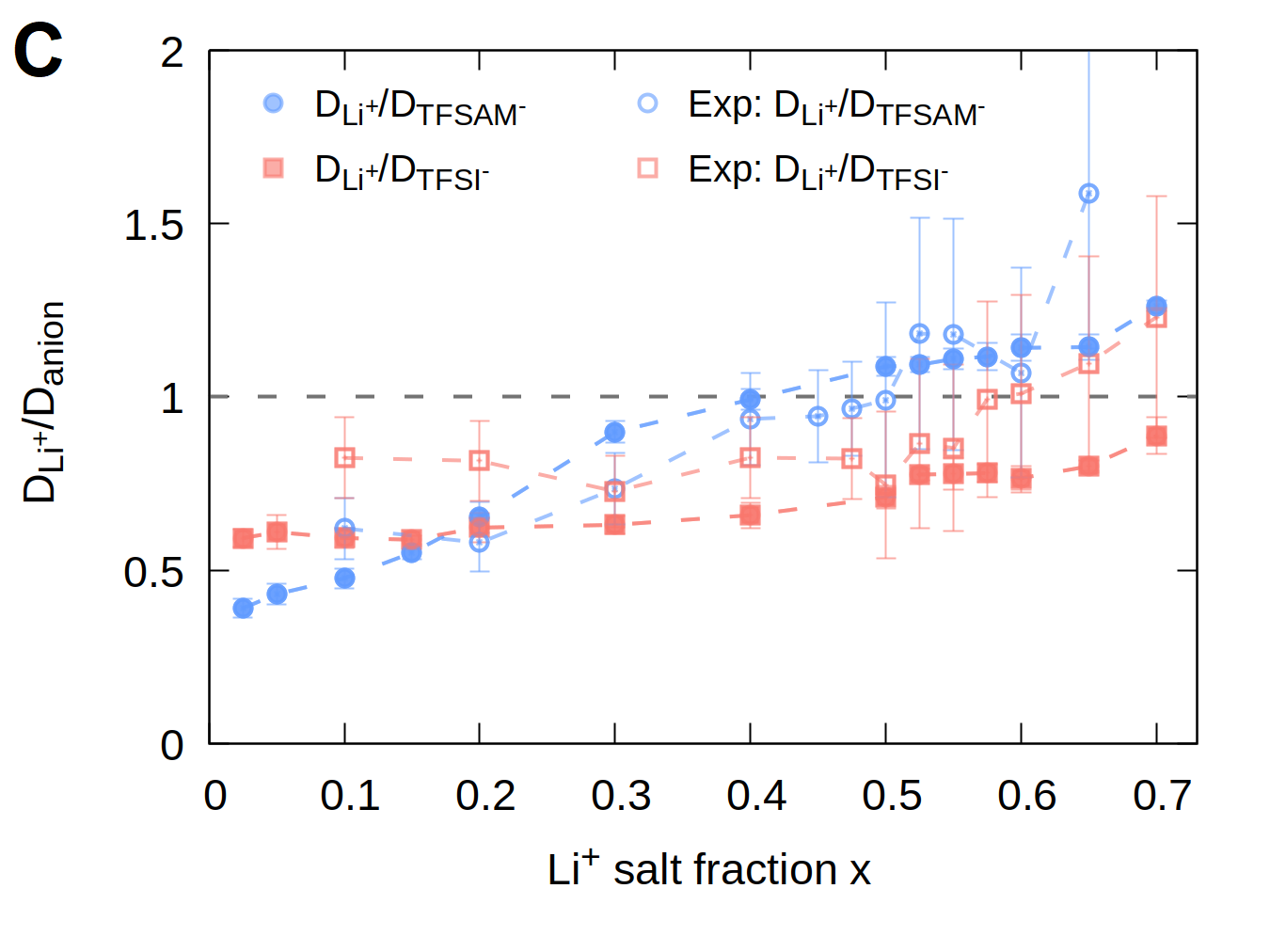}}
  
	\caption{Top and center: Self diffusion coefficients of all ionic species as a function of lithium salt content x. The self diffusion coefficients for the $\text{TFSI}^-$ system at x\,=\,0.15 match results from recent polarizable force field simulations employing a $\text{Pyr}^+_{13}$ cation within a factor of two \cite{li2015effect}. Bottom: Ratio of $\text{Li}^+$ and anion self diffusion coefficients as a function of lithium salt content. The full symbols show the data measured in this work and are compared to the experimental results from N\"urnberg et al.\cite{nuernberg2020}, which are depicted by the open symbols. Reprinted (adapted) with permission from  P. N\"urnberg, E. I. Lozinskaya, A. S. Shaplov and M. Sch\"onhoff, \textit{The Journal of Physical Chemistry B}, 2020, \textbf{124}, 861–870. Copyright 2021 American Chemical Society.} 
	\label{fig:self_diffusion_coefficients}
\end{figure}

To discuss the $\text{Li}^+$ transport characteristics, we first calculated the self-diffusion coefficients of each ionic species according to the Einstein relation:
\begin{equation}
\text{D}_i = \lim_{t \to \infty} \dfrac{\langle \left(\vec{r}_i(t) - \vec{r}_i(0) \right)^2 \rangle}{6t},
\end{equation}
where $\vec{r}_i(t)$ denotes the position of the ion belonging to species $i$ and $\langle .. \rangle$ indicates the ensemble average.
The results are shown in Figure \ref{fig:self_diffusion_coefficients}A and B for both electrolyte mixtures over the entire concentration range under study. As expected, increasing addition of lithium salt causes a slowing down of the overall dynamics, which can be attributed to an increasing electrolyte viscosity \cite{haskins2014computational,self2019transport,chen2018ion}. For moderate salt concentrations up to x\,=\,0.3 the diffusivities rank in both systems as $[\text{cation}_{\text{IL}}]^+ > [\text{anion}]^- > [\text{Li}]^+$, which is in agreement with MD studies investigating the $\text{TFSI}^-$-based mixture as well as experiments \cite{li2015effect,haskins2014computational,nuernberg2020}. 
Whereas for the $\text{TFSI}^-$-containing electrolytes the dynamics slow down continuously, $\text{D}_{\text{Li}^+}$ and $\text{D}_{\text{TFSAM}^-}$ achieve a plateau between x\,=\,0.3\,-\,0.5. The bulk dynamics resume their decline, but the $\text{Li}^+$ diffusion overtakes that of $\text{TFSAM}^-$. Similar observations have been made by N\"urnberg et al. \cite{nuernberg2020}, who reported at elevated salt content a remarkable discontinuance of till then slowing dynamics.
%%%%%%%%%%%%%%%%%%%%%%%%%%%%%%%%%%%%%%%%%%%%%%%%%%%%%%%%%%%%%%%%%%%%%
%%%  self diffusion constants: ratio lithium / anion - experiment
%%%%%%%%%%%%%%%%%%%%%%%%%%%%%%%%%%%%%%%%%%%%%%%%%%%%%%%%%%%%%%%%%%%%%
\noindent To gain insights on the dominant mode of $\text{Li}^+$ transport, which may change over the broad concentration spectrum, N\"urnberg et al. considered the dependence of  $\text{D}_{\text{Li}^+}/\text{D}_{\text{anion}}$ on salt content as plotted in Figure \ref{fig:self_diffusion_coefficients}C. 
Our results confirm the experimental observation that for the $\text{TFSI}^-$-based electrolytes the diffusion ratio is barely sensitive on increasing lithium salt concentrations up to x\,=\,0.5 with a following upward trend towards equally fast $\text{Li}^+$ and $\text{TFSI}^-$ diffusion.
In the $\text{TFSAM}^-$-mixtures, on the other hand, the $\text{Li}$-ions display a steady acceleration relative to $\text{TFSAM}^-$ for increasing x with $\text{D}_{\text{Li}^+}$ faster than $\text{D}_{\text{TFSAM}^-}$ for $\text{x}>0.4$.
N\"urnberg and coworkers put forward the hypothesis that the structural changes of the immediate $\text{Li}^+$ environment induce a transition from a predominantly vehicular diffusion, \textit{i.e.}, a strongly coupled motion of $\text{Li}^+$ and its primary solvation shell, to a structural diffusion mechanism where $\text{Li}^+$ performs hopping-like events between its neighbouring anions. 
The fact that the turning point of the dynamic changes coincides remarkably with the onset of these structural rearrangements is supportive of this interpretation.
%%%%%%%%%%%%%%%%%%%%%%%%%%%%%%%%%%%%%%%%%%%%%%%%%%%%%%%%%%%%%%%%%%%%%
%%%  mean residence times : different coordination sites
%%%%%%%%%%%%%%%%%%%%%%%%%%%%%%%%%%%%%%%%%%%%%%%%%%%%%%%%%%%%%%%%%%%%%
\begin{figure}[ht!]
  \centering
  \subfloat{\includegraphics[width=0.45\textwidth]{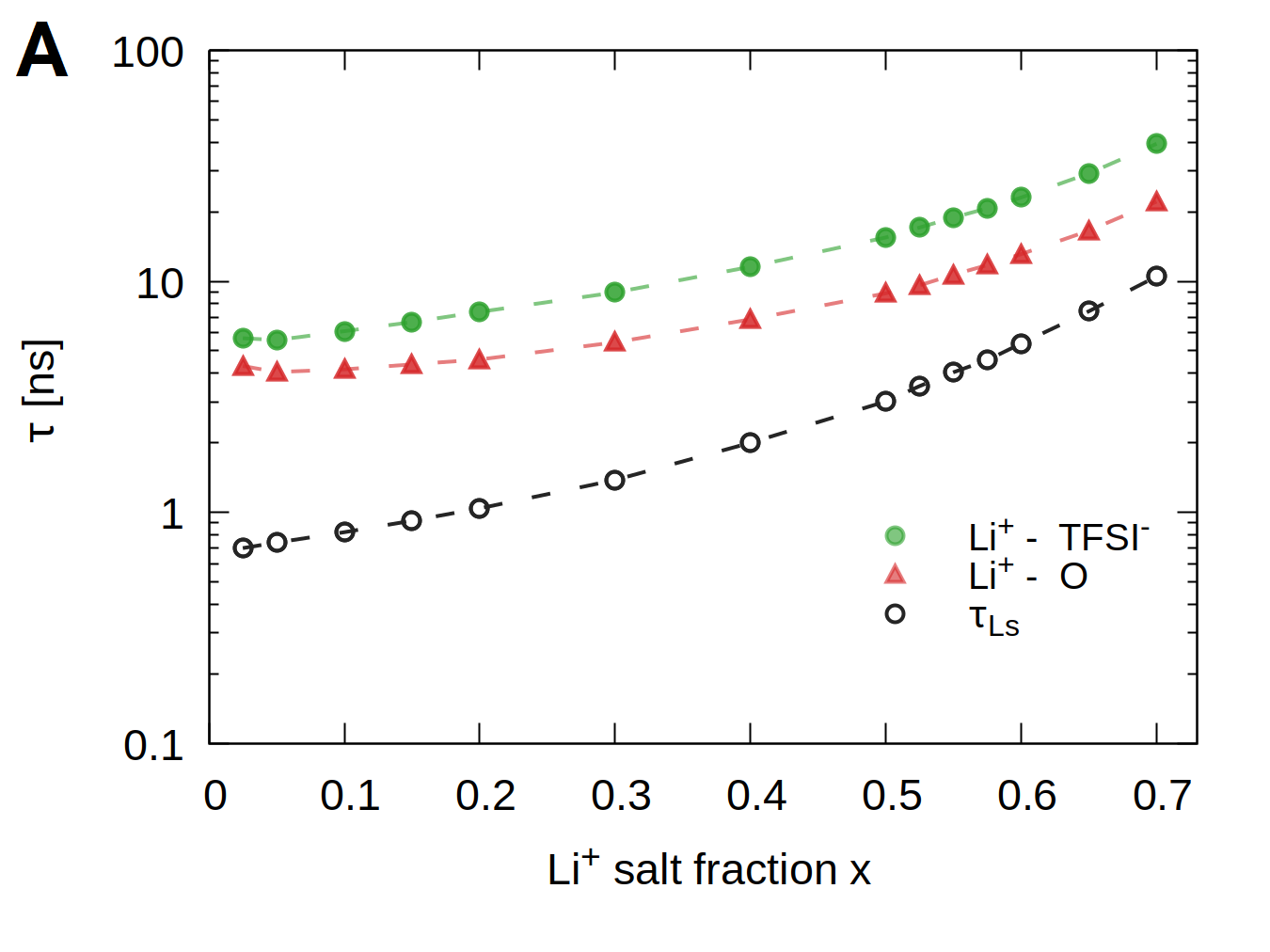}}
  \hfill
  \centering
  \subfloat{\includegraphics[width=0.45\textwidth]{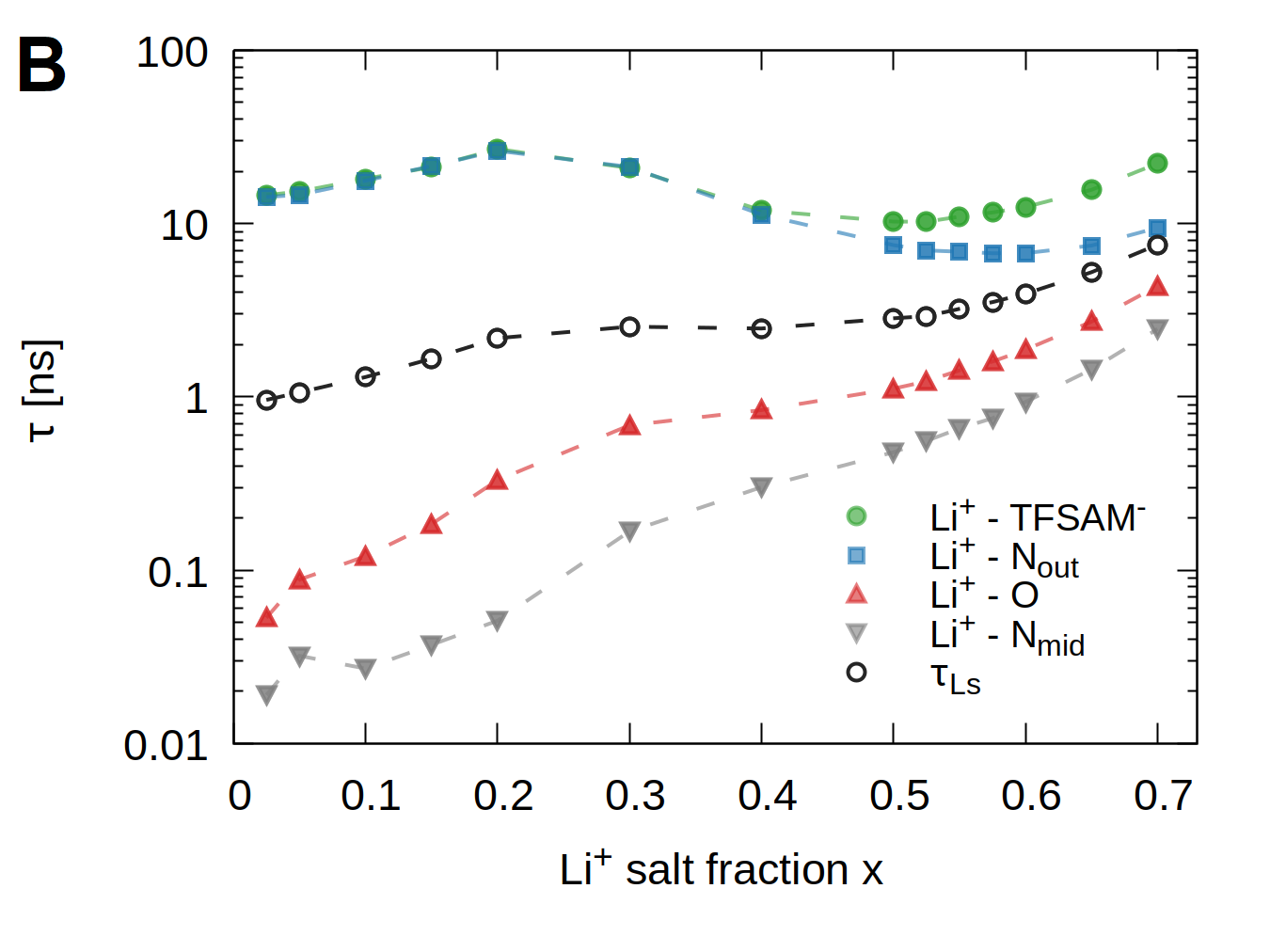}}
  \hfill

  \subfloat{\includegraphics[width=0.42\textwidth]{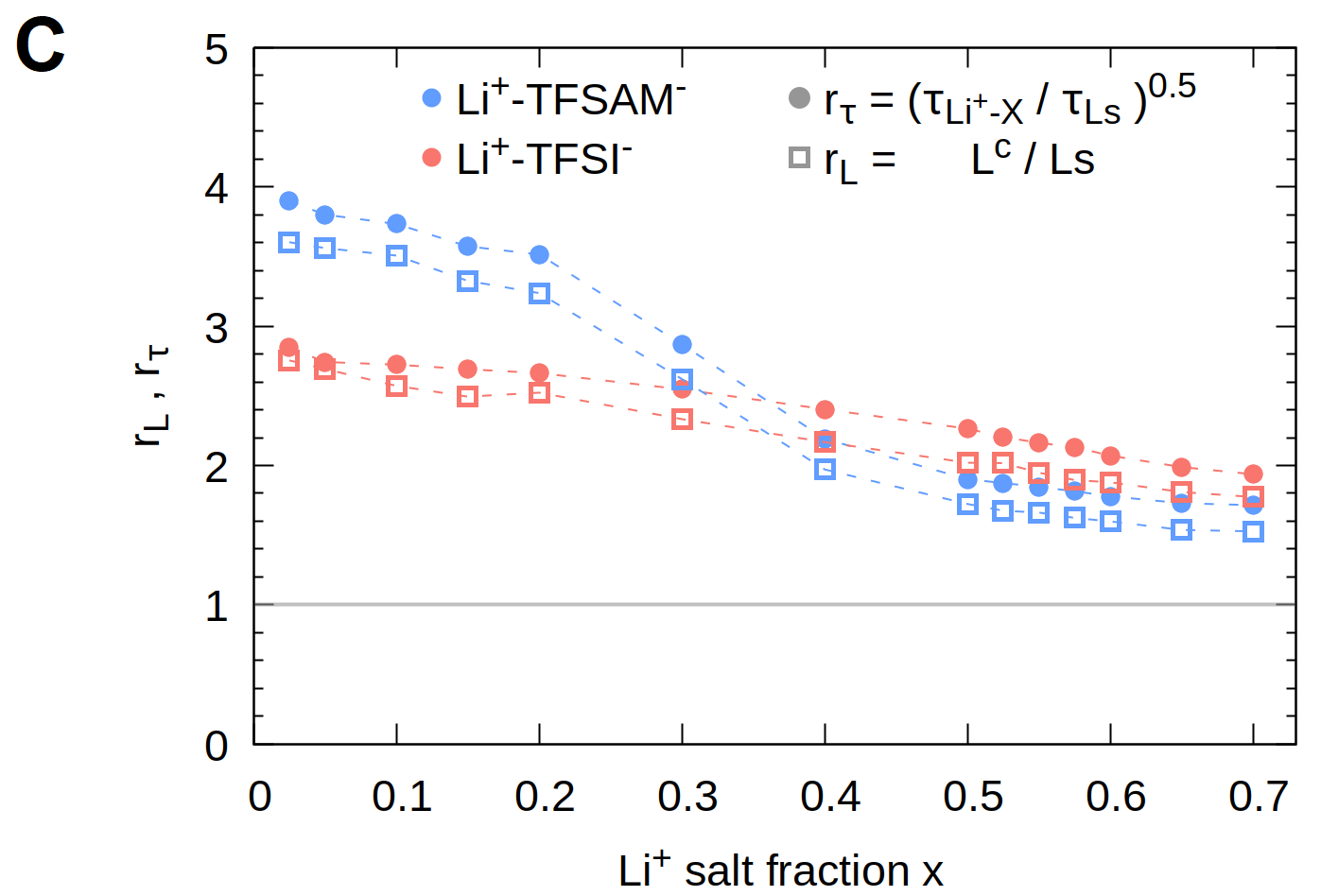} }

  \caption{Top and center: Mean residence times $\tau$ of $\text{Li}^+$ at various atomic binding sites provided by $\text{TFSI}^-$ (top) or $\text{TFSAM}^-$ (center) as a function of lithium salt content x. We note that $\tau_{\text{Li}^+-\text{TFSI}^-}$ for x\,=\,0.15 matches results from recent polarizable force field - simulations employing a $\text{Pyr}_{13}$ cation within a factor of two \cite{li2015effect}. Bottom: Mechanistic interpretation of mean residence times $\tau_{\text{Li}^+-\text{TFSI}^-/\text{TFSAM}^-}$ and $\text{Li}^+$ solvation sphere size Ls in terms of the ratio $\text{r}_{\tau}\,=\,\sqrt{\tau_{\text{Li}^+-\text{anion}}/\tau_{\text{Ls}}}$ as well as  $\text{r}_{\text{L}}\,=\,\text{L}^{\text{c}}/\text{Ls}$.} 
	\label{fig:mean_residence_times_lithium_anions}
\end{figure}

\subsection{Mean residence times} 

One can only surmise the extent of correlated transport from the self-diffusion coefficients, which are ultimately averaged over all ionic species regardless of their binding status, and therefore, strictly speaking, do not provide specific information on the pairwise motion of $\text{Li}^+$ and anion. 
A more precise picture of the $\text{Li}^+$ solvation dynamics can be obtained from the mean residence times $\tau_{\text{Li}^+-X}$, \textit{i.e.}, the average time a $\text{Li}$-ion is attached to species $X$ which is computed from the residence time autocorrelation function (see Supplementary Information section D$^\dag$).
The analysis can be performed for all atomic species to which $\text{Li}^+$ is inclined to bind, \textit{e.g.}, $\text{O}_{\text{TFSI}^-/\text{TFSAM}^-}$ or $\text{N}_{\text{out,TFSAM}^-}$. 
We track the $\text{Li}^+-\text{N}_{\text{mid}}$-bond to measure the average time $\tau_{\text{Li}^+-\text{TFSI}^-/\text{TFSAM}^-}$ that $\text{Li}^+$ spends in the direct vicinity of a distinct anion using the second minimum position in $\text{g}_{\text{Li}^+-\text{N}_{\text{mid}}}$ as a distance cutoff. It is clear from Figure \ref{fig:li_coordination_environment_tfsi_tfsam} that all possible coordination geometries occurring at the respective salt concentration such as mono- and bidentate $\text{TFSI}^-$ binding or direct binding via $\text{O}_{\text{TFSAM}^-}$, $\text{N}_{\text{out,TFSAM}^-}$ or $\text{N}_{\text{mid,TFSAM}^-}$ are thereby contained. 
Figure \ref{fig:mean_residence_times_lithium_anions}A shows that the exchange of the original $\text{TFSI}^-$ solvation environment slows down with increased salt content, which can be deduced from the increasing $\text{Li}^+$ binding time to a distinct $\text{O}_{\text{TFSI}^-}$ atom or a distinct $\text{TFSI}^-$ molecule. 
Most commonly, slowing dynamics in terms of diffusivity in such ionic systems are found to correlate directly with increasing ion pair lifetimes \cite{zhang2015direct,haskins2014computational}. 
Surprisingly, we observe a very unique binding time behaviour for the $\text{TFSAM}^-$-based electrolytes which is displayed in Figure \ref{fig:mean_residence_times_lithium_anions}B.
The structural analysis has demonstrated that at low salt content the $\text{Li}^+$-$\text{TFSAM}^-$ binding is solely mediated via $\text{N}_{\text{out,TFSAM}^-}$ but no direct contacts to $\text{O}_{\text{TFSAM}^-}$ or $\text{N}_{\text{mid,TFSAM}^-}$ atoms are made. Accordingly, we find corresponding mean residence times to be negligibly short-lived and up to two orders in magnitude shorter than $\tau_{\text{Li}^+-\text{N}_{\text{out,TFSAM}^-}}$. 
All residence times increase with increasing salt content as far as x\,$\leq$\,0.2 when $\tau_{\text{Li}^+-\text{TFSAM}^-}$ and $\tau_{\text{Li}^+-\text{N}_{\text{out,TFSAM}^-}}$ exhibit a turning point. 
The average time $\text{Li}^+$ spends in the neighbourhood of $\text{TFSAM}^-$ is reduced significantly, which suggests that $\text{Li}^+$ is bound in weaker configurations.
Concomitant with the structural rearrangement of the primary solvation shell,  $\tau_{\text{Li}^+-\text{O}/\text{N}_{\text{mid}}/\text{N}_{\text{out}}}$ approach the same time scale and thus mirror a competition of these binding sites.

%%%%%%%%%%%%%%%%%%%%%%%%%%%%%%%%%%%%%%%%%%%%%%%%%%%%%%%%%%%%%%%%%%%%%
%%%%%%%%%%%%%%%%%%%%%%%%%%%%%%%%%%%%%%%%%%%%%%%%%%%%%%%%%%%%%%%%%%%%%
%%%%%%%%%%%%%%%%%%%%%%%%%%%%%%%%%%%%%%%%%%%%%%%%%%%%%%%%%%%%%%%%%%%%%
%%% TRANSPORT MECHANISM
%%%%%%%%%%%%%%%%%%%%%%%%%%%%%%%%%%%%%%%%%%%%%%%%%%%%%%%%%%%%%%%%%%%%%
%%%%%%%%%%%%%%%%%%%%%%%%%%%%%%%%%%%%%%%%%%%%%%%%%%%%%%%%%%%%%%%%%%%%%
%%%%%%%%%%%%%%%%%%%%%%%%%%%%%%%%%%%%%%%%%%%%%%%%%%%%%%%%%%%%%%%%%%%%%

\section{Transport mechanism}
Current literature assesses the molecular-scale mechanism of lithium transport in dimensions of either a "vehicular" or "structural" type of diffusion \cite{huang2018solvation,self2019transport,li2012li+,fong2019ion,chen2018ion,haskins2014computational,borodin2018insights,li2015effect,dong2018charge}. 
The terminology is apparently borrowed from the mechanistic description of proton conduction where the conditions are somewhat similar. Due to its large charge density a proton is liable to be absorbed by its environment. If it is attached to a mobile host molecule, the proton diffusion is guided by this vehicular entity (vehicle mechanism) \cite{kreuer1996proton,norbya1990proton,li2019insights,kreuer1982w}. This transport concept contrasts with a repeated jumping along electronegative moieties that are provided by a surrounding parent structure (hopping / Grotthus mechanism) \cite{agmon1995grotthuss,kreuer1996proton,li2019insights}.
In the context of lithium transport, this vehicular concept is often equated with a strongly correlated motion of $\text{Li}^+$ and its shell, where the latter may be constituted by several coordinating molecules. 
The aspect of structural diffusion is less clearly defined but commonly describes the scenario where $\text{Li}^+$ mobility benefits from a frequent exchange of the coordinating molecules, possibly by means of such Grotthus-like hopping events between coordination spheres. \cite{self2019transport,haskins2014computational,molinari2019general,nuernberg2020}\newline
Putting theory into practice, the classification of the dominating mechanism is usually made by balancing a local time or length scale, which is set by a single $\text{Li}^+$-ligand pair, against those established by the diffusivities of $\text{Li}^+$ and ligand. In a variety of recent simulation studies on lithium salt-IL electrolytes, the importance of the vehicular mechanism is estimated by comparing $\text{MSD}_{\text{Li}^+}(\tau_{\text{Li}^+-X})$ against the size of the solvating anion $X$. \cite{haskins2014computational,borodin2006li+,borodin2018insights,li2012li+}
Self and coworkers \cite{self2019transport} systematically studied the $\text{Li}^+$ transport process in super-concentrated lithium salt propylene carbonate (PC) electrolytes, which differ from the IL based electrolytes investigated in this work insofar as $\text{Li}^+$ solvation is achieved not only through anions but also neutral PC molecules. To conclude on the predominant $\text{Li}^+$ transport mechanism in a spectrum from vehicular to structural diffusion, the authors proposed an intuitive comparison of two length scales. The characteristic length scale  
\begin{equation}
\text{L}^c_{\text{Li}^+-\text{X}}\,=\,\sqrt{6\text{D}_{\text{Li}^+}\tau_{\text{Li}^+-X}}
\label{eq:criterion_Lc}
\end{equation}
measures the average distance that $\text{Li}^+$ and ligand $X$ diffuse together and is evaluated against the size of the $X$-based solvation shell $\text{Ls}$. One proposed practical, yet contestable, estimator for $\text{Ls}$ is the  position of the first minimum of $\text{g}_{\text{Li}^+-X_{\text{com}}}$.
We slightly extend this scheme by switching to a temporal perspective and relate the local length scale Ls to a characteristic "self diffusion time" $\tau_{\text{Ls}}$ according to
$\text{MSD}_{\text{Li}^+}\,(\tau_{\text{Ls}})\,=\,\text{Ls}^2$, \textit{i.e.}, the time $\text{Li}^+$ requires to cover the area $\text{Ls}^2$. Please note that we will use $\tau_{\text{Ls}}$, whose concentrations dependence is additionally shown in Figures \ref{fig:mean_residence_times_lithium_anions}A and B, to scale the time dependence of correlative properties later.
By analogy to Self et al. \cite{self2019transport}, the transport criterion can be formulated as
\begin{equation}
\text{r}_{\text{L}}\,\defeq\,\dfrac{\text{L}^c}{\text{Ls}} \qquad \simeq \qquad \text{r}_{\tau}\,\defeq\,\sqrt{\dfrac{\tau_{\text{Li}^+-X}}{\tau_{\text{Ls}}}} \qquad
\begin{cases}
    \qquad  > \quad 1 & \text{vehicular}\\
    \qquad  < \quad 1 & \text{structural}.
    \end{cases} 
\end{equation}
Figure \ref{fig:mean_residence_times_lithium_anions}C shows that an increasing salt content effectuates a drop in the net contribution of the vehicular mode in both electrolyte mixtures. This trend is reported in previous studies \cite{li2012li+,haskins2014computational} and it is reasoned that a relative mobility improvement of alkali metal ions may be causally linked to a concentration induced breakup of their solvation vehicle \cite{nuernberg2020,chen2018ion,forsyth2016novel}. This hypothesis of a structurally conditioned change in transport mechanism is supported by the sudden and steep decline of $\text{r}_{\text{L}/\tau}$ for the $\text{TFSAM}^-$ based mixtures. When exceeding the threshold concentration x\,=\,0.3, for which the primary $\text{Li}^+$ solvation sphere is dramatically reconfigured, the degree of collective $\text{Li}^+$-$\text{TFSAM}^-$ motion falls abruptly below the $\text{TFSI}^-$ analogue. 
As already discussed on basis of the absolute mean residence times, the incipient coordination by less attractive binding sites provided by $\text{TFSAM}^-$ may imply an overall lower binding energy of $\text{Li}^+$ to its solvation cage, which therefore promotes a more efficient renewal of the latter.\newline
Overall, the two common approaches of comparing either characteristic length/time scales or the ratio of self-diffusion coefficients attest to a decreasingly collective nature of the $\text{Li}^+$-solvation complex, although their information content is quite different. On the one hand, we learn from the $\text{r}_{\text{L}/\tau}$ behaviour that the extent to which $\text{Li}^+$ travels with its shell decreases steadily with increasing salt content. However, a change of transport mechanism towards structural diffusion can only be deduced when $\text{r}_{\text{L}/\tau}$ falls below 1 and, naturally, undershooting this threshold is very sensitive to how one specifies Ls. The explanatory power of relative diffusivities, on the other hand, is limited to $\text{D}_{\text{Li}^+}/\text{D}_{\text{anion}} > 1$. In the highly concentrated regime $\text{Li}^+$ diffuses faster than the $\text{TFSAM}^-$ shell anions, which is indeed not compatible with the vehicular picture of a dynamically stable $\text{Li}^+$ shell and strong evidence of structural $\text{Li}^+$ diffusion. Unlike for $\text{r}_{\text{L}/\tau}$, the trend of increasing $\text{D}_{\text{Li}^+}/\text{D}_{\text{anion}} \leq 1$ is open to different interpretations. Apart from the evolution of a new transport mechanism, a decreasing $\text{D}_{\text{anion}}$ could effectuate an increasing diffusivity ratio as well. Referring to the discussion of coordination numbers, a higher lithium salt content correlates with a smaller fraction of $\text{Li}^+$-uncoordinated, \textit{i.e.}, 'free', anions so that $\text{D}_{\text{anion}}$ is slowed down with more anions being integrated in retarding $\text{Li}^+$ solvation shells. In fact, $\text{D}_{\text{Li}^+}/\text{D}_{\text{anion}}=1$ taken alone could be the signature of perfectly vehicular $\text{Li}^+$-shell dynamics. \newline
A key drawback of these two analysis schemes is that both spatial and temporal resolution of how the $\text{Li}^+$-shell disintegrates are not accounted for.
To elucidate the transport mechanism in new and complementary detail, we pose three key questions about the $\text{Li}^+$-shell dynamics:
\begin{itemize}
\item[$\text{Q}_1$] How do anion dynamics couple to the motion of the lithium ion?
\item[$\text{Q}_2$] Do the dynamics of anions, which are coordinated to the same $\text{Li}^+$,  exhibit extra positive correlations among themselves and thus stability, as implied by the picture of a vehicular transport mechanism? 
\item[$\text{Q}_3$] How do anion dynamics respond to multiple lithium coordination at high salt fractions?
\end{itemize}
\begin{figure}[H]
	\centering
	\includegraphics[width=0.4\textwidth]{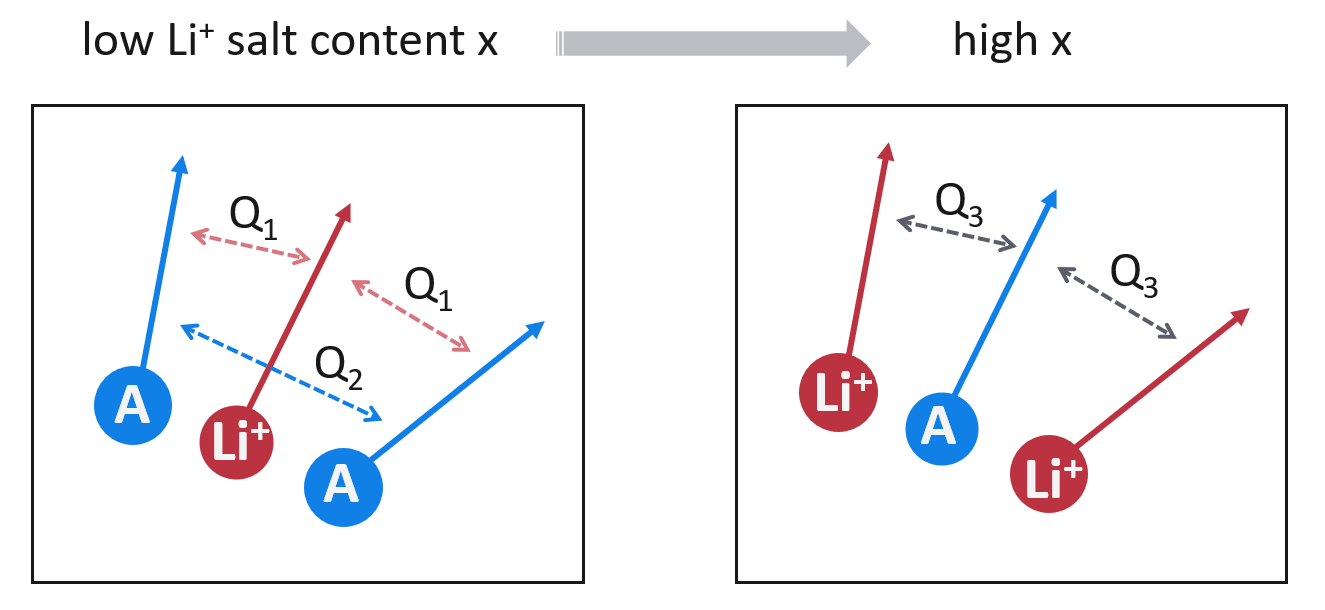}
	%\caption{Key ideas. } 
	%\label{fig:key_questions_transport_mechanism}
\end{figure}
\subsection{Collectivity of $\text{Li}^+$-shell dynamics and Lithium-Coupling Factor $\lambda$} 
To answer these questions, we propose a very simple but systematic characterisation of the (un)coupled dynamics of $\text{Li}^+$ and its anionic neighbourhood. The underlying idea is sketched in Scheme \ref{fig:corr_ij_concept}. We set up an anion neighbour list for every $\text{Li}$-ion based on the procedure that is employed to determine $\tau_{\text{Li}^+-\text{TFSI}^-/\text{TFSAM}^-}$. 
\begin{scheme}[H]
	\centering
	\includegraphics[width=0.5\textwidth]{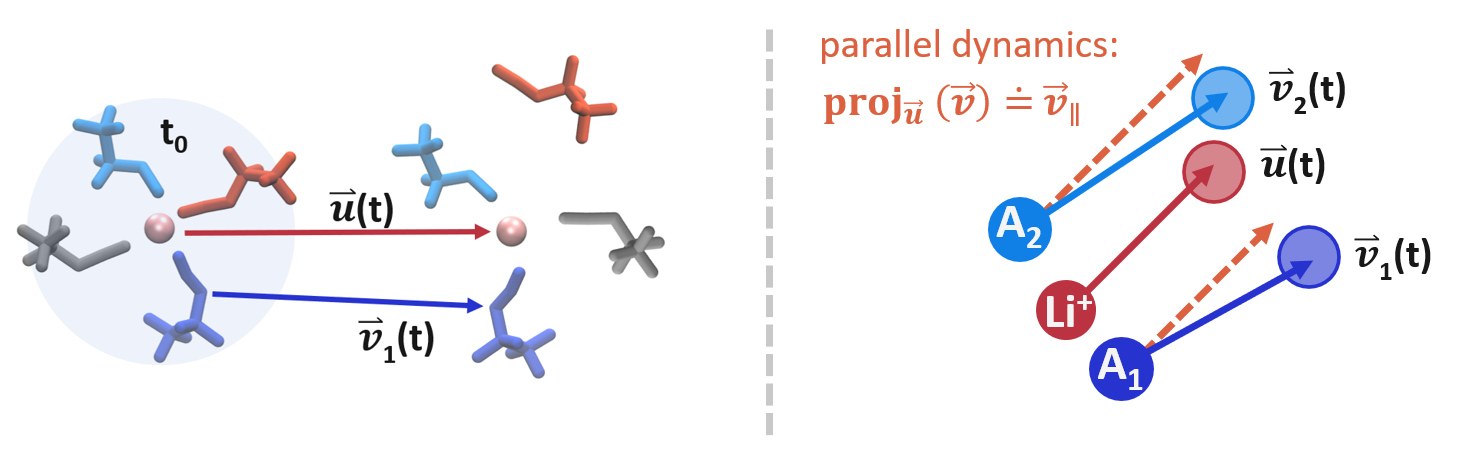}
	\caption{Left: Sketch of a $\text{Li}^+$-solvation complex which is defined at a reference time ${t}_0$ and tracked over time regarding the spatial displacements. 
	Right: To measure the translation $\vec{{v}}_{\parallel j}$ (dashed orange) of the anion in direction of the $\text{Li}$-ion to which it is initially bound, the anion displacement $\vec{{v}}_j({t})$ (blue) is projected on the respective $\text{Li}^+$ displacement $\vec{{u}}({t})$ (red). } 
	\label{fig:corr_ij_concept}
\end{scheme}
\noindent The neighbour list comprises the identities $j$ and positions $\vec{r}^{\,\,i}_{j}({t}_0)$ of the anions binding to the designated $\text{Li}$-ion $i$ at a reference time ${t}_0$. We then track the individual displacements of the $\text{Li}$-ions $\vec{{u}}_{i}({t})\,=\,(\vec{r}_{i}({t})-\vec{r}_{i}({t}_0))$ and the denoted anions $\vec{{v}}^{\,\,i}_{j}({t})\,=\,(\vec{r}^{\,\,i}_{j}({t})-\vec{r}^{\,\,i}_{j}({t}_0))$ as a function of time.
\begin{figure*}[hb!]
	%\centering
	\includegraphics[width=1.0\textwidth]{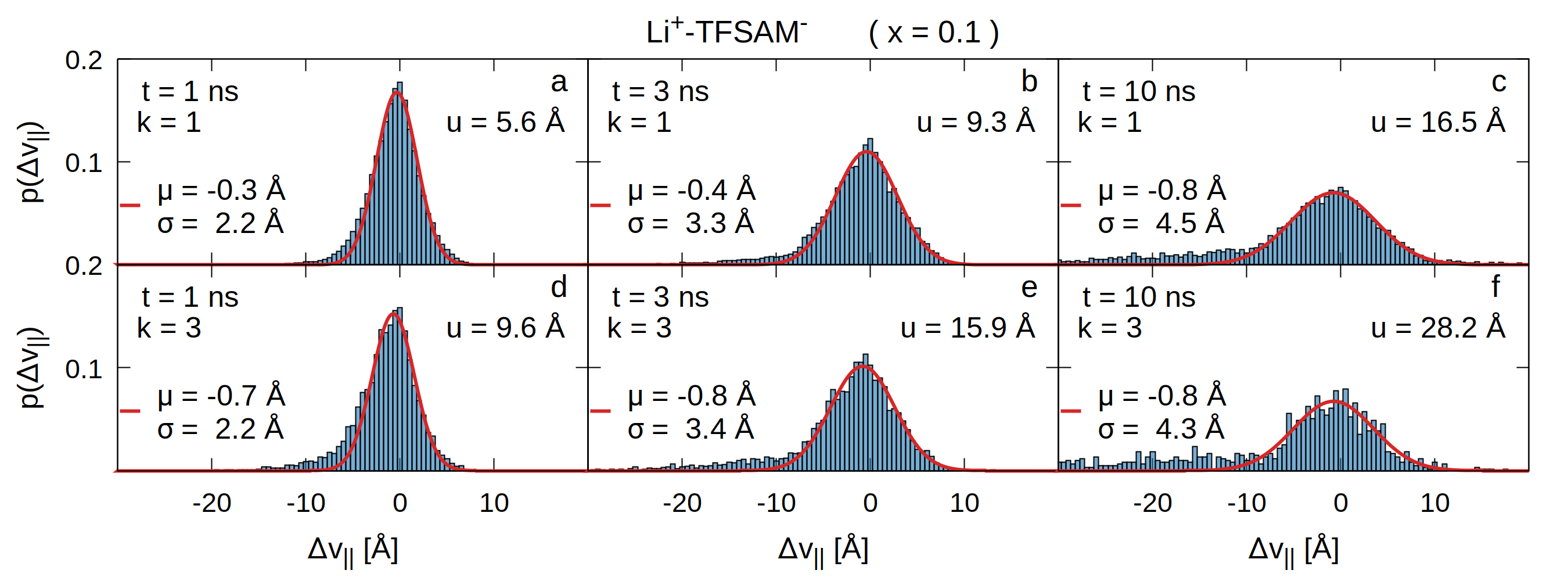}

	\caption{Exemplary for the $\text{TFSAM}^-$-based mixture at a salt fraction x\,=\,0.1: Distributions of the relative distance $\Delta v_{\parallel}$ between anion and $\text{Li}^+$, which are bound at a reference time ${t}_0$, in direction of the $\text{Li}$-ion for various subensembles. The upper panel shows the histograms for the subensemble of $\text{Li}$-ions whose squared displacement $u^2$ matches the $\text{Li}^+$ mean squared displacement $\langle{u}^2\rangle$ at the respective lag times $t=$ 1, 3 and 10\,ns (a, b and c), \textit{i.e.}, $u^2 = k\cdot \langle{u}^2\rangle$ and distance scaling factor $k=1$. The lower panel histograms (d, e and f) belong to the subensemble of $\text{Li}$-ions for which the distance scaling factor $k=3$. The peaks are fitted by a Gaussian function (red) whose expected value $\mu$ and standard deviation $\sigma^2$ are given in the legends. A description of the fitting procedure is provided in the Supplementary Information section F$^\dag$. Please note that the histograms are displayed for a bin width of 0.5\,\angstrom\, for visual reasons but the Gaussian fit is obtained for 0.1\,\angstrom\, binning.} 
	\label{fig:histogram_delta_v_para_Li_TFSAM_x_0_1}
\end{figure*}
\noindent As a starting point for answering question $\text{Q}_1$, we characterise the dynamic collectivity of $\text{Li}^+$ and anion by means of a vector decomposition: we project the anion displacement $\vec{{v}}^{\,\,i}_{j}({t})$ on the displacement $\vec{{u}}_{i}({t})$ of its $\text{Li}$-ion and compute the distance $v^{\,\,i}_{j \parallel}$ which the anion covered parallel to the $\text{Li}$-ion's direction. From this, we can further deduce the change of position $\Delta v_{\parallel}$ of the anion relative to $\text{Li}^+$ in that direction:
\begin{equation}
    \dfrac{ \vec{u}_i\,\cdot\,\vec{v}^{\,\,i}_{j} }{ u_i } = v^{\,\,i}_{j \parallel} \qquad \wedge \qquad  v^{\,\,i}_{j \parallel} - u_i \quad \defeq \quad \Delta v_{\parallel}.
\end{equation}
\noindent In the limiting case of a perfectly coupled motion, the relative displacement is conserved and thus $\Delta v_{\parallel} = 0$. Apart from a natural time dependence, \textit{i.e.}, one expects $\Delta v_{\parallel}$ to scatter increasingly with time as $\text{Li}^+$ and anions become detached, $\Delta v_{\parallel}$ may also depend on the $\text{Li}^+$ displacement $u$. It seems plausible that the dynamic collectivity could be impeded increasingly with larger $u$. To test this assumption, we study the distribution $p(\Delta v_{\parallel})$ for subensembles of $\text{Li}$-ions which covered a specific distance $u$. For a comparison as intuitive as possible, we choose subensembles of $\text{Li}$-ions whose squared displacement $u^2$ corresponds to a multiple $k$ of the mean squared displacement $\langle u^2 \rangle$ at time $t$.\newline
Figure \ref{fig:histogram_delta_v_para_Li_TFSAM_x_0_1} gives an overview of $p(\Delta v_{\parallel})$ as a function of lag time $t$ and $\text{Li}^+$ distance scaling factor $k$ exemplary for the x\,=\,0.1 $\text{TFSAM}^-$-based electrolyte.
In accordance with the previously assessed indicators of strongly coupled $\text{Li}^+$-anion-dynamics, \textit{i.e.}, $\text{r}_{\text{L}/\tau}$, the distributions show a large peak that is centred closely at a relative anion-$\text{Li}^+$ distance $\Delta v_{\parallel} \approx 0\,\,$\angstrom\,. Importantly, one can directly see the disintegration with time. We interpret the developing tail of $p(\Delta v_{\parallel})$ as the superposition of a steadily growing ratio of anions having disengaged from the initial $\text{Li}^+$ solvation shell, whose subsequent dynamics are no longer correlated with the $\text{Li}^+$ reference. However, since the mean residence time $\tau_{\text{Li}^+-\text{TFSAM}^-} = 18.1\,$ns exceeds the longest analyzed lag time $t$ the peak is yet present. To quantify the peak features, i.e., its position $\mu$ and variance $\sigma^2$, and thus the dynamic coupling of $\text{Li}^+$-anion pairs, we empirically fit the distribution by a Gaussian function. To separate the peak from the overlapping signatures of decoupled anions, we adopt a two-step fitting procedure: First, the plain distribution is fitted by a Gaussian which provides us with an educated guess of $\Tilde{\mu}$ and $\Tilde{\sigma}^2$. Then, we refit the distribution but limit the left-hand side of the value range to $-\Tilde{\sigma}+\Tilde{\mu}\,\leq \, \Delta v_{\parallel}$. The final Gaussian fits are plotted in red with $\mu$ and $\sigma$ provided in the legend.
At fixed lag time $t$, we find that $\mu$ is not only different to zero but also shifted towards an increasing anion-$\text{Li}^+$ distance for larger $k$. Interestingly, neither of these two properties is intuitively expected from a vehicular transport concept. 
Assuming that the bound anions move with lithium as a kinetic entity, reorientations within the shell geometry would effect a maximum possible peak width $\sigma$, yet should average to $\mu = 0$ for symmetry reasons. While the peak of coupled dynamics may decrease in height for a decreasing ratio of dynamically preserved $\text{Li}^+$-anion pairs, its unique signature of $\mu\,=\,0$ and constant $\sigma$ after fast local equilibration would not depend on $u$.\newline
We explore the mechanistic background giving rise to this peak behaviour by means of a toy model: A $\text{Li}^+$-anion pair is simplified to two particles which are coupled by a harmonic interaction. The particles' collective and relative dynamics, \textit{i.e.}, $X=(u+v_{\parallel})/2$ and $\Gamma=(u-v_{\parallel})/2$, are both subjected to Gaussian distributions whose statistical parameters are accessible through our analysis of $p(\Delta v_{\parallel})$. As demonstrated in the Supplementary Information section G$^\dag$, this toy model predicts a connection between $\mu$ and $\sigma$:
\begin{equation}
     \mu \quad= \quad - \dfrac{u}{2} \cdot \dfrac{\sigma^2 }{\langle u^2\rangle} \quad\stackrel{\text{for}\, u^2\, =\, k\cdot\langle u^2\rangle}{=} \quad - \dfrac{\sqrt{k}}{2} \cdot \dfrac{\sigma^2 }{\sqrt{\langle u^2\rangle}}.
     \label{eq:mu_delta_v_para_scaling_relation}
\end{equation}
The validity of this relationship with regard to the numerical data is tested in Figure \ref{fig:masterscaling} and we find that the scaling $\mu \propto \sqrt{k}$ as well as the universal interplay of $\mu, \sigma$ and $\langle u^2\rangle$ is surprisingly well-fulfilled. From this, we can conclude that the dynamic complexity of a $\text{Li}^+$-anion pair, and therefore the $\text{Li}^+$ solvation shell, is approximated remarkably well by a harmonic coupling.  
\begin{figure}[htb!]
	\centering
	\includegraphics[width=0.5\textwidth]{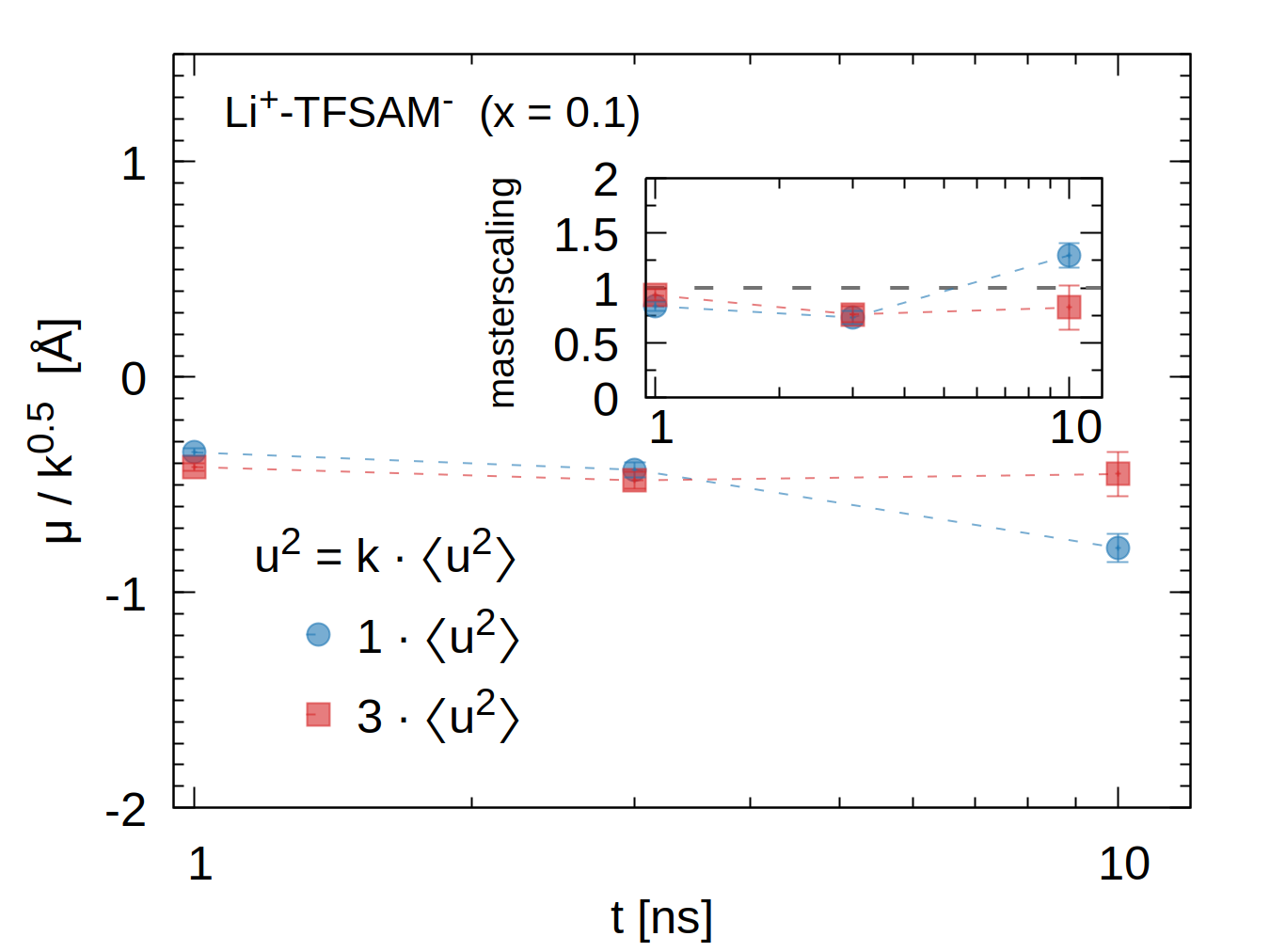}
	\caption{Exemplary for the $\text{TFSAM}^-$-based mixture at a salt fraction x\,=\,0.1: Peak positions $\mu$ of Gaussian fits shown in Figure \ref{fig:histogram_delta_v_para_Li_TFSAM_x_0_1} divided by the square root of the $\text{Li}^+$ distance scaling factor $k$. Inset: Masterscaling of the Gaussian peak parameters $\mu$ and $\sigma^2$ to 1 according to Equation \ref{eq:mu_delta_v_para_scaling_relation} after rearrangement to 1\,=\,$-2\cdot \mu\sqrt{\langle u^2 \rangle}\, / \, \sqrt{k}\sigma^2$.} 
	\label{fig:masterscaling}
\end{figure}
\noindent Turning towards the time evolution of the peak properties, we make two observations: Firstly, we find a slight shifting of the peak position and secondly, a significant broadening of the peak over time. While the peak broadening might be partially attributed to anions exploring the configurational space around $\text{Li}^+$, the behaviour could also be interpreted qualitatively as a gradual weakening of the effective harmonic $\text{Li}^+$-shell interaction. Over time, an increasing share of the initially bound anions has escaped the primary $\text{Li}^+$ neighbourhood as clearly reflected in the steadily growing tail of the distribution. However, since $d\log\sigma^2/d\log t$ is significantly smaller than 1, \textit{i.e.} subdiffusive,  even after detachment these anions will find themselves in a flow field which they share with their previous environment and thus may effectuate a loose yet measurable hydrodynamic coupling to the original $\text{Li}^+$. Overall, it is remarkable that even when $\text{Li}^+$ has travelled beyond the next-neighbour distance (see Figure S1$^\dag$), the harmonic coupling in both subensembles, \textit{e.g.}, comparing Figure \ref{fig:histogram_delta_v_para_Li_TFSAM_x_0_1}c and f, remains equally strong.   \newline
While the residence time autocorrelation (ACF) function (see Equation S2) is a structural measure of the ratio of anions remaining in the $\text{Li}^+$ solvation shell, we may quantify the percentage of dynamically dissociated shell anions from $\int d \Delta v_{\parallel} ( p(\Delta  v_{\parallel} ) - \mathcal{N}(\mu,\sigma^2) ) \,\,\dot{=} \,\,p_{\text{lost}}$, \textit{i.e.}, essentially the tail of $p(\Delta v_{\parallel})$. Figure S15$^\dag$ shows that $p_{\text{lost}}$ increases with time as well as $\text{Li}^+$ displacement $u$, and, furthermore, systematically follows the structural decoupling probed by the ACF (see Figure S7$^\dag$) which underscores the relevance of dynamic coupling beyond the primary solvation shell. 
Please note that an extended data set for both $\text{TFSI}^-$- and $\text{TFSAM}^-$-based electrolytes is provided in the Supplementary Information sections F-H$^\dag$. For the sake of brevity, we discussed the common insights exemplary for the $x\,=\,0.1$ $\text{TFSAM}^-$ mixture.\newline
\begin{scheme}[H]
	\centering
	\includegraphics[width=0.25\textwidth]{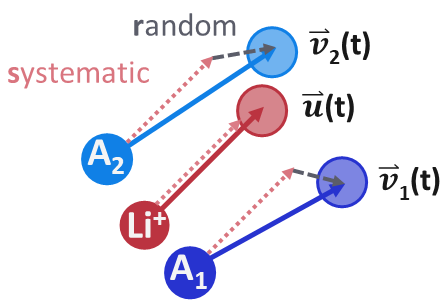}
	\label{fig:sketch_decomposition_lambda}
	\caption{In the coordinate frame of a designated $\text{Li}^+$, the anion displacement $\vec{v}_j$ (blue arrow) is split into a systematic part (dotted pink arrow), which is parallel to the $\text{Li}^+$ displacement $\vec{u}$ (red arrow), and a random motion (dashed grey arrow). As graphically indicated, the  lithium-coupling factor $\lambda$, \textit{e.g.}, $\lambda\,=\,0.85$ in this sketch, corresponds to the ratio of arrow lengths of systematic anion and $\text{Li}^+$ displacement. }
	\label{fig:sketch_decomposition_lambda}
\end{scheme}
\begin{figure*}[ht!]
	\centering
	\includegraphics[width=1.0\textwidth]{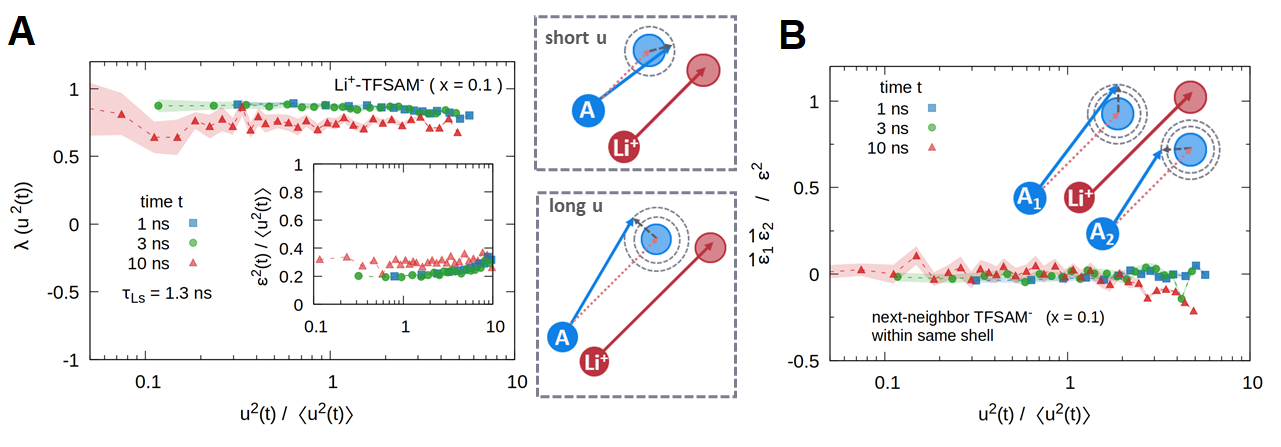}
	\caption{Exemplary for the $\text{TFSAM}^-$-based mixture at a salt fraction x\,=\,0.1: Left: LCF $\lambda$ as a function of $u^2$, which is scaled by the $\text{MSD}_{\text{Li}^+}$ $\langle\,u^{\,2}\,\rangle$ at the respective lag time $t$. Inset: Variance $\epsilon^{\,2}$ of the anion displacement relative to $\text{Li}^+$ trajectory. To allow for a better comparison, $\epsilon^{\,2}$ is scaled by $\text{MSD}_{\text{Li}^+}$ as well. Right: Correlation of the random motion of initially adjacent shell anions $\langle\,\vec{\epsilon}_1 \cdot \vec{\epsilon}_2 \,\rangle /\langle\,\vec{\epsilon}^{\,\,2}\, \rangle $ as a function of $u^2 / \langle\,u^{\,2}\,\rangle$ and the consequential mechanistic picture. Due to rare statistics the fastest 1\textperthousand\, of $\text{Li}^+$ ions are not shown in the plots. } 
	\label{fig:lambda_x2_inset_eps_x2_Li_TFSAM_0_1}
\end{figure*}
\noindent Building on the mechanistic understanding of the coupled $\text{Li}^+$-anion dynamics, we wish to characterise the $\text{Li}^+$ transport properties in their entirety, providing the foundation for a clear comparison of dynamic collectivity across different electrolytes. With this in mind, we make a conceptual switch from the dynamic coupling of distinct $\text{Li}^+$-anion pairs to the ensemble of anions and their collective motion $\langle {v}_{\parallel}\rangle_{u^2}$ in the coordinate frame of  $\text{Li}^+$. \newline 
At a fixed time $t$, we introduce the lithium-coupling factor (LCF) $\lambda$ which expresses the systematic contribution of the anion displacement  $\langle {v}_{\parallel}\rangle_{u^2,t}$ relative to the $\text{Li}^+$ displacement $u$. It may depend on the squared $\text{Li}^+$ displacement $u^2$ and on the chosen time scale $t$. On the most detailed level, we first analyse $\lambda(u^2,t)$ conditioned on a specific time lag $t$ as well as the squared lithium displacement $u^2$.  \newline
In a second step towards a further generalised description, we take the average over all $\text{Li}^+$-anion pairs regardless of the distance covered by a specific $\text{Li}^+$, giving rise to $\lambda(t)$:
\begin{equation}
\lambda(u^{\,2},t)\,=\,\dfrac{\langle\,\vec{u}_i\,\cdot\,\vec{v}^{\,\,i}_{j}\,\rangle_{u^2, t}}{{u}^{\,2}}\,=\,\dfrac{ \langle v_{\parallel}\rangle_{u^2,t}}{{u}}
\quad \wedge \quad
\lambda(t)\,=\,\dfrac{\langle\,\vec{u}_i\,\cdot\,\vec{v}^{\,\,i}_{j}\,\rangle_{t}}{\langle \,\vec{u}^{\,2}\, \rangle_{t}}.
\label{eq:basic_definition_lambda}
\end{equation}
$\langle..\rangle$ denotes the corresponding ensemble average over all $\text{Li}^+$-anion-pairs $ij$ existing at ${t}_0$. For reasons of simplicity, the time dependence of these observables is not explicitly mentioned in parts. Geometrically speaking, the LCF weighs the average projection of $\vec{v}$ on $\vec{u}$ against the $\text{Li}^+$ displacement $u$ and thereby expresses the features of Figure \ref{fig:histogram_delta_v_para_Li_TFSAM_x_0_1} through a number. From a physical viewpoint, the LCF behaves like a correlation coefficient: A strictly coupled motion of $\text{Li}^+$ and its shell anions implies $\lambda\approx 1$ because single leads of either $\vec{u}$ or $\vec{v}$ due to rotational rearrangement within a shell would cancel out over time. When an initially defined neighbourhood disintegrates, the LCF decreases ultimately to zero. 
As visualised in Scheme \ref{fig:sketch_decomposition_lambda}, we may thus decompose the anion dynamics in the reference frame of $\text{Li}^+$ into a systematic part (pink) and random motion (grey):
\begin{equation}
\vec{v}^{\,\,i}_{j}(t) \,\,=\,\, \lambda(u^{\,2},t)\,\cdot\,\vec{u}_{i}(t) \,\,+ \,\,\vec{\epsilon}^{\,\,i}_{j}(u^{\,2},t).
\label{eq:basic_decomposition_y}
\end{equation}
Knowing $\lambda$, we gain additional information on the anion's motional freedom which is measured by the variance:
\begin{equation}\label{eq:basic_definition_eps}
3\cdot\,\epsilon^2  \quad=\quad\langle\,\vec{\epsilon}^{\,\,2}\,\rangle_{u^2,t} \quad = \quad \langle \,\vec{v}^{\,2}\,\rangle_{u^2,t}\,-\,\lambda^2 \langle\, \vec{u}^{\,2}\,\rangle_{u^2,t}.
\end{equation}
Please note that by construction $\langle \, \vec{{u}}_i  \,\vec{\epsilon}^{\,\,i}_{j} \,\rangle\,=\,0$. \newline
In the subsequent analysis we are guided by the initially posed questions 1-3 and examine with increasing generalisation how the dynamic properties of the $\text{Li}^+$-shell-object depend on $\text{Li}^+$ displacement, time and salt concentration. For the sake of brevity, the results are discussed for the example of the $\text{TFSAM}^-$-containing electrolytes, while the complete data set is provided in the Supplementary Information sections I-M$^\dag$. \newline

\subsubsection{Displacement dependence $\lambda({u}^{\,2})$ } Figure \ref{fig:lambda_x2_inset_eps_x2_Li_TFSAM_0_1}A shows that regardless of whether $\text{Li}^+$ covered half or more than five times of $\text{MSD}_{\text{Li}^+}$ at the respective lag time $t$, the average collectivity with its environment remains the same. The finding that $\lambda$ does hardly depend on $u$ is consistent with the predictions of the harmonic toy model. In the limit of $p_{\text{lost}}\,=\,0$, Equation \ref{eq:basic_definition_lambda} implies $\lambda \approx 1 + \mu/u$. As discussed for Figure \ref{fig:masterscaling}, we find $\mu \propto u$ at fixed time $t$ so that consequently $\lambda \approx $ const. for varying $u$. The observation of a strong LCF $\lambda<1$ is thus as a direct consequence of the shifted Gaussian peak position $\mu < 0$. It is remarkable that $\lambda$ maintains the independence of $u$ even at longer lag times $t$ when a noticeable proportion of the anions has already escaped the $\text{Li}^+$-shell as discussed for the tails of $p(\Delta v_{\parallel})$ in Figure \ref{fig:histogram_delta_v_para_Li_TFSAM_x_0_1}, which substantially effectuate the drop of the average coupling strength. On a qualitative level, we attribute the small decline of $\lambda$ for large lithium displacements to the concurrently increasing $p_{\text{lost}}$. \newline
As shown in the inset of Figure \ref{fig:lambda_x2_inset_eps_x2_Li_TFSAM_0_1}A, the independence of $u$ is approximately true for the random dynamics $\epsilon^2$ of the anions as well, which increases only slightly when the designated $\text{Li}^+$ has moved appreciably further. In accordance with the strong LCF, we find that the   $\text{Li}^+$-independent motion is, firstly, small compared to $u^2$ and, secondly, scales with $\langle u^2(t)\rangle$ which further supports the conceptual extension of coupled transport by flow-like properties. Splitting $\vec{\epsilon}$ into contributions parallel $\vec{\epsilon}_{\parallel}$ and orthogonal $\vec{\epsilon}_{\perp}$ to the $\text{Li}^+$ path direction $\vec{u}$, as demonstrated in Figure S19$^\dag$, reveals that the anionic fluctuations relative to the $\text{Li}^+$ trajectory are approximately isotropic with $\vec{\epsilon}_{\perp}^{\,\,2}\,\approx 2\,\cdot\,\vec{\epsilon}_{\parallel}^{\,\,2}$. The random displacement of the anions can therefore be illustrated schematically by spheres which add to the systematic motion in direction of the $\text{Li}^+$ ion as pictured in Figure \ref{fig:lambda_x2_inset_eps_x2_Li_TFSAM_0_1}. \newline
Finally, the vector decomposition into systematic and random dynamics allows us to study the $\text{Li}^+$-independent collectivity of two initially adjacent shell anions by correlating their random motions $\vec{\epsilon}_1$ and $\vec{\epsilon}_2$. As shown in Figure \ref{fig:lambda_x2_inset_eps_x2_Li_TFSAM_0_1}B, the anions exhibit barely a sign of interaction beyond the individual coupling to $\text{Li}^+$, \textit{i.e.}, neither negative correlations due to plausibly repulsive dynamics of likely charges, nor positive correlations that would evidence a stability of the shell environment. We complete our $\text{Li}^+$ transport scheme accordingly by a non-communicating anion environment and conclude that the vehicular transport description is somewhat misleading when taking it at face value.\newline
Last but not least, the absence of displacement dependencies allows us to characterise the dynamic collectivity by a single $\lambda(t)$ and provides the basis for a further simplified model description.\newline
\FloatBarrier

\begin{figure}[ht!]
    \centering
	\includegraphics[width=0.5\textwidth]{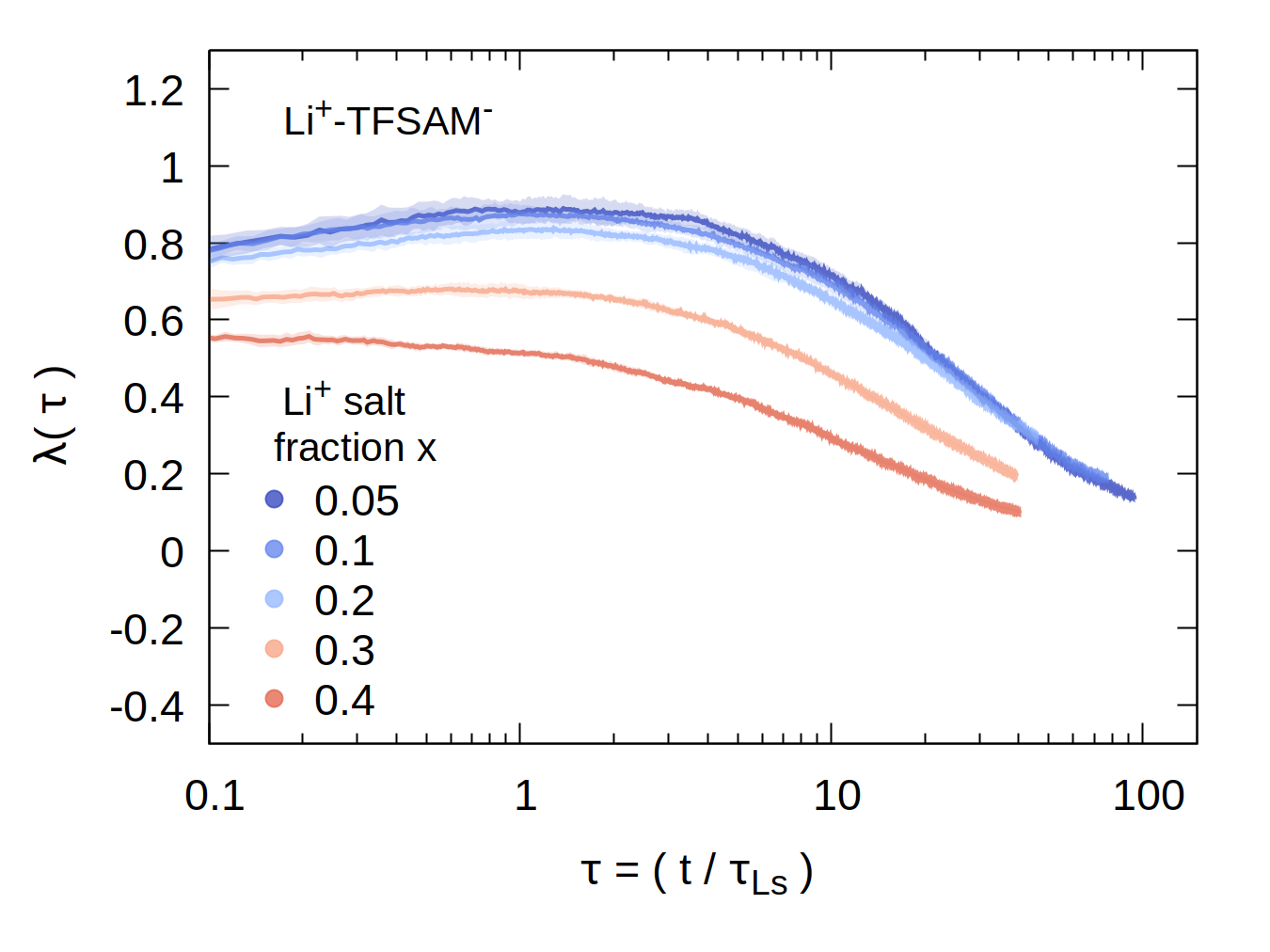}
	\caption{Exemplary for the $\text{TFSAM}^-$-based mixtures for various lithium salt concentrations x: LCF $\lambda$ as a function of $\tau$, which is the time $t$ scaled by the self-diffusion time $\tau_{\text{Ls}}$ corresponding to concentration x.} 
	\label{fig:lambda_correlation_eps1eps2_sketch_TFSAM}
\end{figure}

\subsubsection{Time dependence $\lambda({t})$ } Figure \ref{fig:lambda_correlation_eps1eps2_sketch_TFSAM} shows how the LCF behaves as a function of time. For comparability across different systems, we scale the absolute time $t$ by the mixture-specific self-diffusion time $\tau_{\text{Ls}}$.
To begin with, we find that the temporal decay of $\lambda$ is practically identical for low salt fractions. This indicates that the underlying $\text{Li}^+$ transport principle remains unchanged in this concentration regime. 
As already implied in Figure \ref{fig:lambda_x2_inset_eps_x2_Li_TFSAM_0_1}A, a high LCF of $\lambda \gtrapprox 0.8$ is maintained over multiple $\tau_{\text{Ls}}$ until it steadily decreases, reflecting the dissolution of the initial neighborhood. 
Interestingly, $\lambda(\tau)$ experiences an abrupt downshift when crossing the previously discerned threshold concentration x$\,=\,$0.3. Its decay behaviour is increasingly flattened upon the addition of lithium salt and we find for the high salt concentration regime that $\lambda$ no longer sustains a plateau value as shown in Figure S22$^\dag$.\newline

\subsubsection{Concentration dependence $\lambda(\text{x})$} 
\begin{figure*}
	\centering
	\includegraphics[width=0.95\textwidth]{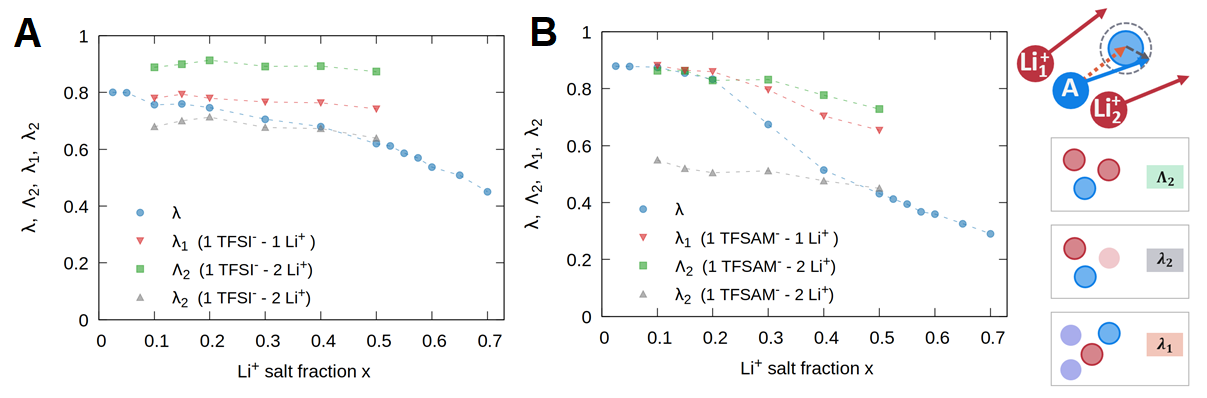}
	\caption{Concentration dependence of LCFs evaluated at  $t\,=\,1\cdot{\tau}_{\text{Ls}}$ for $\text{TFSI}^-$ (left) and $\text{TFSAM}^-$ (right) - based electrolytes. 
The blue symbols show the LCF $\lambda$ averaged over all $\text{Li}^+$-anion pairs. The green data points show the LCF $\Lambda_2$ of one anion to the average displacement $\vec{U}\,=\,\dfrac{1}{2}\left(\vec{u}_{\text{Li}^+_1}+\vec{u}_{\text{Li}^+_2}\right)$ of two initially binding $\text{Li}$-ions as illustrated in the sketches. $\lambda_{1}$ (red) and $\lambda_{2}$ (grey) compare the coupling strengths for the subsets where the anion binds to a single $\text{Li}^+$ ($\lambda_{1}$) or is shared between two $\text{Li}^+$ ($\lambda_{2}$). } 
	\label{fig:lambda_concentration_TFSI_TFSAM_sketch_next_level_principle}
\end{figure*}

On the last level of abstraction, we analyse how $\lambda$ is affected by the salt content. To obtain a systematic overview, we measure $\lambda$ at the characteristic self-diffusion time $\text{t}\,=\,1\cdot\tau_{\text{Ls}}$, which coincides closely with the onset of a diffusive $\text{Li}^+$ motion as shown in Figure S8$^\dag$. We compare the concentration behaviour $\lambda(\text{x})$ of the $\text{TFSI}^-$ versus $\text{TFSAM}^-$-based mixtures in Figure \ref{fig:lambda_concentration_TFSI_TFSAM_sketch_next_level_principle} and make, in accordance with a transport description via $\text{r}_{\text{L}/\tau}$, two key observations: Firstly, we find for both setups that the dynamic collectivity of $\text{Li}^+$ and its anionic environment is progressively suppressed. 
Secondly, the overall reduction is not only greater for the $\text{TFSAM}^-$-mixtures, for which $\lambda$ drops by two thirds, but again seems to be triggered when the salt content exceeds a critical threshold of $\text{x}\,=\,0.3$. 
The aforementioned turning point of the $\text{Li}^+$-shell structure, \textit{e.g.}, the onset of less stable $\text{Li}^+$ binding to the $\text{O}_{\text{TFSAM}^-}$ atoms instead of the more attractive $\text{N}_{\text{out}}$ cyano-group, remains a plausible explanation. The fact that $\lambda_{\text{Li}^+-\text{N}_{\text{out}}}$ is close to a perfect correlation and significantly higher than $\lambda_{\text{Li}^+-\text{O}_{\text{TFSAM}^-}}$ as depicted in Figure S22$^\dag$, further corroborates the understanding that the implementation of such weak coordination sites greatly reduces the dynamic stability of $\text{Li}^+$ environments. We add that for the $\text{TFSI}^-$-based electrolytes, where $\text{Li}^+$ binding can be mediated through oxygen contacts only, the concentration-induced trend towards both monodentate coordination geometries and higher CNs could have a loosening impact on the cohesiveness of the $\text{Li}^+$-shell complex as well.\newline
An explanation for the general decrease of the LCF suggests itself from our adopted modelling perspective: We concluded from \ref{fig:li_coordination_numbers_tfsi_tfsam_distribution}B that with increasing salt concentration a growing ratio of anions must be shared by multiple $\text{Li}$-ions. When an anion is integrated in the solvation shells of two $\text{Li}^+$, how can it maintain systematic pair dynamics? It seems natural that the LCF of a distinct $\text{Li}^+$-anion pair comes at the expense of another and thereby effectuates indirectly a liberation of the lithium ions. \newline
To test this hypothesis, we specifically analyse the LCF for two different subensembles which are sketched in Figure \ref{fig:lambda_concentration_TFSI_TFSAM_sketch_next_level_principle}: Firstly, we select the $\text{Li}^+$-anion pairs where the anion is bound exclusively to a single $\text{Li}^+$ and compute the LCF $\lambda_1$ (red) according to Equation \ref{eq:basic_definition_lambda}. Secondly, we choose the subset of $\text{Li}^+$-anion pairs where the anions are shared by two $\text{Li}^+$ and evaluate their LCF $\lambda_2$ (grey) as per this basic definition. We observe for both electrolyte series that the average $\lambda$ (blue) makes a clear transition between the subsets $\lambda_1$, which is dominant at low x, and $\lambda_2$ which becomes more relevant with increasing x. It is remarkable that both LCFs $\lambda_1$ and $\lambda_2$ remain nearly constant over a broad concentration regime, except for a slight decrease of $\lambda_1$ in the highly concentrated $\text{TFSAM}^-$ electrolytes. The latter could result from a complex structuring of the electrolyte that further impedes the mobility of $\text{Li}^+$-anion pairs. The fact that $\lambda_2$ is considerably smaller than $\lambda_1$ underpins our afore-stated expectation that the systematic anion motion is suppressed through multiple $\text{Li}^+$-binding.\newline
\newpage
\noindent We challenge the strength of our transport observable and generically extend the LCF formalism for the ensemble of anions which have two $\text{Li}^+$ neighbours:
\begin{equation}
\begin{aligned}
\vec{v}_j\,&=\, \Lambda_2\,\cdot\,\vec{U}_i^j\,+\,\vec{\mathcal{E}}_j \qquad \text{with}\quad \vec{U}^{\,j}_i\,=\,\dfrac{1}{2}\left(\vec{u}^{\,\,j}_{\text{Li}^+_{\,1}}+\vec{u}^{\,j}_{\text{Li}^+_{\,2}}\right)\\
\Lambda_2\,&=\,\dfrac{\langle\,\vec{U}^{\,j}_i\vec{v}_j\,\rangle}{\langle\,\vec{U}^2\,\rangle}\qquad \wedge \qquad  \langle \, \vec{\mathcal{E}}^2\,\rangle\,=\, \langle \,\vec{v}^2 \,\rangle - \Lambda^2_2\,\cdot\,\langle \, \vec{U}^2\,\rangle.
\end{aligned}
\label{eq:Lambda2_definintion}
\end{equation}
\noindent In this this approach, we correlate the displacement $\vec{v}_j$ of double coordinating anions with the average displacement vector of the two $\text{Li}^+$ as if they merged to a quasi-lithium particle. Of course, the equal weighting of the $\text{Li}^+$ displacements omits that different binding geometries and sites may entail different coupling strengths. Nevertheless, we find that $\Lambda_2$ (green) recovers the systematic transport characteristics $\lambda_1$ on this higher level of abstraction surprisingly well in terms of concentration dependence and for the $\text{TFSAM}^-$-based mixtures also regarding the absolute value. This means that although the anion motion decreasingly follows that of distinct $\text{Li}^+$, it is coupled in a good approximation to their mean displacement vector. It therefore appears that the $\text{Li}^+$-transport behaviour at high salt concentrations bears significant similarity to the transport mechanism characterised for diluted electrolytes.\newline
The introduced LCFs are observables that are presently only accessible through simulation but not experimentally measurable. However, returning to the ratio of $\text{Li}^+$ and anion diffusion coefficients, whose concentration dependence was the experimentally found indication to postulate a change of $\text{Li}^+$ transport mechanism \cite{nuernberg2020}, we can use our formalism to elucidate why $\text{D}_{\text{Li}^+}/\text{D}_{\text{anion}}$ increases with salt content. \newline After squaring and rearranging the anion dynamics' decomposition for the different subensembles according to Equations \ref{eq:basic_decomposition_y} and \ref{eq:Lambda2_definintion}, the inverse ratio $\text{D}_{\text{anion}}/\text{D}_{\text{Li}^+}$ can be expressed through the observables of systematic and random motion:
\begin{equation}
\begin{split}
\text{subensemble of}\,\, \lambda_1:\, \dfrac{\langle\,\vec{v}^{\,2}\,\rangle}{\langle\,\vec{u}^{\,2}\,\rangle}  &= \lambda_1^2\,+\,\dfrac{\langle\,\vec{\epsilon}^{\,2}\,\rangle}{\langle\,\vec{u}^{\,2}\,\rangle} \\
\text{subensemble of}\,\, \Lambda_2:\,   \dfrac{\langle\,\vec{v}^{\,2}\,\rangle}{\langle\,\vec{u}^{\,2}\,\rangle}  &=  \Lambda_2^2\, \cdot\underbrace{ \dfrac{1}{2}\,\left( 1 + \dfrac{\langle\,\vec{u}_1\vec{u}_2\,\rangle}{\langle\,\vec{u}^{\,2}\,\rangle} \right)}_{<1}\,+\,\dfrac{\langle\,\vec{ \mathcal{E} }^{\,2}\,\rangle}{\langle\,\vec{u}^{\,2}\,\rangle}.
\end{split}
\label{eq:subensemble_d_ratio}
\end{equation}
Please note the special term including the $\text{Li}^+_1-\text{Li}^+_2$ interaction which occurs for the subensemble of $\Lambda_2$. While we found that the coupling strengths $\lambda_1 \lesssim \Lambda_2$ depend only marginally on salt content, it is a priori not known how the terms $\langle\vec{\epsilon}^2\rangle/\langle\vec{u}^2\rangle$, $\langle\vec{\mathcal{E}}^2\rangle/\langle\vec{u}^2\rangle$, and $\langle\vec{u}_1\vec{u}_2\rangle/\langle\vec{u}^2\rangle$, as well as $\langle\vec{u}^2\rangle$ for the two subensembles behave as a function of x. As presented in Figures S26$^\dag$-S29$^\dag$, analysis shows, firstly, that $\langle\vec{u}^2\rangle$ is almost identical for all subensembles and, secondly, that the random motion is even slightly reduced for the double $\text{Li}^+$-bound anions $\langle\vec{\mathcal{E}}^2\rangle/\langle\vec{u}^2\rangle  \lesssim \langle\vec{\epsilon}^2\rangle/\langle\vec{u}^2\rangle$. Thirdly, we find that $\langle\vec{\epsilon}^2\rangle/\langle\vec{u}^2\rangle$, $\langle\vec{\mathcal{E}}^2\rangle/\langle\vec{u}^2\rangle$, and $\langle\vec{u}_1\vec{u}_2\rangle/\langle\vec{u}^2\rangle$ are constant within 15$\,\%$ with respect to x in the regime of moderate salt concentration. The explanatory power of these empirical observations with regard to $\text{D}_{\text{anion}}/\text{D}_{\text{Li}^+}$ lies in the $\text{Li}^+_1-\text{Li}^+_2$ interaction: As to be expected for their cross correlation, we find $\langle\vec{u}_1\vec{u}_2\rangle/\langle\vec{u}^2\rangle < 1$, \textit{e.g.}, $\langle\vec{u}_1\vec{u}_2\rangle/\langle\vec{u}^2\rangle \approx 0.3$ in the $\text{TFSAM}^-$-based electrolytes which is shown in Figure S27$^\dag$. Given the quantitative similarity of the remaining terms and factors stated in Equation \ref{eq:subensemble_d_ratio} for both subensembles, the reluctance of $\text{Li}^+_1-\text{Li}^+_2$ to collective motion effectuates a lower $\langle\vec{v}^2\rangle/\langle\vec{u}^2\rangle$ for the anions coordinating two $\text{Li}^+$. Since the share of the latter increases with salt concentration, the overall $\text{D}_{\text{anion}}/\text{D}_{\text{Li}^+}$ is reduced systematically. Consequently, the emergence of a fundamentally different $\text{Li}^+$ transport mechanism such as hopping is yet a possible but not a necessary explanation for the increasing $\text{D}_{\text{Li}^+}/\text{D}_{\text{anion}}$ ratio.
\newpage
\subsubsection{Dynamic heterogeneity} Even though our preceding analysis does not suggest systematic hopping diffusion of $\text{Li}^+$ in the IL-matrix at elevated salt concentration, its possibility cannot be ruled out so far. A typical signature of such hopping events is the occurrence of humps or additional peaks in the self-part of lithium's van Hove function $\text{G}_{\text{s}}(\text{r,t})$, which probes the displacement distribution within time $t$ \cite{liu2011molecular,liu2021quantitative,mukherji2020hopping}. Figure S30$^\dag$ reveals the absence of such secondary peaks in all electrolyte mixtures despite their strong local structuring which is reflected in $\text{g}_{\text{Li}^+-\text{Li}^+}$ at high salt concentrations (see Figure S1$^\dag$). Furthermore, we find an increasing deviation from an ideal Gaussian behaviour $\text{G}_{\text{0,s}}$. Such dynamic heterogeneities are commonly observed in IL-based materials \cite{liu2011molecular,liu2021quantitative,mukherji2020hopping} and generally quantified via the non-Gaussian parameter $\alpha_2(\text{t})$. We find that the maximum heterogeneity $\alpha_2(\text{t})$ approximately coincides for $\text{Li}^+$, $\text{Pyr}_{14}^+$ and respective anion (see Figure S31$^\dag$), which suggests that their heterogeneous dynamics are coupled and possibly located altogether in faster moving regions.

\section{Conclusions}
In this work, we explored the structural and dynamic properties of lithium salt/binary IL mixtures over a broad range of salt content, employing the intensively studied $\text{TFSI}^-$ anion and the asymmetric analogue $\text{TFSAM}^-$.
Our results confirmed the distinctive structural and dynamic characteristics of the $\text{TFSAM}^-$-based electrolytes which were observed in a recent experimental study. 
At small salt fractions $\text{TFSAM}^-$ coordinated to $\text{Li}^+$ exclusively via the cyano-nitrogen, but when exceeding a threshold concentration of x\,=\,0.3 the $\text{Li}^+$ environment became increasingly diverse. The shorter mean residence times $\tau_{\text{Li}^+-\text{TFSAM}^-}$ were a strong indicator that $\text{Li}^+$ was weaker bound in the emerging configurations. \newline
We further studied the dynamic properties of the $\text{Li}^+$-solvation complex in the framework of generally adopted analysis procedures, \textit{e.g.}, discussion of $\text{D}_{\text{Li}^+}$ relative to $\text{D}_{\text{anion}}$ or $\tau_{\text{Li}^+-\text{anion}}$.
The results gave consistent evidence of strongly collective $\text{Li}^+$-shell motion at small salt fractions in both electrolyte mixtures and showed that the extent of joint $\text{Li}^+$-anion dynamics declines systematically with increasing salt content. \newline
We developed a novel approach that measures the extent to which the dynamics of an initially $\text{Li}^+$-bound anion remain coupled to the respective $\text{Li}^+$, and could analyse key features of the solvation shell which is commonly interpreted as a vehicle guiding $\text{Li}^+$ diffusion. 
Our findings showed that the solvation shell by no means behaved like a stable vehicle. Thus, we suggest to reconsider the description of $\text{Li}^+$ transport in terms of a vehicular mechanism. Instead, our observations highlighted a flow-like motion of $\text{Li}^+$ and its anionic environment and we propose a "coupled diffusion"-conception to be more suitable. \newline
We could discern two different causes of the decreasingly coupled diffusion for the concentrated electrolytes: On the one hand, an increased anion-sharing between $\text{Li}$-ions weakened the dynamic coupling strength of an individual $\text{Li}^+$-anion pair. On the other hand, the novel $\text{TFSAM}^-$ anion demonstrated how the incorporation of different binding sites, among which $\text{Li}^+$ showed clear interaction preferences, had a further loosening effect in the emerging ion network.
In the light of the structure-dynamics relationship elucidated in this work, tuning the $\text{Li}^+$ coordination environment appears to be a promising path for a tailored electrolyte design.\newline
We believe that our methodology to characterise the collective dynamics of a central particle and its environment can easily be applied to similar research areas where transport phenomena are studied. \newline

\section*{Conflicts of interest}
There are no conflicts to declare.

\section*{Acknowledgements}
Analysis and simulations have been performed on the computing cluster PALMA2 at the University of M\"unster. We thankfully acknowledge the financial support from MWIDE NRW as part of the "GrEEn" project (funding code: 313-W044A).

%%%END OF MAIN TEXT%%%

%The \balance command can be used to balance the columns on the final page if desired. It should be placed anywhere within the first column of the last page.

\balance

%If notes are included in your references you can change the title from 'References' to 'Notes and references' using the following command:
%\renewcommand\refname{Notes and references}

%%%REFERENCES%%%
%\bibliography{rsc} %You need to replace "rsc" on this line with the name of your .bib file
\bibliography{literature} %You need to replace "rsc" on this line with the name of your .bib file

\providecommand*{\mcitethebibliography}{\thebibliography}
\csname @ifundefined\endcsname{endmcitethebibliography}
{\let\endmcitethebibliography\endthebibliography}{}
\begin{mcitethebibliography}{58}
\providecommand*{\natexlab}[1]{#1}
\providecommand*{\mciteSetBstSublistMode}[1]{}
\providecommand*{\mciteSetBstMaxWidthForm}[2]{}
\providecommand*{\mciteBstWouldAddEndPuncttrue}
  {\def\EndOfBibitem{\unskip.}}
\providecommand*{\mciteBstWouldAddEndPunctfalse}
  {\let\EndOfBibitem\relax}
\providecommand*{\mciteSetBstMidEndSepPunct}[3]{}
\providecommand*{\mciteSetBstSublistLabelBeginEnd}[3]{}
\providecommand*{\EndOfBibitem}{}
\mciteSetBstSublistMode{f}
\mciteSetBstMaxWidthForm{subitem}
{(\emph{\alph{mcitesubitemcount}})}
\mciteSetBstSublistLabelBeginEnd{\mcitemaxwidthsubitemform\space}
{\relax}{\relax}

\bibitem[Eftekhari \emph{et~al.}(2016)Eftekhari, Liu, and
  Chen]{eftekhari2016different}
A.~Eftekhari, Y.~Liu and P.~Chen, \emph{Journal of Power Sources}, 2016,
  \textbf{334}, 221--239\relax
\mciteBstWouldAddEndPuncttrue
\mciteSetBstMidEndSepPunct{\mcitedefaultmidpunct}
{\mcitedefaultendpunct}{\mcitedefaultseppunct}\relax
\EndOfBibitem
\bibitem[Ghandi(2014)]{ghandi2014review}
K.~Ghandi, \emph{Green and Sustainable Chemistry}, 2014, \textbf{2014},
  44--53\relax
\mciteBstWouldAddEndPuncttrue
\mciteSetBstMidEndSepPunct{\mcitedefaultmidpunct}
{\mcitedefaultendpunct}{\mcitedefaultseppunct}\relax
\EndOfBibitem
\bibitem[Gali{\'n}ski \emph{et~al.}(2006)Gali{\'n}ski, Lewandowski, and
  St{\k{e}}pniak]{galinski2006ionic}
M.~Gali{\'n}ski, A.~Lewandowski and I.~St{\k{e}}pniak, \emph{Electrochimica
  acta}, 2006, \textbf{51}, 5567--5580\relax
\mciteBstWouldAddEndPuncttrue
\mciteSetBstMidEndSepPunct{\mcitedefaultmidpunct}
{\mcitedefaultendpunct}{\mcitedefaultseppunct}\relax
\EndOfBibitem
\bibitem[Armand \emph{et~al.}(2011)Armand, Endres, MacFarlane, Ohno, and
  Scrosati]{armand2011ionic}
M.~Armand, F.~Endres, D.~R. MacFarlane, H.~Ohno and B.~Scrosati, in
  \emph{Materials For Sustainable Energy: A Collection of Peer-Reviewed
  Research and Review Articles from Nature Publishing Group}, World Scientific,
  2011, pp. 129--137\relax
\mciteBstWouldAddEndPuncttrue
\mciteSetBstMidEndSepPunct{\mcitedefaultmidpunct}
{\mcitedefaultendpunct}{\mcitedefaultseppunct}\relax
\EndOfBibitem
\bibitem[MacFarlane \emph{et~al.}(2014)MacFarlane, Tachikawa, Forsyth, Pringle,
  Howlett, Elliott, Davis, Watanabe, Simon, and Angell]{macfarlane2014energy}
D.~R. MacFarlane, N.~Tachikawa, M.~Forsyth, J.~M. Pringle, P.~C. Howlett, G.~D.
  Elliott, J.~H. Davis, M.~Watanabe, P.~Simon and C.~A. Angell, \emph{Energy \&
  Environmental Science}, 2014, \textbf{7}, 232--250\relax
\mciteBstWouldAddEndPuncttrue
\mciteSetBstMidEndSepPunct{\mcitedefaultmidpunct}
{\mcitedefaultendpunct}{\mcitedefaultseppunct}\relax
\EndOfBibitem
\bibitem[Elia \emph{et~al.}(2016)Elia, Ulissi, Jeong, Passerini, and
  Hassoun]{elia2016exceptional}
G.~A. Elia, U.~Ulissi, S.~Jeong, S.~Passerini and J.~Hassoun, \emph{Energy \&
  Environmental Science}, 2016, \textbf{9}, 3210--3220\relax
\mciteBstWouldAddEndPuncttrue
\mciteSetBstMidEndSepPunct{\mcitedefaultmidpunct}
{\mcitedefaultendpunct}{\mcitedefaultseppunct}\relax
\EndOfBibitem
\bibitem[Yoon \emph{et~al.}(2013)Yoon, Howlett, Best, Forsyth, and
  Macfarlane]{yoon2013fast}
H.~Yoon, P.~Howlett, A.~S. Best, M.~Forsyth and D.~R. Macfarlane, \emph{Journal
  of the Electrochemical Society}, 2013, \textbf{160}, A1629\relax
\mciteBstWouldAddEndPuncttrue
\mciteSetBstMidEndSepPunct{\mcitedefaultmidpunct}
{\mcitedefaultendpunct}{\mcitedefaultseppunct}\relax
\EndOfBibitem
\bibitem[Wilken \emph{et~al.}(2015)Wilken, Xiong, Scheers, Jacobsson, and
  Johansson]{wilken2015ionic}
S.~Wilken, S.~Xiong, J.~Scheers, P.~Jacobsson and P.~Johansson, \emph{Journal
  of Power Sources}, 2015, \textbf{275}, 935--942\relax
\mciteBstWouldAddEndPuncttrue
\mciteSetBstMidEndSepPunct{\mcitedefaultmidpunct}
{\mcitedefaultendpunct}{\mcitedefaultseppunct}\relax
\EndOfBibitem
\bibitem[Lewandowski and {\'S}widerska-Mocek(2009)]{lewandowski2009ionic}
A.~Lewandowski and A.~{\'S}widerska-Mocek, \emph{Journal of Power Sources},
  2009, \textbf{194}, 601--609\relax
\mciteBstWouldAddEndPuncttrue
\mciteSetBstMidEndSepPunct{\mcitedefaultmidpunct}
{\mcitedefaultendpunct}{\mcitedefaultseppunct}\relax
\EndOfBibitem
\bibitem[Rogers and Seddon(2003)]{rogers2003ionic}
R.~D. Rogers and K.~R. Seddon, \emph{Science}, 2003, \textbf{302},
  792--793\relax
\mciteBstWouldAddEndPuncttrue
\mciteSetBstMidEndSepPunct{\mcitedefaultmidpunct}
{\mcitedefaultendpunct}{\mcitedefaultseppunct}\relax
\EndOfBibitem
\bibitem[Zhou \emph{et~al.}(2011)Zhou, Boyle, Malpezzi, Mele, Shin, Passerini,
  and Henderson]{zhou2011phase}
Q.~Zhou, P.~D. Boyle, L.~Malpezzi, A.~Mele, J.-H. Shin, S.~Passerini and W.~A.
  Henderson, \emph{Chemistry of Materials}, 2011, \textbf{23}, 4331--4337\relax
\mciteBstWouldAddEndPuncttrue
\mciteSetBstMidEndSepPunct{\mcitedefaultmidpunct}
{\mcitedefaultendpunct}{\mcitedefaultseppunct}\relax
\EndOfBibitem
\bibitem[Brinkk\"otter \emph{et~al.}(2017)Brinkk\"otter, Lozinskaya, Ponkratov,
  Vlasov, Rosenwinkel, Malyshkina, Vygodskii, Shaplov, and
  Sch\"onhoff]{brinkkotter2017influence}
M.~Brinkk\"otter, E.~I. Lozinskaya, D.~O. Ponkratov, P.~S. Vlasov, M.~P.
  Rosenwinkel, I.~A. Malyshkina, Y.~Vygodskii, A.~S. Shaplov and
  M.~Sch\"onhoff, \emph{Electrochimica Acta}, 2017, \textbf{237},
  237--247\relax
\mciteBstWouldAddEndPuncttrue
\mciteSetBstMidEndSepPunct{\mcitedefaultmidpunct}
{\mcitedefaultendpunct}{\mcitedefaultseppunct}\relax
\EndOfBibitem
\bibitem[Giffin \emph{et~al.}(2017)Giffin, Moretti, Jeong, and
  Passerini]{giffin2017decoupling}
G.~A. Giffin, A.~Moretti, S.~Jeong and S.~Passerini, \emph{Journal of Power
  Sources}, 2017, \textbf{342}, 335--341\relax
\mciteBstWouldAddEndPuncttrue
\mciteSetBstMidEndSepPunct{\mcitedefaultmidpunct}
{\mcitedefaultendpunct}{\mcitedefaultseppunct}\relax
\EndOfBibitem
\bibitem[Reber \emph{et~al.}(2020)Reber, Takenaka, K\"uhnel, Yamada, and
  Battaglia]{reber2020impact}
D.~Reber, N.~Takenaka, R.-S. K\"uhnel, A.~Yamada and C.~Battaglia, \emph{The
  Journal of Physical Chemistry Letters}, 2020, \textbf{11}, 4720--4725\relax
\mciteBstWouldAddEndPuncttrue
\mciteSetBstMidEndSepPunct{\mcitedefaultmidpunct}
{\mcitedefaultendpunct}{\mcitedefaultseppunct}\relax
\EndOfBibitem
\bibitem[Shaplov \emph{et~al.}(2015)Shaplov, Lozinskaya, Vlasov, Morozova,
  Antonov, Aubert, Armand, and Vygodskii]{shaplov2015new}
A.~S. Shaplov, E.~I. Lozinskaya, P.~S. Vlasov, S.~M. Morozova, D.~Y. Antonov,
  P.-H. Aubert, M.~Armand and Y.~S. Vygodskii, \emph{Electrochimica Acta},
  2015, \textbf{175}, 254--260\relax
\mciteBstWouldAddEndPuncttrue
\mciteSetBstMidEndSepPunct{\mcitedefaultmidpunct}
{\mcitedefaultendpunct}{\mcitedefaultseppunct}\relax
\EndOfBibitem
\bibitem[Hoffknecht \emph{et~al.}(2017)Hoffknecht, Drews, He, and
  Paillard]{hoffknecht2017tfsam}
J.-P. Hoffknecht, M.~Drews, X.~He and E.~Paillard, \emph{Electrochimica Acta},
  2017, \textbf{250}, 25--34\relax
\mciteBstWouldAddEndPuncttrue
\mciteSetBstMidEndSepPunct{\mcitedefaultmidpunct}
{\mcitedefaultendpunct}{\mcitedefaultseppunct}\relax
\EndOfBibitem
\bibitem[Marczewski \emph{et~al.}(2014)Marczewski, Stanje, Hanzu, Wilkening,
  and Johansson]{marczewski2014ionic}
M.~J. Marczewski, B.~Stanje, I.~Hanzu, M.~Wilkening and P.~Johansson,
  \emph{Physical Chemistry Chemical Physics}, 2014, \textbf{16},
  12341--12349\relax
\mciteBstWouldAddEndPuncttrue
\mciteSetBstMidEndSepPunct{\mcitedefaultmidpunct}
{\mcitedefaultendpunct}{\mcitedefaultseppunct}\relax
\EndOfBibitem
\bibitem[Girard \emph{et~al.}(2015)Girard, Hilder, Zhu, Nucciarone, Whitbread,
  Zavorine, Moser, Forsyth, Macfarlane, and Howlett]{girard2015electrochemical}
G.~M. Girard, M.~Hilder, H.~Zhu, D.~Nucciarone, K.~Whitbread, S.~Zavorine,
  M.~Moser, M.~Forsyth, D.~R. Macfarlane and P.~C. Howlett, \emph{Physical
  Chemistry Chemical Physics}, 2015, \textbf{17}, 8706--8713\relax
\mciteBstWouldAddEndPuncttrue
\mciteSetBstMidEndSepPunct{\mcitedefaultmidpunct}
{\mcitedefaultendpunct}{\mcitedefaultseppunct}\relax
\EndOfBibitem
\bibitem[N\"urnberg \emph{et~al.}(2020)N\"urnberg, Lozinskaya, Shaplov, and
  Sch\"onhoff]{nuernberg2020}
P.~N\"urnberg, E.~I. Lozinskaya, A.~S. Shaplov and M.~Sch\"onhoff, \emph{The
  Journal of Physical Chemistry B}, 2020, \textbf{124}, 861--870\relax
\mciteBstWouldAddEndPuncttrue
\mciteSetBstMidEndSepPunct{\mcitedefaultmidpunct}
{\mcitedefaultendpunct}{\mcitedefaultseppunct}\relax
\EndOfBibitem
\bibitem[{Van Der Spoel} \emph{et~al.}(2005){Van Der Spoel}, Lindahl, Hess,
  Groenhof, Mark, and Berendsen]{VanDerSpoel2005}
D.~{Van Der Spoel}, E.~Lindahl, B.~Hess, G.~Groenhof, A.~E. Mark and H.~J.
  Berendsen, \emph{Journal of Computational Chemistry}, 2005, \textbf{26},
  1701--1718\relax
\mciteBstWouldAddEndPuncttrue
\mciteSetBstMidEndSepPunct{\mcitedefaultmidpunct}
{\mcitedefaultendpunct}{\mcitedefaultseppunct}\relax
\EndOfBibitem
\bibitem[P{\'a}ll \emph{et~al.}(2015)P{\'a}ll, Abraham, Kutzner, Hess, and
  Lindahl]{Pall2015}
S.~P{\'a}ll, M.~J. Abraham, C.~Kutzner, B.~Hess and E.~Lindahl, Solving
  Software Challenges for Exascale, Cham, 2015, pp. 3--27\relax
\mciteBstWouldAddEndPuncttrue
\mciteSetBstMidEndSepPunct{\mcitedefaultmidpunct}
{\mcitedefaultendpunct}{\mcitedefaultseppunct}\relax
\EndOfBibitem
\bibitem[Abraham \emph{et~al.}(2015)Abraham, Murtola, Schulz, P{\'{a}}ll,
  Smith, Hess, and Lindah]{Abraham2015}
M.~J. Abraham, T.~Murtola, R.~Schulz, S.~P{\'{a}}ll, J.~C. Smith, B.~Hess and
  E.~Lindah, \emph{SoftwareX}, 2015, \textbf{1-2}, 19--25\relax
\mciteBstWouldAddEndPuncttrue
\mciteSetBstMidEndSepPunct{\mcitedefaultmidpunct}
{\mcitedefaultendpunct}{\mcitedefaultseppunct}\relax
\EndOfBibitem
\bibitem[Berendsen \emph{et~al.}(1995)Berendsen, van~der Spoel, and van
  Drunen]{Berendsen1995}
H.~J. Berendsen, D.~van~der Spoel and R.~van Drunen, \emph{Computer Physics
  Communications}, 1995, \textbf{91}, 43--56\relax
\mciteBstWouldAddEndPuncttrue
\mciteSetBstMidEndSepPunct{\mcitedefaultmidpunct}
{\mcitedefaultendpunct}{\mcitedefaultseppunct}\relax
\EndOfBibitem
\bibitem[Gouveia \emph{et~al.}(2017)Gouveia, Bernardes, Tom{\'e}, Lozinskaya,
  Vygodskii, Shaplov, Lopes, and Marrucho]{gouveia2017ionic}
A.~S. Gouveia, C.~E. Bernardes, L.~C. Tom{\'e}, E.~I. Lozinskaya, Y.~S.
  Vygodskii, A.~S. Shaplov, J.~N.~C. Lopes and I.~M. Marrucho, \emph{Physical
  Chemistry Chemical Physics}, 2017, \textbf{19}, 29617--29624\relax
\mciteBstWouldAddEndPuncttrue
\mciteSetBstMidEndSepPunct{\mcitedefaultmidpunct}
{\mcitedefaultendpunct}{\mcitedefaultseppunct}\relax
\EndOfBibitem
\bibitem[{Canongia Lopes} and P\'adua(2012)]{CanongiaLopes2012}
J.~N. {Canongia Lopes} and A.~A. P\'adua, \emph{Theoretical Chemistry
  Accounts}, 2012, \textbf{131}, 1--11\relax
\mciteBstWouldAddEndPuncttrue
\mciteSetBstMidEndSepPunct{\mcitedefaultmidpunct}
{\mcitedefaultendpunct}{\mcitedefaultseppunct}\relax
\EndOfBibitem
\bibitem[{Canongia Lopes} \emph{et~al.}(2004){Canongia Lopes}, Deschamps, and
  P\'adua]{JoseN.CanongiaLopes2004}
J.~N. {Canongia Lopes}, J.~Deschamps and A.~A.~H. P\'adua, \emph{The Journal of
  Physical Chemistry B}, 2004, \textbf{108}, 2038--2047\relax
\mciteBstWouldAddEndPuncttrue
\mciteSetBstMidEndSepPunct{\mcitedefaultmidpunct}
{\mcitedefaultendpunct}{\mcitedefaultseppunct}\relax
\EndOfBibitem
\bibitem[Lopes and P\'adua(2004)]{Lopes2004}
J.~N. Lopes and A.~A. P\'adua, \emph{Journal of Physical Chemistry B}, 2004,
  \textbf{108}, 16893--16898\relax
\mciteBstWouldAddEndPuncttrue
\mciteSetBstMidEndSepPunct{\mcitedefaultmidpunct}
{\mcitedefaultendpunct}{\mcitedefaultseppunct}\relax
\EndOfBibitem
\bibitem[Shimizu \emph{et~al.}(2010)Shimizu, Almantariotis, {Costa Gomes},
  P{\'{a}}dua, and {Canongia Lopes}]{Shimizu2010}
K.~Shimizu, D.~Almantariotis, M.~F. {Costa Gomes}, A.~A. P{\'{a}}dua and J.~N.
  {Canongia Lopes}, \emph{Journal of Physical Chemistry B}, 2010, \textbf{114},
  3592--3600\relax
\mciteBstWouldAddEndPuncttrue
\mciteSetBstMidEndSepPunct{\mcitedefaultmidpunct}
{\mcitedefaultendpunct}{\mcitedefaultseppunct}\relax
\EndOfBibitem
\bibitem[Self \emph{et~al.}(2019)Self, Fong, and Persson]{self2019transport}
J.~Self, K.~D. Fong and K.~A. Persson, \emph{ACS Energy Letters}, 2019,
  \textbf{4}, 2843--2849\relax
\mciteBstWouldAddEndPuncttrue
\mciteSetBstMidEndSepPunct{\mcitedefaultmidpunct}
{\mcitedefaultendpunct}{\mcitedefaultseppunct}\relax
\EndOfBibitem
\bibitem[Molinari \emph{et~al.}(2019)Molinari, Mailoa, and
  Kozinsky]{molinari2019general}
N.~Molinari, J.~P. Mailoa and B.~Kozinsky, \emph{The journal of physical
  chemistry letters}, 2019, \textbf{10}, 2313--2319\relax
\mciteBstWouldAddEndPuncttrue
\mciteSetBstMidEndSepPunct{\mcitedefaultmidpunct}
{\mcitedefaultendpunct}{\mcitedefaultseppunct}\relax
\EndOfBibitem
\bibitem[Molinari and Kozinsky(2020)]{molinari2020chelation}
N.~Molinari and B.~Kozinsky, \emph{The Journal of Physical Chemistry B}, 2020,
  \textbf{124}, 2676--2684\relax
\mciteBstWouldAddEndPuncttrue
\mciteSetBstMidEndSepPunct{\mcitedefaultmidpunct}
{\mcitedefaultendpunct}{\mcitedefaultseppunct}\relax
\EndOfBibitem
\bibitem[Thum \emph{et~al.}(2020)Thum, Heuer, Shimizu, and
  Lopes]{thum2020solvate}
A.~Thum, A.~Heuer, K.~Shimizu and J.~N.~C. Lopes, \emph{Physical Chemistry
  Chemical Physics}, 2020, \textbf{22}, 525--535\relax
\mciteBstWouldAddEndPuncttrue
\mciteSetBstMidEndSepPunct{\mcitedefaultmidpunct}
{\mcitedefaultendpunct}{\mcitedefaultseppunct}\relax
\EndOfBibitem
\bibitem[Huang \emph{et~al.}(2018)Huang, Louren\c{c}o, Costa, Zhang, Maginn,
  and Gurkan]{huang2018solvation}
Q.~Huang, T.~C. Louren\c{c}o, L.~T. Costa, Y.~Zhang, E.~J. Maginn and
  B.~Gurkan, \emph{The Journal of Physical Chemistry B}, 2018, \textbf{123},
  516--527\relax
\mciteBstWouldAddEndPuncttrue
\mciteSetBstMidEndSepPunct{\mcitedefaultmidpunct}
{\mcitedefaultendpunct}{\mcitedefaultseppunct}\relax
\EndOfBibitem
\bibitem[Nos{\'{e}} and Klein(1983)]{Nose1983}
S.~Nos{\'{e}} and M.~Klein, \emph{Molecular Physics}, 1983, \textbf{50},
  1055--1076\relax
\mciteBstWouldAddEndPuncttrue
\mciteSetBstMidEndSepPunct{\mcitedefaultmidpunct}
{\mcitedefaultendpunct}{\mcitedefaultseppunct}\relax
\EndOfBibitem
\bibitem[Nos{\'{e}}(1984)]{Nose1984}
S.~Nos{\'{e}}, \emph{Molecular Physics}, 1984, \textbf{52}, 255--268\relax
\mciteBstWouldAddEndPuncttrue
\mciteSetBstMidEndSepPunct{\mcitedefaultmidpunct}
{\mcitedefaultendpunct}{\mcitedefaultseppunct}\relax
\EndOfBibitem
\bibitem[Hoover(1985)]{Hoover1985}
W.~G. Hoover, \emph{Physical Review A}, 1985, \textbf{31}, 1695--1697\relax
\mciteBstWouldAddEndPuncttrue
\mciteSetBstMidEndSepPunct{\mcitedefaultmidpunct}
{\mcitedefaultendpunct}{\mcitedefaultseppunct}\relax
\EndOfBibitem
\bibitem[Parrinello and Rahman(1981)]{Parrinello1981}
M.~Parrinello and A.~Rahman, \emph{Journal of Applied Physics}, 1981,
  \textbf{52}, 7182--7190\relax
\mciteBstWouldAddEndPuncttrue
\mciteSetBstMidEndSepPunct{\mcitedefaultmidpunct}
{\mcitedefaultendpunct}{\mcitedefaultseppunct}\relax
\EndOfBibitem
\bibitem[Borodin \emph{et~al.}(2018)Borodin, Giffin, Moretti, Haskins, Lawson,
  Henderson, and Passerini]{borodin2018insights}
O.~Borodin, G.~A. Giffin, A.~Moretti, J.~B. Haskins, J.~W. Lawson, W.~A.
  Henderson and S.~Passerini, \emph{The Journal of Physical Chemistry C}, 2018,
  \textbf{122}, 20108--20121\relax
\mciteBstWouldAddEndPuncttrue
\mciteSetBstMidEndSepPunct{\mcitedefaultmidpunct}
{\mcitedefaultendpunct}{\mcitedefaultseppunct}\relax
\EndOfBibitem
\bibitem[Li \emph{et~al.}(2015)Li, Borodin, Smith, and Bedrov]{li2015effect}
Z.~Li, O.~Borodin, G.~D. Smith and D.~Bedrov, \emph{The Journal of Physical
  Chemistry B}, 2015, \textbf{119}, 3085--3096\relax
\mciteBstWouldAddEndPuncttrue
\mciteSetBstMidEndSepPunct{\mcitedefaultmidpunct}
{\mcitedefaultendpunct}{\mcitedefaultseppunct}\relax
\EndOfBibitem
\bibitem[Kubisiak \emph{et~al.}(2019)Kubisiak, Wr{\'o}bel, and
  Eilmes]{kubisiak2019molecular}
P.~Kubisiak, P.~Wr{\'o}bel and A.~Eilmes, \emph{The Journal of Physical
  Chemistry B}, 2019, \textbf{124}, 413--421\relax
\mciteBstWouldAddEndPuncttrue
\mciteSetBstMidEndSepPunct{\mcitedefaultmidpunct}
{\mcitedefaultendpunct}{\mcitedefaultseppunct}\relax
\EndOfBibitem
\bibitem[Li \emph{et~al.}(2012)Li, Smith, and Bedrov]{li2012li+}
Z.~Li, G.~D. Smith and D.~Bedrov, \emph{The Journal of Physical Chemistry B},
  2012, \textbf{116}, 12801--12809\relax
\mciteBstWouldAddEndPuncttrue
\mciteSetBstMidEndSepPunct{\mcitedefaultmidpunct}
{\mcitedefaultendpunct}{\mcitedefaultseppunct}\relax
\EndOfBibitem
\bibitem[Monteiro \emph{et~al.}(2008)Monteiro, Bazito, Siqueira, Ribeiro, and
  Torresi]{monteiro2008transport}
M.~J. Monteiro, F.~F. Bazito, L.~J. Siqueira, M.~C. Ribeiro and R.~M. Torresi,
  \emph{The Journal of Physical Chemistry B}, 2008, \textbf{112},
  2102--2109\relax
\mciteBstWouldAddEndPuncttrue
\mciteSetBstMidEndSepPunct{\mcitedefaultmidpunct}
{\mcitedefaultendpunct}{\mcitedefaultseppunct}\relax
\EndOfBibitem
\bibitem[Haskins \emph{et~al.}(2014)Haskins, Bennett, Wu, Hernandez, Borodin,
  Monk, Bauschlicher~Jr, and Lawson]{haskins2014computational}
J.~B. Haskins, W.~R. Bennett, J.~J. Wu, D.~M. Hernandez, O.~Borodin, J.~D.
  Monk, C.~W. Bauschlicher~Jr and J.~W. Lawson, \emph{The Journal of Physical
  Chemistry B}, 2014, \textbf{118}, 11295--11309\relax
\mciteBstWouldAddEndPuncttrue
\mciteSetBstMidEndSepPunct{\mcitedefaultmidpunct}
{\mcitedefaultendpunct}{\mcitedefaultseppunct}\relax
\EndOfBibitem
\bibitem[Chen \emph{et~al.}(2018)Chen, Howlett, and Forsyth]{chen2018ion}
F.~Chen, P.~Howlett and M.~Forsyth, \emph{The Journal of Physical Chemistry C},
  2018, \textbf{122}, 105--114\relax
\mciteBstWouldAddEndPuncttrue
\mciteSetBstMidEndSepPunct{\mcitedefaultmidpunct}
{\mcitedefaultendpunct}{\mcitedefaultseppunct}\relax
\EndOfBibitem
\bibitem[Lesch \emph{et~al.}(2014)Lesch, Jeremias, Moretti, Passerini, Heuer,
  and Borodin]{lesch2014combined}
V.~Lesch, S.~Jeremias, A.~Moretti, S.~Passerini, A.~Heuer and O.~Borodin,
  \emph{The Journal of Physical Chemistry B}, 2014, \textbf{118},
  7367--7375\relax
\mciteBstWouldAddEndPuncttrue
\mciteSetBstMidEndSepPunct{\mcitedefaultmidpunct}
{\mcitedefaultendpunct}{\mcitedefaultseppunct}\relax
\EndOfBibitem
\bibitem[Zhang and Maginn(2015)]{zhang2015direct}
Y.~Zhang and E.~J. Maginn, \emph{The Journal of Physical Chemistry Letters},
  2015, \textbf{6}, 700--705\relax
\mciteBstWouldAddEndPuncttrue
\mciteSetBstMidEndSepPunct{\mcitedefaultmidpunct}
{\mcitedefaultendpunct}{\mcitedefaultseppunct}\relax
\EndOfBibitem
\bibitem[Fong \emph{et~al.}(2019)Fong, Self, Diederichsen, Wood, McCloskey, and
  Persson]{fong2019ion}
K.~D. Fong, J.~Self, K.~M. Diederichsen, B.~M. Wood, B.~D. McCloskey and K.~A.
  Persson, \emph{ACS central science}, 2019, \textbf{5}, 1250--1260\relax
\mciteBstWouldAddEndPuncttrue
\mciteSetBstMidEndSepPunct{\mcitedefaultmidpunct}
{\mcitedefaultendpunct}{\mcitedefaultseppunct}\relax
\EndOfBibitem
\bibitem[Dong and Bedrov(2018)]{dong2018charge}
D.~Dong and D.~Bedrov, \emph{The Journal of Physical Chemistry B}, 2018,
  \textbf{122}, 9994--10004\relax
\mciteBstWouldAddEndPuncttrue
\mciteSetBstMidEndSepPunct{\mcitedefaultmidpunct}
{\mcitedefaultendpunct}{\mcitedefaultseppunct}\relax
\EndOfBibitem
\bibitem[Kreuer(1996)]{kreuer1996proton}
K.-D. Kreuer, \emph{Chemistry of materials}, 1996, \textbf{8}, 610--641\relax
\mciteBstWouldAddEndPuncttrue
\mciteSetBstMidEndSepPunct{\mcitedefaultmidpunct}
{\mcitedefaultendpunct}{\mcitedefaultseppunct}\relax
\EndOfBibitem
\bibitem[Norbya(1990)]{norbya1990proton}
T.~Norbya, \emph{Solid State Ionics}, 1990, \textbf{40}, 857--862\relax
\mciteBstWouldAddEndPuncttrue
\mciteSetBstMidEndSepPunct{\mcitedefaultmidpunct}
{\mcitedefaultendpunct}{\mcitedefaultseppunct}\relax
\EndOfBibitem
\bibitem[Li \emph{et~al.}(2019)Li, Yin, Zheng, Sui, Zhou, Chen, and
  Zhu]{li2019insights}
Q.~Li, Q.~Yin, Y.-S. Zheng, Z.-J. Sui, X.-G. Zhou, D.~Chen and Y.-A. Zhu,
  \emph{Langmuir}, 2019, \textbf{35}, 9962--9969\relax
\mciteBstWouldAddEndPuncttrue
\mciteSetBstMidEndSepPunct{\mcitedefaultmidpunct}
{\mcitedefaultendpunct}{\mcitedefaultseppunct}\relax
\EndOfBibitem
\bibitem[Kreuer(1982)]{kreuer1982w}
K.~Kreuer, \emph{Angew. Chem. Int. Ed. Eng}, 1982, \textbf{21}, 208--211\relax
\mciteBstWouldAddEndPuncttrue
\mciteSetBstMidEndSepPunct{\mcitedefaultmidpunct}
{\mcitedefaultendpunct}{\mcitedefaultseppunct}\relax
\EndOfBibitem
\bibitem[Agmon(1995)]{agmon1995grotthuss}
N.~Agmon, \emph{Chemical Physics Letters}, 1995, \textbf{244}, 456--462\relax
\mciteBstWouldAddEndPuncttrue
\mciteSetBstMidEndSepPunct{\mcitedefaultmidpunct}
{\mcitedefaultendpunct}{\mcitedefaultseppunct}\relax
\EndOfBibitem
\bibitem[Borodin \emph{et~al.}(2006)Borodin, Smith, and
  Henderson]{borodin2006li+}
O.~Borodin, G.~D. Smith and W.~Henderson, \emph{The Journal of Physical
  Chemistry B}, 2006, \textbf{110}, 16879--16886\relax
\mciteBstWouldAddEndPuncttrue
\mciteSetBstMidEndSepPunct{\mcitedefaultmidpunct}
{\mcitedefaultendpunct}{\mcitedefaultseppunct}\relax
\EndOfBibitem
\bibitem[Forsyth \emph{et~al.}(2016)Forsyth, Yoon, Chen, Zhu, MacFarlane,
  Armand, and Howlett]{forsyth2016novel}
M.~Forsyth, H.~Yoon, F.~Chen, H.~Zhu, D.~R. MacFarlane, M.~Armand and P.~C.
  Howlett, \emph{The Journal of Physical Chemistry C}, 2016, \textbf{120},
  4276--4286\relax
\mciteBstWouldAddEndPuncttrue
\mciteSetBstMidEndSepPunct{\mcitedefaultmidpunct}
{\mcitedefaultendpunct}{\mcitedefaultseppunct}\relax
\EndOfBibitem
\bibitem[Liu and Maginn(2011)]{liu2011molecular}
H.~Liu and E.~Maginn, \emph{The Journal of Chemical Physics}, 2011,
  \textbf{135}, 124507\relax
\mciteBstWouldAddEndPuncttrue
\mciteSetBstMidEndSepPunct{\mcitedefaultmidpunct}
{\mcitedefaultendpunct}{\mcitedefaultseppunct}\relax
\EndOfBibitem
\bibitem[Liu \emph{et~al.}(2021)Liu, Luo, Sokolov, and
  Paddison]{liu2021quantitative}
H.~Liu, X.~Luo, A.~P. Sokolov and S.~J. Paddison, \emph{The Journal of Physical
  Chemistry B}, 2021, \textbf{125}, 372--381\relax
\mciteBstWouldAddEndPuncttrue
\mciteSetBstMidEndSepPunct{\mcitedefaultmidpunct}
{\mcitedefaultendpunct}{\mcitedefaultseppunct}\relax
\EndOfBibitem
\bibitem[Mukherji \emph{et~al.}(2020)Mukherji, Avula, Kumar, and
  Balasubramanian]{mukherji2020hopping}
S.~Mukherji, N.~V. Avula, R.~Kumar and S.~Balasubramanian, \emph{The Journal of
  Physical Chemistry Letters}, 2020, \textbf{11}, 9613--9620\relax
\mciteBstWouldAddEndPuncttrue
\mciteSetBstMidEndSepPunct{\mcitedefaultmidpunct}
{\mcitedefaultendpunct}{\mcitedefaultseppunct}\relax
\EndOfBibitem
\end{mcitethebibliography}


\providecommand{\latin}[1]{#1}
\makeatletter
\providecommand{\doi}
  {\begingroup\let\do\@makeother\dospecials
  \catcode`\{=1 \catcode`\}=2 \doi@aux}
\providecommand{\doi@aux}[1]{\endgroup\texttt{#1}}
\makeatother
\providecommand*\mcitethebibliography{\thebibliography}
\csname @ifundefined\endcsname{endmcitethebibliography}
  {\let\endmcitethebibliography\endthebibliography}{}
\begin{mcitethebibliography}{37}
\providecommand*\natexlab[1]{#1}
\providecommand*\mciteSetBstSublistMode[1]{}
\providecommand*\mciteSetBstMaxWidthForm[2]{}
\providecommand*\mciteBstWouldAddEndPuncttrue
  {\def\EndOfBibitem{\unskip.}}
\providecommand*\mciteBstWouldAddEndPunctfalse
  {\let\EndOfBibitem\relax}
\providecommand*\mciteSetBstMidEndSepPunct[3]{}
\providecommand*\mciteSetBstSublistLabelBeginEnd[3]{}
\providecommand*\EndOfBibitem{}
\mciteSetBstSublistMode{f}
\mciteSetBstMaxWidthForm{subitem}{(\alph{mcitesubitemcount})}
\mciteSetBstSublistLabelBeginEnd
  {\mcitemaxwidthsubitemform\space}
  {\relax}
  {\relax}

\bibitem[{Van Der Spoel} \latin{et~al.}(2005){Van Der Spoel}, Lindahl, Hess,
  Groenhof, Mark, and Berendsen]{VanDerSpoel2005}
{Van Der Spoel},~D.; Lindahl,~E.; Hess,~B.; Groenhof,~G.; Mark,~A.~E.;
  Berendsen,~H.~J. {GROMACS: Fast, flexible, and free}. \emph{Journal of
  Computational Chemistry} \textbf{2005}, \emph{26}, 1701--1718\relax
\mciteBstWouldAddEndPuncttrue
\mciteSetBstMidEndSepPunct{\mcitedefaultmidpunct}
{\mcitedefaultendpunct}{\mcitedefaultseppunct}\relax
\EndOfBibitem
\bibitem[P{\'a}ll \latin{et~al.}(2015)P{\'a}ll, Abraham, Kutzner, Hess, and
  Lindahl]{Pall2015}
P{\'a}ll,~S.; Abraham,~M.~J.; Kutzner,~C.; Hess,~B.; Lindahl,~E. Tackling
  Exascale Software Challenges in Molecular Dynamics Simulations with GROMACS.
  Solving Software Challenges for Exascale. Cham, 2015; pp 3--27\relax
\mciteBstWouldAddEndPuncttrue
\mciteSetBstMidEndSepPunct{\mcitedefaultmidpunct}
{\mcitedefaultendpunct}{\mcitedefaultseppunct}\relax
\EndOfBibitem
\bibitem[Abraham \latin{et~al.}(2015)Abraham, Murtola, Schulz, P{\'{a}}ll,
  Smith, Hess, and Lindah]{Abraham2015}
Abraham,~M.~J.; Murtola,~T.; Schulz,~R.; P{\'{a}}ll,~S.; Smith,~J.~C.;
  Hess,~B.; Lindah,~E. {Gromacs: High performance molecular simulations through
  multi-level parallelism from laptops to supercomputers}. \emph{SoftwareX}
  \textbf{2015}, \emph{1-2}, 19--25\relax
\mciteBstWouldAddEndPuncttrue
\mciteSetBstMidEndSepPunct{\mcitedefaultmidpunct}
{\mcitedefaultendpunct}{\mcitedefaultseppunct}\relax
\EndOfBibitem
\bibitem[Berendsen \latin{et~al.}(1995)Berendsen, van~der Spoel, and van
  Drunen]{Berendsen1995}
Berendsen,~H.~J.; van~der Spoel,~D.; van Drunen,~R. {GROMACS: A message-passing
  parallel molecular dynamics implementation}. \emph{Computer Physics
  Communications} \textbf{1995}, \emph{91}, 43--56\relax
\mciteBstWouldAddEndPuncttrue
\mciteSetBstMidEndSepPunct{\mcitedefaultmidpunct}
{\mcitedefaultendpunct}{\mcitedefaultseppunct}\relax
\EndOfBibitem
\bibitem[Jorgensen \latin{et~al.}(1996)Jorgensen, Maxwell, and
  Tirado-Rives]{WilliamL.Jorgensen1996}
Jorgensen,~W.~L.; Maxwell,~D.~S.; Tirado-Rives,~J. {Development and testing of
  the OPLS all-atom force field on conformational energetics and properties of
  organic liquids}. \emph{Journal of the American Chemical Society}
  \textbf{1996}, \emph{118}, 11225--11236\relax
\mciteBstWouldAddEndPuncttrue
\mciteSetBstMidEndSepPunct{\mcitedefaultmidpunct}
{\mcitedefaultendpunct}{\mcitedefaultseppunct}\relax
\EndOfBibitem
\bibitem[Gouveia \latin{et~al.}(2017)Gouveia, Bernardes, Tom{\'e}, Lozinskaya,
  Vygodskii, Shaplov, Lopes, and Marrucho]{gouveia2017ionic}
Gouveia,~A.~S.; Bernardes,~C.~E.; Tom{\'e},~L.~C.; Lozinskaya,~E.~I.;
  Vygodskii,~Y.~S.; Shaplov,~A.~S.; Lopes,~J. N.~C.; Marrucho,~I.~M. Ionic
  liquids with anions based on fluorosulfonyl derivatives: from asymmetrical
  substitutions to a consistent force field model. \emph{Physical Chemistry
  Chemical Physics} \textbf{2017}, \emph{19}, 29617--29624\relax
\mciteBstWouldAddEndPuncttrue
\mciteSetBstMidEndSepPunct{\mcitedefaultmidpunct}
{\mcitedefaultendpunct}{\mcitedefaultseppunct}\relax
\EndOfBibitem
\bibitem[{Canongia Lopes} and P\'adua(2012){Canongia Lopes}, and
  P\'adua]{CanongiaLopes2012}
{Canongia Lopes},~J.~N.; P\'adua,~A.~A. {CL\&P: A generic and systematic force
  field for ionic liquids modeling}. \emph{Theoretical Chemistry Accounts}
  \textbf{2012}, \emph{131}, 1--11\relax
\mciteBstWouldAddEndPuncttrue
\mciteSetBstMidEndSepPunct{\mcitedefaultmidpunct}
{\mcitedefaultendpunct}{\mcitedefaultseppunct}\relax
\EndOfBibitem
\bibitem[{Canongia Lopes} \latin{et~al.}(2004){Canongia Lopes}, Deschamps, and
  P\'adua]{JoseN.CanongiaLopes2004}
{Canongia Lopes},~J.~N.; Deschamps,~J.; P\'adua,~A. A.~H. {Modeling Ionic
  Liquids Using a Systematic All-Atom Force Field}. \emph{The Journal of
  Physical Chemistry B} \textbf{2004}, \emph{108}, 2038--2047\relax
\mciteBstWouldAddEndPuncttrue
\mciteSetBstMidEndSepPunct{\mcitedefaultmidpunct}
{\mcitedefaultendpunct}{\mcitedefaultseppunct}\relax
\EndOfBibitem
\bibitem[Lopes and P\'adua(2004)Lopes, and P\'adua]{Lopes2004}
Lopes,~J.~N.; P\'adua,~A.~A. {Molecular force field for ionic liquids composed
  of triflate or bistriflylimide anions}. \emph{Journal of Physical Chemistry
  B} \textbf{2004}, \emph{108}, 16893--16898\relax
\mciteBstWouldAddEndPuncttrue
\mciteSetBstMidEndSepPunct{\mcitedefaultmidpunct}
{\mcitedefaultendpunct}{\mcitedefaultseppunct}\relax
\EndOfBibitem
\bibitem[Shimizu \latin{et~al.}(2010)Shimizu, Almantariotis, {Costa Gomes},
  P{\'{a}}dua, and {Canongia Lopes}]{Shimizu2010}
Shimizu,~K.; Almantariotis,~D.; {Costa Gomes},~M.~F.; P{\'{a}}dua,~A.~A.;
  {Canongia Lopes},~J.~N. {Molecular force field for ionic liquids V:
  Hydroxyethylimidazolium, dimethoxy-2methylimidazolium, and
  fluoroalkylimidazolium cations and Bis(fluorosulfonyl)amide,
  perfluoroalkanesulfonylamide, and fluoroalkylfluorophosphate anions}.
  \emph{Journal of Physical Chemistry B} \textbf{2010}, \emph{114},
  3592--3600\relax
\mciteBstWouldAddEndPuncttrue
\mciteSetBstMidEndSepPunct{\mcitedefaultmidpunct}
{\mcitedefaultendpunct}{\mcitedefaultseppunct}\relax
\EndOfBibitem
\bibitem[Doherty \latin{et~al.}(2017)Doherty, Zhong, Gathiaka, Li, and
  Acevedo]{doherty2017revisiting}
Doherty,~B.; Zhong,~X.; Gathiaka,~S.; Li,~B.; Acevedo,~O. Revisiting OPLS force
  field parameters for ionic liquid simulations. \emph{Journal of chemical
  theory and computation} \textbf{2017}, \emph{13}, 6131--6145\relax
\mciteBstWouldAddEndPuncttrue
\mciteSetBstMidEndSepPunct{\mcitedefaultmidpunct}
{\mcitedefaultendpunct}{\mcitedefaultseppunct}\relax
\EndOfBibitem
\bibitem[Bhargava and Balasubramanian(2007)Bhargava, and
  Balasubramanian]{Bhargava2007}
Bhargava,~B.~L.; Balasubramanian,~S. {Refined potential model for atomistic
  simulations of ionic liquid [bmim][PF6]}. \emph{The Journal of Chemical
  Physics} \textbf{2007}, \emph{127}, 114510\relax
\mciteBstWouldAddEndPuncttrue
\mciteSetBstMidEndSepPunct{\mcitedefaultmidpunct}
{\mcitedefaultendpunct}{\mcitedefaultseppunct}\relax
\EndOfBibitem
\bibitem[Chaban \latin{et~al.}(2011)Chaban, Voroshylova, and
  Kalugin]{Chaban2011}
Chaban,~V.~V.; Voroshylova,~I.~V.; Kalugin,~O.~N. {A new force field model for
  the simulation of transport properties of imidazolium-based ionic liquids}.
  \emph{Physical Chemistry Chemical Physics} \textbf{2011}, \emph{13},
  7910\relax
\mciteBstWouldAddEndPuncttrue
\mciteSetBstMidEndSepPunct{\mcitedefaultmidpunct}
{\mcitedefaultendpunct}{\mcitedefaultseppunct}\relax
\EndOfBibitem
\bibitem[Self \latin{et~al.}(2019)Self, Fong, and Persson]{self2019transport}
Self,~J.; Fong,~K.~D.; Persson,~K.~A. Transport in superconcentrated LiPF6 and
  LiBF4/propylene carbonate electrolytes. \emph{ACS Energy Letters}
  \textbf{2019}, \emph{4}, 2843--2849\relax
\mciteBstWouldAddEndPuncttrue
\mciteSetBstMidEndSepPunct{\mcitedefaultmidpunct}
{\mcitedefaultendpunct}{\mcitedefaultseppunct}\relax
\EndOfBibitem
\bibitem[Fong \latin{et~al.}(2019)Fong, Self, Diederichsen, Wood, McCloskey,
  and Persson]{fong2019ion}
Fong,~K.~D.; Self,~J.; Diederichsen,~K.~M.; Wood,~B.~M.; McCloskey,~B.~D.;
  Persson,~K.~A. Ion transport and the true transference number in nonaqueous
  polyelectrolyte solutions for lithium ion batteries. \emph{ACS central
  science} \textbf{2019}, \emph{5}, 1250--1260\relax
\mciteBstWouldAddEndPuncttrue
\mciteSetBstMidEndSepPunct{\mcitedefaultmidpunct}
{\mcitedefaultendpunct}{\mcitedefaultseppunct}\relax
\EndOfBibitem
\bibitem[Molinari \latin{et~al.}(2019)Molinari, Mailoa, Craig, Christensen, and
  Kozinsky]{molinari2019transport}
Molinari,~N.; Mailoa,~J.~P.; Craig,~N.; Christensen,~J.; Kozinsky,~B. Transport
  anomalies emerging from strong correlation in ionic liquid electrolytes.
  \emph{Journal of Power Sources} \textbf{2019}, \emph{428}, 27--36\relax
\mciteBstWouldAddEndPuncttrue
\mciteSetBstMidEndSepPunct{\mcitedefaultmidpunct}
{\mcitedefaultendpunct}{\mcitedefaultseppunct}\relax
\EndOfBibitem
\bibitem[Molinari \latin{et~al.}(2019)Molinari, Mailoa, and
  Kozinsky]{molinari2019general}
Molinari,~N.; Mailoa,~J.~P.; Kozinsky,~B. General trend of a negative Li
  effective charge in ionic liquid electrolytes. \emph{The journal of physical
  chemistry letters} \textbf{2019}, \emph{10}, 2313--2319\relax
\mciteBstWouldAddEndPuncttrue
\mciteSetBstMidEndSepPunct{\mcitedefaultmidpunct}
{\mcitedefaultendpunct}{\mcitedefaultseppunct}\relax
\EndOfBibitem
\bibitem[Molinari and Kozinsky(2020)Molinari, and
  Kozinsky]{molinari2020chelation}
Molinari,~N.; Kozinsky,~B. Chelation-Induced Reversal of Negative Cation
  Transference Number in Ionic Liquid Electrolytes. \emph{The Journal of
  Physical Chemistry B} \textbf{2020}, \emph{124}, 2676--2684\relax
\mciteBstWouldAddEndPuncttrue
\mciteSetBstMidEndSepPunct{\mcitedefaultmidpunct}
{\mcitedefaultendpunct}{\mcitedefaultseppunct}\relax
\EndOfBibitem
\bibitem[Huang \latin{et~al.}(2018)Huang, Louren\c{c}o, Costa, Zhang, Maginn,
  and Gurkan]{huang2018solvation}
Huang,~Q.; Louren\c{c}o,~T.~C.; Costa,~L.~T.; Zhang,~Y.; Maginn,~E.~J.;
  Gurkan,~B. Solvation Structure and Dynamics of Li+ in Ternary Ionic
  Liquid--Lithium Salt Electrolytes. \emph{The Journal of Physical Chemistry B}
  \textbf{2018}, \emph{123}, 516--527\relax
\mciteBstWouldAddEndPuncttrue
\mciteSetBstMidEndSepPunct{\mcitedefaultmidpunct}
{\mcitedefaultendpunct}{\mcitedefaultseppunct}\relax
\EndOfBibitem
\bibitem[Thum \latin{et~al.}(2020)Thum, Heuer, Shimizu, and
  Lopes]{thum2020solvate}
Thum,~A.; Heuer,~A.; Shimizu,~K.; Lopes,~J. N.~C. Solvate ionic liquids based
  on lithium bis (trifluoromethanesulfonyl) imide--glyme systems: coordination
  in MD simulations with scaled charges. \emph{Physical Chemistry Chemical
  Physics} \textbf{2020}, \emph{22}, 525--535\relax
\mciteBstWouldAddEndPuncttrue
\mciteSetBstMidEndSepPunct{\mcitedefaultmidpunct}
{\mcitedefaultendpunct}{\mcitedefaultseppunct}\relax
\EndOfBibitem
\bibitem[Lemkul \latin{et~al.}(2016)Lemkul, Huang, Roux, and
  MacKerell~Jr]{lemkul2016empirical}
Lemkul,~J.~A.; Huang,~J.; Roux,~B.; MacKerell~Jr,~A.~D. An empirical
  polarizable force field based on the classical drude oscillator model:
  development history and recent applications. \emph{Chemical reviews}
  \textbf{2016}, \emph{116}, 4983--5013\relax
\mciteBstWouldAddEndPuncttrue
\mciteSetBstMidEndSepPunct{\mcitedefaultmidpunct}
{\mcitedefaultendpunct}{\mcitedefaultseppunct}\relax
\EndOfBibitem
\bibitem[Salomon-Ferrer \latin{et~al.}(2013)Salomon-Ferrer, Case, and
  Walker]{salomon2013overview}
Salomon-Ferrer,~R.; Case,~D.~A.; Walker,~R.~C. An overview of the Amber
  biomolecular simulation package. \emph{Wiley Interdisciplinary Reviews:
  Computational Molecular Science} \textbf{2013}, \emph{3}, 198--210\relax
\mciteBstWouldAddEndPuncttrue
\mciteSetBstMidEndSepPunct{\mcitedefaultmidpunct}
{\mcitedefaultendpunct}{\mcitedefaultseppunct}\relax
\EndOfBibitem
\bibitem[Li \latin{et~al.}(2019)Li, Bouchal, Mendez-Morales, Rollet, Rizzi,
  Le~Vot, Favier, Rotenberg, Borodin, and Fontaine]{li2019transport}
Li,~Z.; Bouchal,~R.; Mendez-Morales,~T.; Rollet,~A.-L.; Rizzi,~C.; Le~Vot,~S.;
  Favier,~F.; Rotenberg,~B.; Borodin,~O.; Fontaine,~O. Transport Properties of
  Li-TFSI Water-in-Salt Electrolytes. \emph{The Journal of Physical Chemistry
  B} \textbf{2019}, \emph{123}, 10514--10521\relax
\mciteBstWouldAddEndPuncttrue
\mciteSetBstMidEndSepPunct{\mcitedefaultmidpunct}
{\mcitedefaultendpunct}{\mcitedefaultseppunct}\relax
\EndOfBibitem
\bibitem[N\"urnberg \latin{et~al.}(2020)N\"urnberg, Lozinskaya, Shaplov, and
  Sch\"onhoff]{nuernberg2020}
N\"urnberg,~P.; Lozinskaya,~E.~I.; Shaplov,~A.~S.; Sch\"onhoff,~M. Li
  Coordination of a Novel Asymmetric Anion in Ionic Liquid-in-Li Salt
  Electrolytes. \emph{The Journal of Physical Chemistry B} \textbf{2020},
  \emph{124}, 861--870\relax
\mciteBstWouldAddEndPuncttrue
\mciteSetBstMidEndSepPunct{\mcitedefaultmidpunct}
{\mcitedefaultendpunct}{\mcitedefaultseppunct}\relax
\EndOfBibitem
\bibitem[Martinez \latin{et~al.}(2009)Martinez, Andrade, Birgin, and
  Mart{\'{i}}nez]{Martinez2009}
Martinez,~L.; Andrade,~R.; Birgin,~E.~G.; Mart{\'{i}}nez,~J.~M. {PACKMOL: A
  package for building initial configurations for molecular dynamics
  simulations}. \emph{Journal of Computational Chemistry} \textbf{2009},
  \emph{30}, 2157--2164\relax
\mciteBstWouldAddEndPuncttrue
\mciteSetBstMidEndSepPunct{\mcitedefaultmidpunct}
{\mcitedefaultendpunct}{\mcitedefaultseppunct}\relax
\EndOfBibitem
\bibitem[Berendsen \latin{et~al.}(1984)Berendsen, Postma, van Gunsteren,
  DiNola, and Haak]{Berendsen1984}
Berendsen,~H. J.~C.; Postma,~J. P.~M.; van Gunsteren,~W.~F.; DiNola,~A.;
  Haak,~J.~R. {Molecular dynamics with coupling to an external bath}. \emph{The
  Journal of Chemical Physics} \textbf{1984}, \emph{81}, 3684--3690\relax
\mciteBstWouldAddEndPuncttrue
\mciteSetBstMidEndSepPunct{\mcitedefaultmidpunct}
{\mcitedefaultendpunct}{\mcitedefaultseppunct}\relax
\EndOfBibitem
\bibitem[Bussi \latin{et~al.}(2007)Bussi, Donadio, and Parrinello]{Bussi2007}
Bussi,~G.; Donadio,~D.; Parrinello,~M. {Canonical sampling through velocity
  rescaling}. \emph{The Journal of Chemical Physics} \textbf{2007}, \emph{126},
  014101\relax
\mciteBstWouldAddEndPuncttrue
\mciteSetBstMidEndSepPunct{\mcitedefaultmidpunct}
{\mcitedefaultendpunct}{\mcitedefaultseppunct}\relax
\EndOfBibitem
\bibitem[Parrinello and Rahman(1981)Parrinello, and Rahman]{Parrinello1981}
Parrinello,~M.; Rahman,~A. {Polymorphic transitions in single crystals: A new
  molecular dynamics method}. \emph{Journal of Applied Physics} \textbf{1981},
  \emph{52}, 7182--7190\relax
\mciteBstWouldAddEndPuncttrue
\mciteSetBstMidEndSepPunct{\mcitedefaultmidpunct}
{\mcitedefaultendpunct}{\mcitedefaultseppunct}\relax
\EndOfBibitem
\bibitem[Nos{\'{e}} and Klein(1983)Nos{\'{e}}, and Klein]{Nose1983}
Nos{\'{e}},~S.; Klein,~M. {Constant pressure molecular dynamics for molecular
  systems}. \emph{Molecular Physics} \textbf{1983}, \emph{50}, 1055--1076\relax
\mciteBstWouldAddEndPuncttrue
\mciteSetBstMidEndSepPunct{\mcitedefaultmidpunct}
{\mcitedefaultendpunct}{\mcitedefaultseppunct}\relax
\EndOfBibitem
\bibitem[Nos{\'{e}}(1984)]{Nose1984}
Nos{\'{e}},~S. {A molecular dynamics method for simulations in the canonical
  ensemble}. \emph{Molecular Physics} \textbf{1984}, \emph{52}, 255--268\relax
\mciteBstWouldAddEndPuncttrue
\mciteSetBstMidEndSepPunct{\mcitedefaultmidpunct}
{\mcitedefaultendpunct}{\mcitedefaultseppunct}\relax
\EndOfBibitem
\bibitem[Hoover(1985)]{Hoover1985}
Hoover,~W.~G. {Canonical dynamics: Equilibrium phase-space distributions}.
  \emph{Physical Review A} \textbf{1985}, \emph{31}, 1695--1697\relax
\mciteBstWouldAddEndPuncttrue
\mciteSetBstMidEndSepPunct{\mcitedefaultmidpunct}
{\mcitedefaultendpunct}{\mcitedefaultseppunct}\relax
\EndOfBibitem
\bibitem[Hess \latin{et~al.}(1997)Hess, Bekker, Berendsen, and
  Fraaije]{Hess1997}
Hess,~B.; Bekker,~H.; Berendsen,~H. J.~C.; Fraaije,~J. G. E.~M. {LINCS: A
  linear constraint solver for molecular simulations}. \emph{Journal of
  Computational Chemistry} \textbf{1997}, \emph{18}, 1463--1472\relax
\mciteBstWouldAddEndPuncttrue
\mciteSetBstMidEndSepPunct{\mcitedefaultmidpunct}
{\mcitedefaultendpunct}{\mcitedefaultseppunct}\relax
\EndOfBibitem
\bibitem[Hess(2007)]{Hess*2007}
Hess,~B. {P LINCS:A Parallel Linear Constraint Solver for Molecular
  Simulation}. \textbf{2007}, \relax
\mciteBstWouldAddEndPunctfalse
\mciteSetBstMidEndSepPunct{\mcitedefaultmidpunct}
{}{\mcitedefaultseppunct}\relax
\EndOfBibitem
\bibitem[Michaud-Agrawal \latin{et~al.}(2011)Michaud-Agrawal, Denning, Woolf,
  and Beckstein]{michaud2011mdanalysis}
Michaud-Agrawal,~N.; Denning,~E.~J.; Woolf,~T.~B.; Beckstein,~O. MDAnalysis: a
  toolkit for the analysis of molecular dynamics simulations. \emph{Journal of
  computational chemistry} \textbf{2011}, \emph{32}, 2319--2327\relax
\mciteBstWouldAddEndPuncttrue
\mciteSetBstMidEndSepPunct{\mcitedefaultmidpunct}
{\mcitedefaultendpunct}{\mcitedefaultseppunct}\relax
\EndOfBibitem
\bibitem[Gowers \latin{et~al.}(2019)Gowers, Linke, Barnoud, Reddy, Melo,
  Seyler, Domanski, Dotson, Buchoux, and Kenney]{gowers2019mdanalysis}
Gowers,~R.~J.; Linke,~M.; Barnoud,~J.; Reddy,~T. J.~E.; Melo,~M.~N.;
  Seyler,~S.~L.; Domanski,~J.; Dotson,~D.~L.; Buchoux,~S.; Kenney,~I.~M.
  \emph{MDAnalysis: a Python package for the rapid analysis of molecular
  dynamics simulations}; 2019\relax
\mciteBstWouldAddEndPuncttrue
\mciteSetBstMidEndSepPunct{\mcitedefaultmidpunct}
{\mcitedefaultendpunct}{\mcitedefaultseppunct}\relax
\EndOfBibitem
\bibitem[Zhang \latin{et~al.}(2020)Zhang, Nasrabadi, Aryal, and
  Ganesan]{zhang2020mechanisms}
Zhang,~Z.; Nasrabadi,~A.~T.; Aryal,~D.; Ganesan,~V. Mechanisms of Ion Transport
  in Lithium Salt-Doped Polymeric Ionic Liquid Electrolytes.
  \emph{Macromolecules} \textbf{2020}, \emph{53}, 6995--7008\relax
\mciteBstWouldAddEndPuncttrue
\mciteSetBstMidEndSepPunct{\mcitedefaultmidpunct}
{\mcitedefaultendpunct}{\mcitedefaultseppunct}\relax
\EndOfBibitem
\end{mcitethebibliography}

\bibliographystyle{rsc} %the RSC's .bst file

\end{document}

\textbf{A: Simulation protocol }

%software package
All-atomistic molecular dynamics (MD) simulations of lithium salt-ionic liquid (IL) mixtures were performed using the software package GROMACS (version 2018.8) \cite{VanDerSpoel2005,Pall2015,Abraham2015,Berendsen1995}. 
%force field
The atomic interactions were parameterized according to well-established OPLS-AA\cite{WilliamL.Jorgensen1996}-derived CL\&P force field developed by Canongia Lopes and Padua specifically for modeling ILs \cite{gouveia2017ionic,CanongiaLopes2012,JoseN.CanongiaLopes2004,Lopes2004,Shimizu2010}.
Electronic polarization and charge transfer effects were accounted for in a mean field sense via rescaling the atomic point charges, which is the prevalent practice when relying on non-polarizable force fields to study ionic liquids \cite{doherty2017revisiting,Bhargava2007,Chaban2011,self2019transport,fong2019ion,molinari2019transport,molinari2019general,molinari2020chelation,huang2018solvation,thum2020solvate}, because a more accurate treatment of the electronic polarizability by means of Drude oscillators \cite{lemkul2016empirical} or induced point dipoles \cite{salomon2013overview} comes at a great computational cost. 
Very recent studies on lithium salt-ionic liquid mixtures \cite{molinari2019general,molinari2020chelation,thum2020solvate,huang2018solvation} and a variety of lithium salt containing electrolytes \cite{li2019transport,self2019transport,fong2019ion} have demonstrated successfully the ability of non-polarizable force fields, employing scaled partial charges, to capture and confirm experimental observations. 
In this work the the atomic point charges of all species were uniformly scaled down by a factor of 0.8 \cite{molinari2019general,molinari2020chelation,thum2020solvate,huang2018solvation,doherty2017revisiting}.
%systems
We study structural and dynamical properties of the $\text{Li}^+_{x}-\text{Pyr}_{14, (1-x)}^+-\text{TFSI}^-$ and $\text{Li}^+_x-\text{Pyr}_{14, (1-x)}^+-\text{TFSAM}^-$ electrolytes in the same concentration range from $x\,=\,0.0$ (neat IL) to $x\,=\,0.7$ as experimentally reported in \cite{nuernberg2020}.

The initial configurations were created with the PACKMOL software\cite{Martinez2009} which randomly distributed 1000 ion pairs, corresponding to the respective lithium salt to ionic liquid ratio, in a cubic box. 
The systems were first exposed to an energy minimization and then pre-equilibrated under NPT conditions for 40 ns at a high temperature of 500\,K controlling pressure via a Berendsen barostat (relaxation time constant $\tau_P\,=\,5.0\,$ps, compressibility of $4.5\times 10^{-5}\,$bar) coupled to a reference pressure of 1 bar and temperature via a velocity-rescale thermostat (relaxation time constant $\tau_T\,=\,1.0\,$ps)\cite{Berendsen1984,Bussi2007}. 
Then the systems were cooled down to 400\,K and equilibrated for another 100\,ns. 
In the subsequent production run of 400\,ns duration, that was used for data acquisition, pressure and temperature were coupled to an extended Parrinello-Rahman and Nos\'e-Hoover ensemble using the same relaxation time constants as before\cite{Parrinello1981,Nose1983,Nose1984,Hoover1985}.
The equations of motion were solved via the leap-frog algorithm at a time step of 2\,fs. The center of mass of the system was repositioned every simulation step. Furthermore, cutoffs for the long range electrostatic and the van der Waals interactions were both set to 1.4\,nm and the linear constraint solver (LINCS) was employed to constrain the hydrogen bonds \cite{Hess1997,Hess*2007}. 
The simulation trajectories were analyzed with custom scripts supported by the Python library MDAnalysis \cite{michaud2011mdanalysis,gowers2019mdanalysis}.

%%%%%%%%%%%%%%%%%%%%%%%%%%%%%%%%%%%%%%%%%%%%%%%%%%%%%%%%%%%%%%%%%%%%%%%%%%%%%%%%
%%%%%%%%%%%%%%%%%%%%%%%%%%%%%%%%%%%%%%%%%%%%%%%%%%%%%%%%%%%%%%%%%%%%%%%%%%%%%%%%
%%%%%%%%%%%%%%%%%%%%%%%%%%%%%%%%%%%%%%%%%%%%%%%%%%%%%%%%%%%%%%%%%%%%%%%%%%%%%%%%
%%%%%%%%%%%%%%%%%%%%%%%%%%%%%%%%%%%%%%%%%%%%%%%%%%%%%%%%%%%%%%%%%%%%%%%%%%%%%%%%
%%%%%%%%%%%%%%%%%%%%%%%%%%%%%%%%%%%%%%%%%%%%%%%%%%%%%%%%%%%%%%%%%%%%%%%%%%%%%%%%
%%%%%%%%%%%%%%%%%%%%%%%%%%%%%%%%%%%%%%%%%%%%%%%%%%%%%%%%%%%%%%%%%%%%%%%%%%%%%%%%
\newpage
\textbf{B: Structural properties}

%%%%%%%%%%%%%%%%%%%%%%%%%%%%%%%%%%%%%%%%%%%%%%%%%%%%%%%%%%%%%%%%%%%%%%%%%%%%%%%%
%%%%   RDF
The radial distribution functions $g_{\text{Li}^+\text{-X}}$ are computed according to
\begin{equation}
g_{ab}(r) = \dfrac{V}{4\pi r^2N_aN_b} \sum_{i=1}^{N_a}\sum_{j=1}^{N_b} \langle \delta \left( |\vec{\text{r}}_i - \vec{\text{r}}_j| - r \right) \rangle,
\end{equation}
where $N_a/V$ and $N_b/V$ denotes the average number density of species $a$ and species $b$ with $V$ being the volume of the simulation box. The brackets $\langle .. \rangle$ indicate the ensemble average.

\begin{figure}[H]
  \centering
  \subfloat{\includegraphics[width=0.5\textwidth]{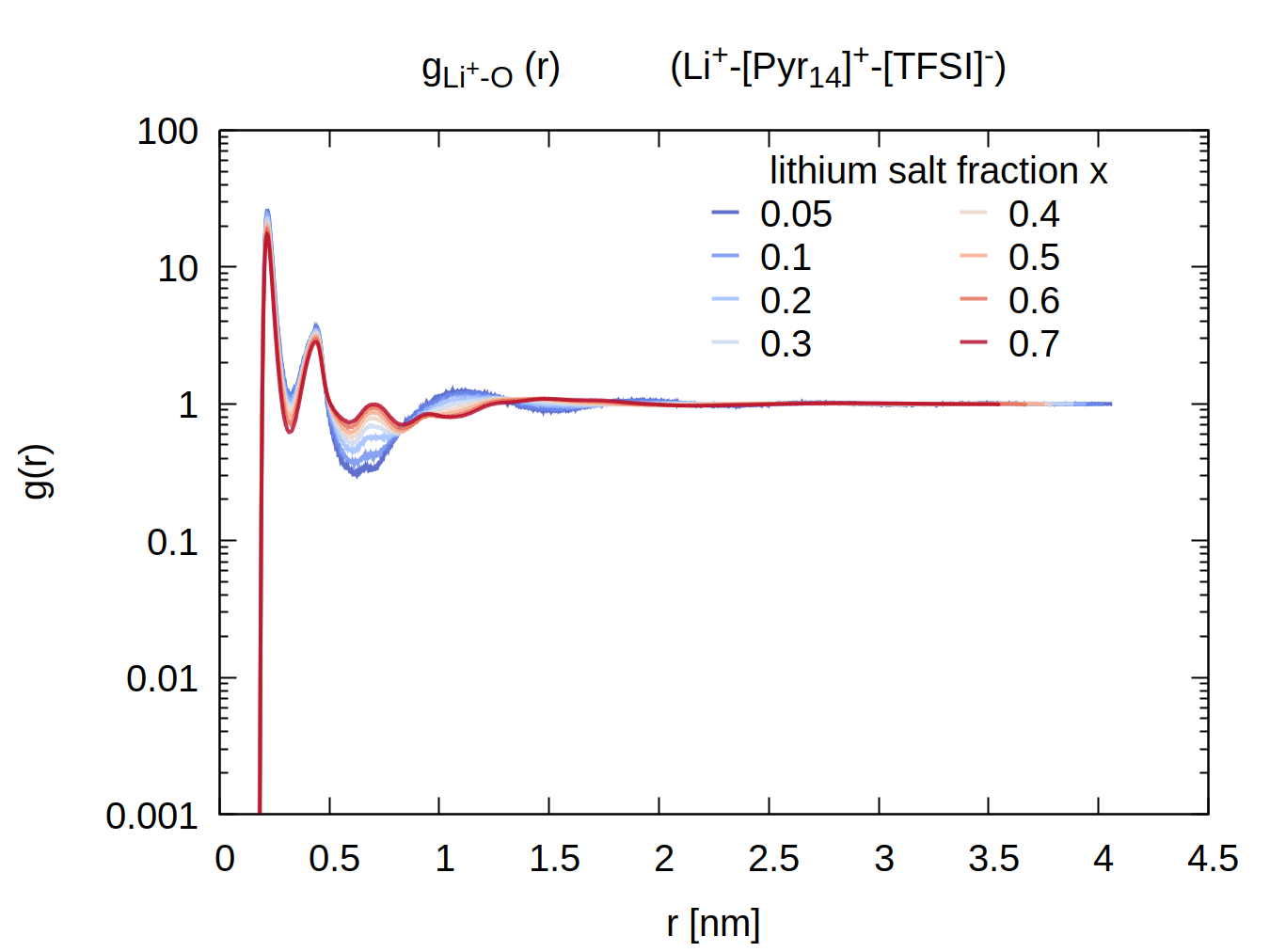}\label{fig:rdf_li_os_linlog}}
  \hfill
  \subfloat{\includegraphics[width=0.5\textwidth]{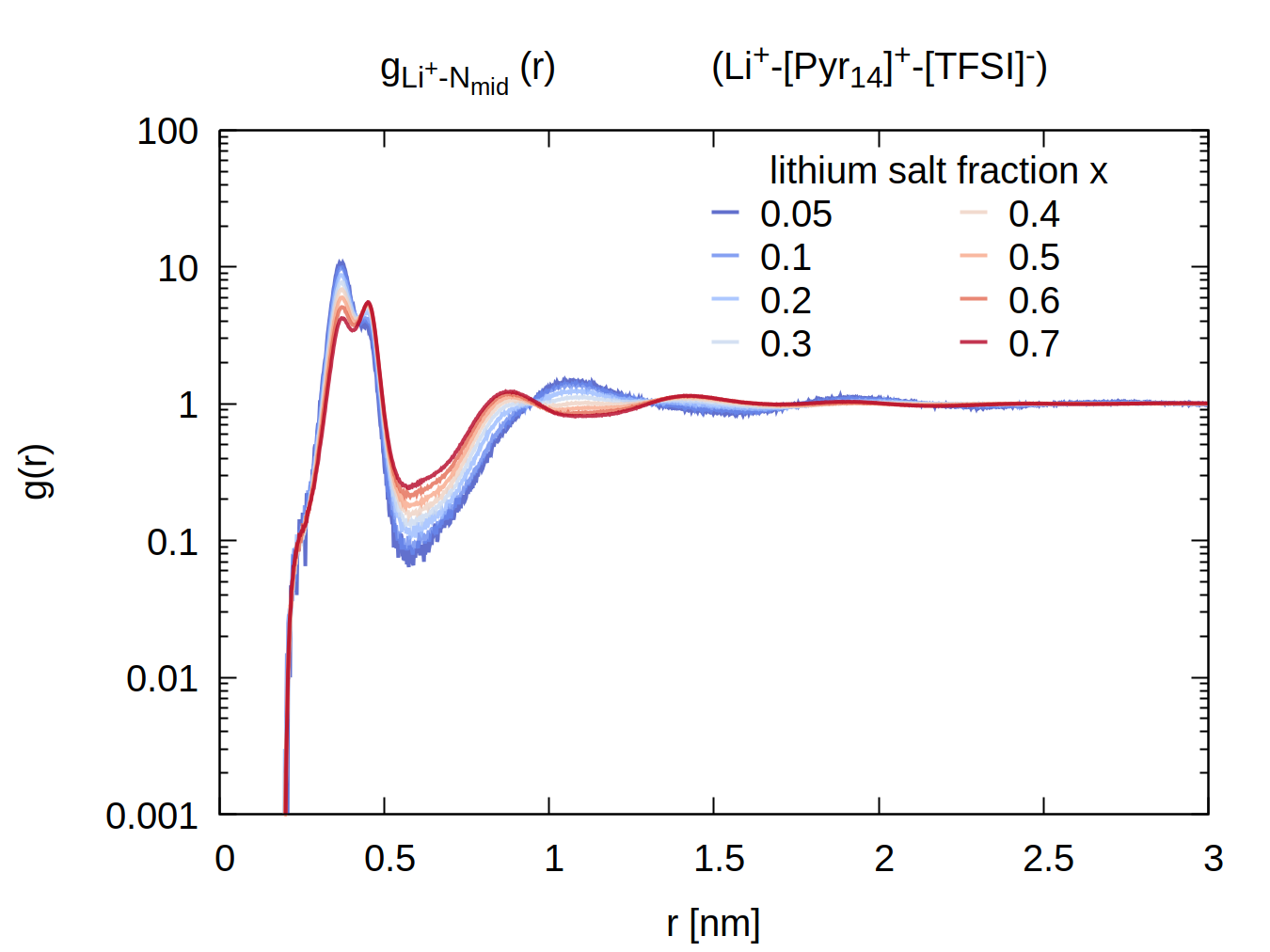}\label{fig:rdf_li_ni_linlog}}

\end{figure}

\begin{figure}[H]
  \centering
  \subfloat{\includegraphics[width=0.5\textwidth]{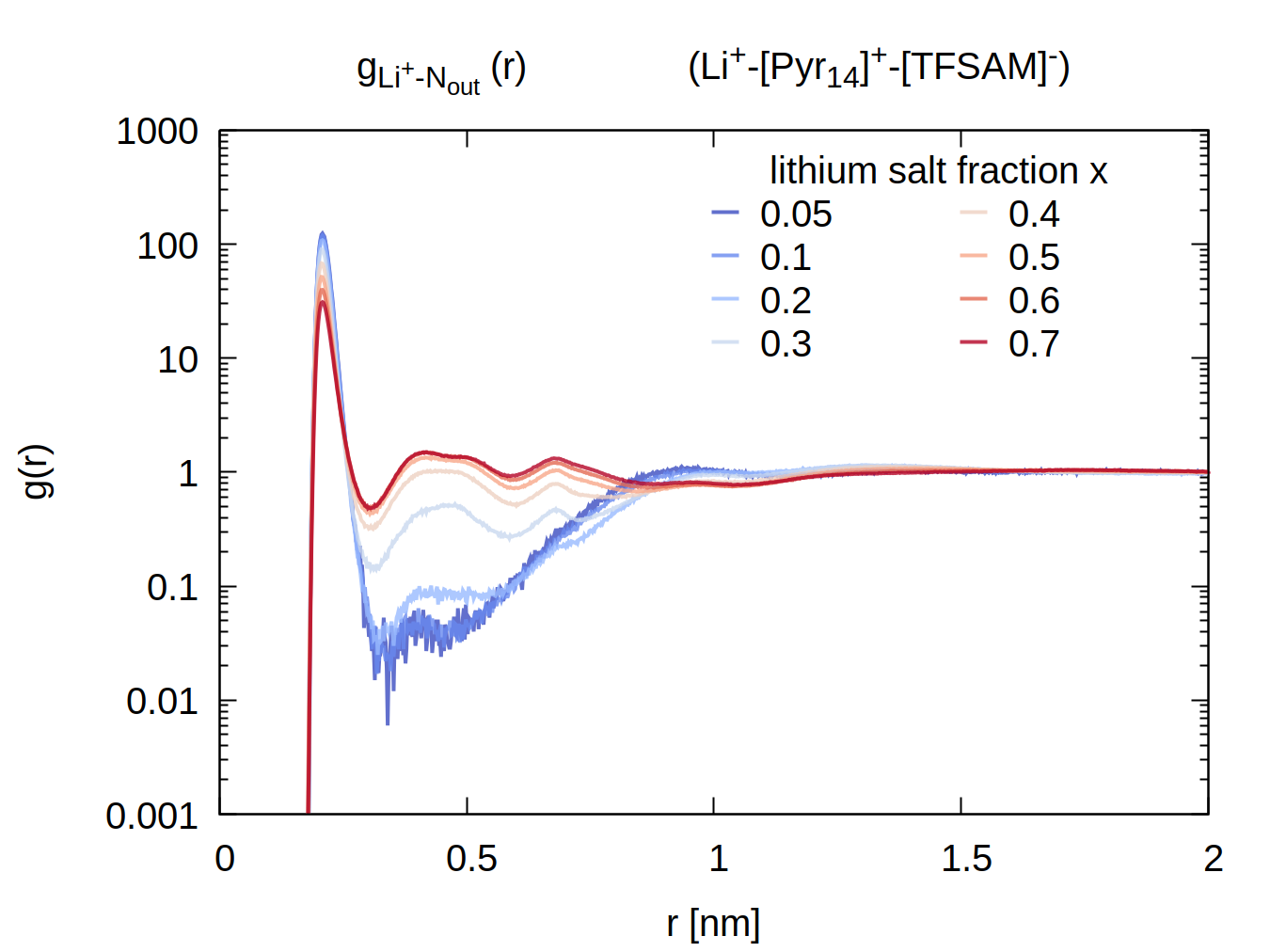}\label{fig:rdf_li_nc_linlog}}
  \hfill
  \subfloat{\includegraphics[width=0.5\textwidth]{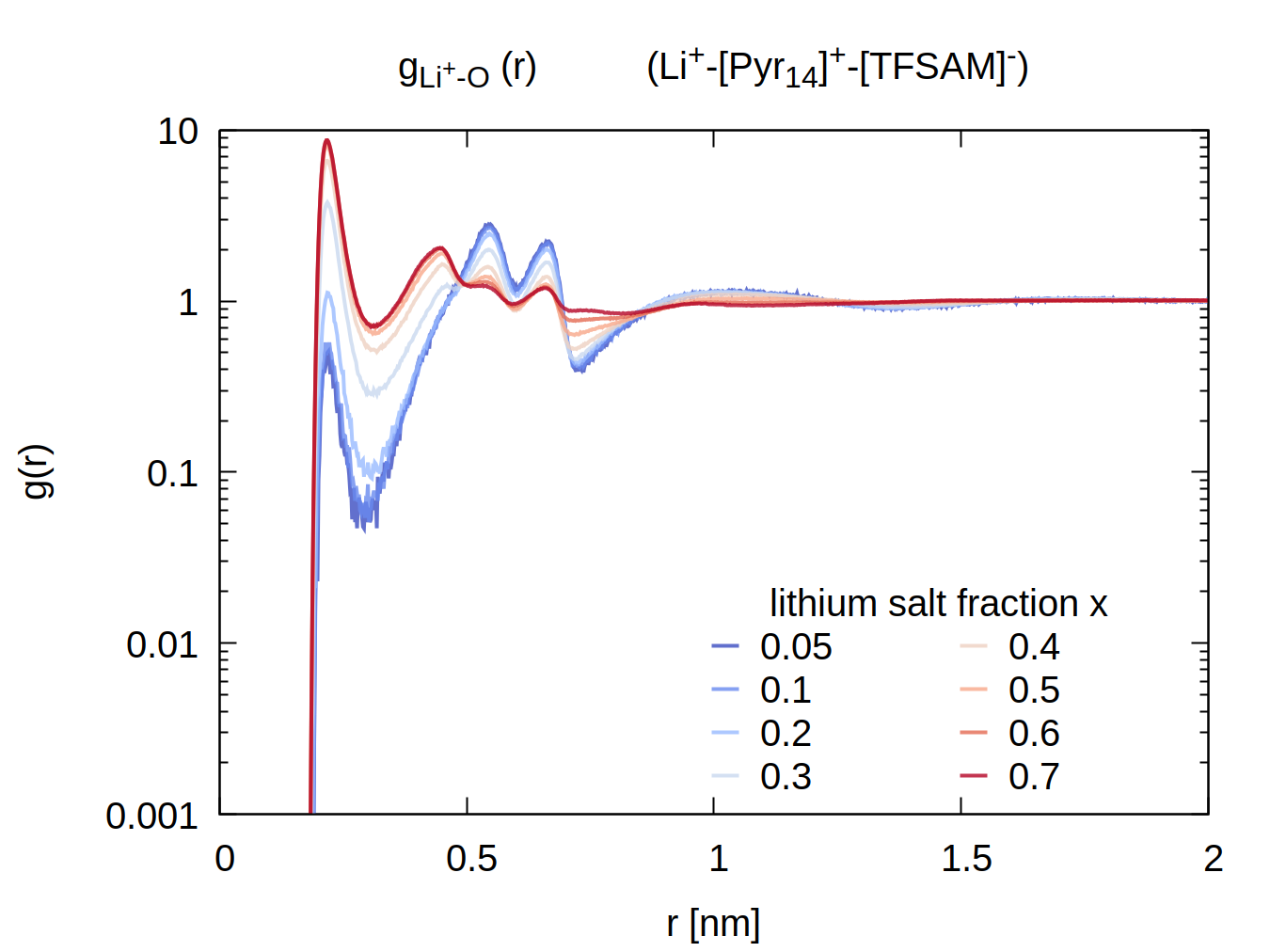}\label{fig:rdf_li_op_linlog}}

\end{figure}

\begin{figure}[H]
  \centering
  \subfloat{\includegraphics[width=0.5\textwidth]{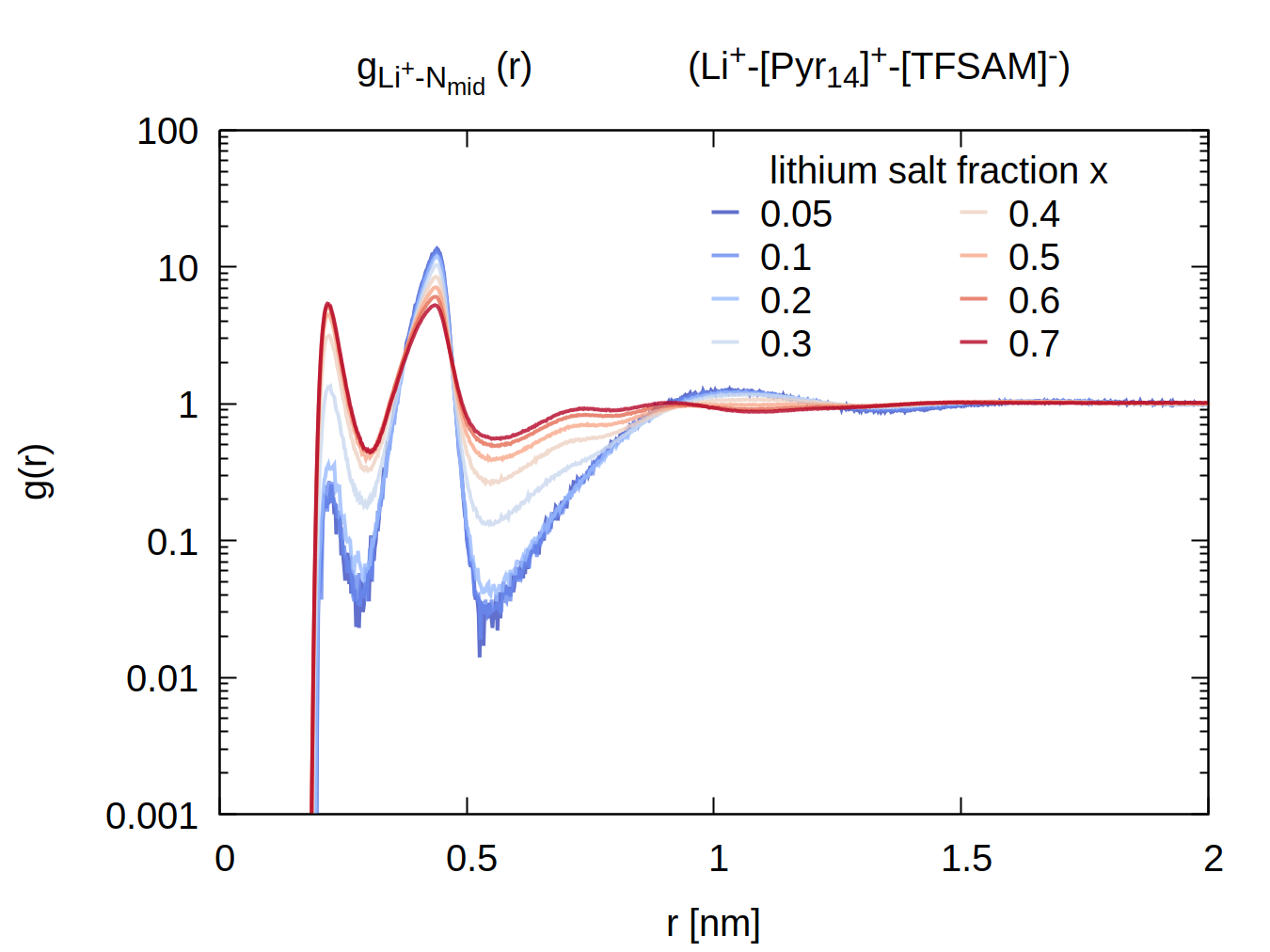}\label{fig:rdf_li_nh_linlog}}
  \hfill

\end{figure}

\begin{figure}[H]
  \centering
  \subfloat{\includegraphics[width=0.5\textwidth]{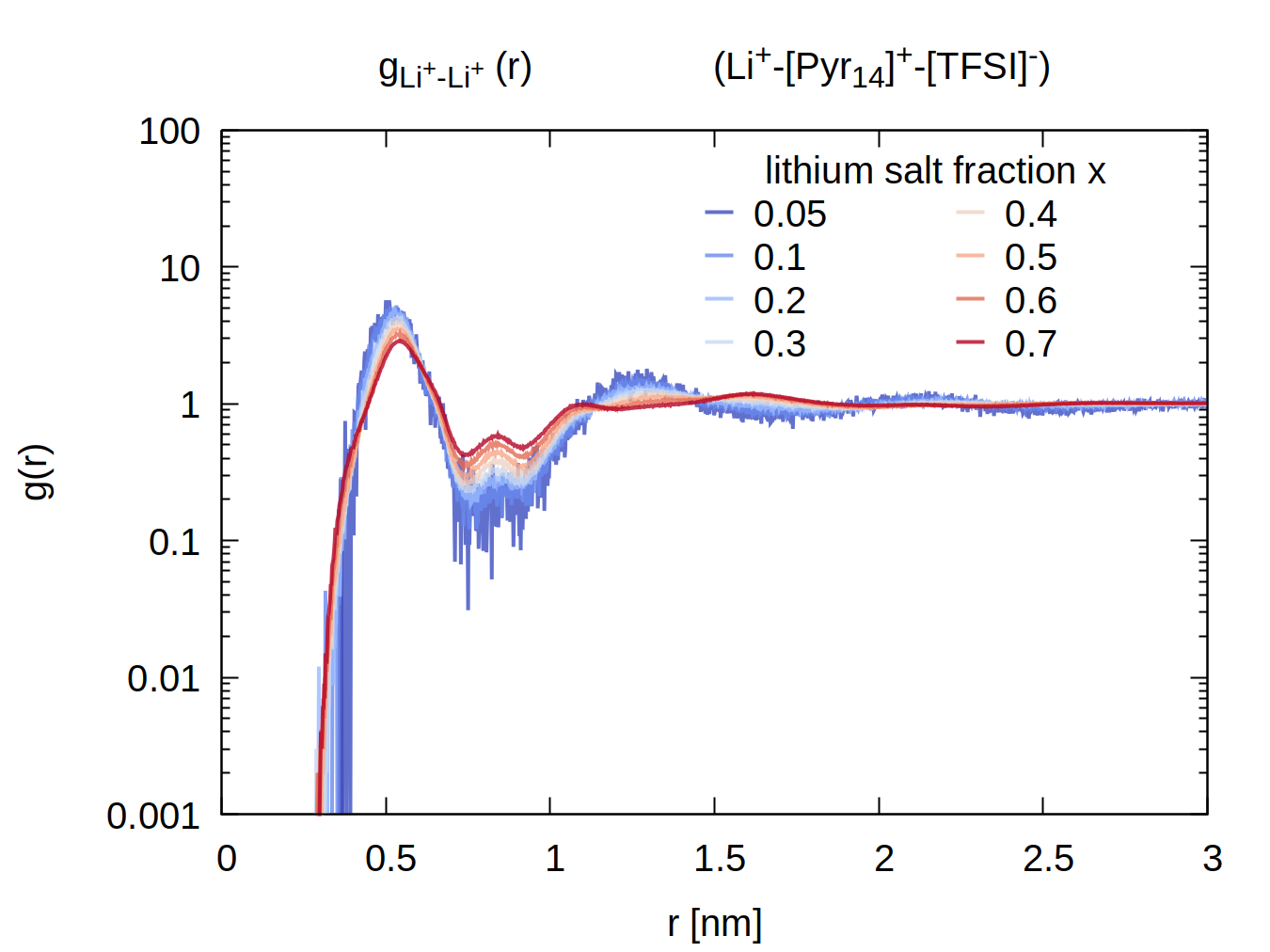}\label{fig:rdf_li_li_TFSI_linlog}}
  \hfill
  \subfloat{\includegraphics[width=0.5\textwidth]{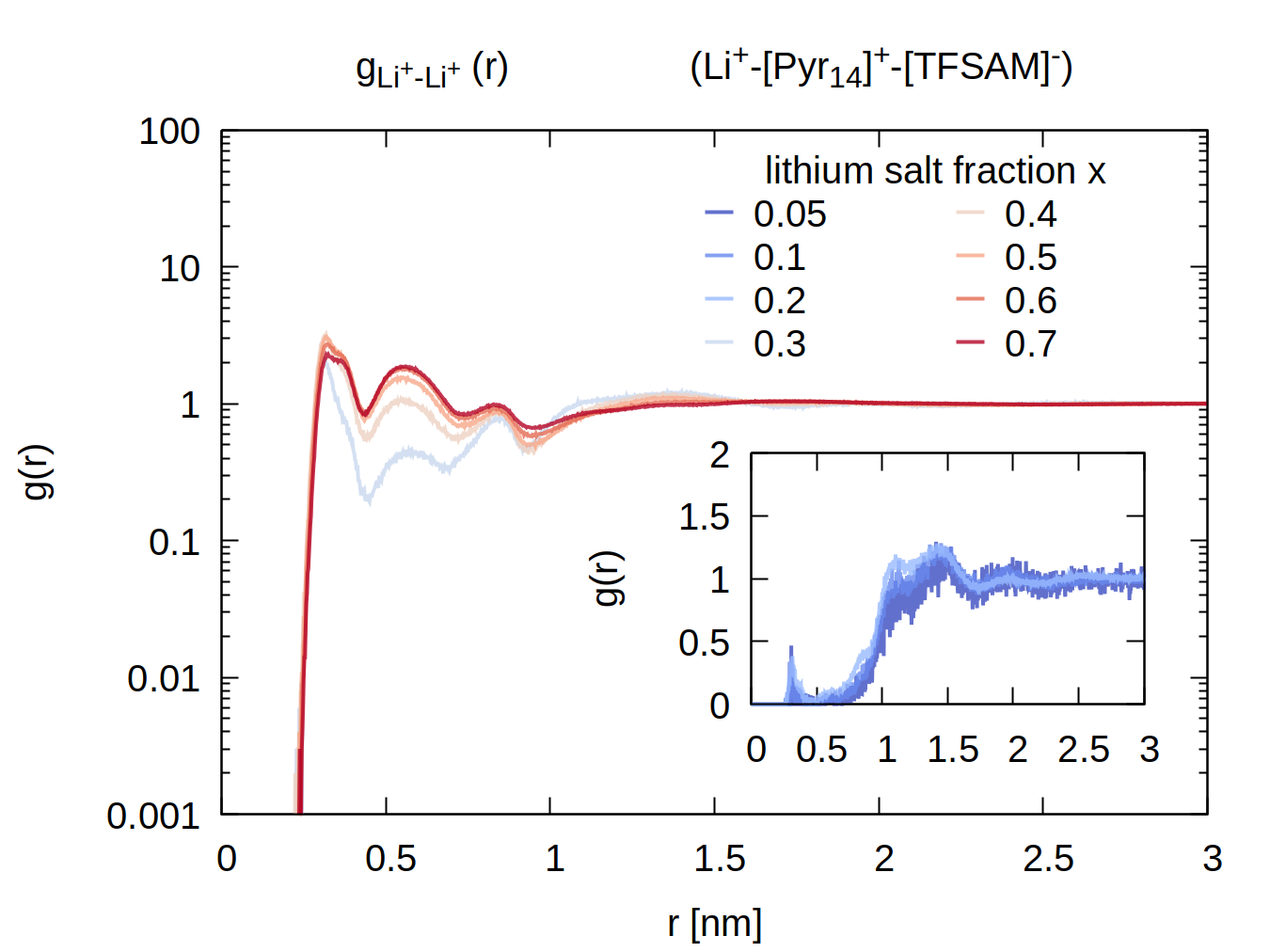}\label{fig:rdf_li_li_TFSAM_linlog}}
  \caption{Overview of radial distribution functions $\text{g}_{\text{Li}^+-X}$(r) between $\text{Li}^+$ ions and \newline nitrogen / oxygen binding sites provided by $\text{TFSI}^-$  and $\text{TFSAM}^-$ as well as $\text{g}_{\text{Li}^+-\text{Li}^+}$(r) in both electrolyte series on a log scale.  }
\end{figure}

\begin{figure}[H]
  \centering
  \subfloat{\includegraphics[width=0.5\textwidth]{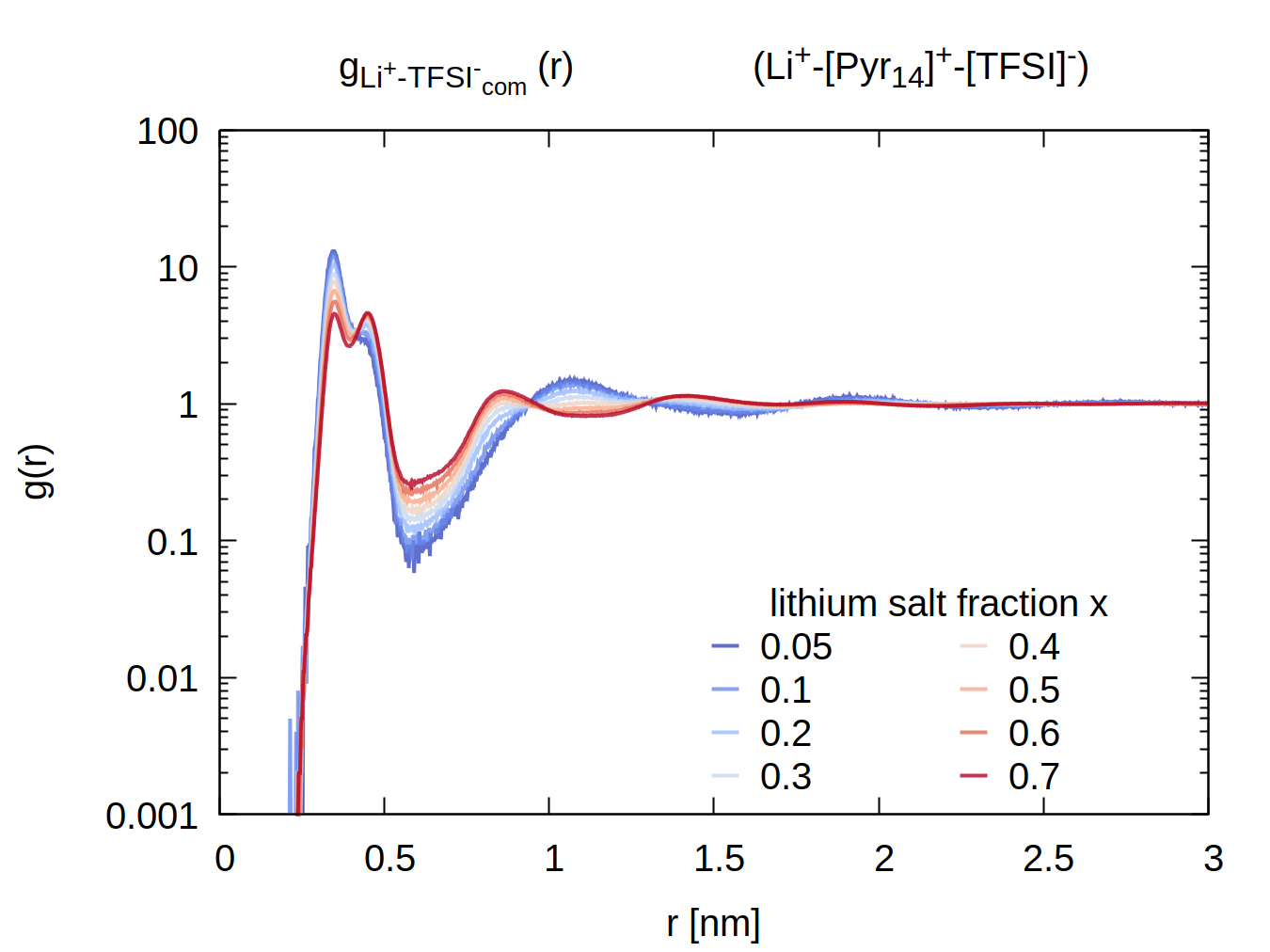}\label{fig:rdf_li_tfsi_com_linlog}}
  \hfill
  \subfloat{\includegraphics[width=0.5\textwidth]{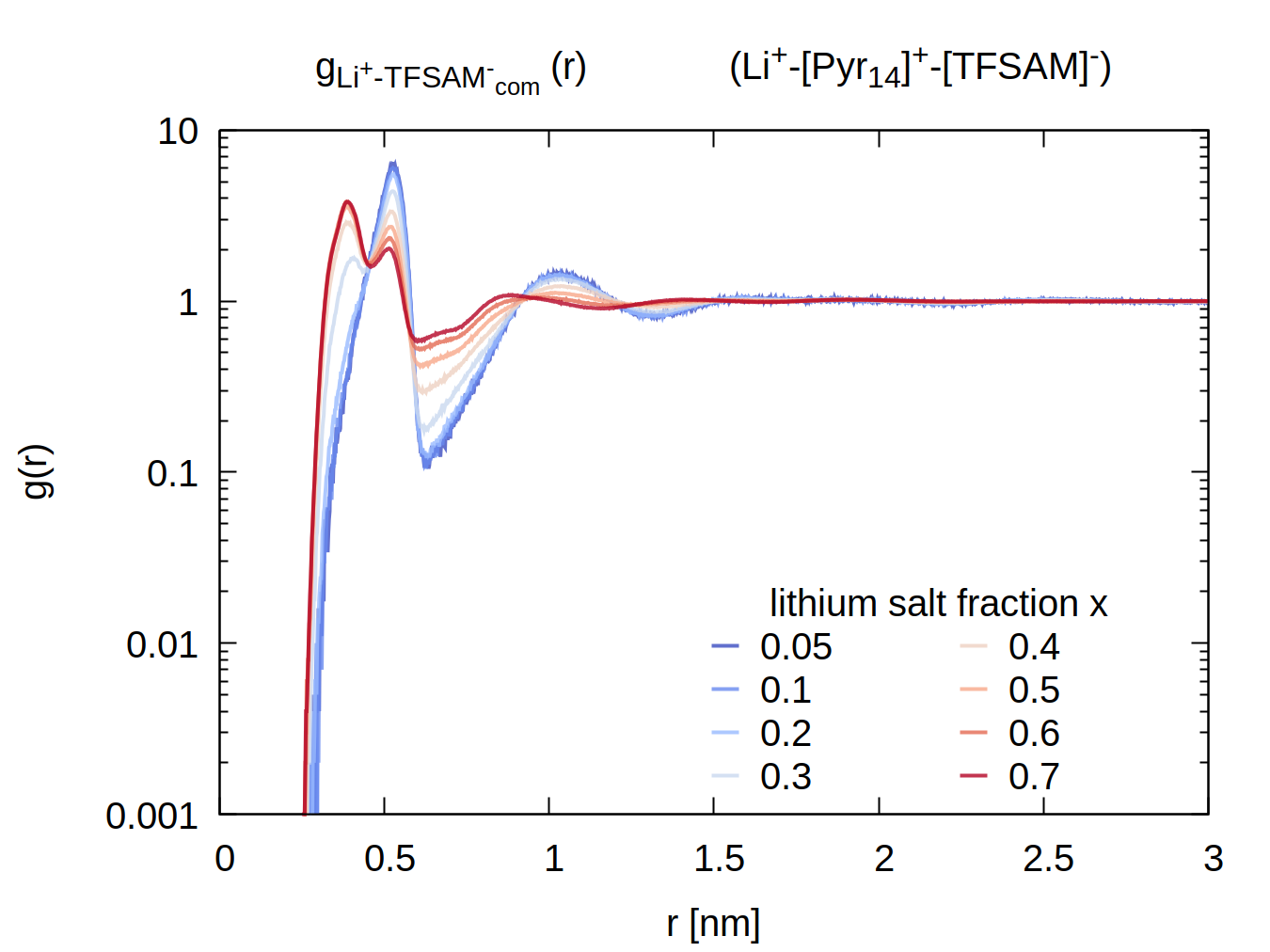}\label{fig:rdf_li_tfsam_com_linlog}}
  \caption{Overview of lithium - anion(com) radial distribution functions $\text{g}_{\text{Li}^+-\text{anion}_{\text{com}}}$(r) on a log scale. The global minimum position is employed as the solvation shell size Ls in analogy to the procedure introduced by Self et al.\cite{self2019transport}}.
\end{figure}

%%%%%%%%%%%%%%%%%%%%%%%%%%%%%%%%%%%%%%%%%%%%%%%%%%%%%%%%%%%%%%%%%%%%%%%%%%%%%%%%
%%%%   CUMULATIVE NUMBERS

\begin{figure}[H]
  \centering
  \subfloat{\includegraphics[width=0.5\textwidth]{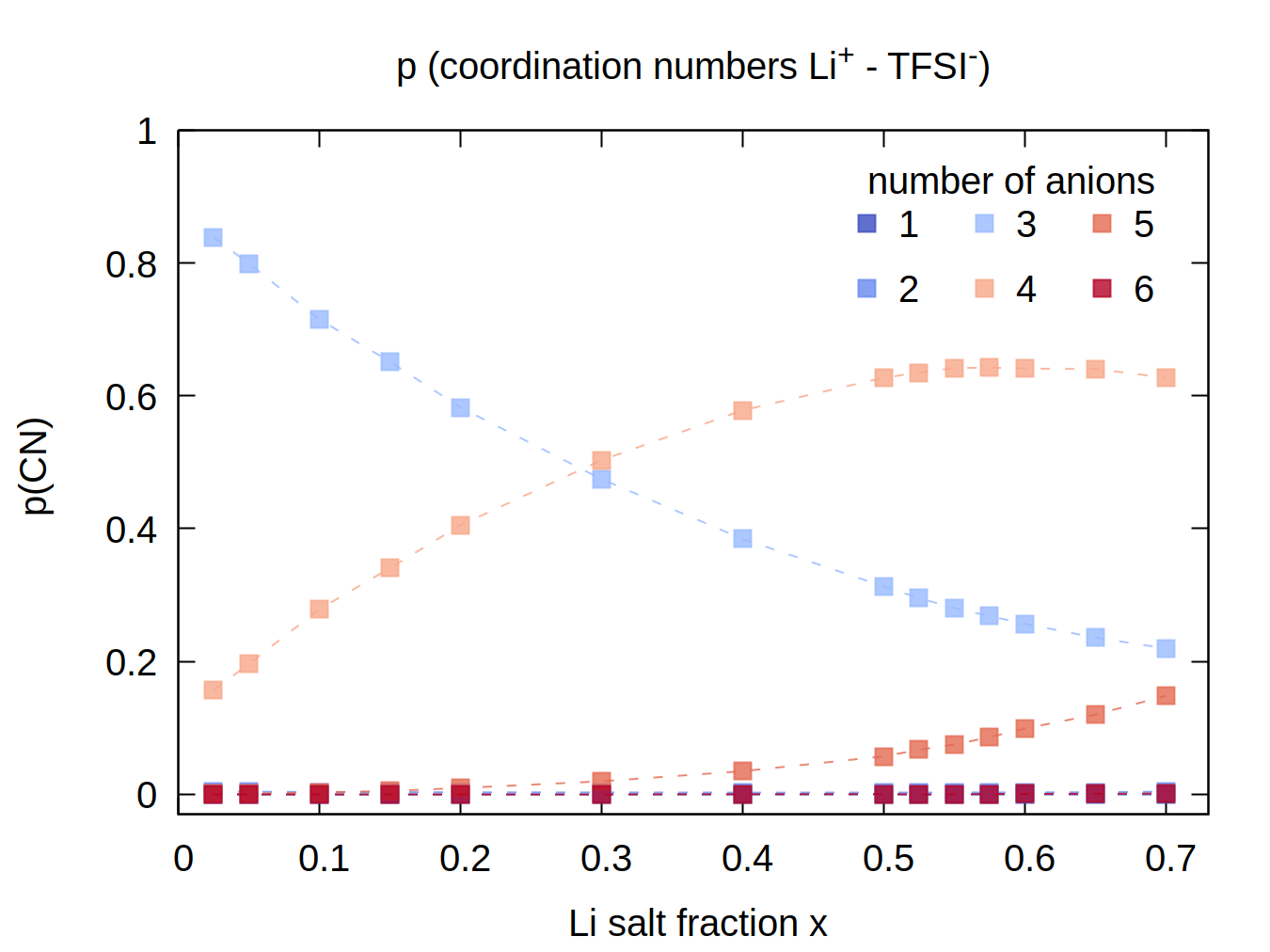}\label{fig:li_coordination_distribution_tfsi}}
  \hfill
  \subfloat{\includegraphics[width=0.5\textwidth]{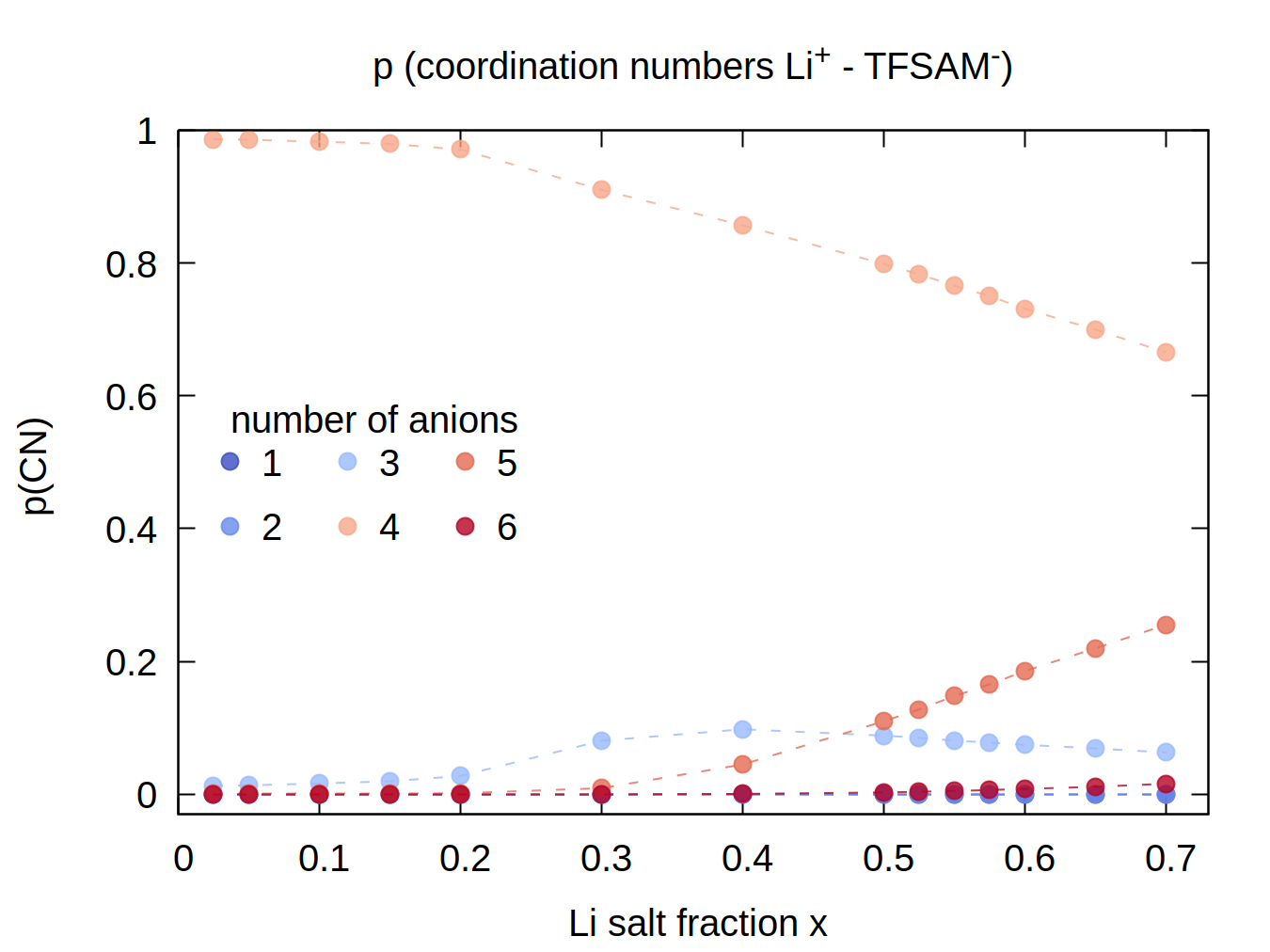}\label{fig:li_coordination_distribution_tfsam}}
  \caption{Probability distribution of $\text{Li}^+$-anion coordination numbers in the $\text{TFSI}^-$ (left) and $\text{TFSAM}^-$ (right) -based electrolytes as a function of lithium salt content x.}
  \label{fig:li_coordination_distribution_tfsi_tfsam}
\end{figure}

\begin{figure}[H]
  \centering
  \includegraphics[width=0.7\textwidth]{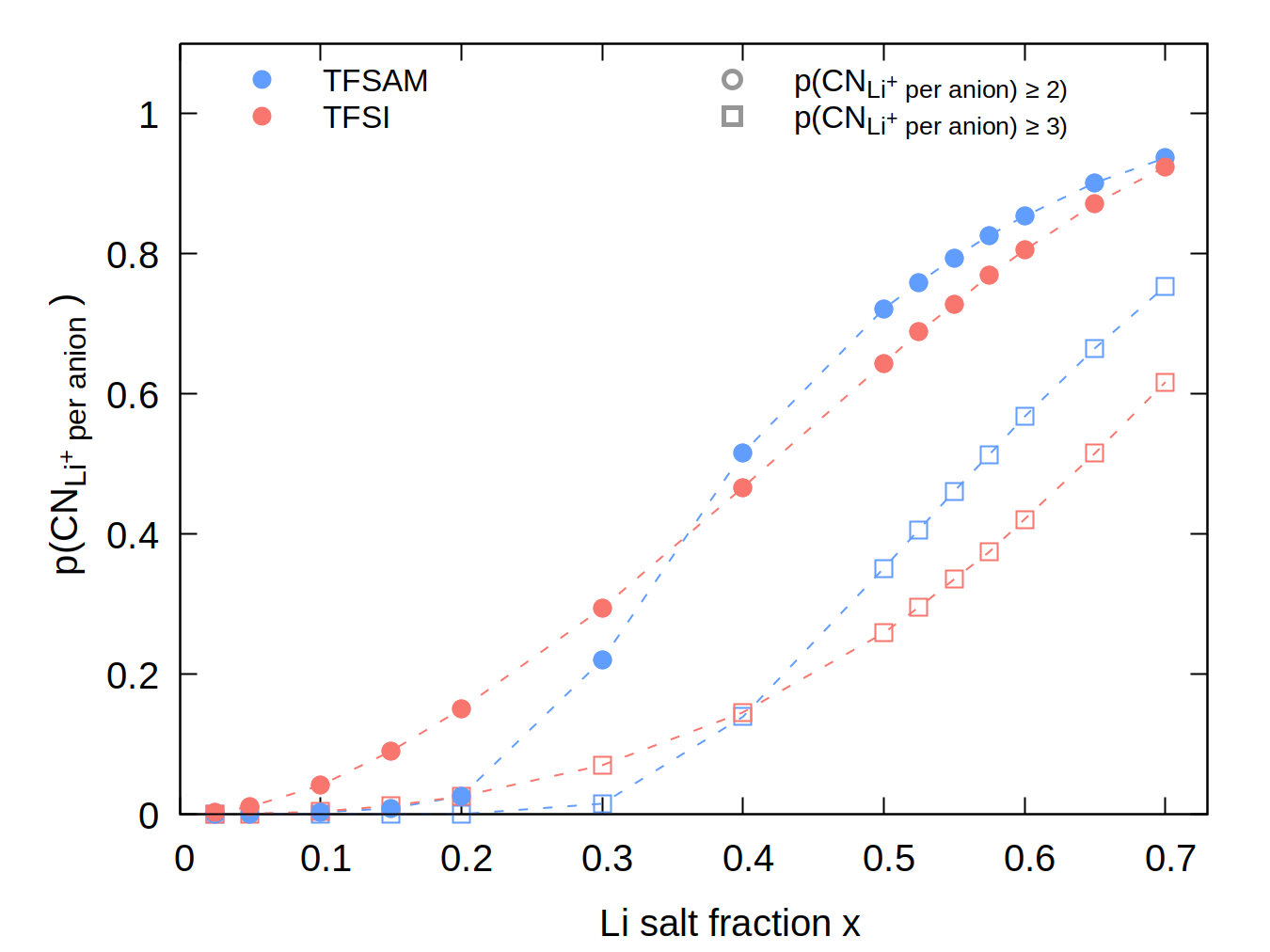}
  \caption{Probability distribution of lithium neighbors per $\text{TFSI}^-$/$\text{TFSAM}^-$  p($\text{CN}_{\text{Li}^+ \text{per anion}}$) as a function of lithium salt content x.}
  \label{fig:coordination_of_anions_via_lithium_p_CN}
\end{figure}

%%%%%%%%%%%%%%%%%%%%%%%%%%%%%%%%%%%%%%%%%%%%%%%%%%%%%%%%%%%%%%%%%%%%%%%%%%%%%%%%
%%%%%%%%%%%%%%%%%%%%%%%%%%%%%%%%%%%%%%%%%%%%%%%%%%%%%%%%%%%%%%%%%%%%%%%%%%%%%%%%
%%%% Polymer monomer angle orientation:

%%%%%%%%%%%%%%%%%%%%%%%%%%%%%%%%%%%%%%%%%%%%%%%%%%%%%%%%%%%%%%%%%%%%%%%%%%%%%%%%
%%%%%%%%%%%%%%%%%%%%%%%%%%%%%%%%%%%%%%%%%%%%%%%%%%%%%%%%%%%%%%%%%%%%%%%%%%%%%%%%

\newpage
\textbf{C: Mean squared displacements}

\begin{figure}[H]
  \centering
  \subfloat{\includegraphics[width=0.5\textwidth]{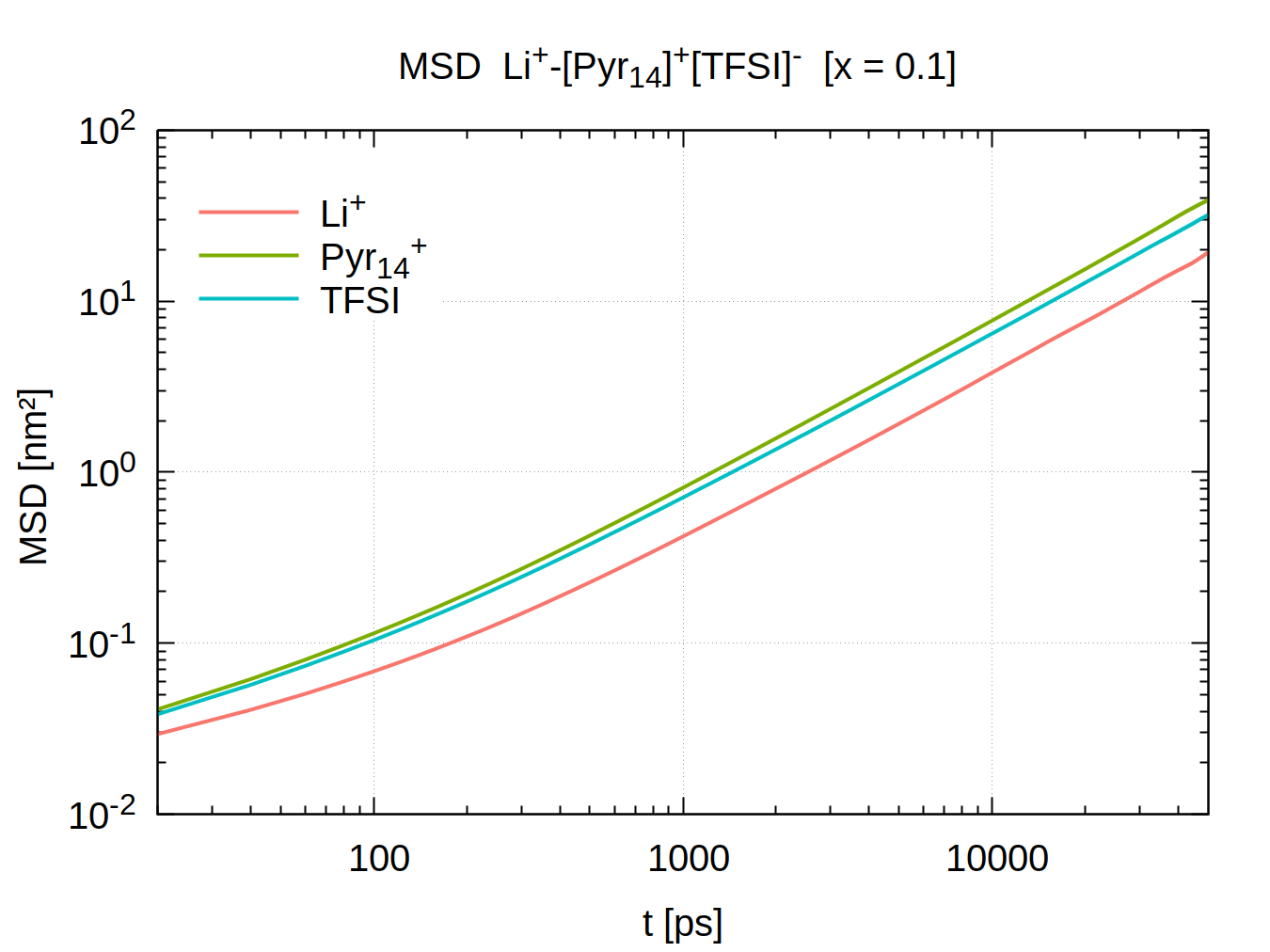}\label{fig:msds_tfsi_x_0_1}}
  \hfill
  \subfloat{\includegraphics[width=0.5\textwidth]{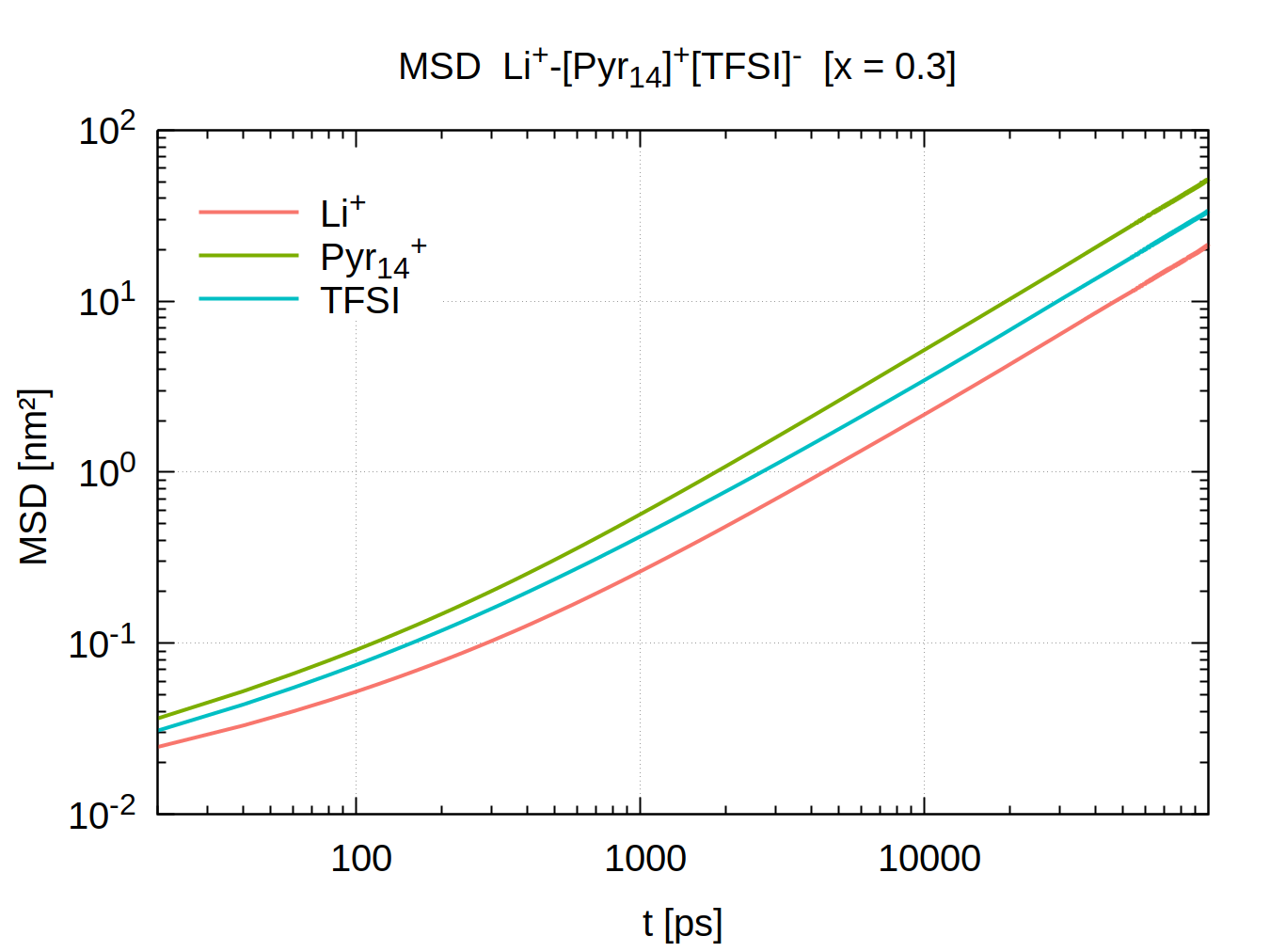}\label{fig:msds_tfsi_x_0_3}}
  \hfill
  \centering
  \subfloat{\includegraphics[width=0.5\textwidth]{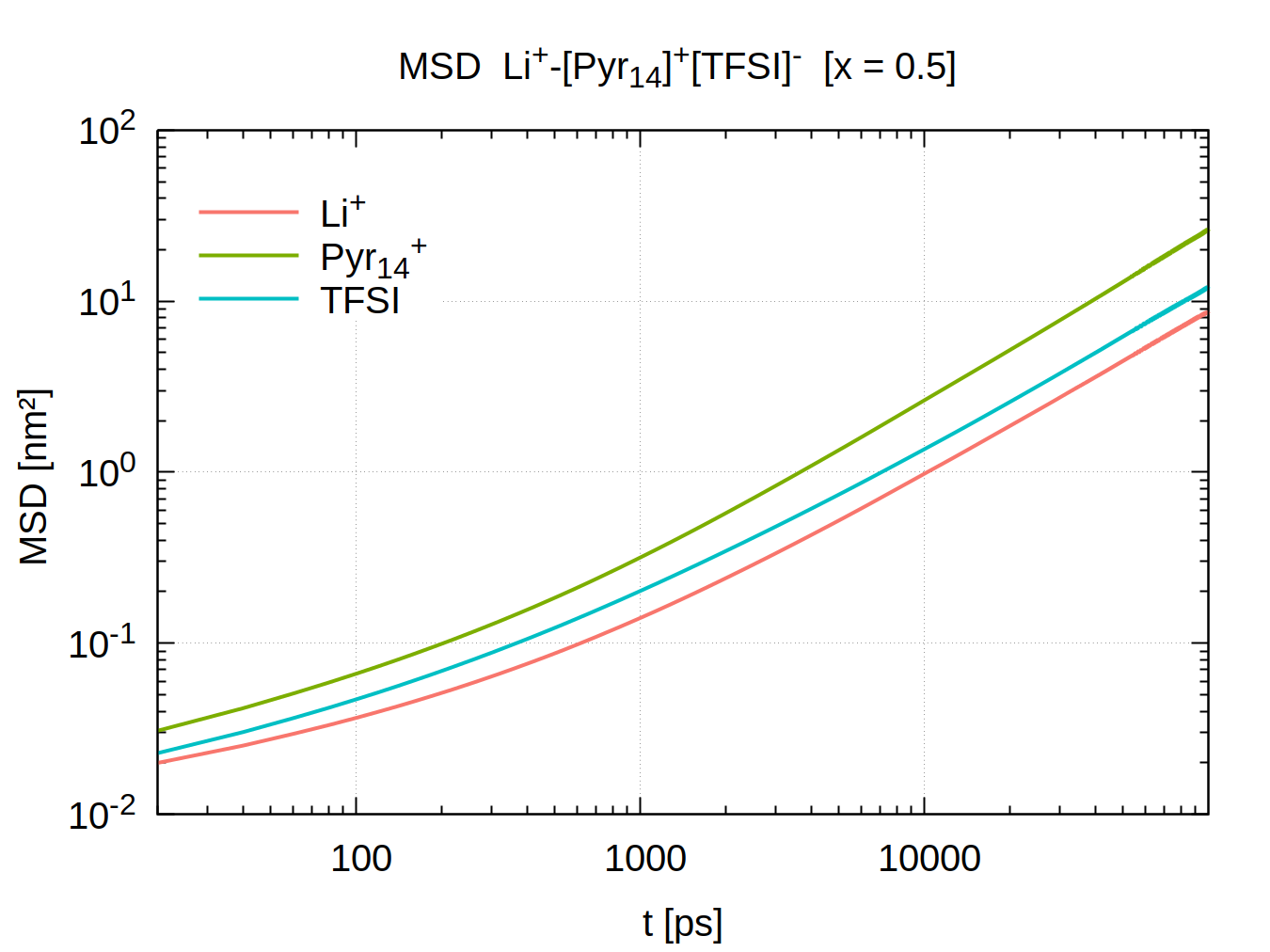}\label{fig:msds_tfsi_x_0_1}}
  \hfill
  \subfloat{\includegraphics[width=0.5\textwidth]{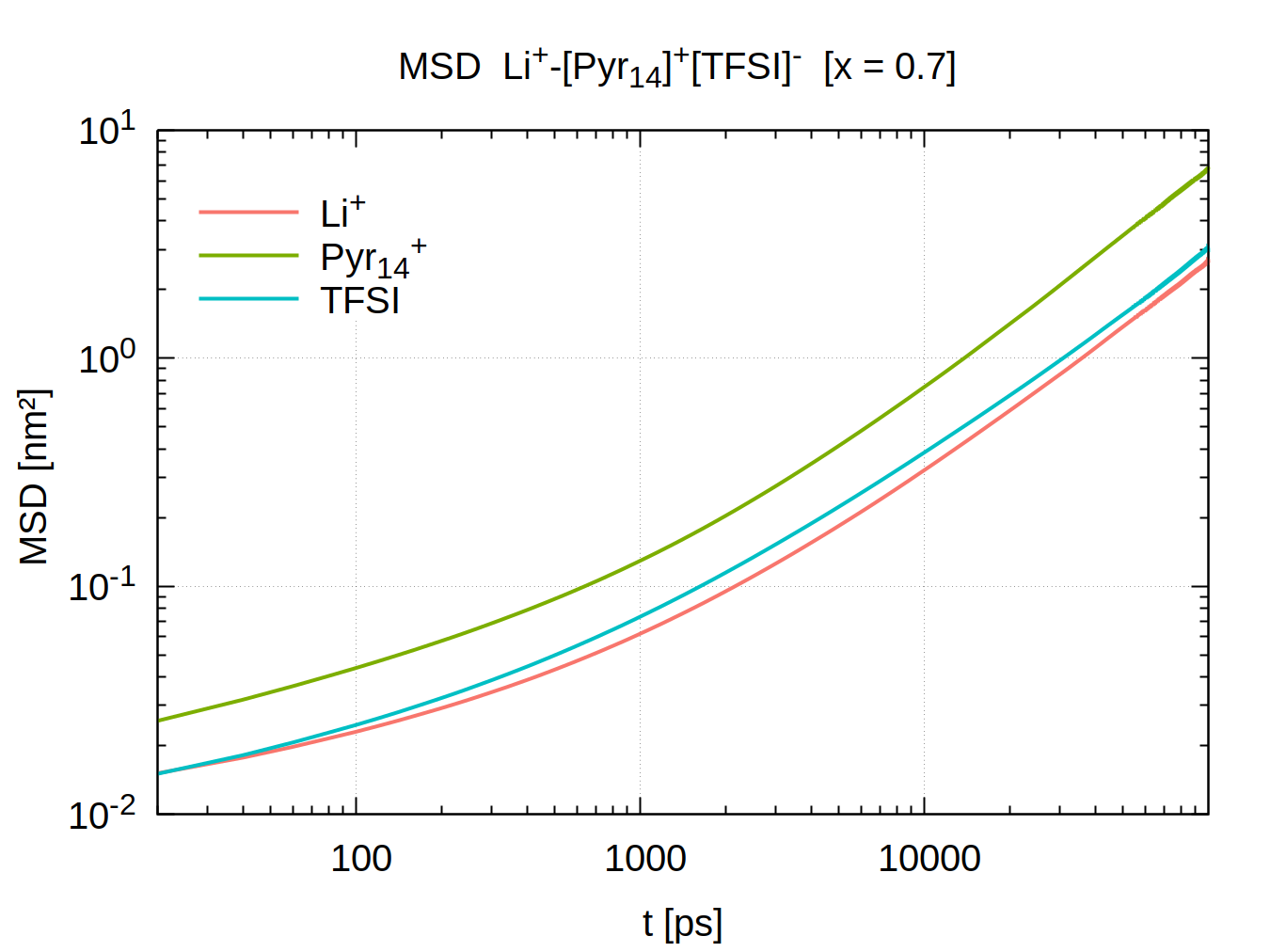}\label{fig:msds_tfsi_x_0_3}}
  \caption{Example mean squared displacements of $\text{Li}^+, \text{Pyr}_{14}^+$ and $\text{TFSI}^-$ as a function of time for lithium salt contents x=0.1 and x=0.3 (top) as well as x=0.5 and x=0.7 (bottom).}
\end{figure}

\begin{figure}[H]
  \centering
  \subfloat{\includegraphics[width=0.5\textwidth]{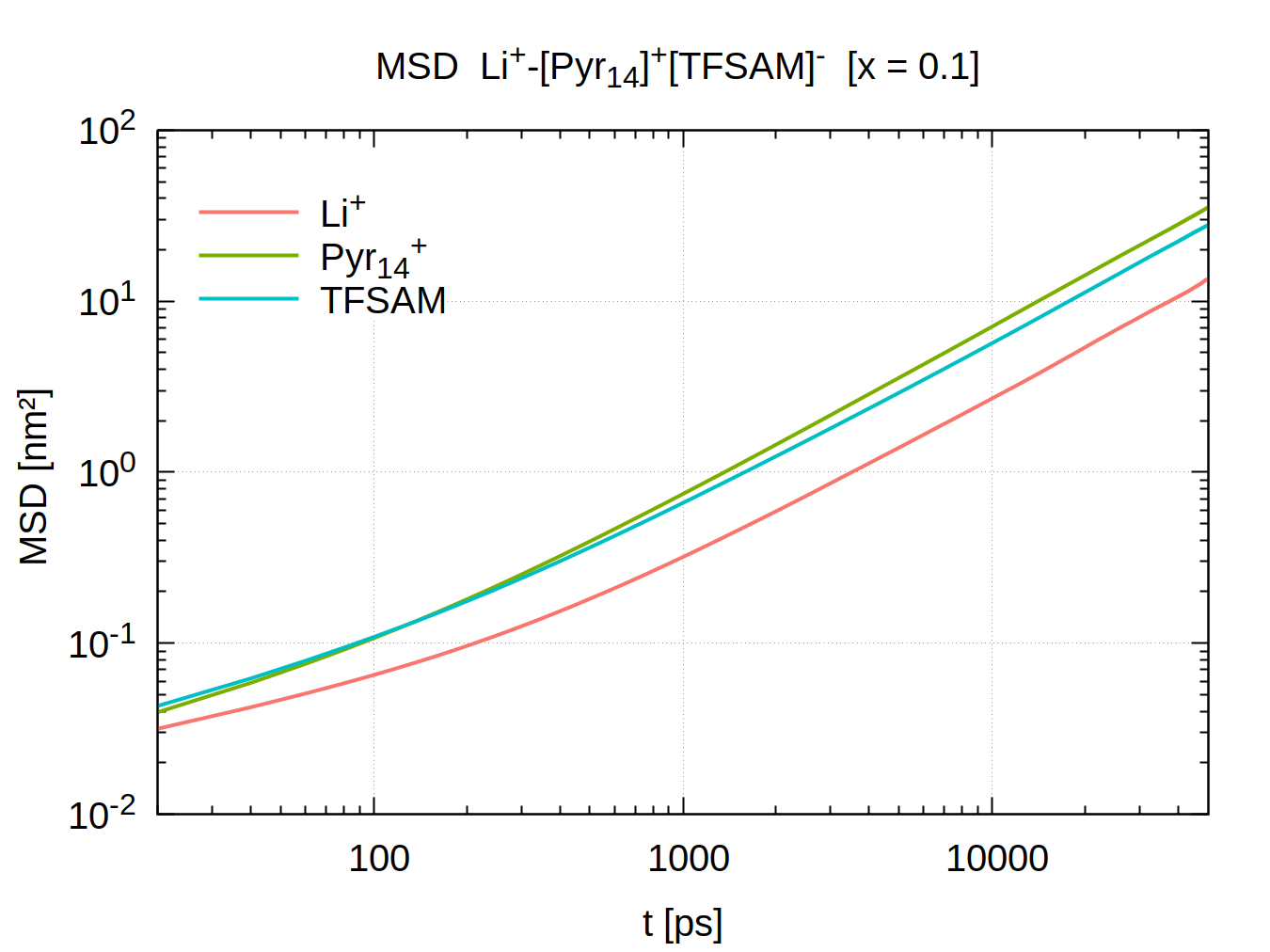}\label{fig:msds_tfsam_x_0_1}}
  \hfill
  \subfloat{\includegraphics[width=0.5\textwidth]{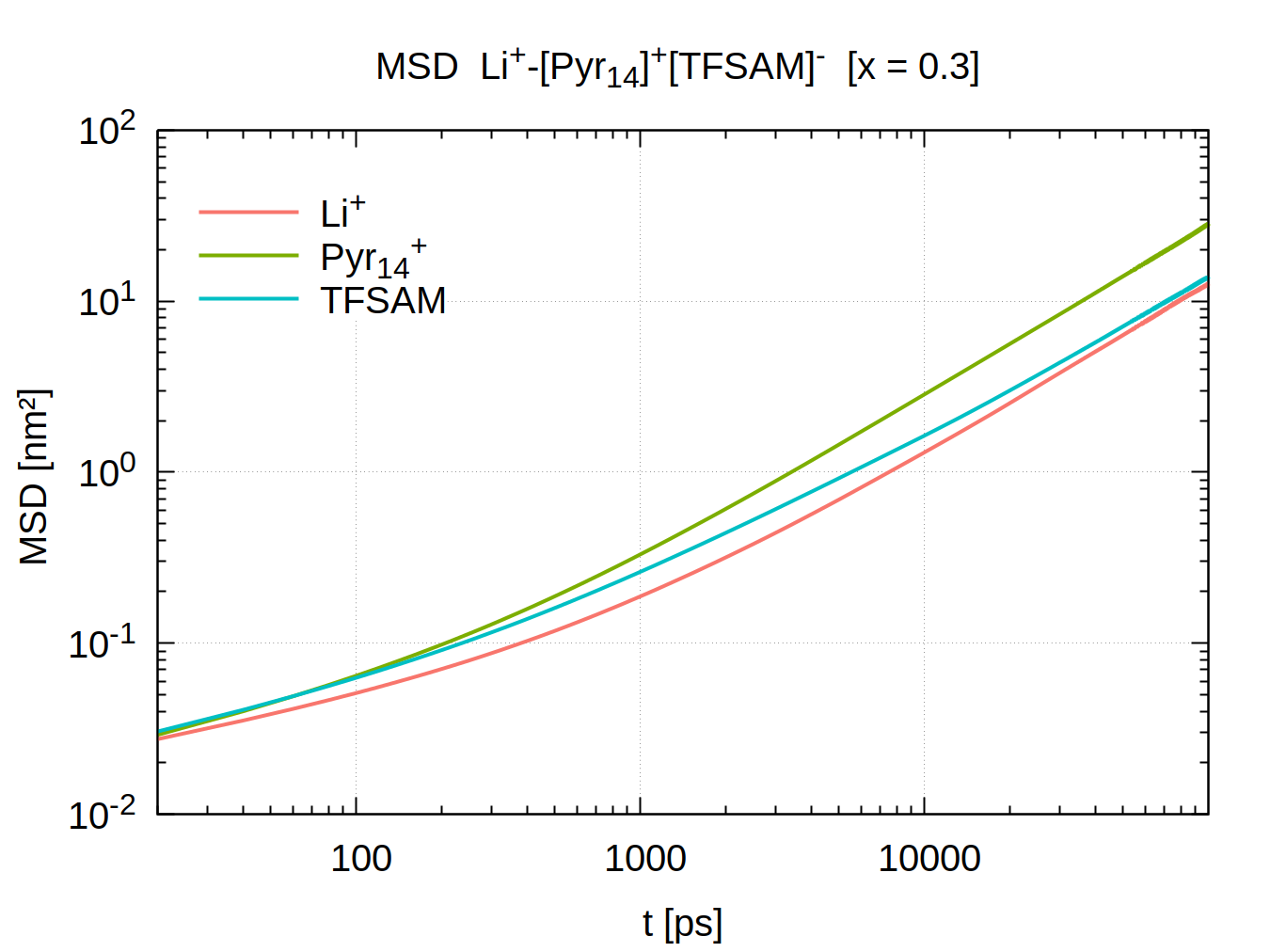}\label{fig:msds_tfsam_x_0_3}}
  \hfill
  \centering
  \subfloat{\includegraphics[width=0.5\textwidth]{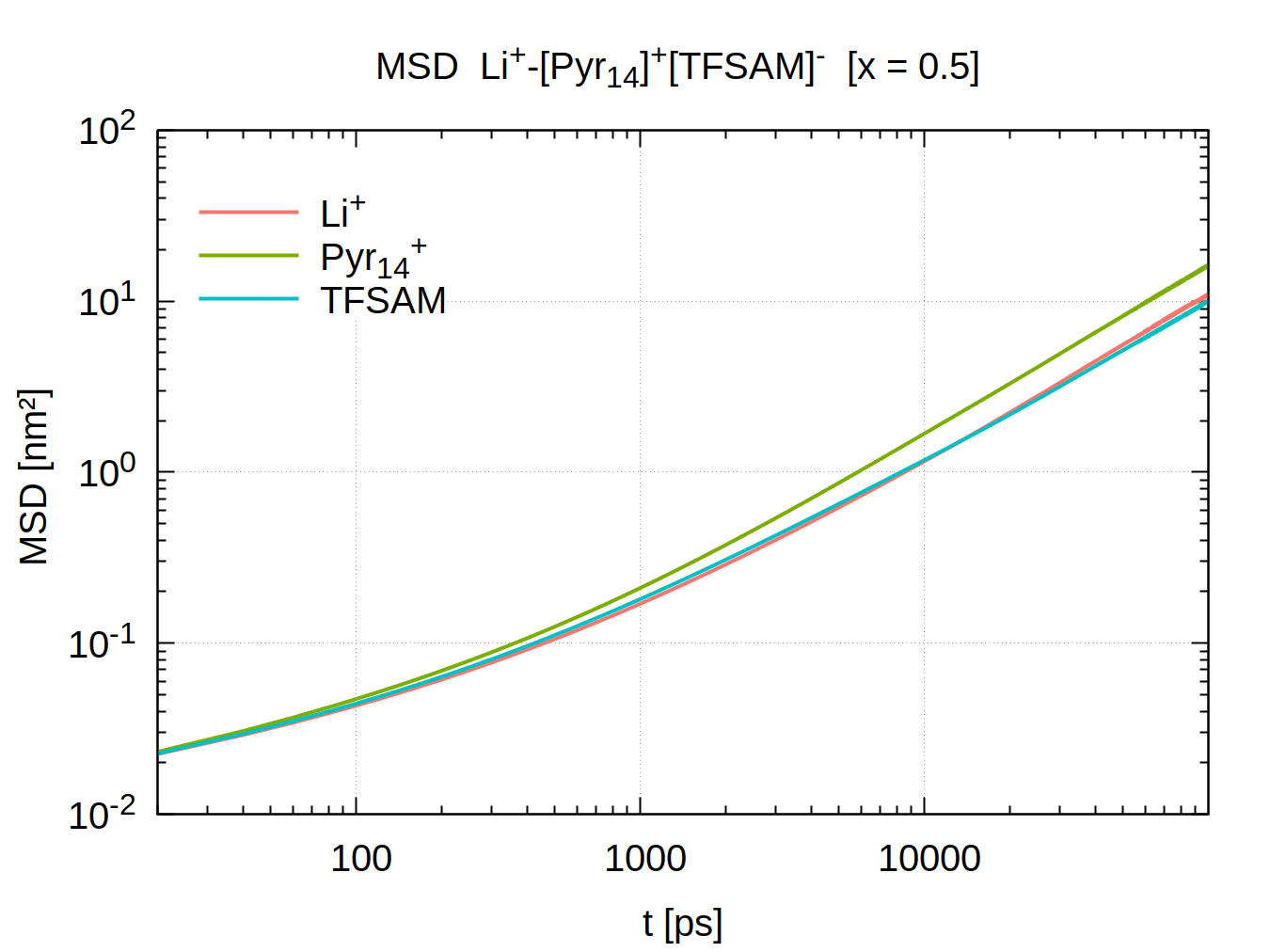}\label{fig:msds_tfsam_x_0_1}}
  \hfill
  \subfloat{\includegraphics[width=0.5\textwidth]{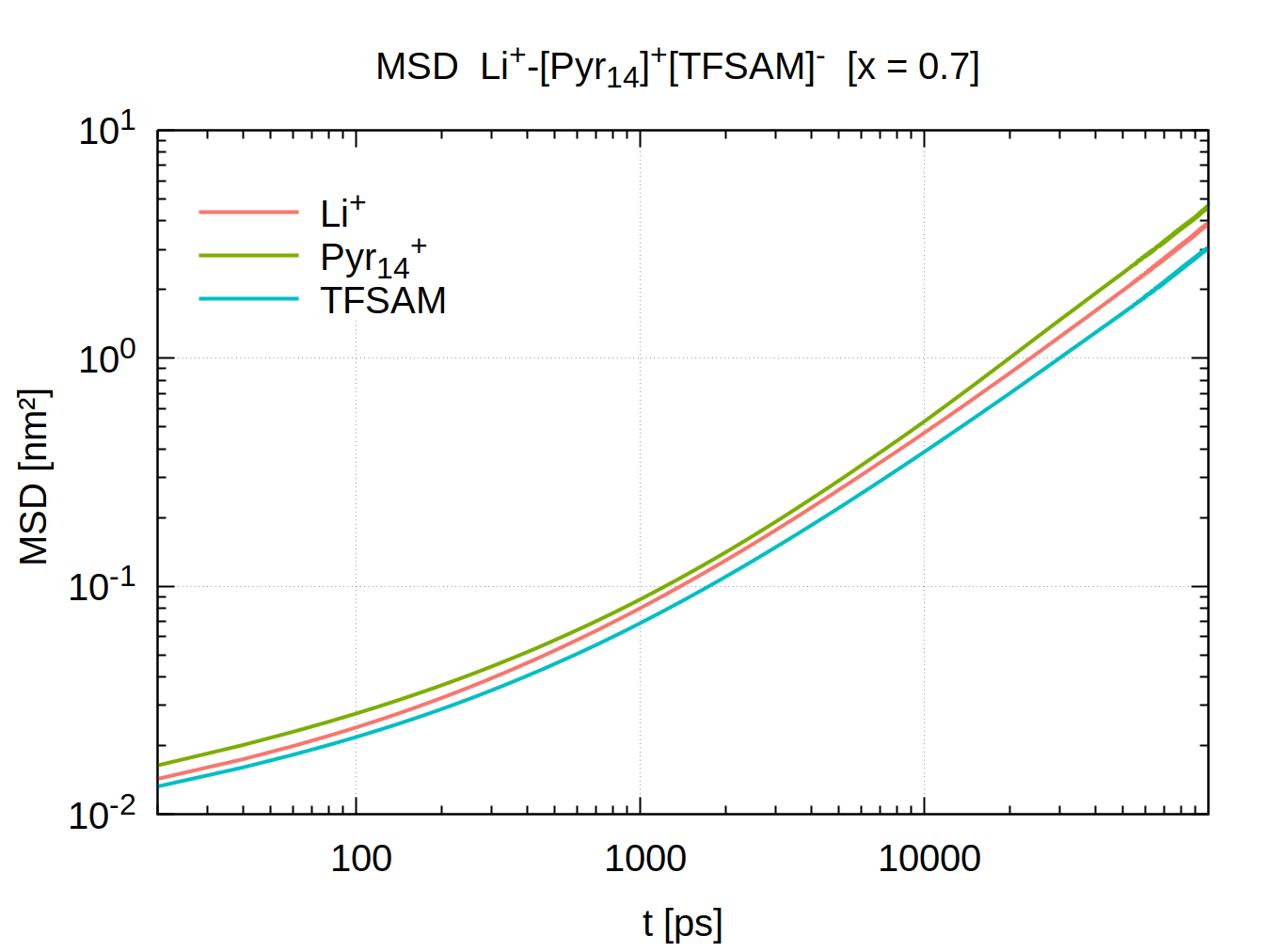}\label{fig:msds_tfsam_x_0_3}}
  \caption{Example mean squared displacements of $\text{Li}^+, \text{Pyr}_{14}^+$ and $\text{TFSAM}^-$ as a function of time for lithium salt contents x=0.1 and x=0.3 (top) as well as x=0.5 and x=0.7 (bottom).}
\end{figure}

\newpage

\textbf{D: Mean residence times}\newline
The mean residence times shown in the main part of the manuscript are computed from the residence time autocorrelation function (ACF) as in reference \cite{zhang2020mechanisms}  :
\begin{equation}
\text{ACF}_{ij}(t) = \dfrac{ \Biggl\langle H_{ij}(t')H_{ij}(t'+t) \Biggr\rangle}{\Biggl\langle H_{ij}(t')H_{ij}(t
) \Biggr\rangle},
\end{equation}
where $H_{ij}$ evaluates to 1 when species $j$ is found within the assigned cutoff to species $i$ and otherwise set to 0. If not mentioned otherwise, the first minimum position of the radial distribution function $\text{g}_{\text{Li}^+-X}(r)$ is employed as the cutoff distance to determine present $\text{Li}^+-X$ binding. The brackets $\langle .. \rangle$ denote the ensemble average over all pairs $ij$ and time origins $t'$.
The ACF is then fitted by a stretched exponential $f(t)$:
\begin{equation}
f(t) = \exp\left( -(t/\tau')^{\beta}\right),
\end{equation}
where $\beta$ and $\tau'$ are the fitting parameters.
The mean residence time $\langle\tau\rangle$ is obtained from the integral:
\begin{equation}
\langle\tau\rangle = \int_{0}^{\infty} dt \exp\left( -(t/\tau')^{\beta}\right) = \dfrac{\tau'}{\beta}\Gamma(1/\beta)
\end{equation}

\begin{figure}[H]
  \centering
  \subfloat{\includegraphics[width=0.5\textwidth]{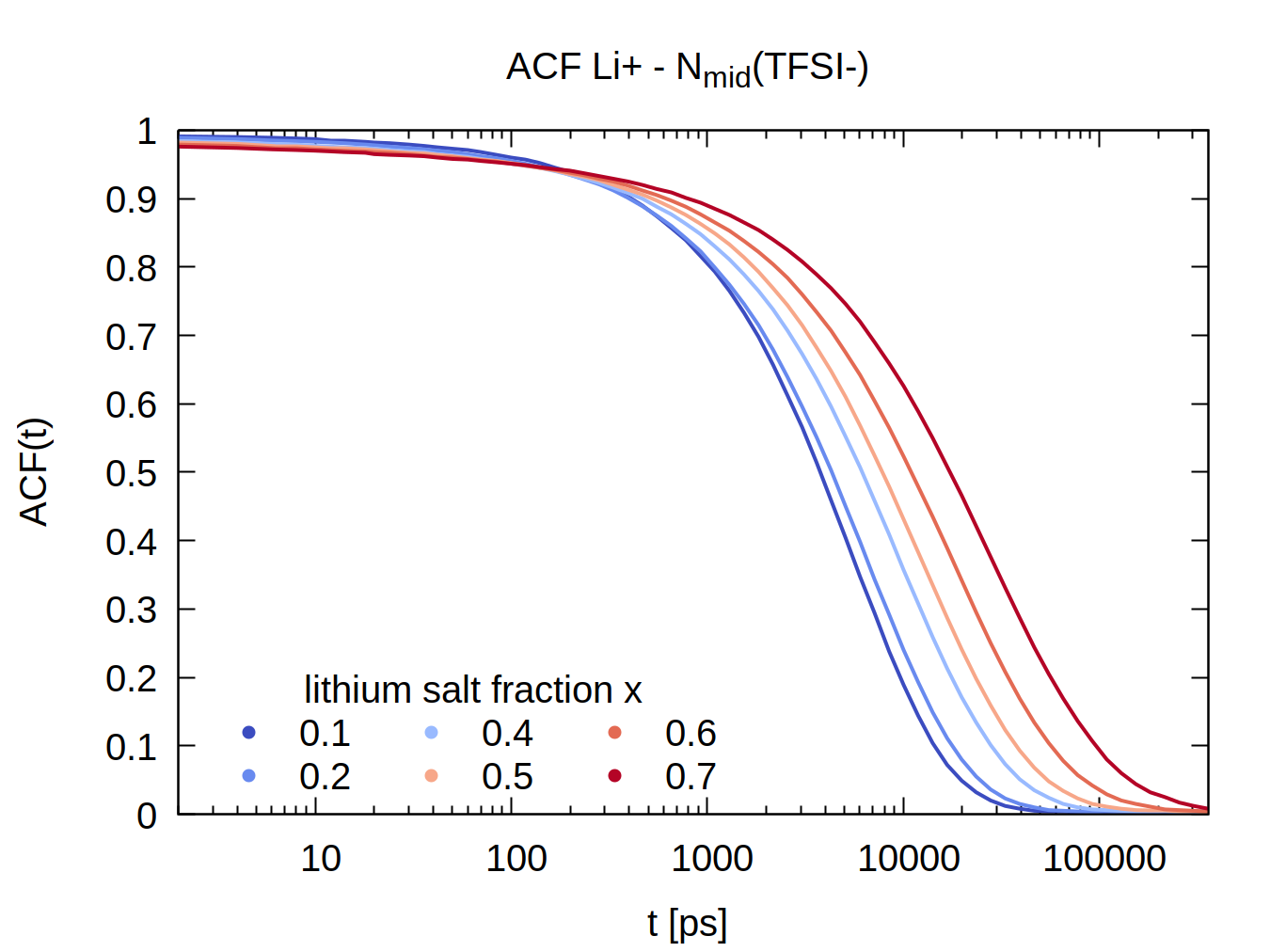}\label{fig:acf_li_n_tfsi}}
  \hfill
  \subfloat{\includegraphics[width=0.5\textwidth]{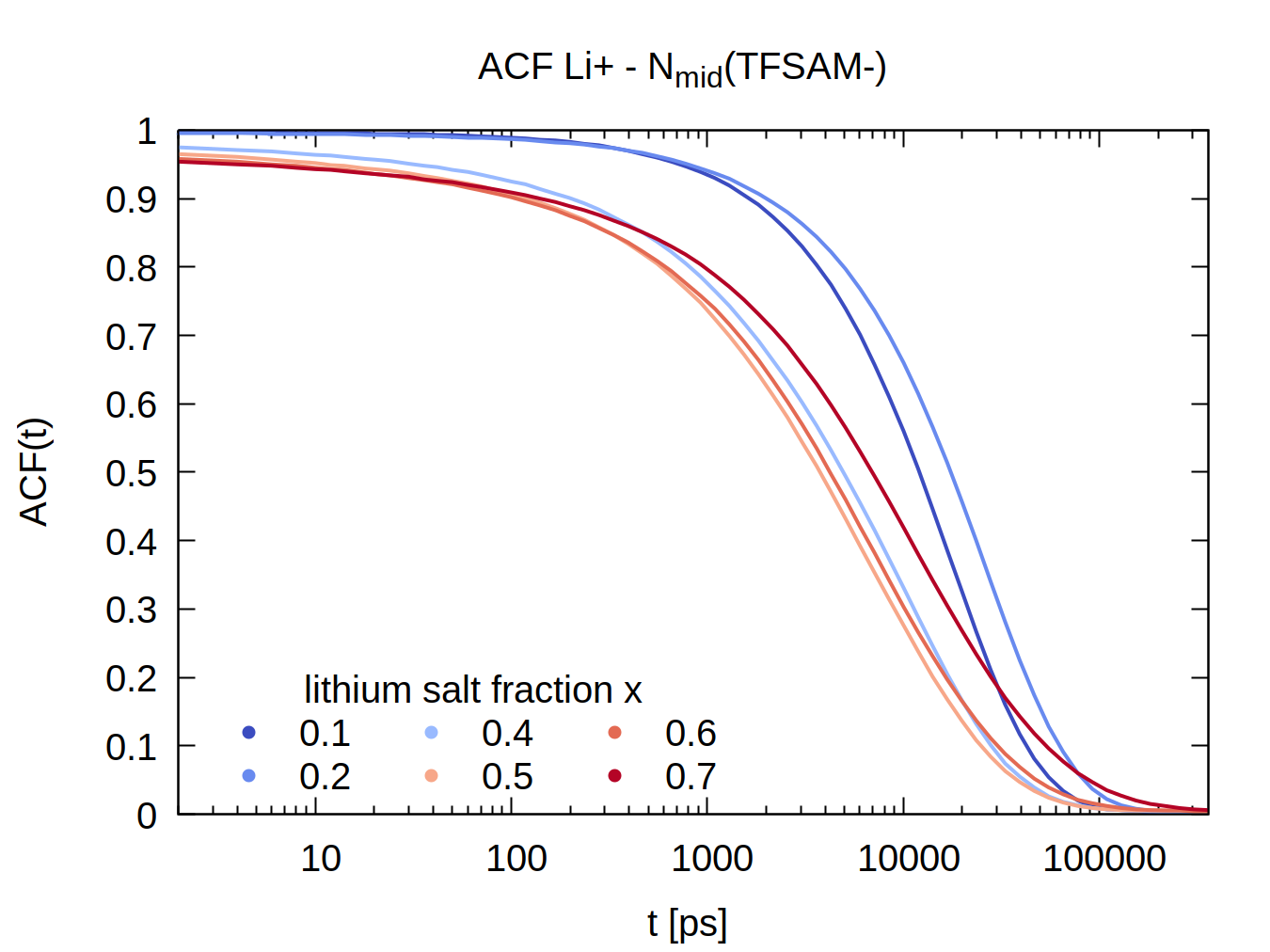}\label{fig:acf_li_n_max_tfsam}}
  \caption{Exemplary overview of the concentration dependence of the autocorrelation functions of $\text{Li}^+-\text{TFSI}^-$ and $\text{Li}^+-\text{TFSAM}^-$ measured via $\text{N}_{\text{mid}}$ employing cutoffs of 5.5\,\angstrom\,\,.}
\end{figure}

\newpage

\textbf{E: Deviation of $\text{MSD}_{\text{Li}^+}(\tau_{\text{Ls}})$ from diffusive dynamics} \newline

\begin{figure}[H]
  \centering
  \includegraphics[width=0.7\textwidth]{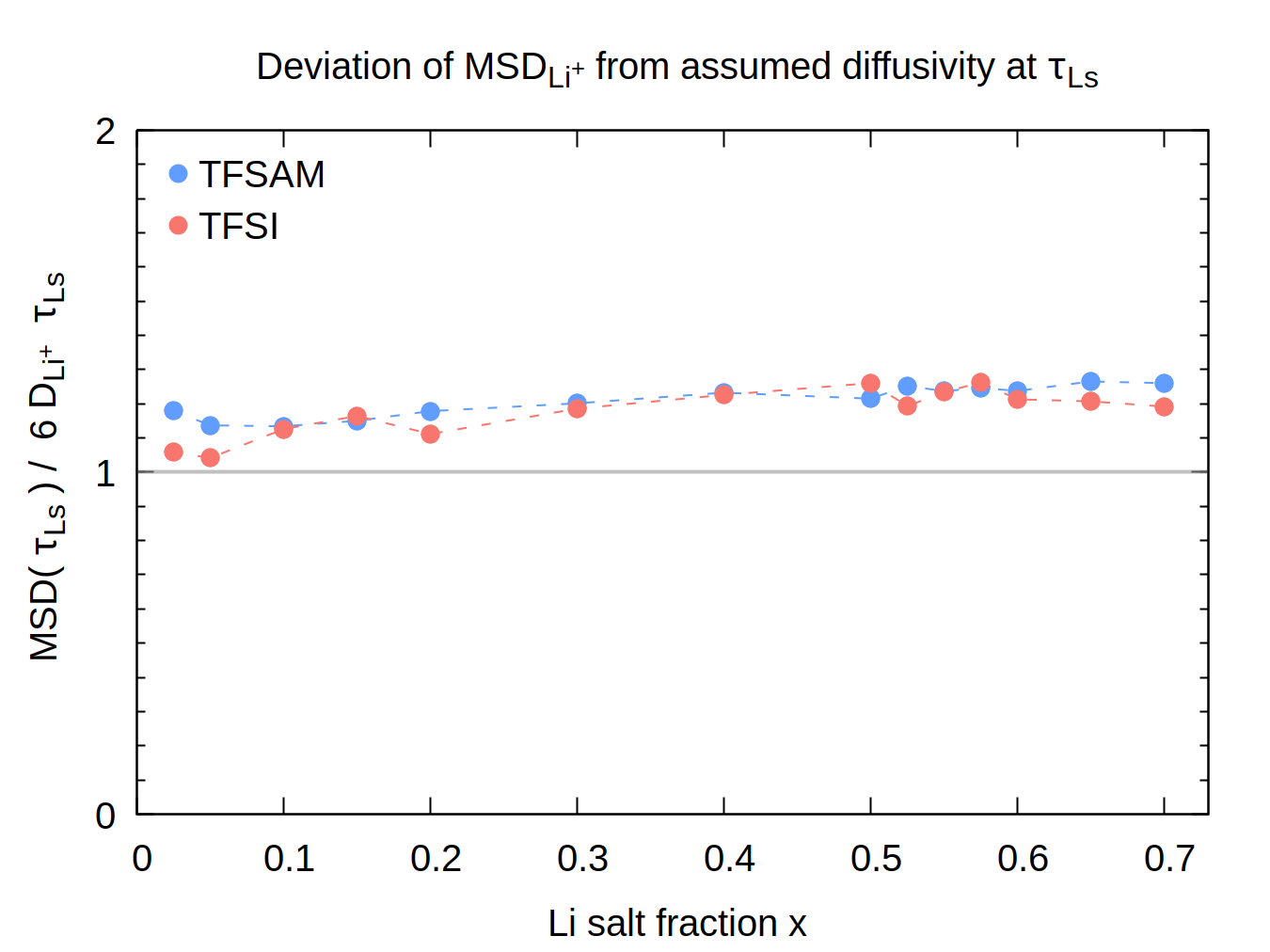}
    \caption{Evaluating the deviation from purely diffusive lithium dynamics at critical time scale $\tau_{\text{Ls}}$, i.e. $\text{MSD}_{\text{Li}^+}(\text{t})\,=\,6\text{D}_{\text{Li}^+}\text{t}$ at times $\tau_{\text{Ls}}$.}
\label{fig:msd_vs_6D_tau_Ls}
\end{figure}

%%%%%%%%%%%%%%%%%%%%%%%%%%%%%%%%%%%%%%%%%%%%%%%%%%%%%%%%%%%%%%%%%%%%%%%%%%%%%%%%%%
%%%%%%%%%%%%%%%%%%%%%%%%%%%%%%%%%%%%%%%%%%%%%%%%%%%%%%%%%%%%%%%%%%%%%%%%%%%%%%%%%%
%%%%%%%%%%%%%%%%%%%%%%%%%%%%%%%%%%%%%%%%%%%%%%%%%%%%%%%%%%%%%%%%%%%%%%%%%%%%%%%%%%
%%%%%%%%%%%%%%%%%%%%%%%%%%%%%%%%%%%%%%%%%%%%%%%%%%%%%%%%%%%%%%%%%%%%%%%%%%%%%%%%%%
%%%%%%%%%%%%%%%%%%%%%%%%%%%%%%%%%%%%%%%%%%%%%%%%%%%%%%%%%%%%%%%%%%%%%%%%%%%%%%%%%%
%%%%%%%%%%%%%%%%%%%%%%%%%%%%%%%%%%%%%%%%%%%%%%%%%%%%%%%%%%%%%%%%%%%%%%%%%%%%%%%%%%
%%%%%%%%%%%%%%%%%%%%%%%%%%%%%%%%%%%%%%%%%%%%%%%%%%%%%%%%%%%%%%%%%%%%%%%%%%%%%%%%%%
%%%%%%%%%%%%%%%%%%%%%%%%%%%%%%%%%%%%%%%%%%%%%%%%%%%%%%%%%%%%%%%%%%%%%%%%%%%%%%%%%%
%%%%%%%%%%%%%%%%%%%%%%%%%%%%%%%%%%%%%%%%%%%%%%%%%%%%%%%%%%%%%%%%%%%%%%%%%%%%%%%%%%
%%%%%%%%%%%%%%%%%%%%%%%%%%%%%%%%%%%%%%%%%%%%%%%%%%%%%%%%%%%%%%%%%%%%%%%%%%%%%%%%%%
%%%%%%%%%%%%%%%%%%%%%%%%%%%%%%%%%%%%%%%%%%%%%%%%%%%%%%%%%%%%%%%%%%%%%%%%%%%%%%%%%%
%%%%%%%%%%%%%%%%%%%%%%%%%%%%%%%%%%%%%%%%%%%%%%%%%%%%%%%%%%%%%%%%%%%%%%%%%%%%%%%%%%

\newpage
\textbf{F: $p(\Delta v_{\parallel})$ for various lag times and lithium subensembles}\newline

\begin{figure}[H]
  \centering
  \subfloat{\includegraphics[width=1.0\textwidth]{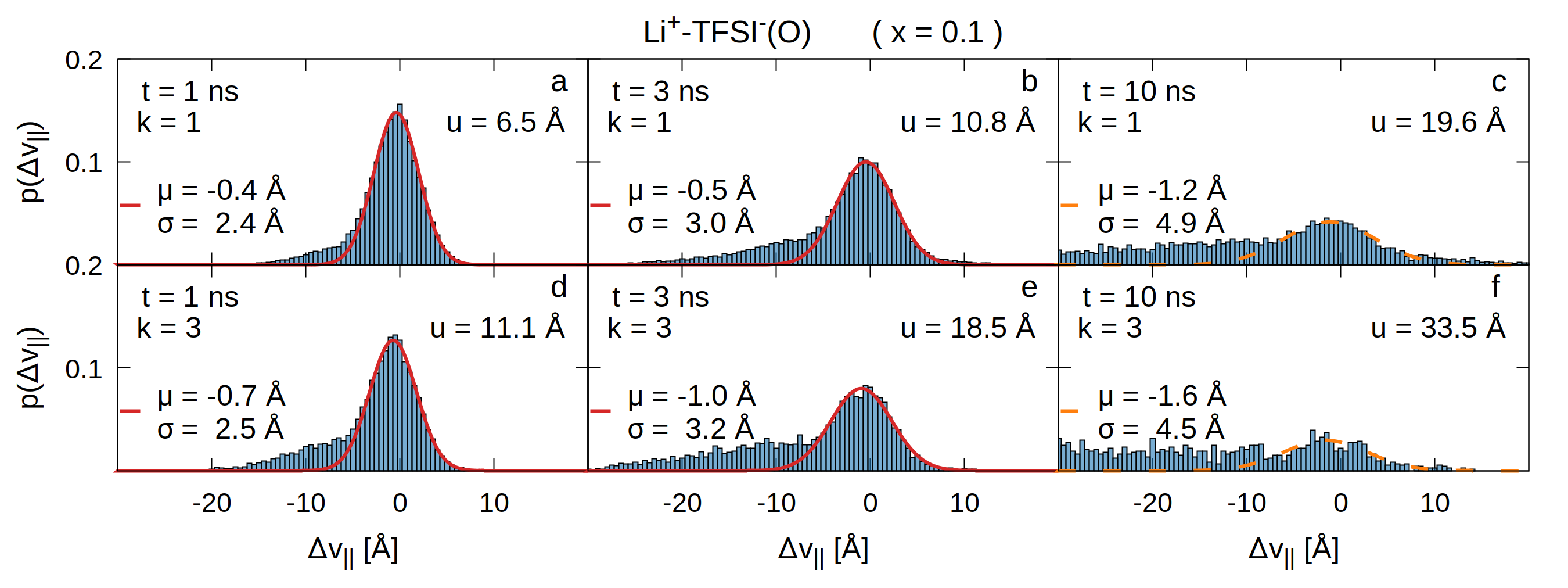}}
  \hfill
  \subfloat{\includegraphics[width=1.0\textwidth]{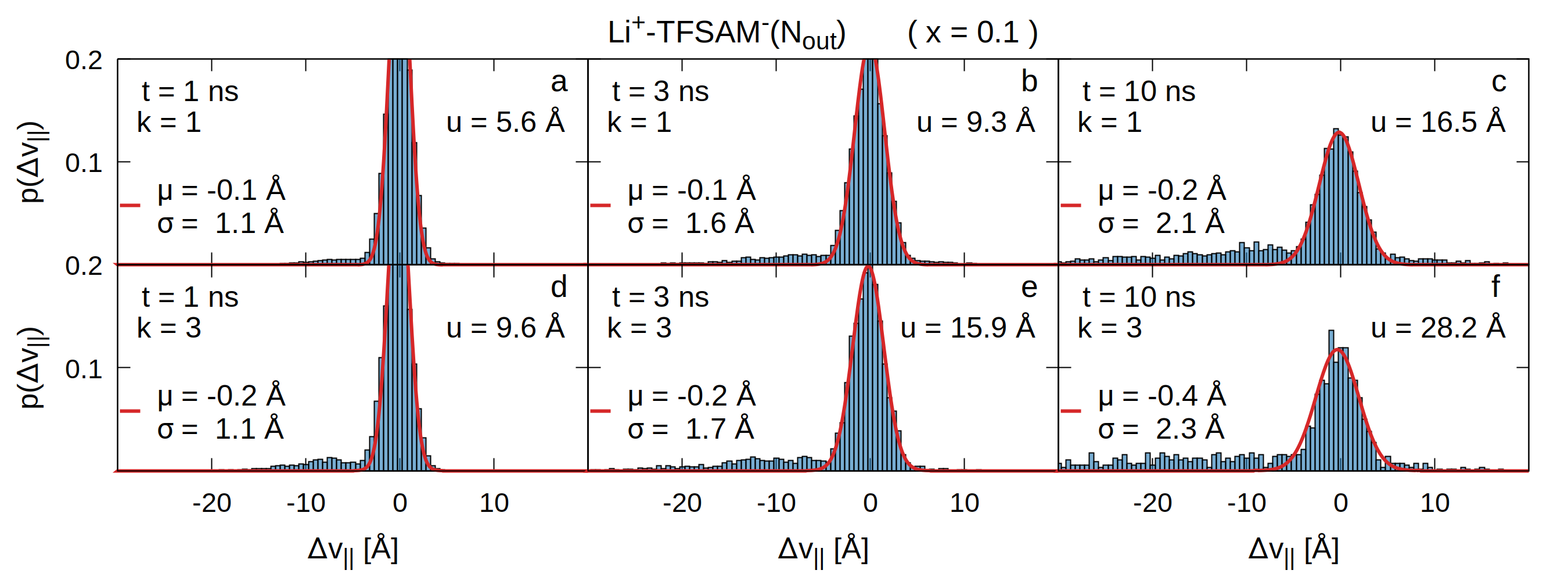}}
 \caption{Distributions $p(\Delta v_{\parallel})$ of $\text{TFSI}^-$ and $\text{TFSAM}^-$ for the subensembles of $u^2=k\cdot\langle u^2\rangle$ with $k$\,=\,1 (a,b and c) and 3 (d, e and f) for various lag times $t$ and salt concentrations x using an upper threshold tolerance $u^2_{\text{r}} = 1.5\cdot k \cdot\langle u^2 \rangle$.}
 
 \label{fig:delta_v_para_tfsam_tfsi}
\end{figure}

The panels in Figure S\ref{fig:delta_v_para_tfsam_tfsi} show the distributions $p(\Delta v_{\parallel})$ of $\text{TFSI}^-$ and $\text{TFSAM}^-$ relative distances in direction of the lithium displacement for salt concentrations x\,=\,0.05, 0.1, 0.2 and 0.5. The upper panels (a, b and c) display the histograms obtained for the anions whose designated lithium ions exhibit a squared displacement equal to the lithium mean squared displacement at the corresponding time $t$, i.e. $u^2 = 1 \cdot \langle u^2 \rangle$. The lower panels (d, e and f) measure $p(\Delta v_{\parallel})$ for the subensemble of lithium ions which achieved a squared displacement $u^2 = 3 \cdot \langle u^2 \rangle$. \\
In order to extent the statistical analysis of the subensembles to a larger data set, we introduce a tolerance interval $\{ u^2_{\text{l}} \leq u^2 \leq u^2_{\text{r}}\}$, whose upper and lower boundaries $u^2_{\text{l}}$ and $u^2_{\text{r}}$ are set in such a way that the sampled average $u^2$ of the subensemble corresponds to the target $k\cdot\langle u^2\rangle$ (see Table S\ref{table:threshold_values}). Because the lithium ions' individual squared displacements are normally distributed the threshold boundaries cannot be chosen symmetrically.\\
To characterize the peak positions and widths, we proceed according to the following protocol:
\begin{enumerate}
    \item $\Delta v_{\parallel}$ is measured according to Equation 5 in the main manuscript and discretized employing a bin width of 0.1\,\angstrom\, and normalized. For reasons of visual appearance the histograms shown in the panels employ a bin width of 0.5\,\angstrom.
    
    \item The obtained histogram $p(\Delta v_{\parallel})$ is empirically fitted by a Gaussian function \newline $g(x)\,=\,\Tilde{a}\cdot\exp\left( \dfrac{(x-\Tilde{\mu})^2}{2\Tilde{\sigma}^2}\right)$ \newline with the amplitude $\Tilde{a}$, the expected value $\Tilde{\mu}$ and variance $\Tilde{\sigma}^2$.
    
    \item Since anions naturally decouple from the lithium ion's dynamics upon detachment, it seems plausible that the distributions exhibit a skew towards negative $\Delta v_{\parallel}$ values. We find that an increasing amount of initial lithium-anion pairs has separated over time and is reflected in a growing tail of $p(\Delta v_{\parallel})$. To separate the peak features belonging to the coupled lithium-anion dynamics from the overlapping distribution of dissociating dynamics, we restrict the left-hand side of the fit interval to $\left[-\Tilde{\sigma}+\Tilde{\mu},\infty \right)$. The peak is refitted by a Gaussian function (red) $g(x)\,=\,{a}\cdot\exp\left( \dfrac{(x-{\mu})^2}{2{\sigma}^2}\right)$ with the amplitude $a$, the expected value ${\mu}$ and variance ${\sigma}^2$. 
    
    \item Due to the significantly shorter mean residence times $\tau_{\text{Li}^+-\text{TFSI}^-}$ compared to  $\tau_{\text{Li}^+-\text{TFSAM}^-}$ as discussed in the main manuscript, $p(\Delta v_{\parallel})$ is more disintegrated for $\text{TFSI}^-$ at the longest analysed lag time of $t=10\,$ns. Since the fit protocol step (3) fails to expose the peak originating from yet retained coupled dynamics, we fitted the peak manually (dashed orange).
\end{enumerate}

\begin{table}[h!]
\centering
\begin{tabular}{||c c c c||} 
 \hline
 $k$ &  $u^2$  & $u^2_{\text{l}}$ & $u^2_{\text{r}}$ \\[0.5ex] 
 \hline\hline
 1 &  $1 \,\cdot\, \langle u^2 \rangle $ & $u^2\,/\,1.33$ & $u^2\,\cdot\,1.3$ \\ 
 3 &  $3 \,\cdot\, \langle u^2 \rangle $ & $u^2\,/\,1.22$ & $u^2\,\cdot\,1.3$ \\ 
 1 &  $1 \,\cdot\, \langle u^2 \rangle $ & $u^2\,/\,1.58$ & $u^2\,\cdot\,1.5$ \\ 
 3 &  $3 \,\cdot\, \langle u^2 \rangle $ & $u^2\,/\,1.31$ & $u^2\,\cdot\,1.5$ \\ [1ex] 
 \hline
\end{tabular}
\caption{Overview of numerically determined lower boundaries $u^2_{\text{l}}$ for a given distance scaling factor $k$ and upper tolerance thresholds $u^2_{\text{r}}$ of either 1.3 or 1.5, \textit{i.e.}, 30 or 50 percent tolerance for lithium ions covering a squared distance larger than the target $u^2$.}
\label{table:threshold_values}
\end{table}
At the very bottom of Figure S\ref{fig:delta_v_para_tfsam_tfsi}, $p(\Delta v_{\parallel})$ is additionally shown for the explicit lithium binding sites provided by the respective anion, \textit{i.e.}, $\text{TFSI}^-(\text{O})$ and  $\text{TFSAM}^-(\text{N}_{\text{out}})$ at x\,=\,0.1.\\

\begin{figure}[H]
  \centering
  \subfloat{\includegraphics[width=0.9\textwidth]{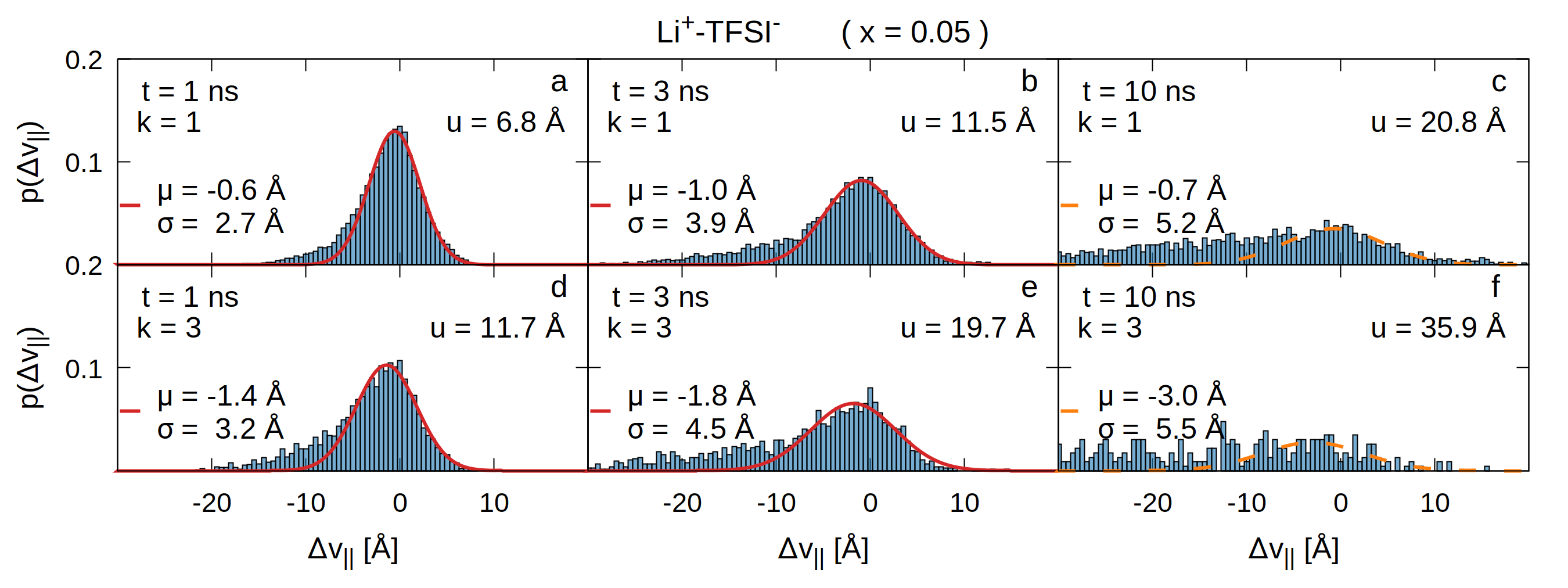}}
  \hfill
  \subfloat{\includegraphics[width=0.9\textwidth]{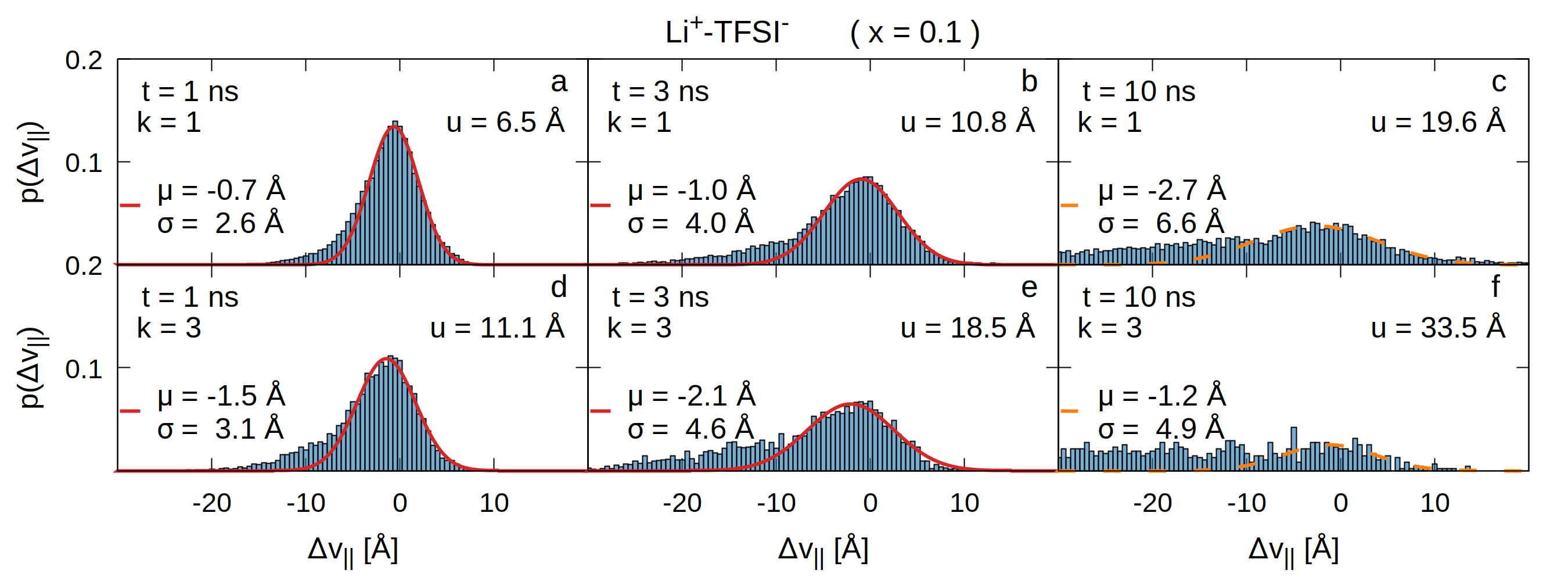}}
  \hfill
  \centering
  \subfloat{\includegraphics[width=0.9\textwidth]{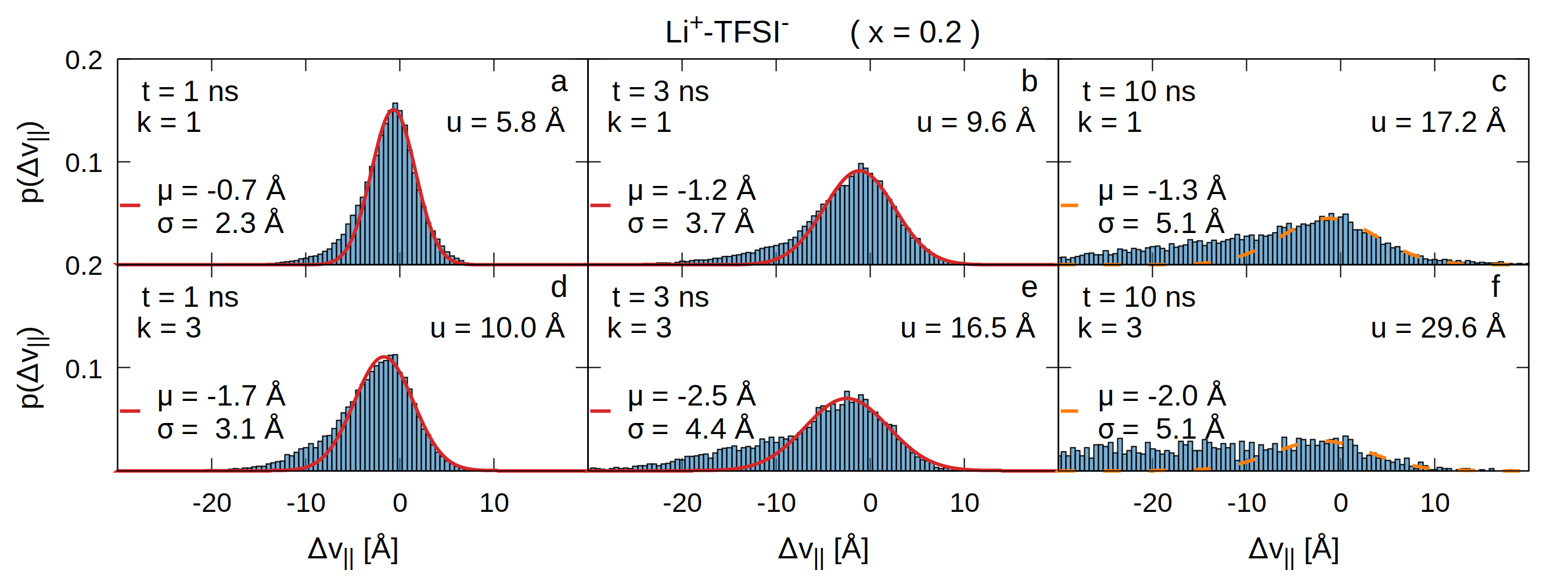}}
  \hfill
  \subfloat{\includegraphics[width=0.9\textwidth]{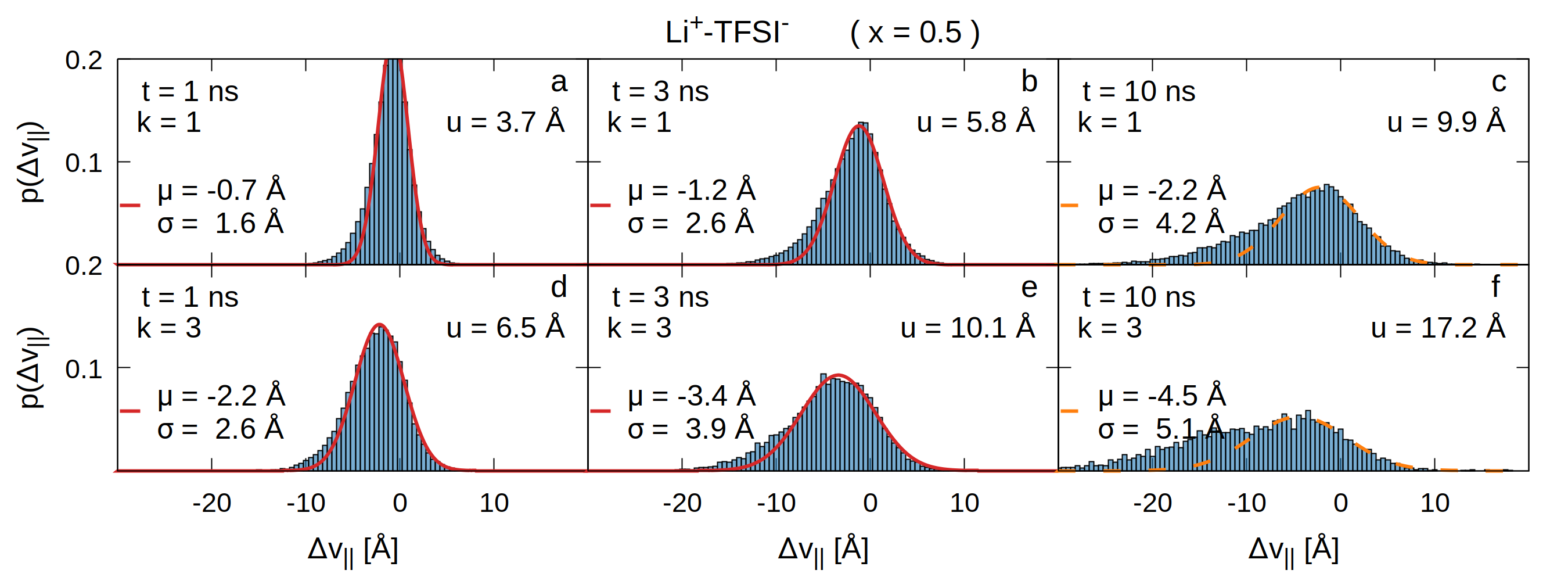}}
\end{figure}

\begin{figure}[H]
  \centering
  \subfloat{\includegraphics[width=0.9\textwidth]{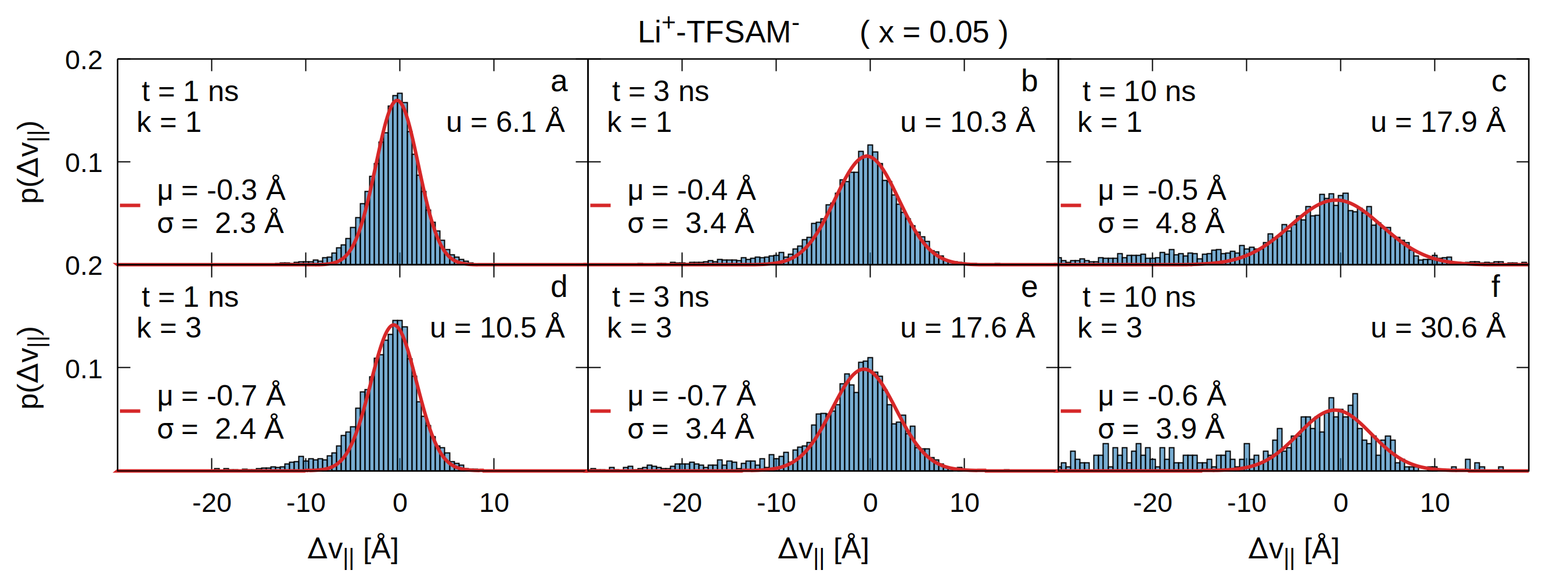}}
  \hfill
  \subfloat{\includegraphics[width=0.9\textwidth]{figures/SI_PANEL_delta_v_para_1ns_to_10ns_1_3_k_v_para_TFSAM_NH_x_0_1_binning_tolerance_1_5.png}}
  \hfill
  \centering
  \subfloat{\includegraphics[width=0.9\textwidth]{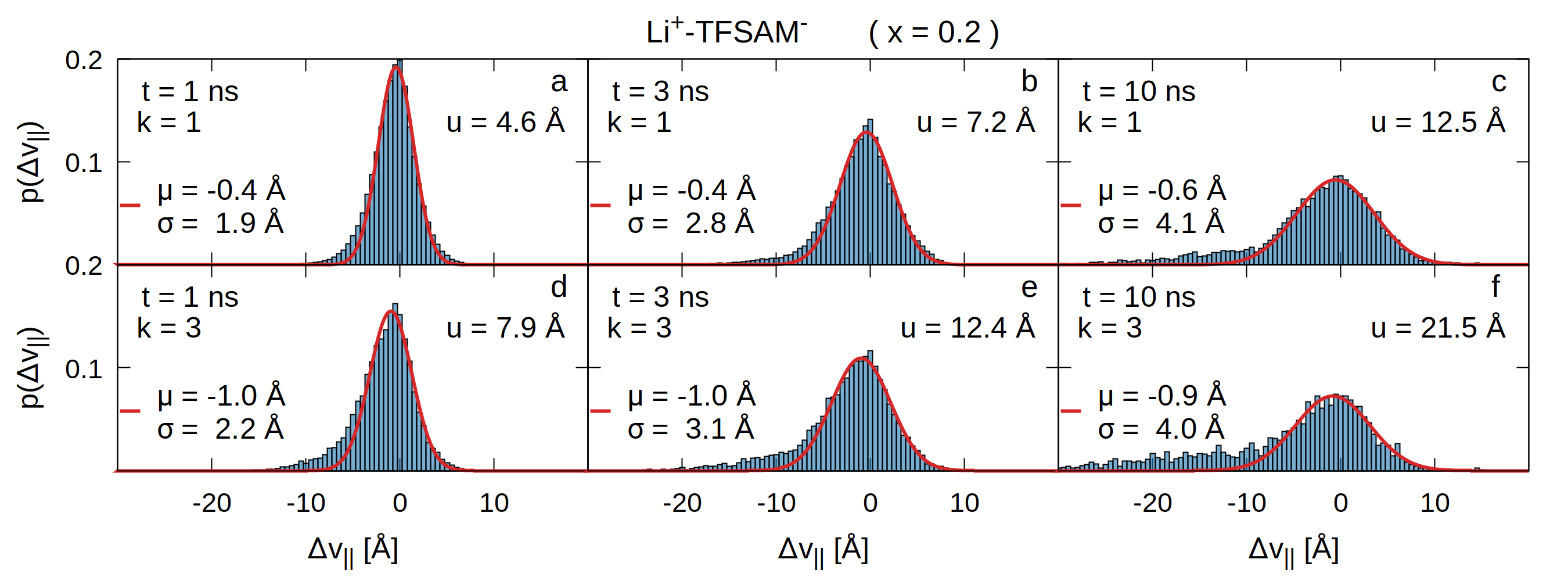}}
  \hfill
  \subfloat{\includegraphics[width=0.9\textwidth]{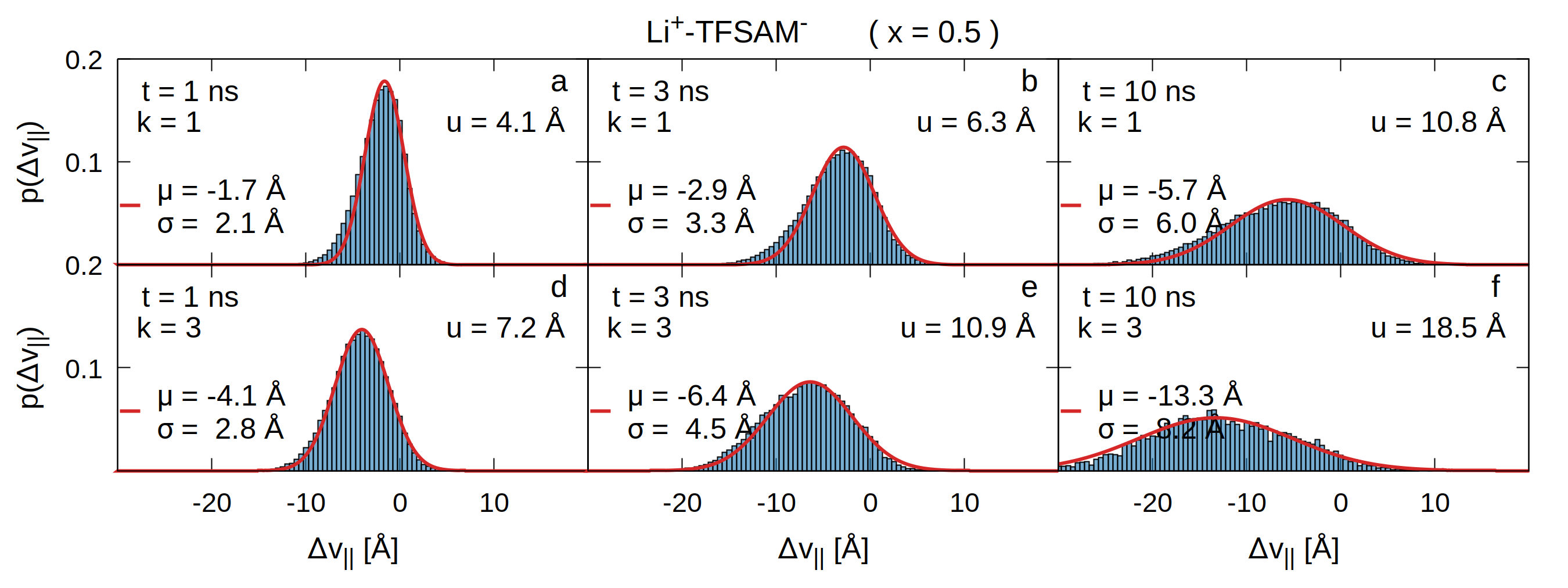}}
 \end{figure}

\begin{figure}[H]

  \centering
  \subfloat{\includegraphics[width=0.9\textwidth]{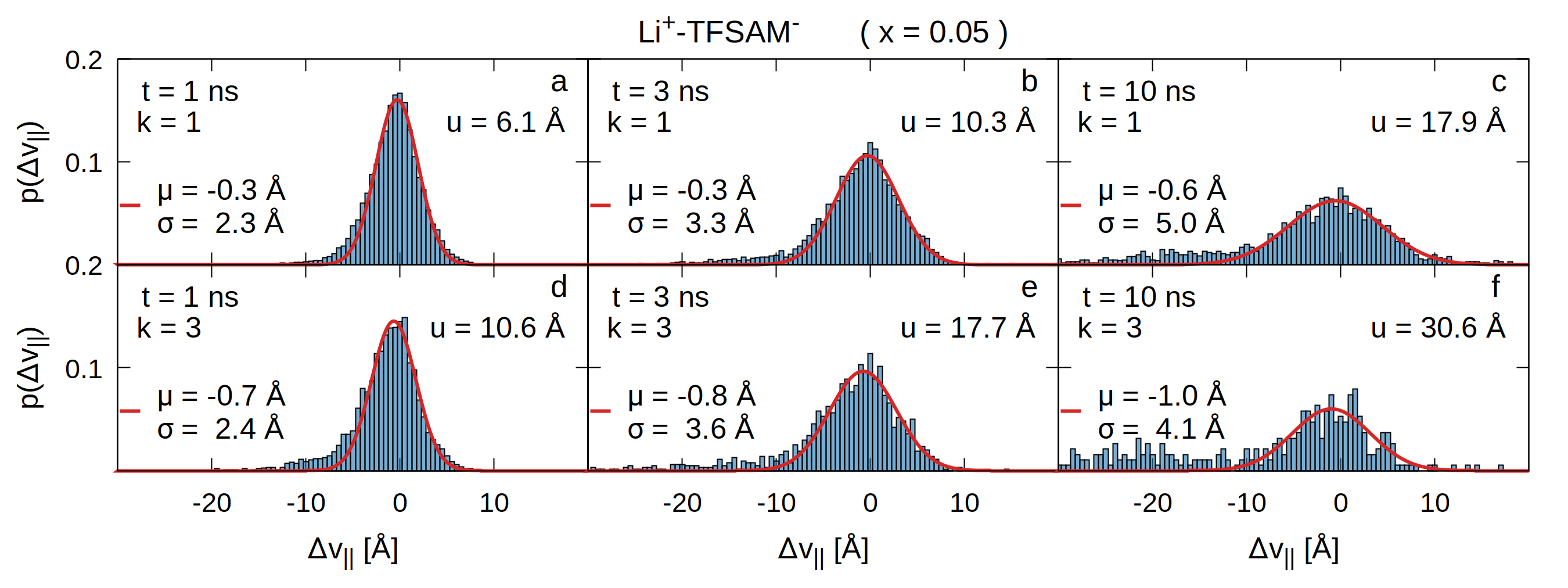}}
  \hfill
  \subfloat{\includegraphics[width=0.9\textwidth]{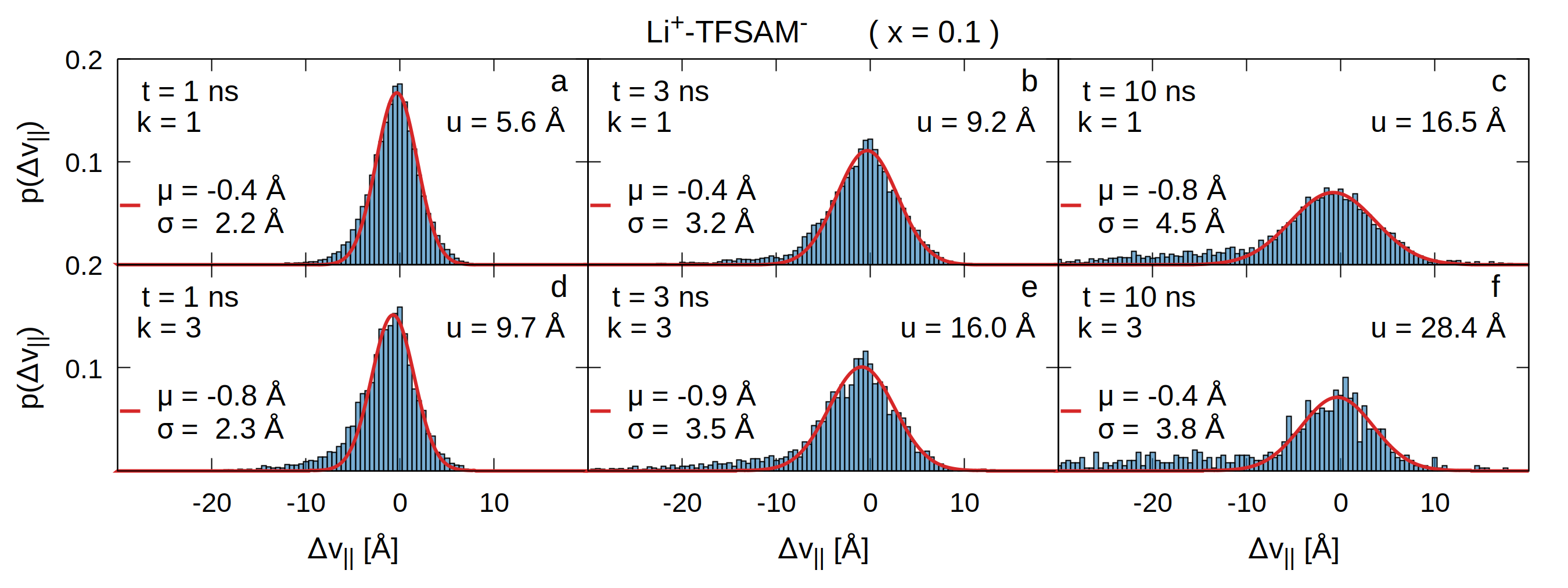}}
  \hfill
  \centering
  \subfloat{\includegraphics[width=0.9\textwidth]{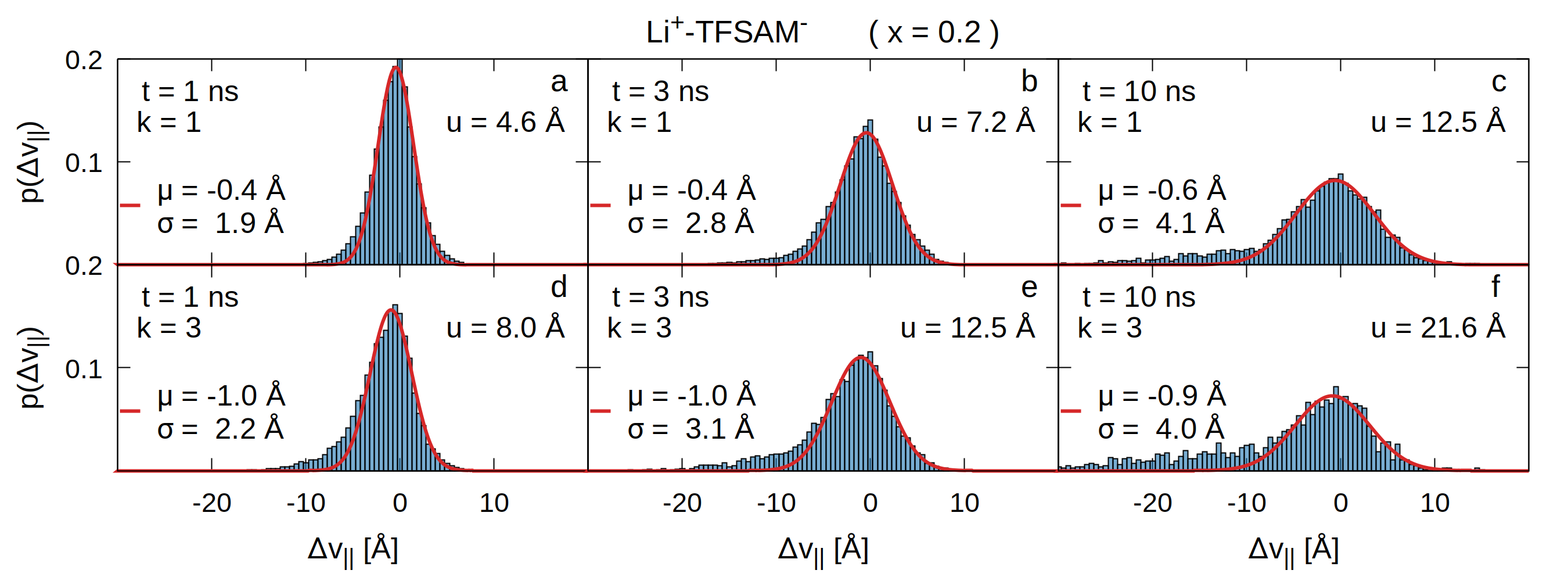}}

 \caption{Distributions $p(\Delta v_{\parallel})$ of $\text{TFSAM}^-$ at a salt concentration x\,=\,0.1 for the subensembles of $u^2=k\cdot\langle u^2\rangle$ with $k$\,=\,1 (a, b and c) and 3 (d, e and f) using an upper threshold tolerance $u^2_{\text{r}} = 1.3\cdot k \cdot\langle u^2 \rangle$.}
 
 \label{fig:delta_v_para_tfsam_tolerance_1_3}
\end{figure}

%%%%%%%%%%%%%%%%%%%%%%%%%%%%%%%%%%%%%%%%%%%%%%%%%%%%%%%%%%%%%%%%%%%%%%%%%%%%%%%%%%
%%%%%%%%%%%%%%%%%%%%%%%%%%%%%%%%%%%%%%%%%%%%%%%%%%%%%%%%%%%%%%%%%%%%%%%%%%%%%%%%%%
%%%%%%%%%%%%%%%%%%%%%%%%%%%%%%%%%%%%%%%%%%%%%%%%%%%%%%%%%%%%%%%%%%%%%%%%%%%%%%%%%%
%%%%%%%%%%%%%%%%%%%%%%%%%%%%%%%%%%%%%%%%%%%%%%%%%%%%%%%%%%%%%%%%%%%%%%%%%%%%%%%%%%
%%%%%%%%%%%%%%%%%%%%%%%%%%%%%%%%%%%%%%%%%%%%%%%%%%%%%%%%%%%%%%%%%%%%%%%%%%%%%%%%%%
%%%%%%%%%%%%%%%%%%%%%%%%%%%%%%%%%%%%%%%%%%%%%%%%%%%%%%%%%%%%%%%%%%%%%%%%%%%%%%%%%%

\newpage

\textbf{G: Relationship of Gaussian peak parameters $\mu$ and $\sigma^2$ : A particle-spring-model approach}

With the aim to understand why the Gaussian peak centers $\mu$ are not positioned at $\Delta v_{\parallel}=0$\,\angstrom\, but increasingly shifted for a larger distance scaling factor $k$, we propose a simple thought experiment:\\
We idealize a lithium-anion-pair to behave like two particles that are coupled through a harmonic interaction, \textit{i.e.}, connected by a spring. Since we analyse the anion dynamics in the reference frame of the lithium ion, the coupled dynamics reduce, firstly, to one spatial dimension and, secondly, can be conveniently expressed by the system's collective and relative displacements 
\begin{equation}
    \begin{aligned}
        X &= \dfrac{1}{2}\cdot \left(u + v_{\parallel} \right) \\
        \Gamma &= \dfrac{1}{2}\cdot \left(u - v_{\parallel} \right) .
    \end{aligned}
\end{equation}
The collective coordinate $X$ thus describes the diffusive motion of the coupled particles whereas the relative coordinate $\Gamma$ measures the fluctuations of the particles' relative positions.
Assuming a normal distribution for both, \textit{i.e.}, $\mathcal{N}(\mu_{X},\sigma_{X}^2)$ and $\mathcal{N}(\mu_{\Gamma},\sigma_{\Gamma}^2)$, we can deduce the conditional probability distribution $P(v_{\parallel}-u|u)$ that we sampled in the histograms $p(\Delta v_{\parallel})$, starting with $P(v_{\parallel}|u) = P(u,v_{\parallel})/P(u)$:
\begin{equation}
    \begin{aligned}
        P(u,v_{\parallel}) &\propto \mathcal{N}(\mu_{X},\sigma_{X}^2) \cdot \mathcal{N}(\mu_{\Gamma},\sigma_{\Gamma}^2) \\
        &\propto \exp\left(  -\dfrac{1}{2}\dfrac{(u+v_{\parallel})^2}{4\sigma_{X}^2}\right) \cdot \exp\left( -\dfrac{1}{2}\dfrac{(u-v_{\parallel})^2}{4\sigma_{\Gamma}^2}   \right) \\
        &\propto \exp\left( -\dfrac{1}{2}\cdot \dfrac{1}{ 4\cdot AB / (A+B)^2} \left[   v_{\parallel} - \underbrace{ \dfrac{A-B}{A+B}\cdot u }_{\mu_{v_{\parallel}}}    \right]^2  \right)\cdot \exp(..) \quad \text{with}\quad A =\sigma_{X}^2 \quad B = \sigma_{\Gamma}^2   
    \end{aligned}
\end{equation}
Therefore, it holds for $\mu \dot{=} \mu_{\Delta v_{\parallel}}$:
\begin{equation}
    \begin{aligned}
     \mu &= \mu_{v_{\parallel}} - u \\
                                &= \dfrac{ \sigma_{X}^2  -\sigma_{\Gamma}^2 }{\sigma_{X}^2  +\sigma_{\Gamma}^2}\cdot u - u \\
                                &= -2 \cdot \dfrac{\sigma_{\Gamma}^2}{\sigma_{\Gamma}^2+\sigma_{X}^2 } \cdot u
     \end{aligned}
     \label{eq:mu_delta_v_raw}
\end{equation}
The variances $\sigma_{X}^2$ and $\sigma_{\Gamma}^2$ are related to the observables $\langle u^2\rangle$ and $\sigma^2 \dot{=}\,\sigma^2_{\Delta v_{\parallel}}$ which are accessible through our analysis:
\begin{equation}
    \sigma_{u}^2 = \langle u^2 \rangle - \underbrace{\langle u \rangle^2}_{0} \doteq \sigma_{X}^2 + \sigma_{\Gamma}^2 \qquad\land\qquad \sigma_{\Gamma}^2  = \dfrac{1}{4}\sigma^2.
\end{equation}
Consequently, we can rewrite Equation S\ref{eq:mu_delta_v_raw}:
\begin{equation}
    \begin{aligned}
     \mu &= - \dfrac{u}{2} \cdot \dfrac{\sigma^2 }{\langle u^2\rangle} \\
     &= - \dfrac{\sqrt{k}}{2} \cdot \dfrac{\sigma^2 }{\sqrt{\langle u^2\rangle}} \qquad \text{for} \qquad u^2\,=\,k\cdot \langle u^2\rangle.
     \end{aligned}
     \label{eq:mu_delta_v_para_scaling_relation}
\end{equation}
The expression "masterscaling" in the main manuscript refers to a simple rearrangement of Equation S\ref{eq:mu_delta_v_para_scaling_relation} :
\begin{equation}
    1 = - 2 \cdot \dfrac{ \mu \sqrt{\langle u^2\rangle}   }{\sqrt{k} \sigma^2}.
    \label{eq:masterscaling_peak_parameters}
\end{equation}

\begin{figure}[H]
  \centering
  \subfloat{\includegraphics[width=0.5\textwidth]{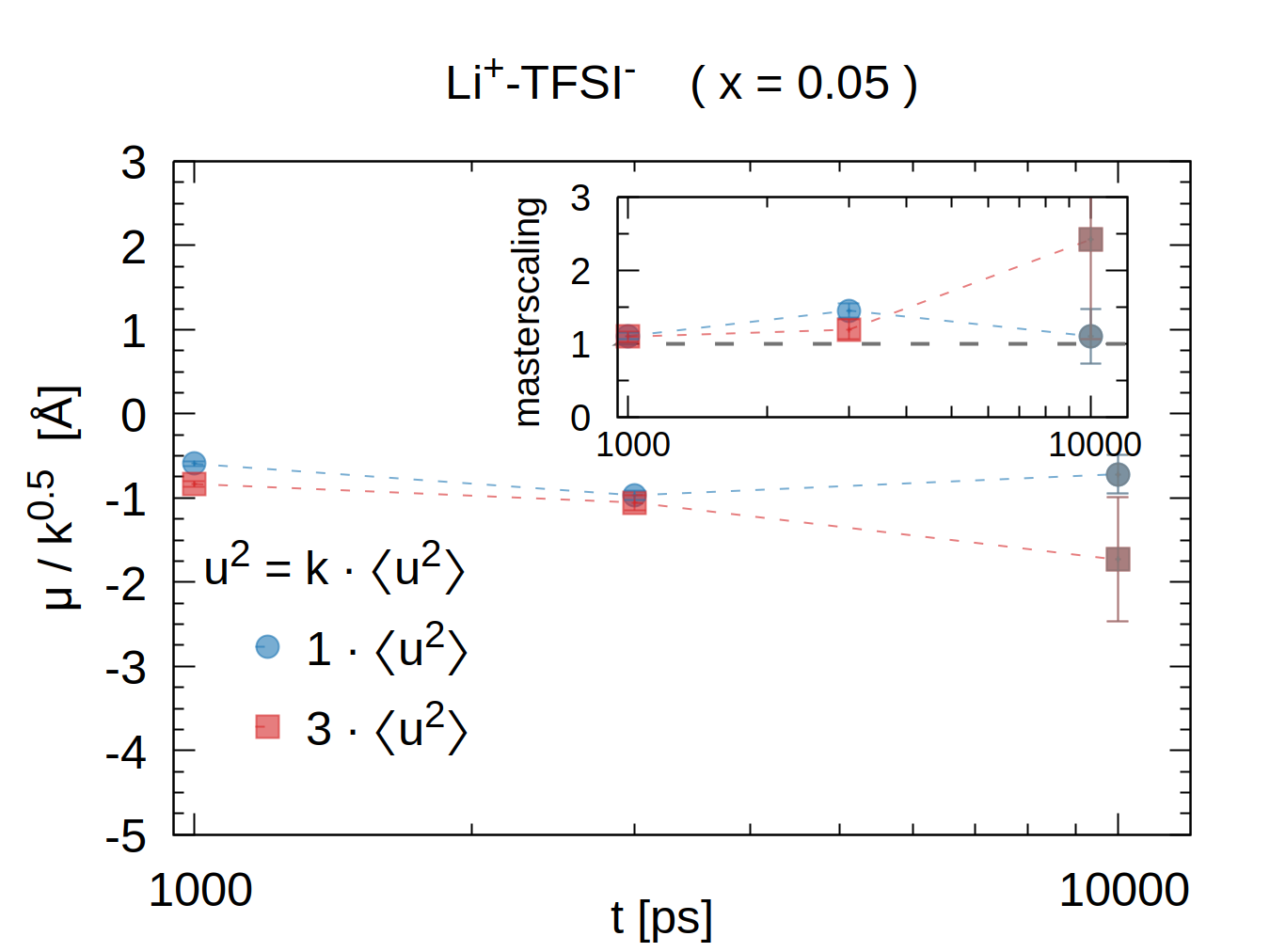}}
  \hfill
  \subfloat{\includegraphics[width=0.5\textwidth]{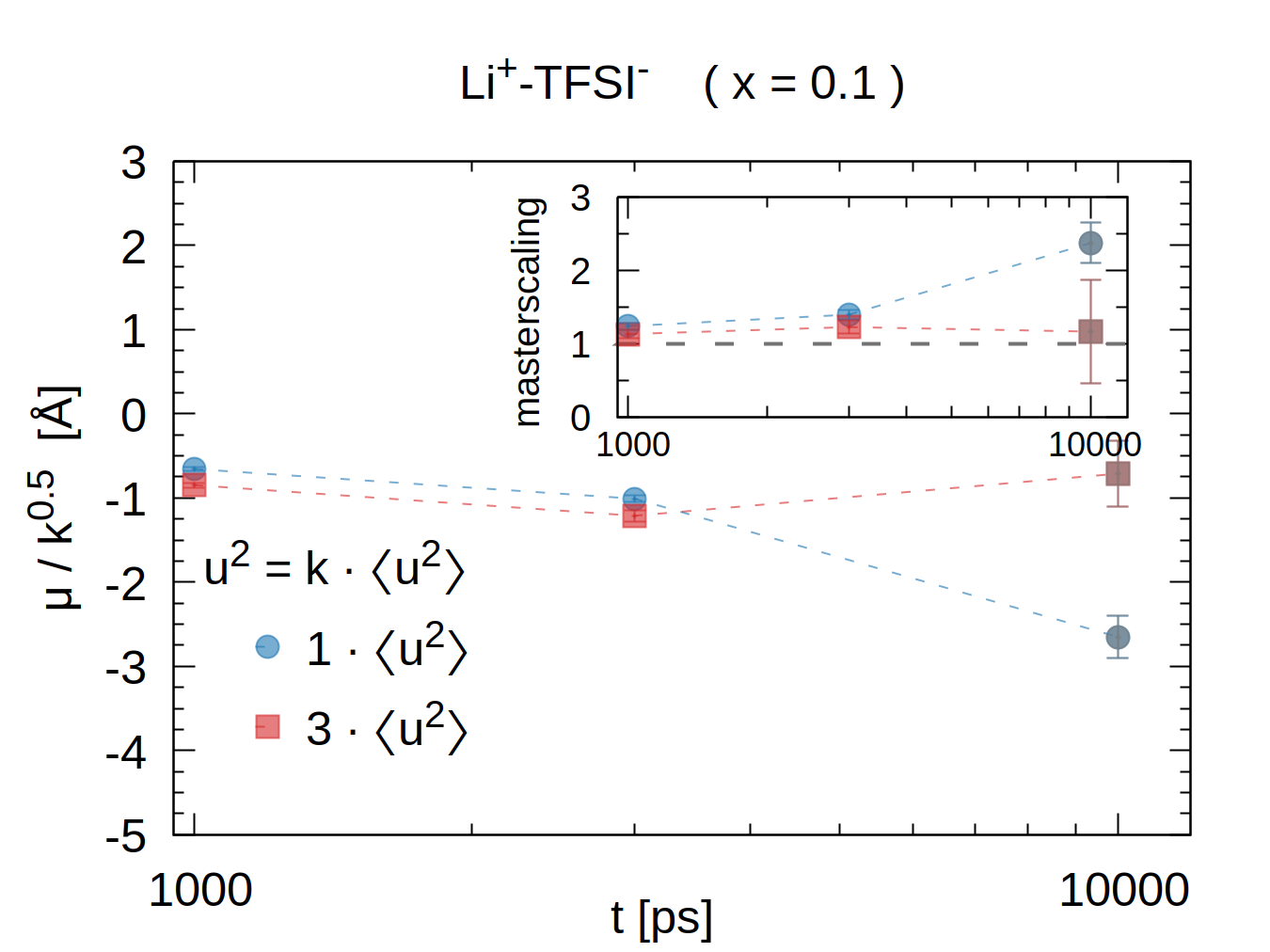}}
  \hfill
  \centering
  \subfloat{\includegraphics[width=0.5\textwidth]{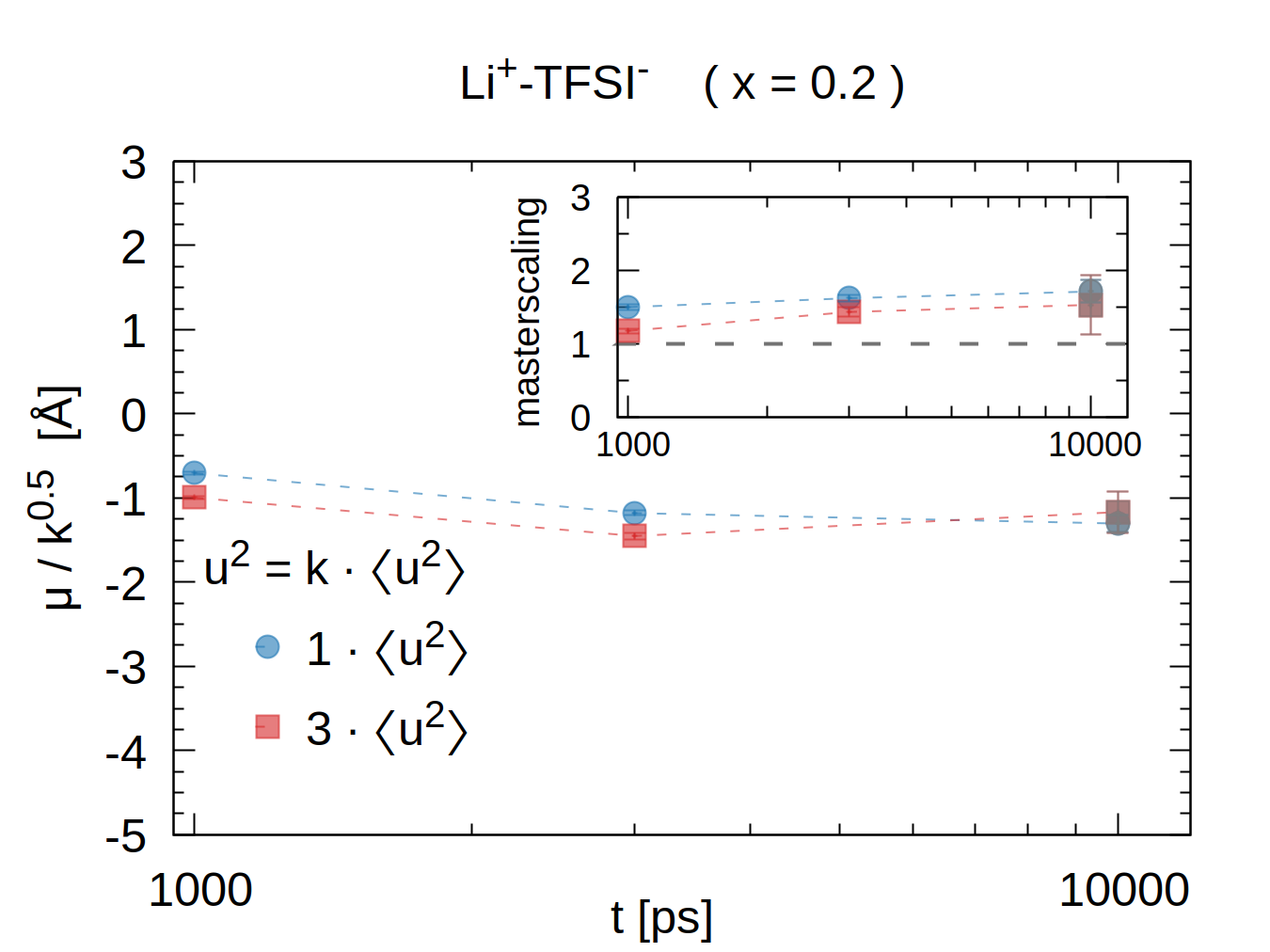}}
  \hfill
  \subfloat{\includegraphics[width=0.5\textwidth]{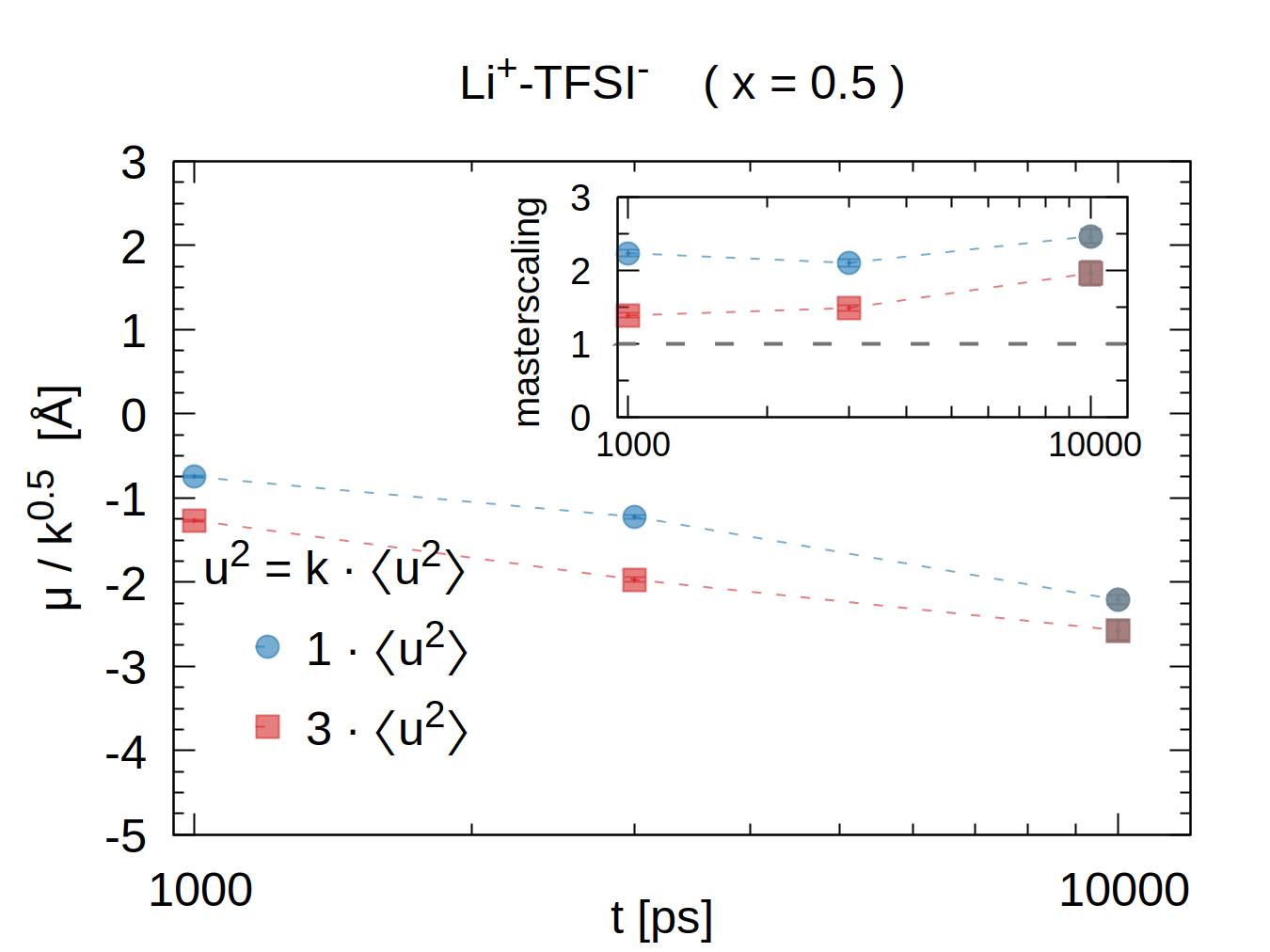}}
\end{figure}

\begin{figure}[H]
  \centering
  \subfloat{\includegraphics[width=0.5\textwidth]{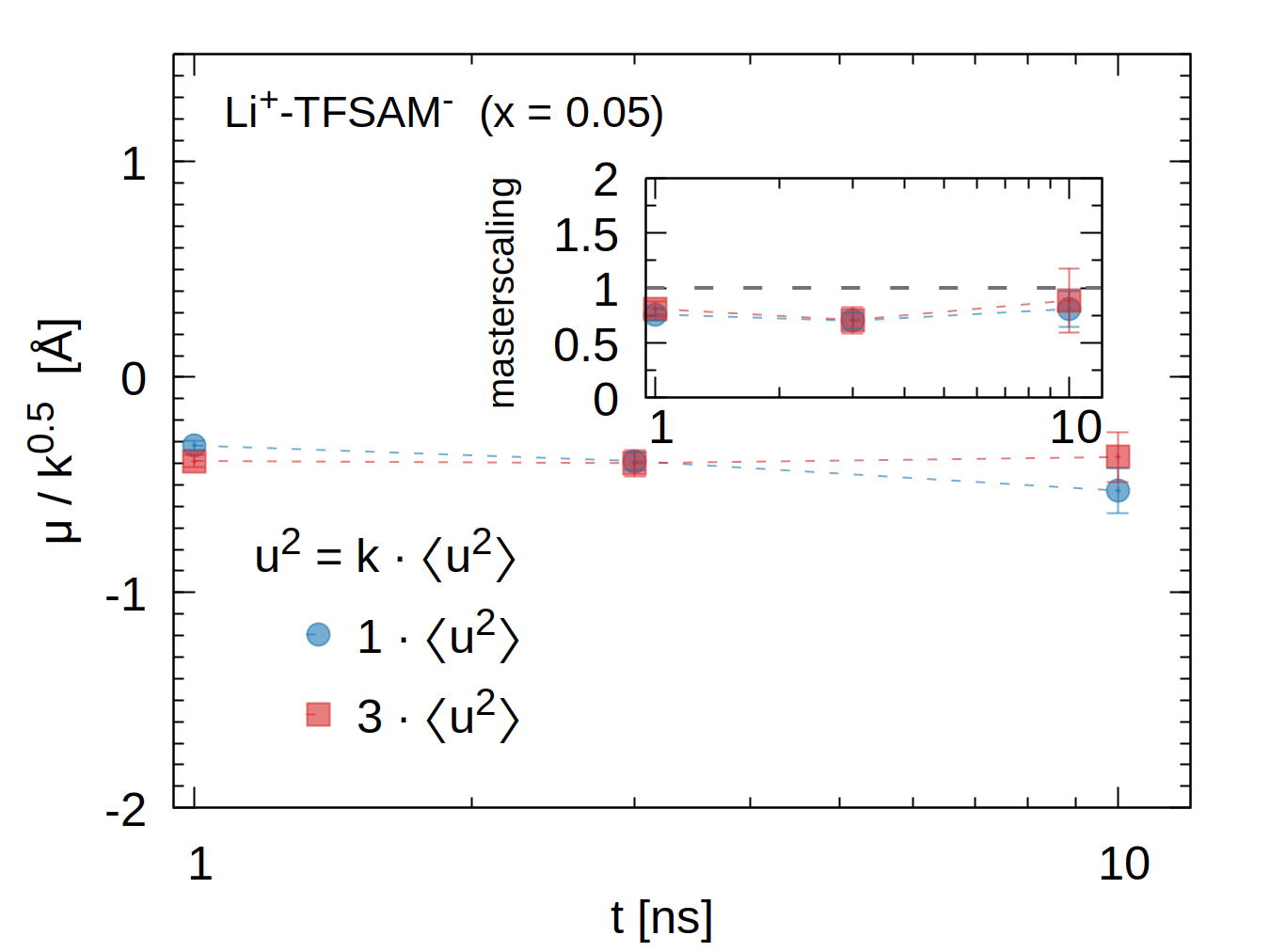}}
  \hfill
  \subfloat{\includegraphics[width=0.5\textwidth]{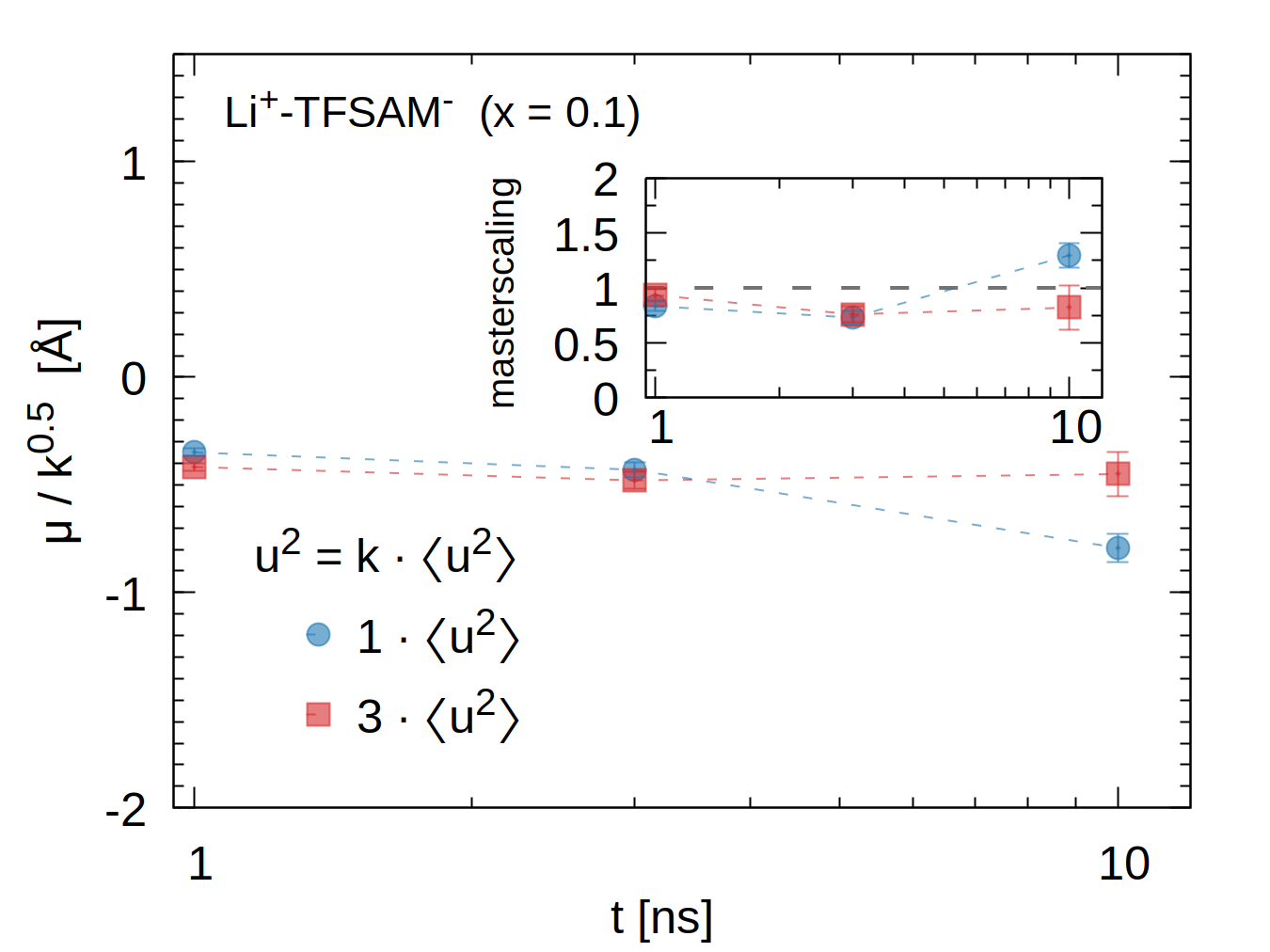}}
  \hfill
  \centering
  \subfloat{\includegraphics[width=0.5\textwidth]{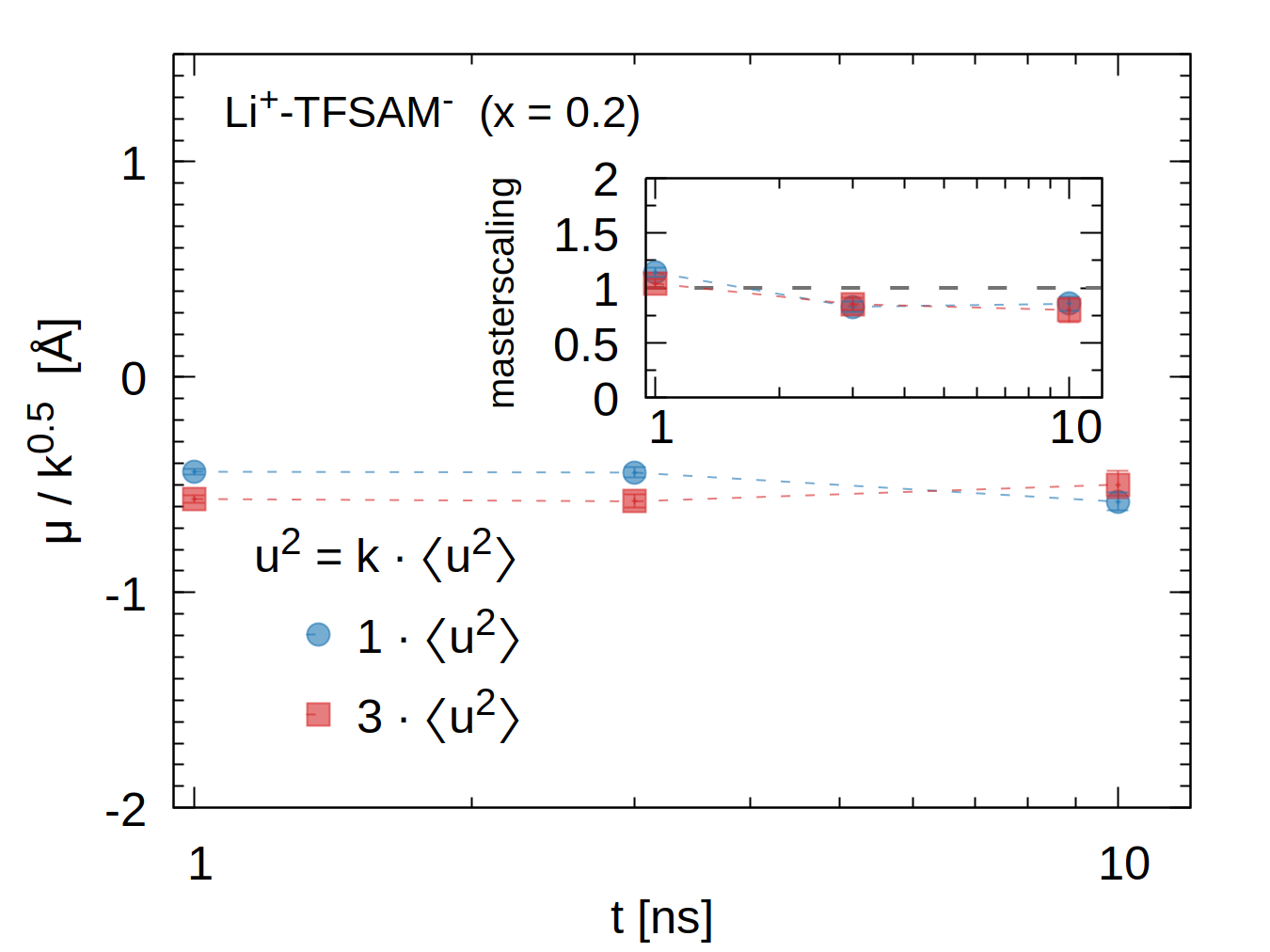}}
  \hfill
  \subfloat{\includegraphics[width=0.5\textwidth]{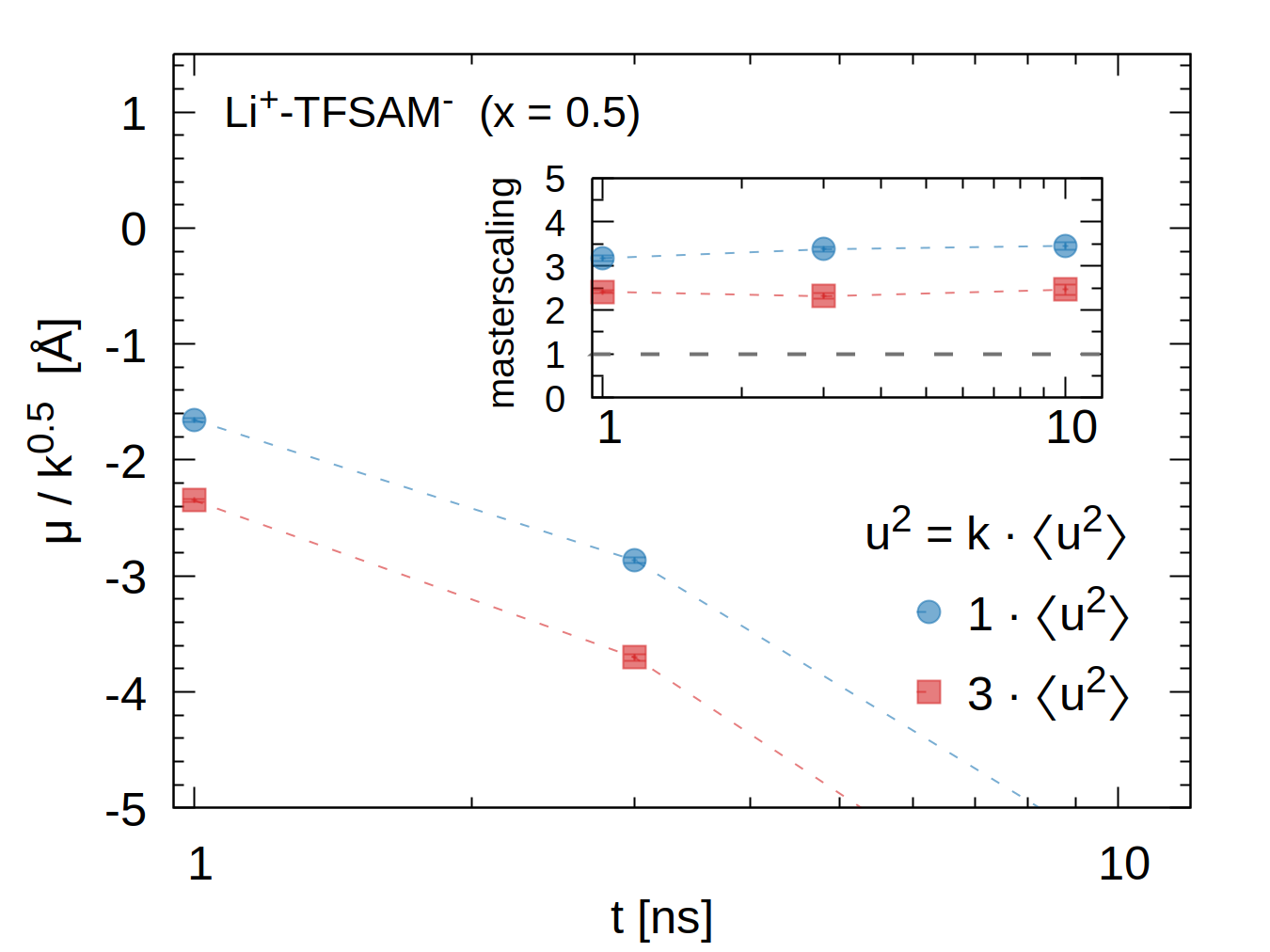}}

 \caption{Gaussian peak positions $\mu(t)$ divided by the square root of the distance scaling factor $k$. Inset: Masterscaling of the Gaussian peak parameters $\mu$ and $\sigma^2$ to 1 according to Equation S\ref{eq:masterscaling_peak_parameters}. The data points relying on the manually performed Gaussian fits are highlighted in grey. The data points are based on measurements employing an upper threshold tolerance $u^2_{\text{r}} = 1.5\cdot k \cdot\langle u^2 \rangle$. }
 \label{fig:masterscaling_delta_v_para_tfsam_tfsi}
\end{figure}

\newpage
\begin{figure}[H]
  \centering
  \subfloat{\includegraphics[width=0.5\textwidth]{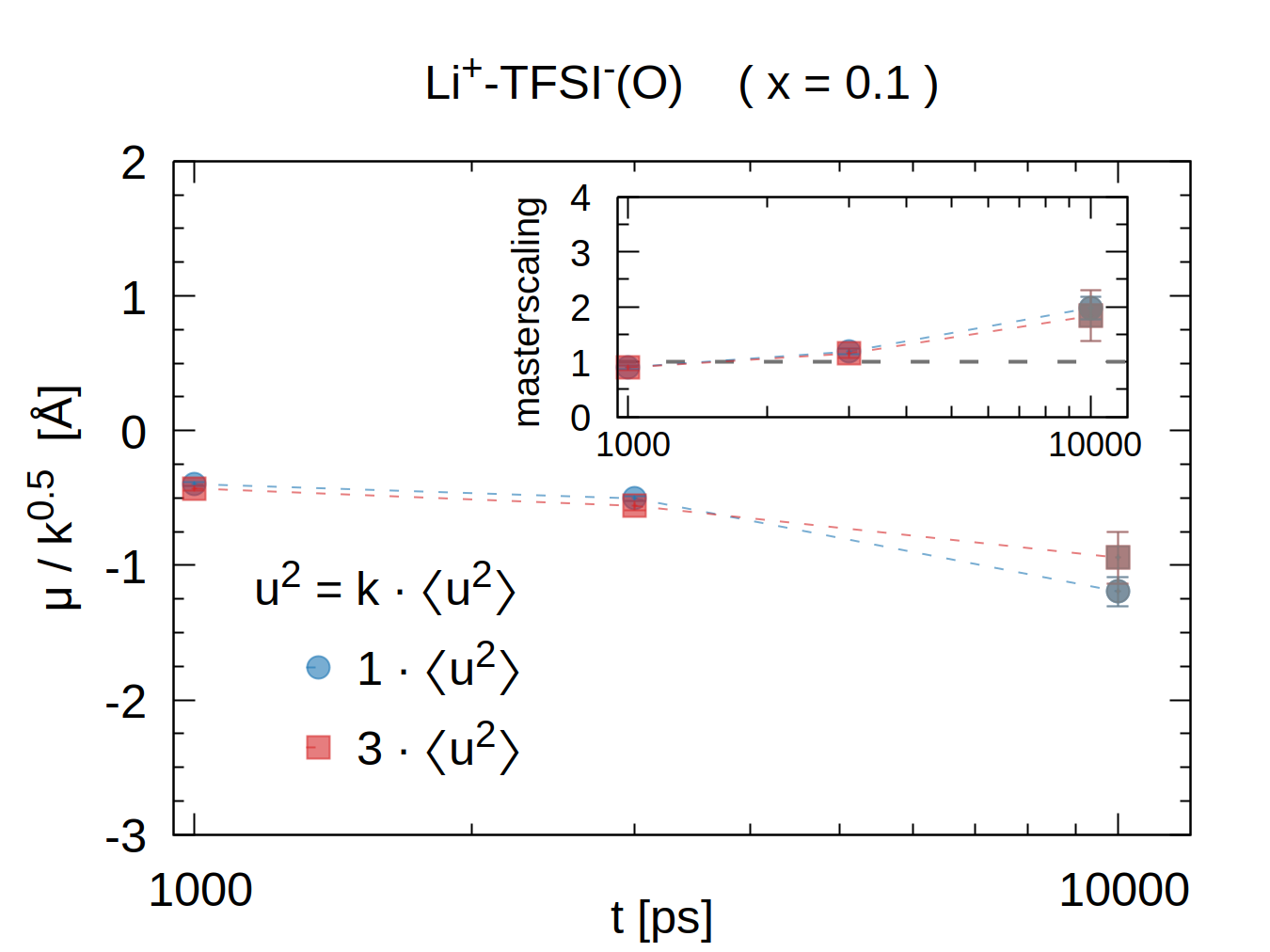}}
  \hfill
  \subfloat{\includegraphics[width=0.5\textwidth]{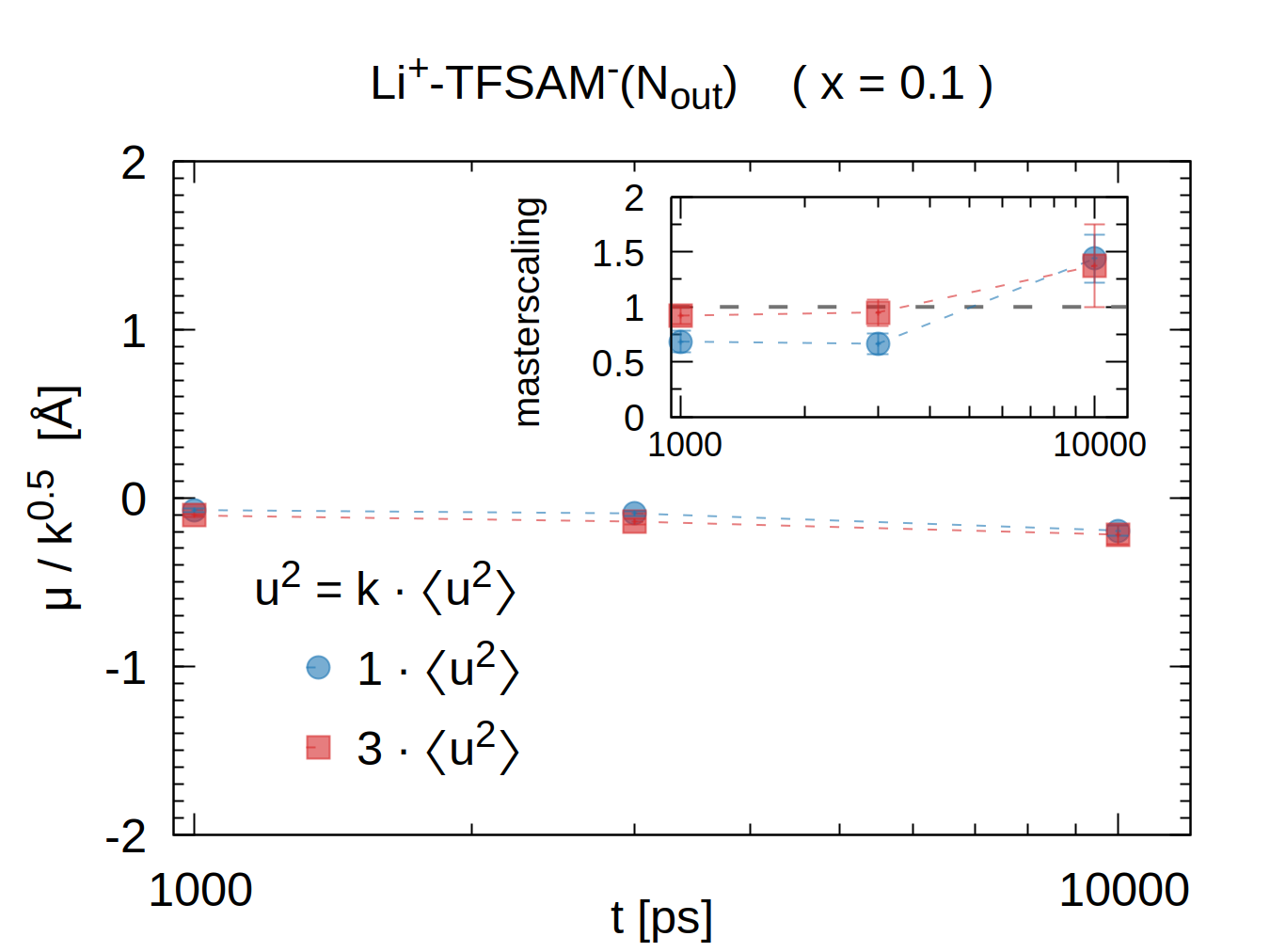}}
  \caption{Scaling relations exemplary for the direct anion binding sites $\text{TFSI}^-(\text{O})$ and $\text{TFSAM}^-(\text{N}_{\text{mid}})$ at a low salt content of x\,=\,0.1: Gaussian peak positions $\mu(t)$ divided by the square root of the distance scaling factor $k$. Inset: Masterscaling of the Gaussian peak parameters $\mu$ and $\sigma^2$ to 1 according to Equation S\ref{eq:masterscaling_peak_parameters}. The data points relying on the manually performed Gaussian fits are highlighted in grey. The data points are measured using an upper threshold tolerance $u^2_{\text{r}} = 1.5\cdot k \cdot\langle u^2 \rangle$. }
 \label{fig:masterscaling_delta_v_para_tfsam_n_out_tfsi_o}
\end{figure}

\begin{figure}[H]
  \centering
  \subfloat{\includegraphics[width=0.5\textwidth]{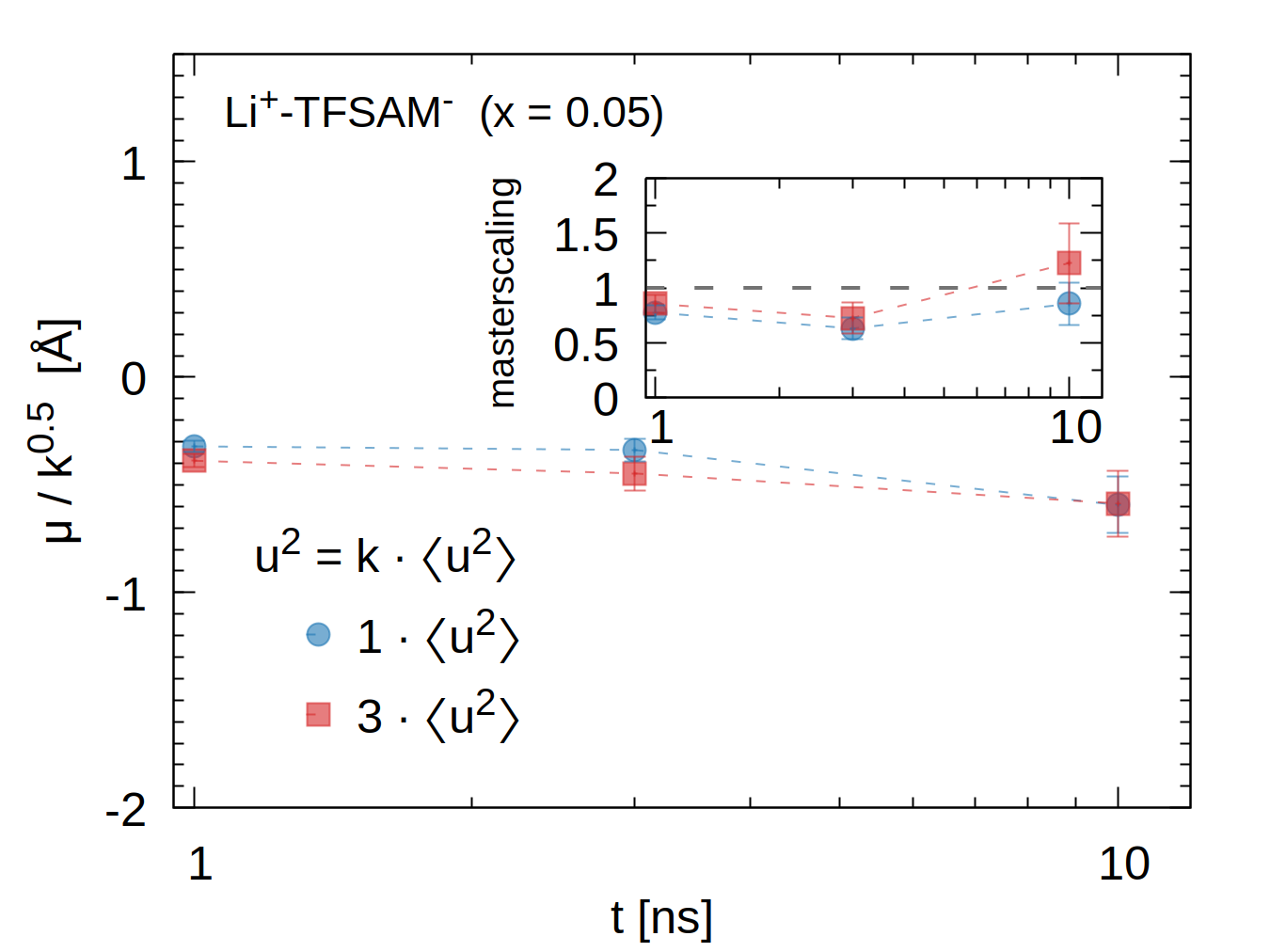}}
  \hfill
  \subfloat{\includegraphics[width=0.5\textwidth]{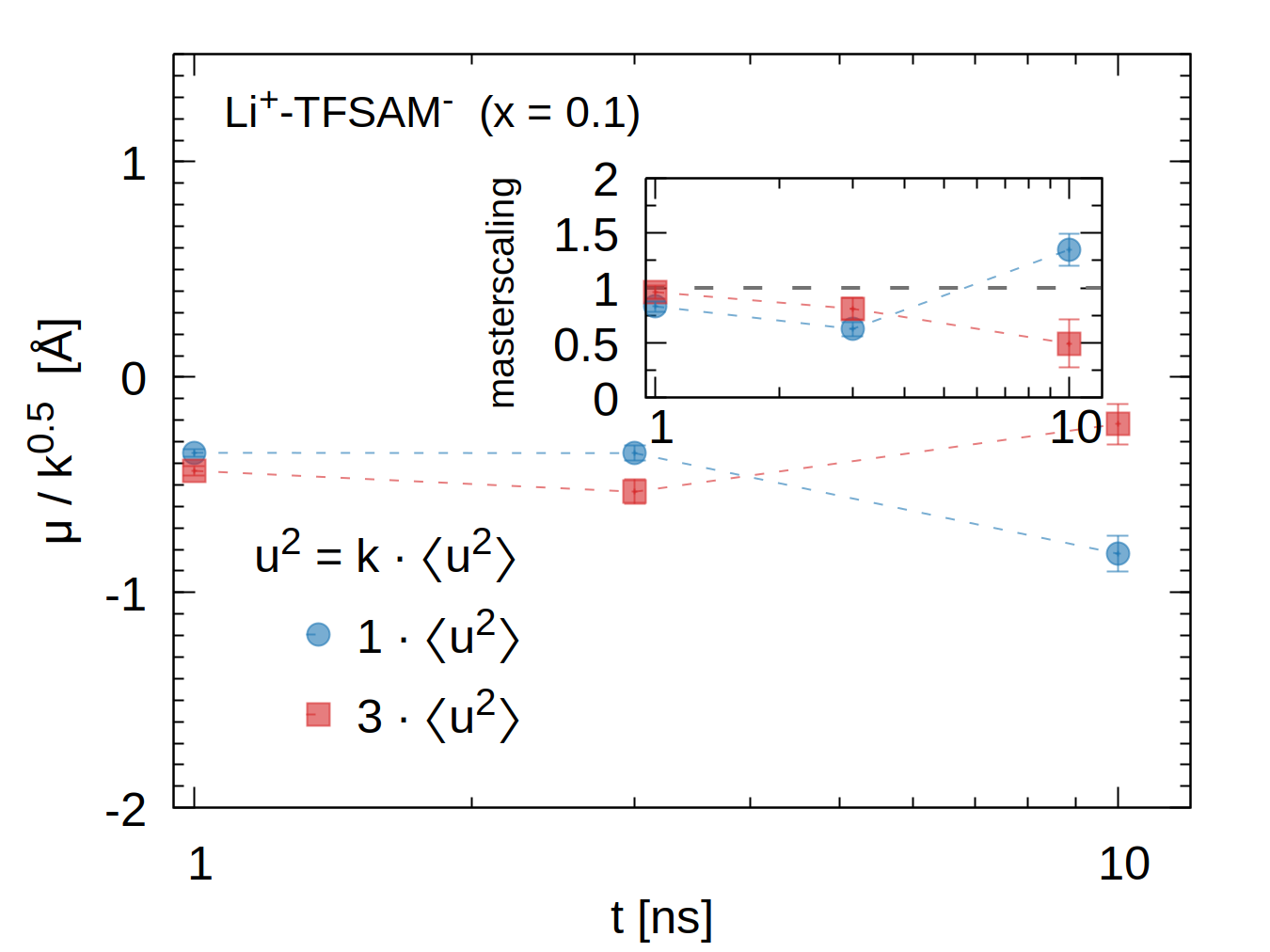}}

 \caption{Scaling relations exemplary for the $\text{TFSAM}^-$-based mixtures for x\,=\,0.05 and 0.1 using an upper threshold tolerance $u^2_{\text{r}} = 1.3\cdot k \cdot\langle u^2 \rangle$ .}
 \label{fig:masterscaling_delta_v_para_tfsam_tfsi_tolerance_1_3}
\end{figure}

%%%%%%%%%%%%%%%%%%%%%%%%%%%%%%%%%%%%%%%%%%%%%%%%%%%%%%%%%%%%%%%%%%%%%%%%%%%%%%%%%%
%%%%%%%%%%%%%%%%%%%%%%%%%%%%%%%%%%%%%%%%%%%%%%%%%%%%%%%%%%%%%%%%%%%%%%%%%%%%%%%%%%
%%%%%%%%%%%%%%%%%%%%%%%%%%%%%%%%%%%%%%%%%%%%%%%%%%%%%%%%%%%%%%%%%%%%%%%%%%%%%%%%%%
%%%%%%%%%%%%%%%%%%%%%%%%%%%%%%%%%%%%%%%%%%%%%%%%%%%%%%%%%%%%%%%%%%%%%%%%%%%%%%%%%%
%%%%%%%%%%%%%%%%%%%%%%%%%%%%%%%%%%%%%%%%%%%%%%%%%%%%%%%%%%%%%%%%%%%%%%%%%%%%%%%%%%
%%%%%%%%%%%%%%%%%%%%%%%%%%%%%%%%%%%%%%%%%%%%%%%%%%%%%%%%%%%%%%%%%%%%%%%%%%%%%%%%%%
%%%%%%%%%%%%%%%%%%%%%%%%%%%%%%%%%%%%%%%%%%%%%%%%%%%%%%%%%%%%%%%%%%%%%%%%%%%%%%%%%%
%%%%%%%%%%%%%%%%%%%%%%%%%%%%%%%%%%%%%%%%%%%%%%%%%%%%%%%%%%%%%%%%%%%%%%%%%%%%%%%%%%
%%%%%%%%%%%%%%%%%%%%%%%%%%%%%%%%%%%%%%%%%%%%%%%%%%%%%%%%%%%%%%%%%%%%%%%%%%%%%%%%%%
\newpage
\textbf{H: Quantification of dynamically decoupled anions $p_{\text{lost}}$} \newline

The ratio of dynamically decoupled anions is estimated from the relative displacement distribution by subtracting the Gaussian peak fit $\mathcal{N}(\mu, \sigma^2) $ from the histogram data $p(\Delta v_{\parallel})$ and summing up the remaining counts: $\int d\Delta v_{\parallel}( p(\Delta v_{\parallel})-\mathcal{N}(\mu,\sigma^2) ) \,\,\dot{=} \,\,p_{\text{lost}}$. The part of the histogram which is attributed to $p_{\text{lost}}$ is highlighted in blue.
\begin{figure}[H]
  \centering
  \subfloat{\includegraphics[width=1.0\textwidth]{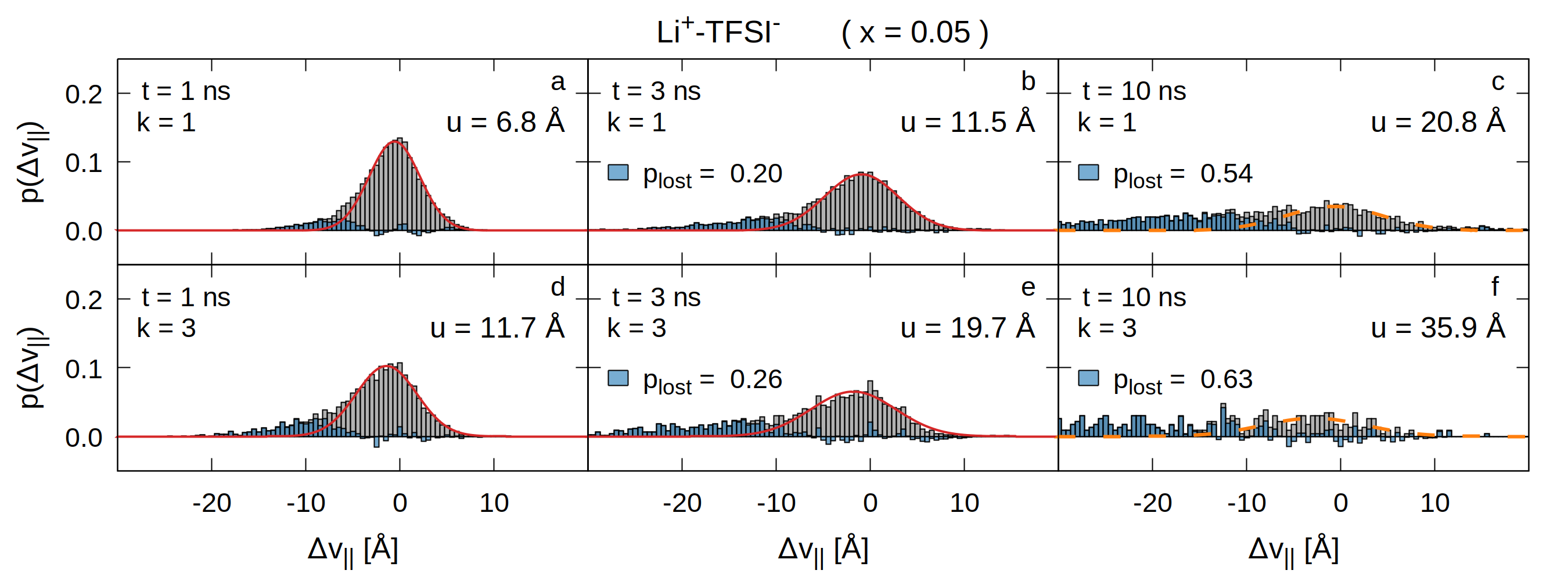}}
  \hfill
  \subfloat{\includegraphics[width=1.0\textwidth]{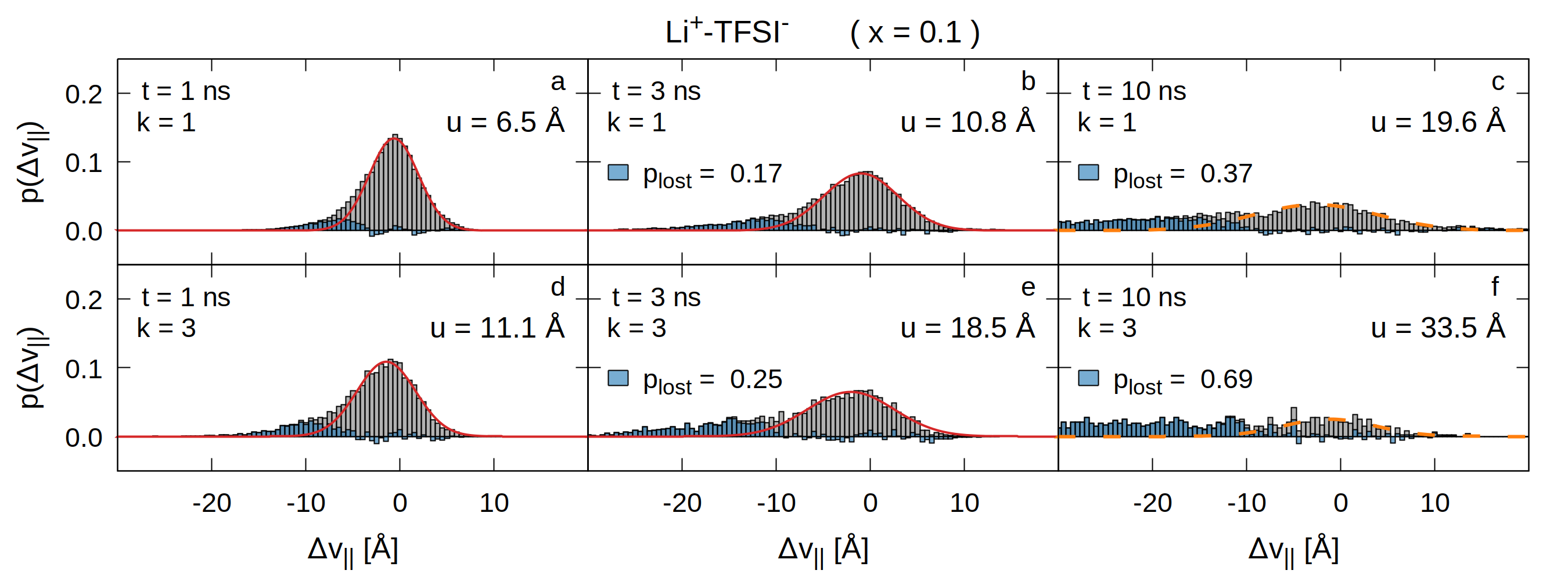}}
  \end{figure}
  \begin{figure}[H]
  \centering
  \subfloat{\includegraphics[width=1.0\textwidth]{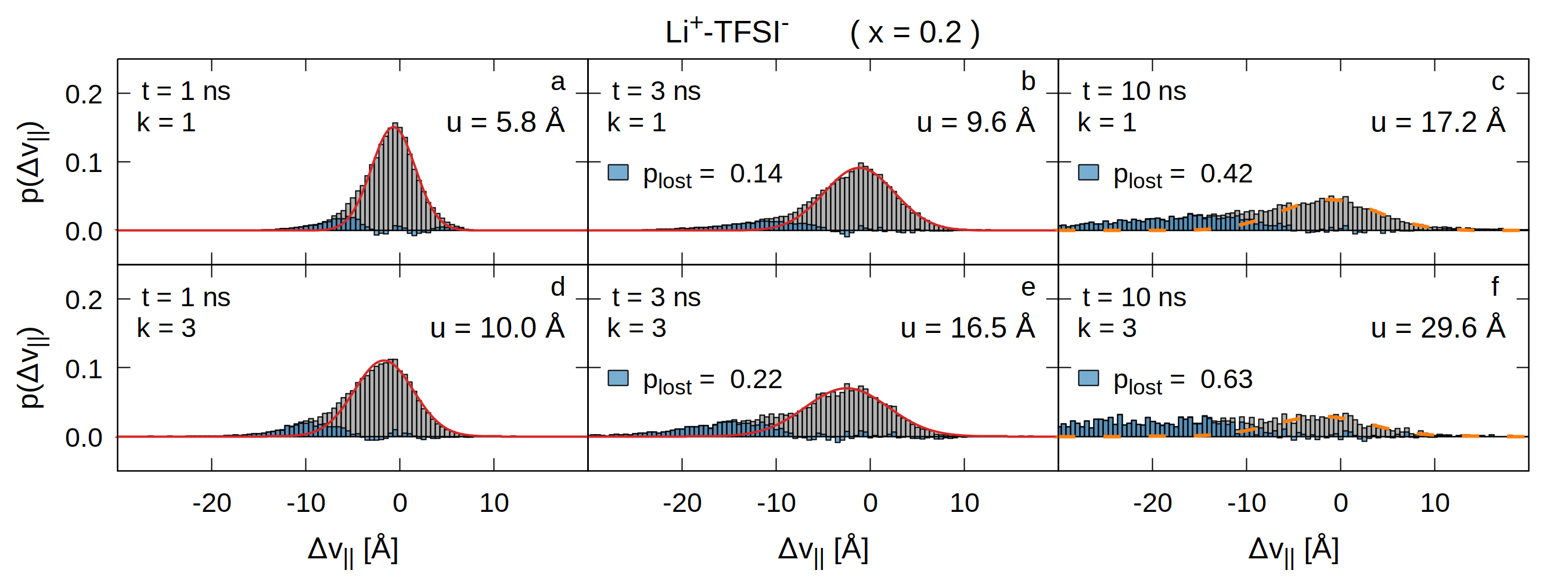}}
  \end{figure}
  \begin{figure}[H]
  \centering
  \subfloat{\includegraphics[width=1.0\textwidth]{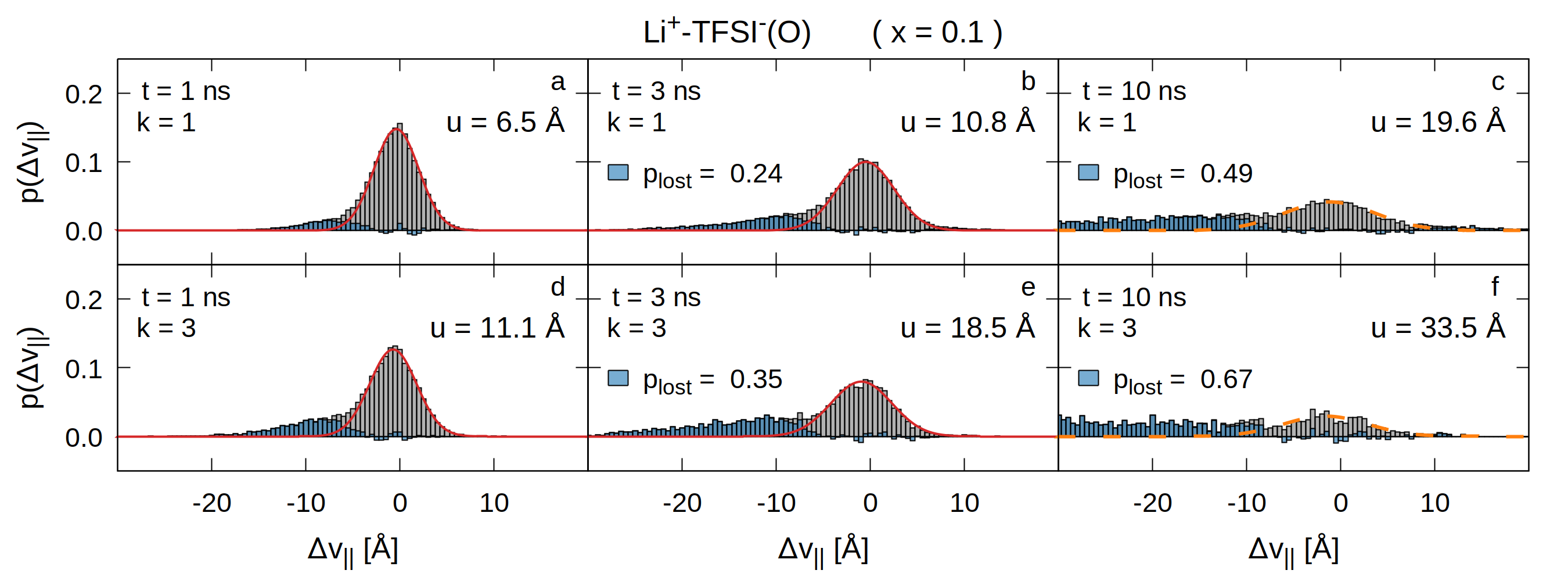}}
  
 \caption{Estimation of the amount $p_{\text{lost}}$ of dynamically decoupled $\text{TFSI}^-$ shell anions at different lag times $t$, lithium squared displacements $k\cdot\langle u^2\rangle$ and various salt contents x. Since the distributions of coupled (Gaussian peak) and decoupled (tail) dynamics overlap considerably at the shortest analysed lag time of $t\,=\,1\,$ns, a precise quantitative estimate of $p_{\text{lost}}$ is not feasible through this procedure. }
 \label{fig:panels_v_para_lost_TFSI}
\end{figure}

\begin{figure}[H]
  \centering
  \subfloat{\includegraphics[width=1.0\textwidth]{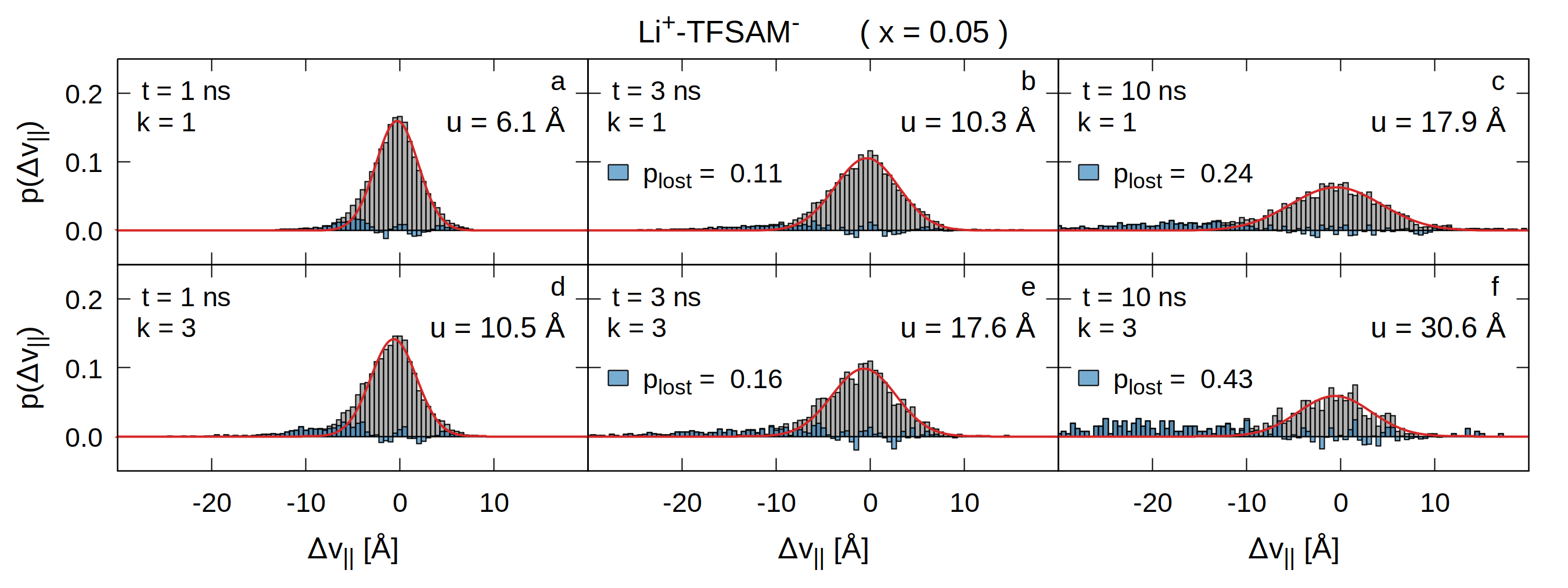}}
  \hfill
  \subfloat{\includegraphics[width=1.0\textwidth]{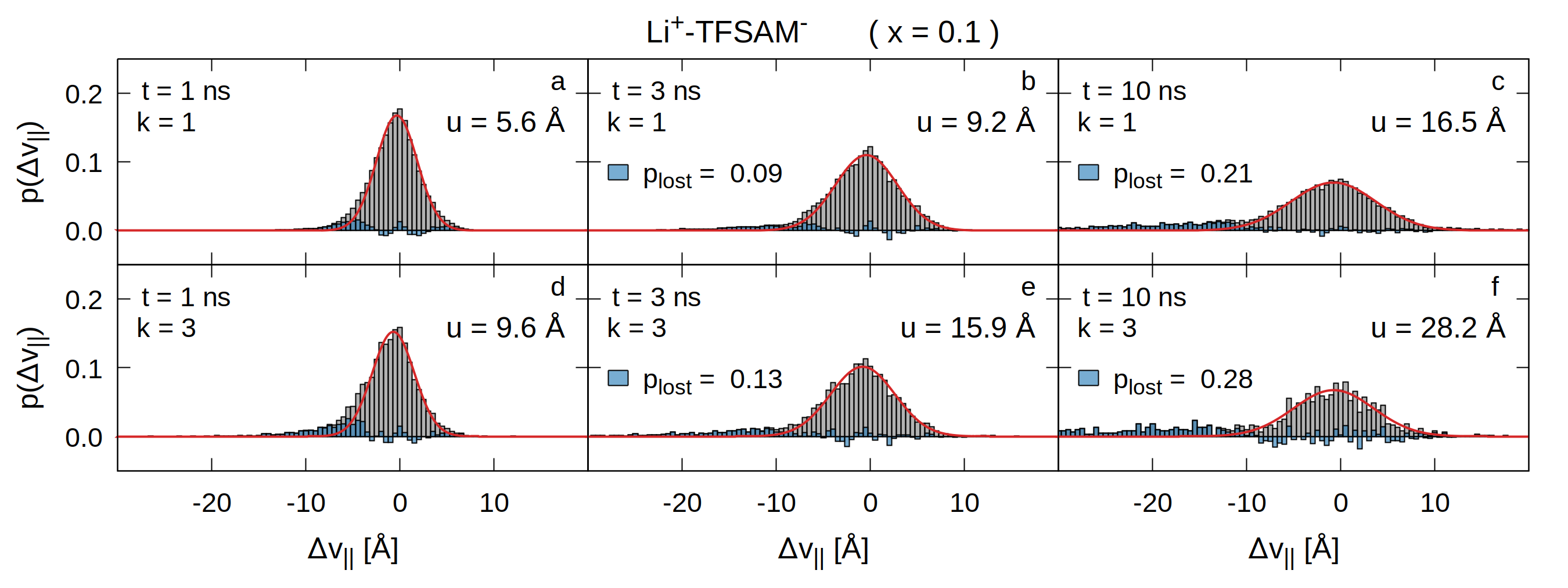}}
    \end{figure}
  \begin{figure}[H]
  \centering

  \subfloat{\includegraphics[width=1.0\textwidth]{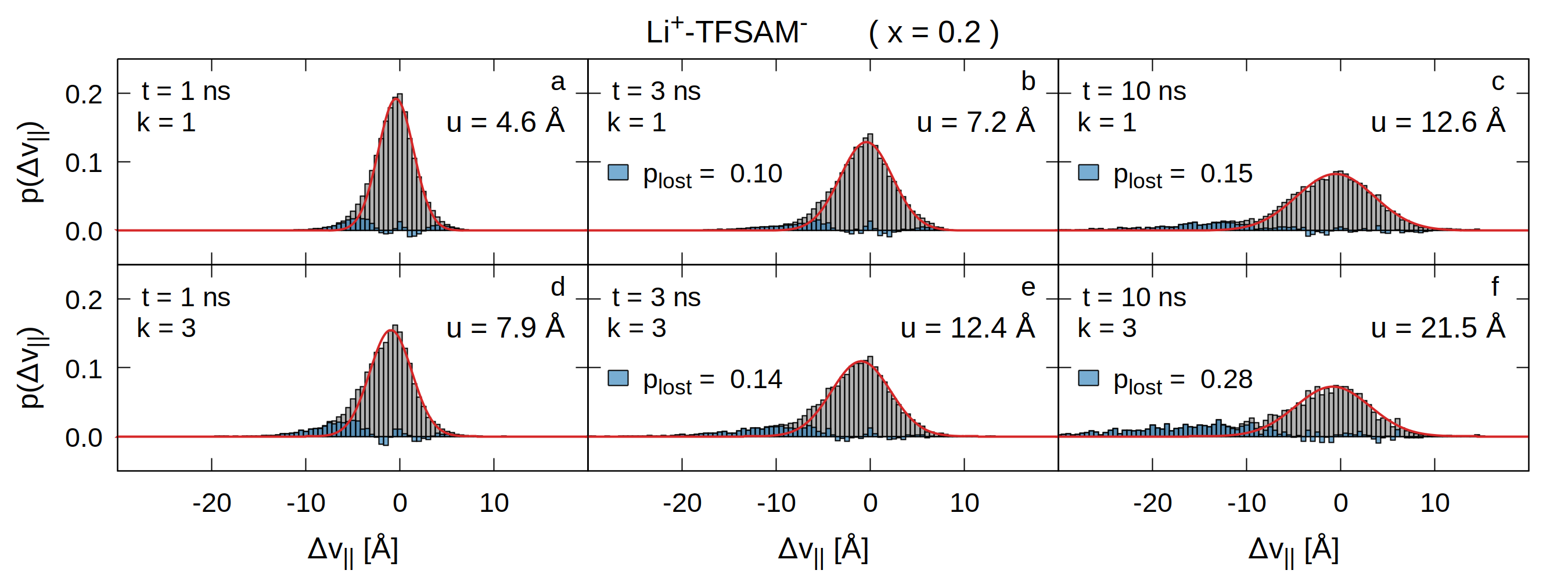}}
  \hfill
  \subfloat{\includegraphics[width=1.0\textwidth]{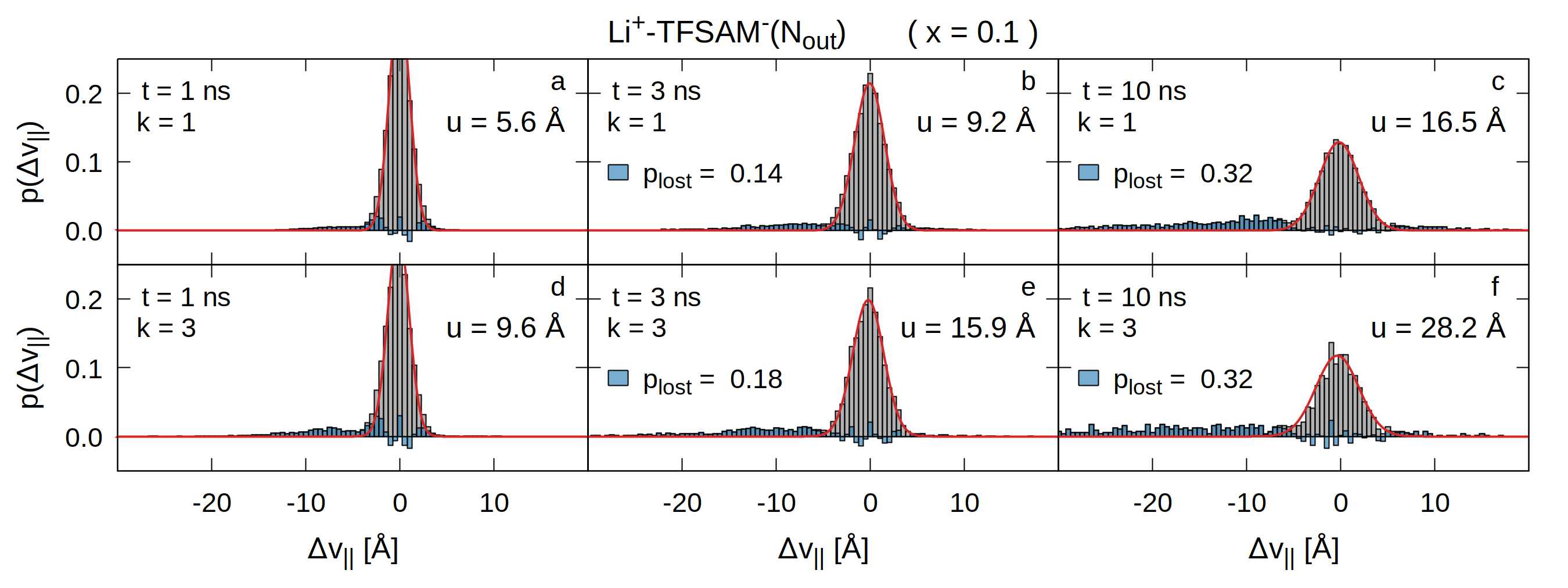}}
  
 \caption{Estimation of the amount $p_{\text{lost}}$ of dynamically decoupled $\text{TFSAM}^-$ shell anions at different lag times $t$, lithium squared displacements $k\cdot\langle u^2\rangle$ and various salt contents x. Since the distributions of coupled (Gaussian peak) and decoupled (tail) dynamics overlap considerably at the shortest analysed lag time of $t\,=\,1\,$ns, a precise quantitative estimate of $p_{\text{lost}}$ is not feasible through this procedure.}
 \label{fig:panels_v_para_lost_TFSAM}
\end{figure}

%%%%%%%%%%%%%%%%%%%%%%%%%%%%%%%%%%%%%%%%%%%%%%%%%%%%%%%%%%%%%%%%%%%%%%%%%%%%%%%%%%
%%%%%%%%%%%%%%%%%%%%%%%%%%%%%%%%%%%%%%%%%%%%%%%%%%%%%%%%%%%%%%%%%%%%%%%%%%%%%%%%%%
%%%%%%%%%%%%%%%%%%%%%%%%%%%%%%%%%%%%%%%%%%%%%%%%%%%%%%%%%%%%%%%%%%%%%%%%%%%%%%%%%%
%%%%%%%%%%%%%%%%%%%%%%%%%%%%%%%%%%%%%%%%%%%%%%%%%%%%%%%%%%%%%%%%%%%%%%%%%%%%%%%%%%
%%%%%%%%%%%%%%%%%%%%%%%%%%%%%%%%%%%%%%%%%%%%%%%%%%%%%%%%%%%%%%%%%%%%%%%%%%%%%%%%%%
%%%%%%%%%%%%%%%%%%%%%%%%%%%%%%%%%%%%%%%%%%%%%%%%%%%%%%%%%%%%%%%%%%%%%%%%%%%%%%%%%%
%%%%%%%%%%%%%%%%%%%%%%%%%%%%%%%%%%%%%%%%%%%%%%%%%%%%%%%%%%%%%%%%%%%%%%%%%%%%%%%%%%
%%%%%%%%%%%%%%%%%%%%%%%%%%%%%%%%%%%%%%%%%%%%%%%%%%%%%%%%%%%%%%%%%%%%%%%%%%%%%%%%%%
%%%%%%%%%%%%%%%%%%%%%%%%%%%%%%%%%%%%%%%%%%%%%%%%%%%%%%%%%%%%%%%%%%%%%%%%%%%%%%%%%%
%%%%%%%%%%%%%%%%%%%%%%%%%%%%%%%%%%%%%%%%%%%%%%%%%%%%%%%%%%%%%%%%%%%%%%%%%%%%%%%%%%
%%%%%%%%%%%%%%%%%%%%%%%%%%%%%%%%%%%%%%%%%%%%%%%%%%%%%%%%%%%%%%%%%%%%%%%%%%%%%%%%%%
%%%%%%%%%%%%%%%%%%%%%%%%%%%%%%%%%%%%%%%%%%%%%%%%%%%%%%%%%%%%%%%%%%%%%%%%%%%%%%%%%%
%%%%%%%%%%%%%%%%%%%%%%%%%%%%%%%%%%%%%%%%%%%%%%%%%%%%%%%%%%%%%%%%%%%%%%%%%%%%%%%%%%
%%%%%%%%%%%%%%%%%%%%%%%%%%%%%%%%%%%%%%%%%%%%%%%%%%%%%%%%%%%%%%%%%%%%%%%%%%%%%%%%%%
%%%%%%%%%%%%%%%%%%%%%%%%%%%%%%%%%%%%%%%%%%%%%%%%%%%%%%%%%%%%%%%%%%%%%%%%%%%%%%%%%%

\newpage
\textbf{I: LCF $\lambda$ as a function of squared lithium displacement $u^2(t)$}

The $\text{Li}^+$ coupling factor $\lambda$, which measures the extent to which an initially lithium-bound anion follows the dynamics of this very $\text{Li}^+$, is binned according to the squared displacement $ u^{\,2}$ of this $\text{Li}^+$. The data sets are generated by averaging over multiple individual blocks, \textit{e.g.}, the 400\,ns trajectories are divided into 130 blocks to evaluate the lag time $\text{t}\,=\,3$\,ns.

\begin{figure}[H]
  \centering
  \subfloat{\includegraphics[width=0.5\textwidth]{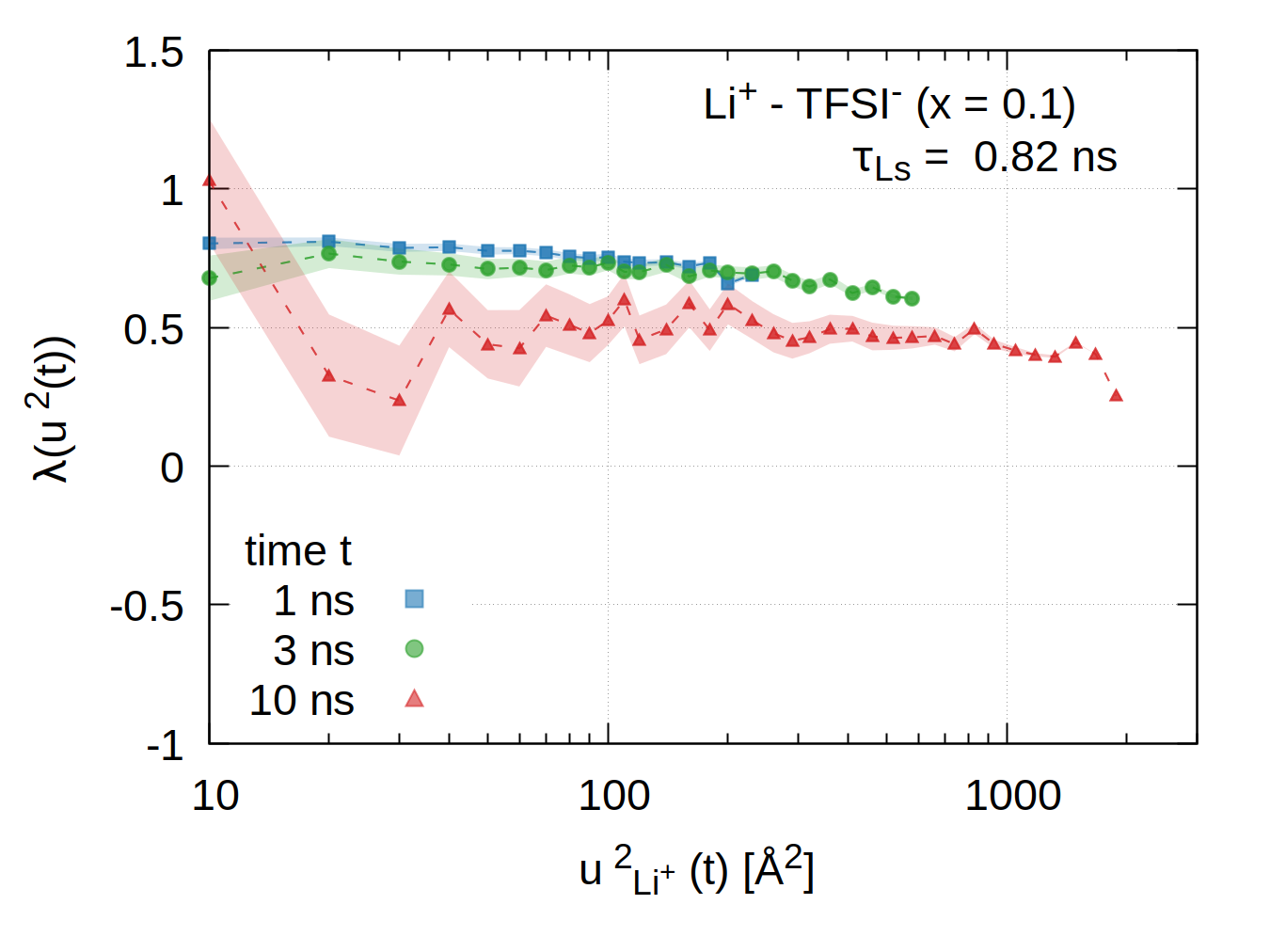}}
  \hfill
  \subfloat{\includegraphics[width=0.5\textwidth]{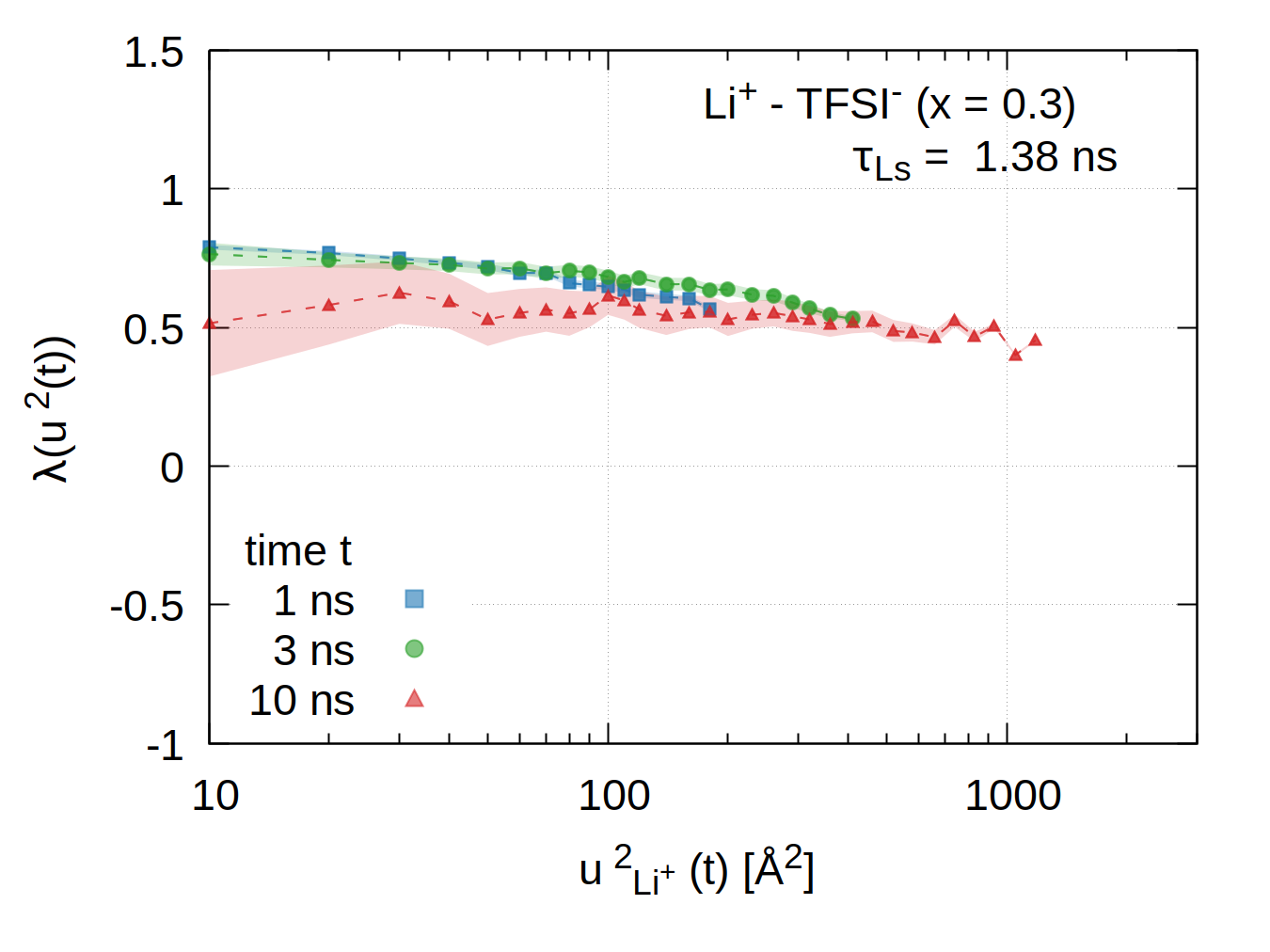}}
\end{figure}

\begin{figure}[H]
  \centering
  \subfloat{\includegraphics[width=0.5\textwidth]{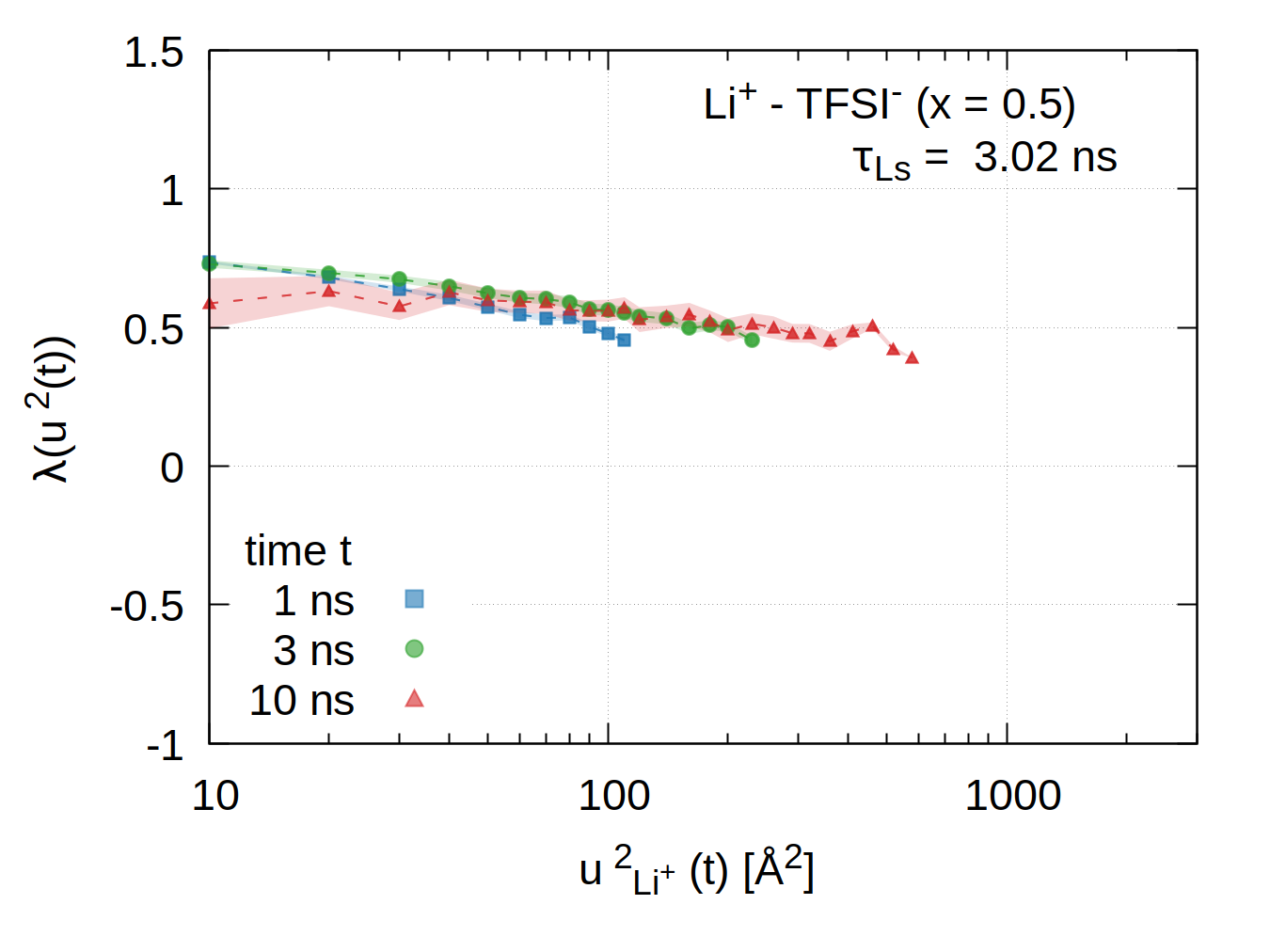}}
  \hfill
  \subfloat{\includegraphics[width=0.5\textwidth]{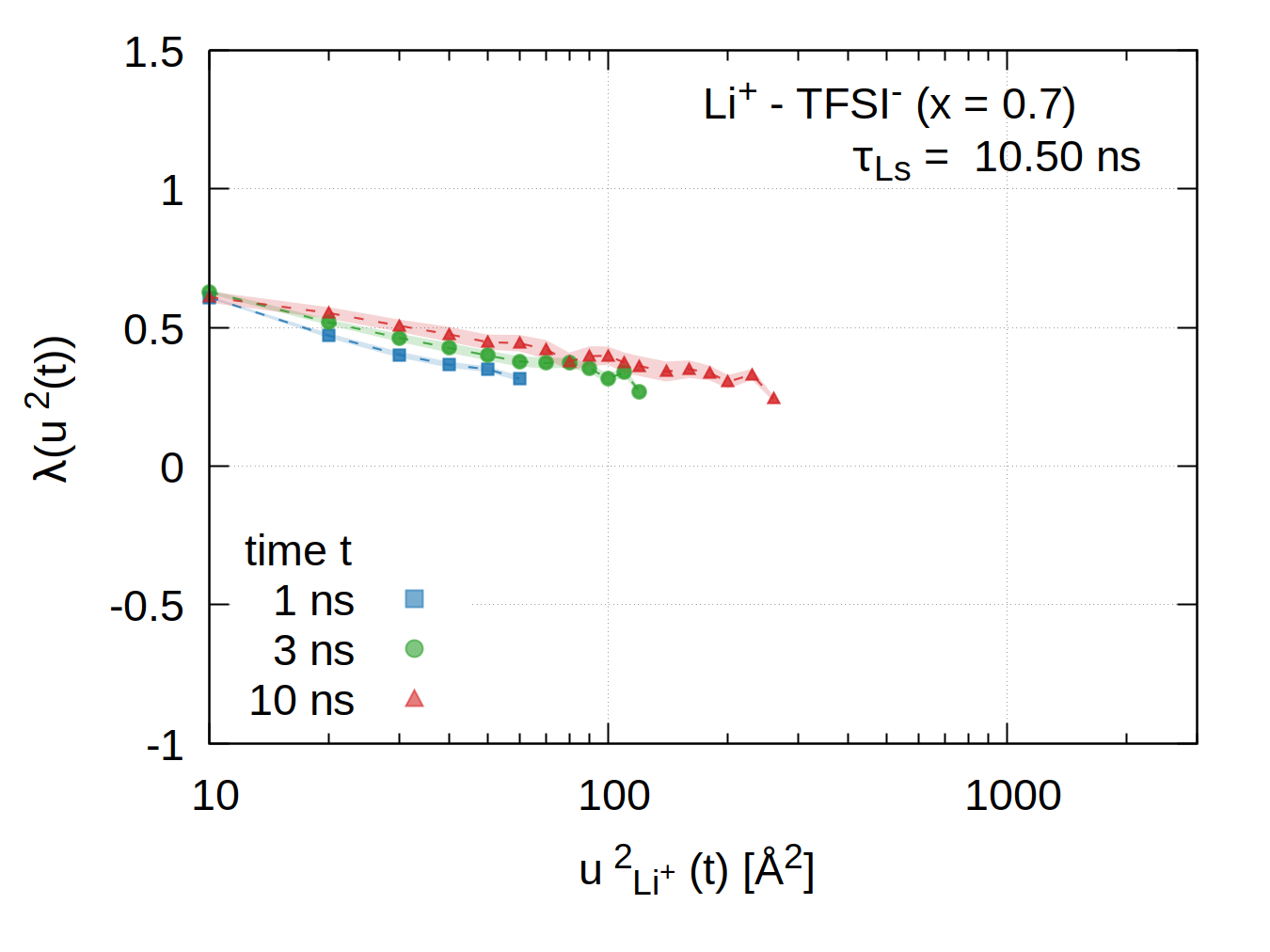}}

 \caption{Coupling of anion motion to lithium dynamics measured via $\lambda$ as a function of $ u^2(\text{t})_{\text{Li}^+}$ exemplary for the lithium salt fractions x\,=\,0.1, 0.3, 0.5 and 0.7 in the $\text{TFSI}^-$-containing mixtures.}
 
 \label{fig:lambda_x2_tfsi}
\end{figure}

\begin{figure}[H]
  \centering
  \subfloat{\includegraphics[width=0.5\textwidth]{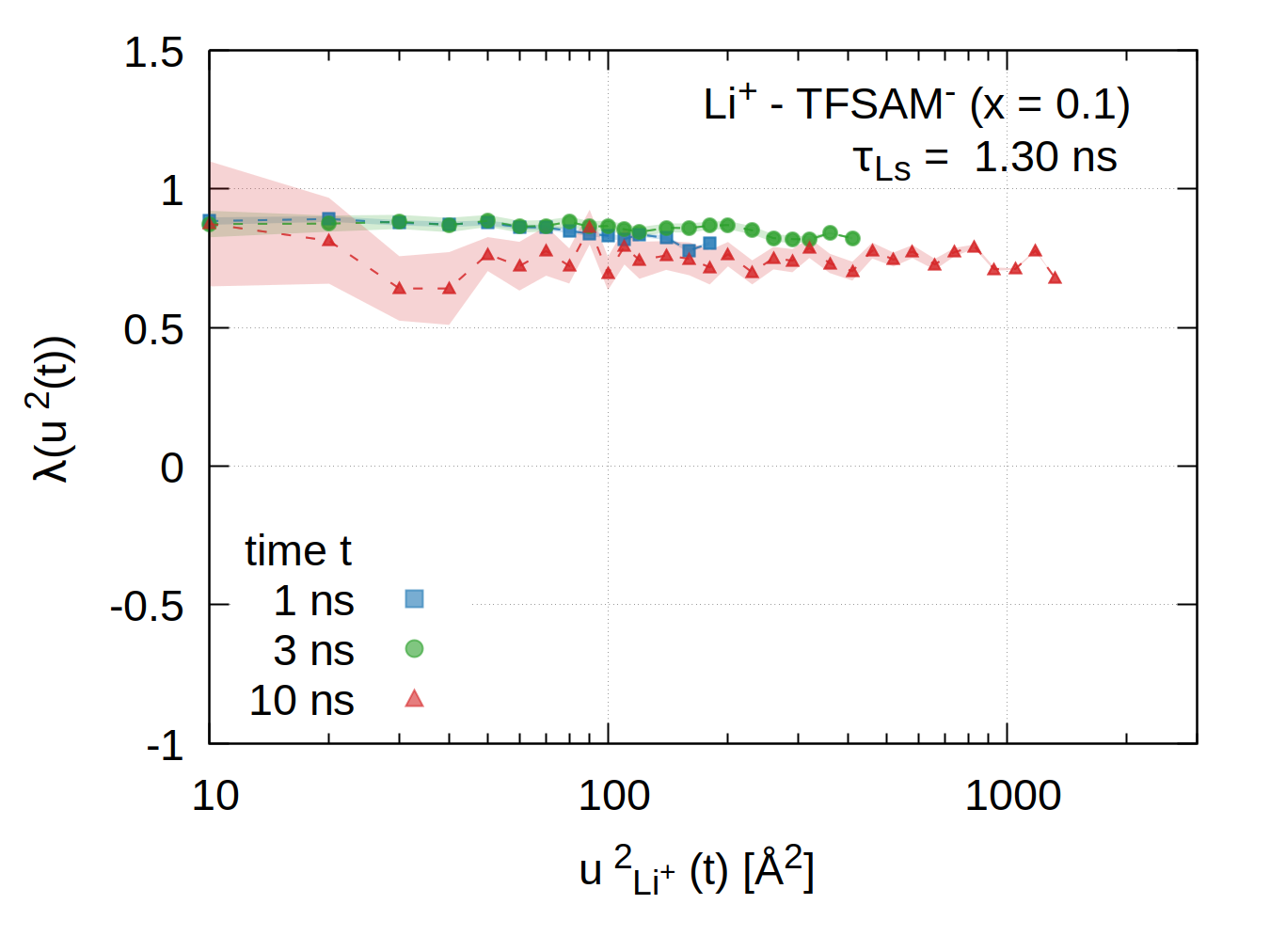}}
  \hfill
  \subfloat{\includegraphics[width=0.5\textwidth]{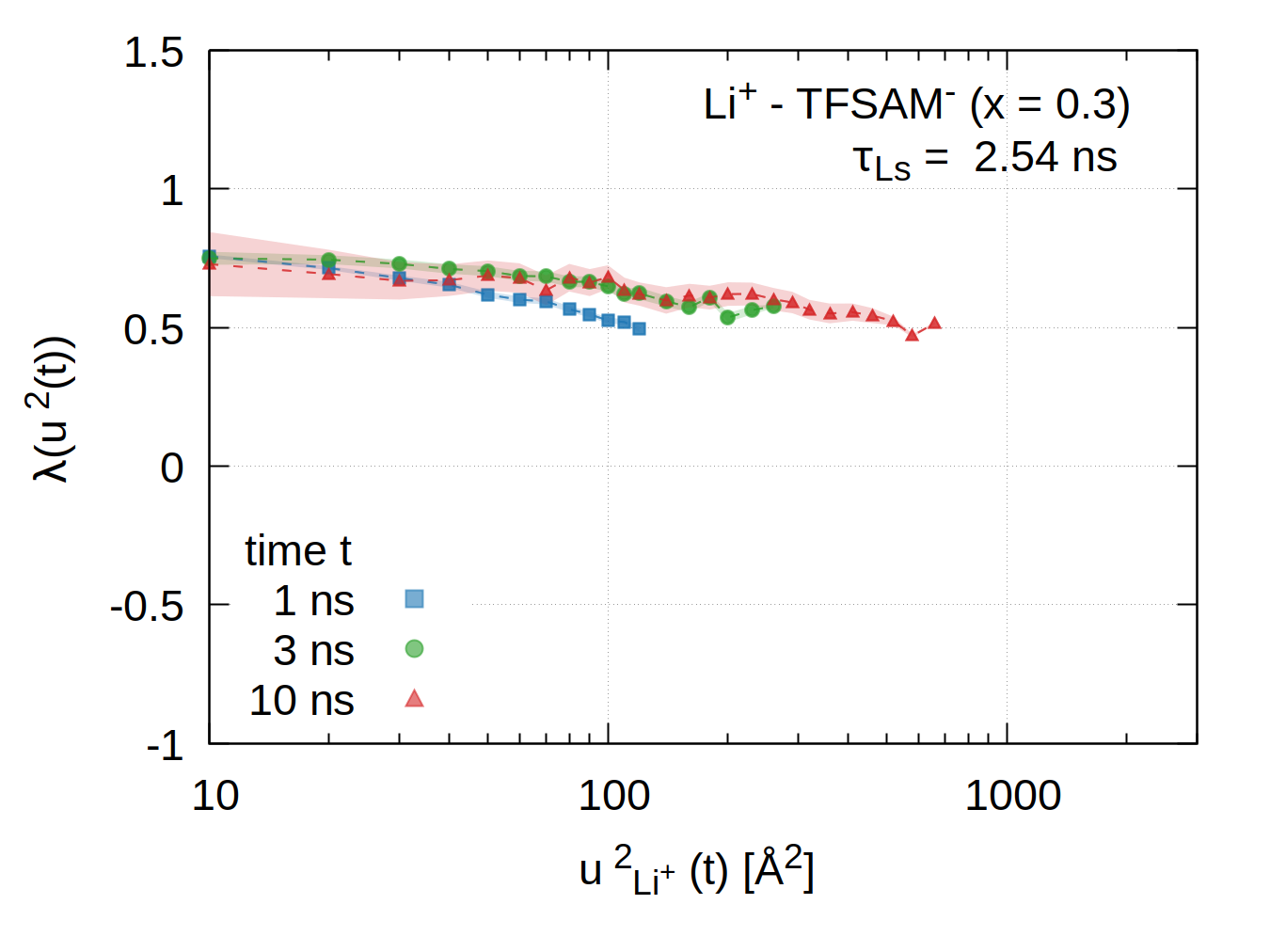}}
\end{figure}

\begin{figure}[H]
  \centering
  \subfloat{\includegraphics[width=0.5\textwidth]{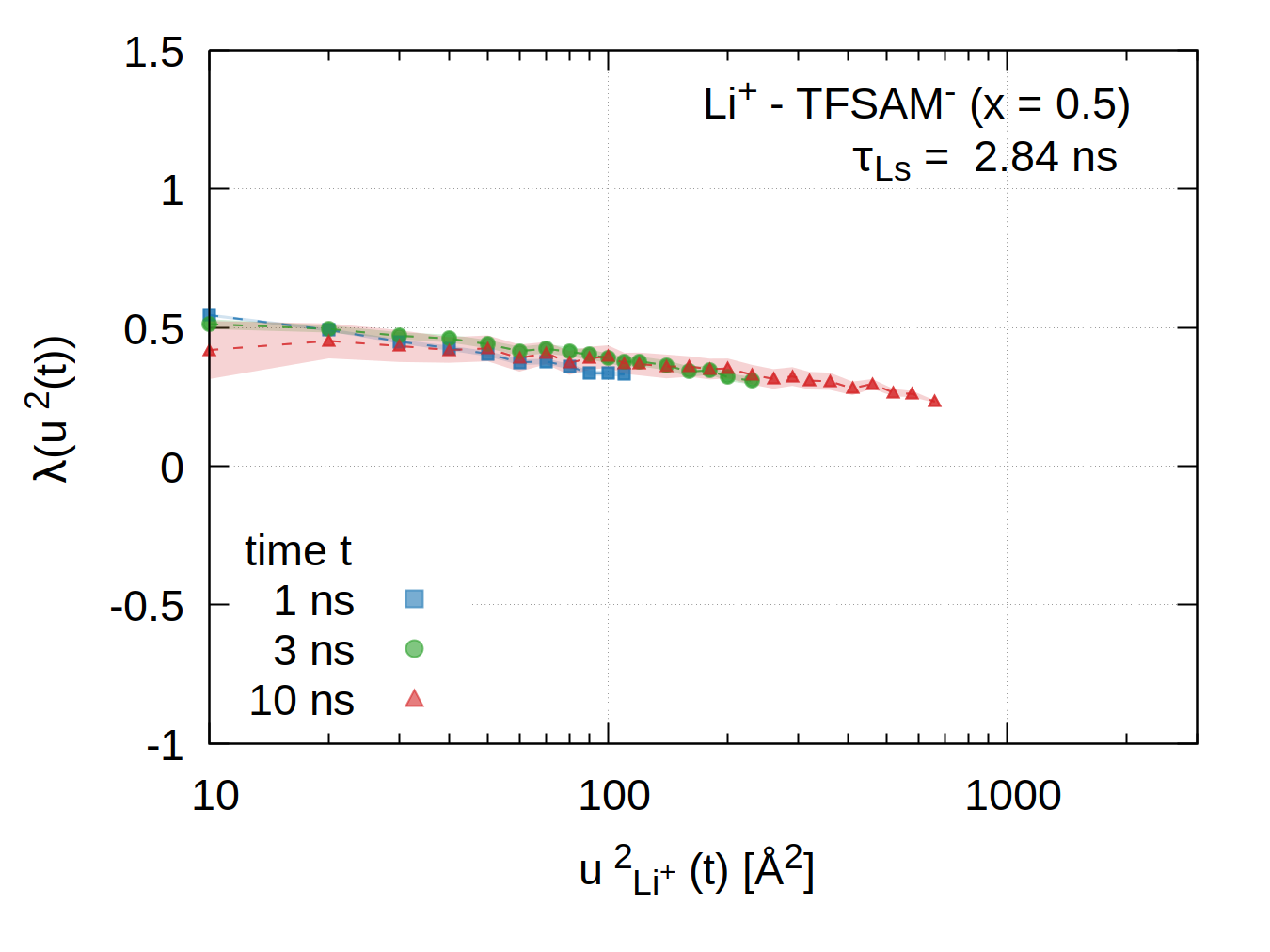}}
  \hfill
  \subfloat{\includegraphics[width=0.5\textwidth]{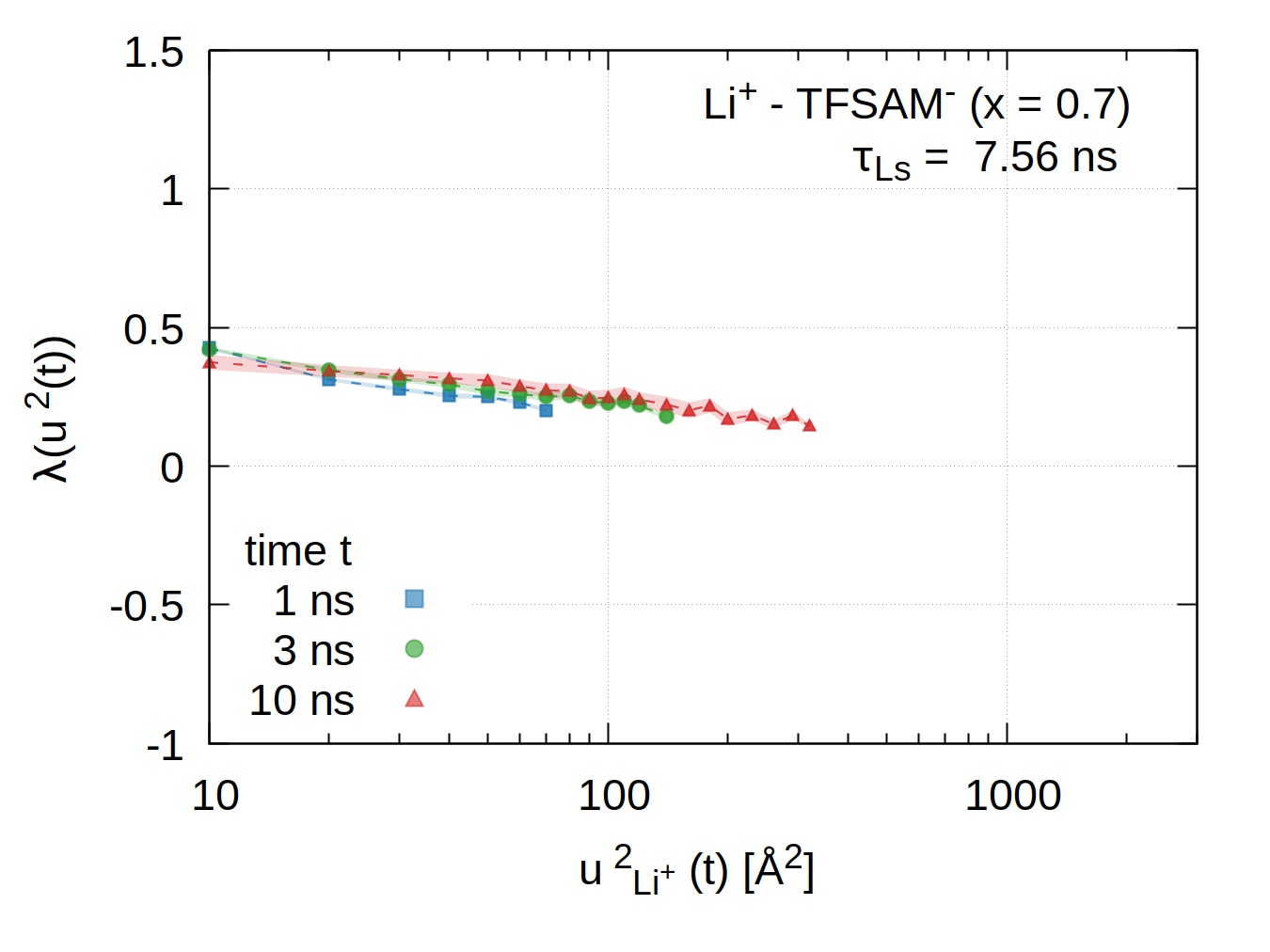}}

  \caption{Coupling of anion motion to lithium dynamics measured via $\lambda$ as a function of $u^2(\text{t})_{\text{Li}^+}$ exemplary for the lithium salt fractions x\,=\,0.1, 0.3, 0.5 and 0.7 in the $\text{TFSAM}^-$-containing mixtures.}
    \label{fig:lambda_x2_tfsam}

\end{figure}

%%%%%%%%%%%%%%%%%%%%%%%%%%%%%%%%%%%%%%%%%%%%%%%%%%%%%%%%%%%%%%%%%%%%%%%%%%%%%%%%%%
%%%%%%%%%%%%%%%%%%%%%%%%%%%%%%%%%%%%%%%%%%%%%%%%%%%%%%%%%%%%%%%%%%%%%%%%%%%%%%%%%%
%%%%%%%%%%%%%%%%%%%%%%%%%%%%%%%%%%%%%%%%%%%%%%%%%%%%%%%%%%%%%%%%%%%%%%%%%%%%%%%%%%
%%%%%%%%%%%%%%%%%%%%%%%%%%%%%%%%%%%%%%%%%%%%%%%%%%%%%%%%%%%%%%%%%%%%%%%%%%%%%%%%%%
%%%%%%%%%%%%%%%%%%%%%%%%%%%%%%%%%%%%%%%%%%%%%%%%%%%%%%%%%%%%%%%%%%%%%%%%%%%%%%%%%%
%%%%%%%%%%%%%%%%%%%%%%%%%%%%%%%%%%%%%%%%%%%%%%%%%%%%%%%%%%%%%%%%%%%%%%%%%%%%%%%%%%
%%%%%%%%%%%%%%%%%%%%%%%%%%%%%%%%%%%%%%%%%%%%%%%%%%%%%%%%%%%%%%%%%%%%%%%%%%%%%%%%%%
%%%%%%%%%%%%%%%%%%%%%%%%%%%%%%%%%%%%%%%%%%%%%%%%%%%%%%%%%%%%%%%%%%%%%%%%%%%%%%%%%%
%%%%%%%%%%%%%%%%%%%%%%%%%%%%%%%%%%%%%%%%%%%%%%%%%%%%%%%%%%%%%%%%%%%%%%%%%%%%%%%%%%
%%%%%%%%%%%%%%%%%%%%%%%%%%%%%%%%%%%%%%%%%%%%%%%%%%%%%%%%%%%%%%%%%%%%%%%%%%%%%%%%%%
%%%%%%%%%%%%%%%%%%%%%%%%%%%%%%%%%%%%%%%%%%%%%%%%%%%%%%%%%%%%%%%%%%%%%%%%%%%%%%%%%%
%%%%%%%%%%%%%%%%%%%%%%%%%%%%%%%%%%%%%%%%%%%%%%%%%%%%%%%%%%%%%%%%%%%%%%%%%%%%%%%%%%

\newpage
\textbf{J: $\vec{\epsilon}^2$, $\vec{\epsilon}_{\parallel}^2$ and $\vec{\epsilon}_{\perp}^2$ as a function of squared lithium displacement $u^2(t)$}

One may split the random motion $\vec{\epsilon}$ of an anion into contributions parallel and orthogonal to the lithium path direction $\hat{r}\,=\,\vec{u}\,/\,|\,\vec{u}\,|$, which can be computed as

\begin{equation}
\begin{aligned}
  \vec{\epsilon} &=\, \vec{v} -  \lambda_{u^{\,2}} \cdot \vec{u}\\
  \vec{\epsilon}_{\parallel} &=\, \left(\vec{\epsilon}\cdot \hat{r} \right)\,\cdot \, \hat{r}\\
  \vec{\epsilon}_{\perp} &=\,\vec{\epsilon}\,-\,\vec{\epsilon}_{\parallel} . 
\end{aligned}
\end{equation}
$\lambda_{u^{\,2}}$ corresponds to the definition in Equation 7 in the main manuscript \newline
$\lambda(u^{\,2},t)\,=\,\dfrac{\langle\,\vec{u}_i\,\cdot\,\vec{v}^{\,\,i}_{j}\,\rangle_{u^2, t}}{{u}^{\,2}}\,=\,\dfrac{\langle v_{\parallel}\rangle_{u^2,t}}{u}$, \textit{i.e.}, the data shown in Figures S\ref{fig:lambda_x2_tfsi} and S\ref{fig:lambda_x2_tfsam}.

\begin{figure}[H]
  \centering
  \subfloat{\includegraphics[width=0.5\textwidth]{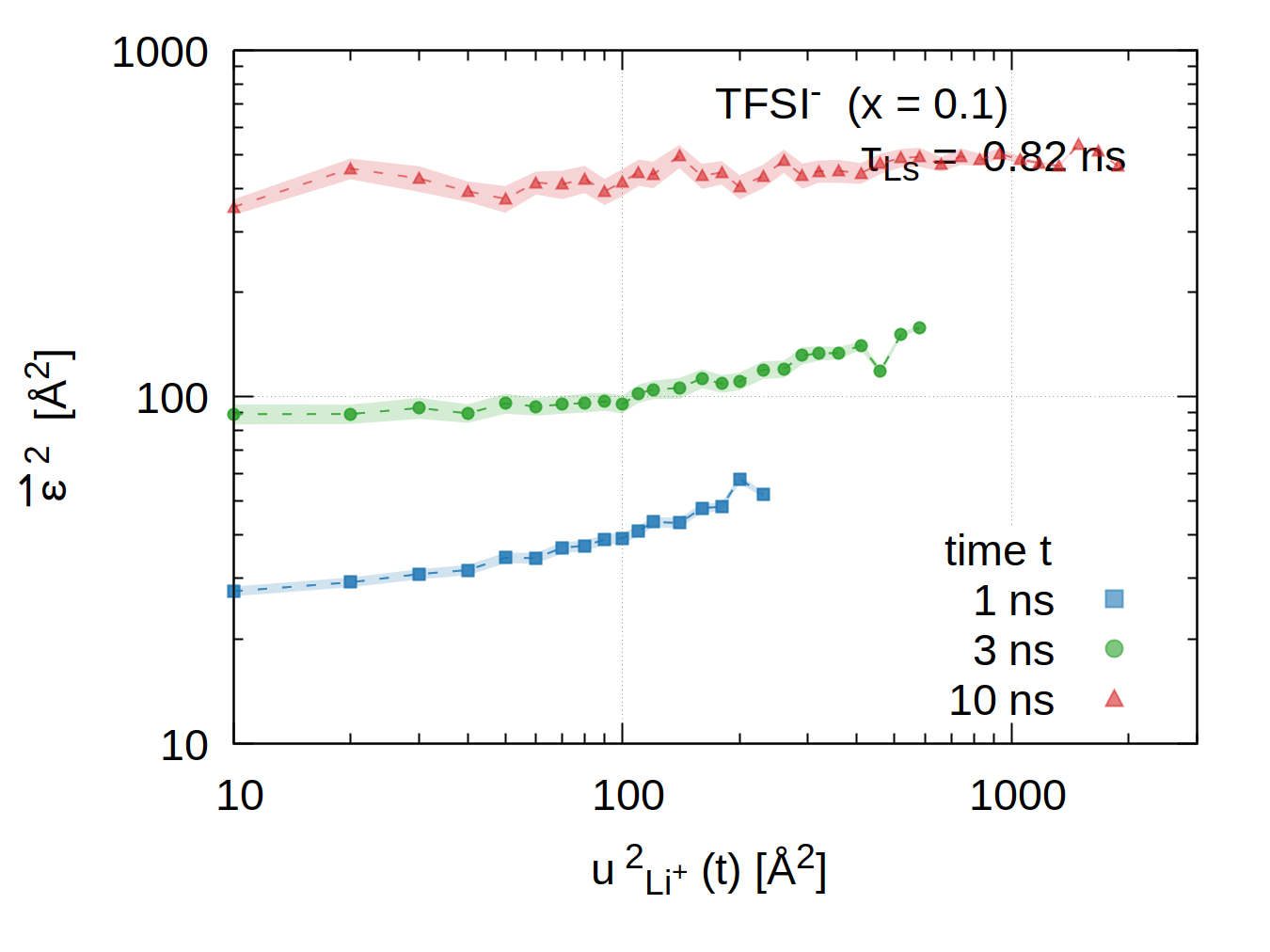}}
  \hfill
  \subfloat{\includegraphics[width=0.5\textwidth]{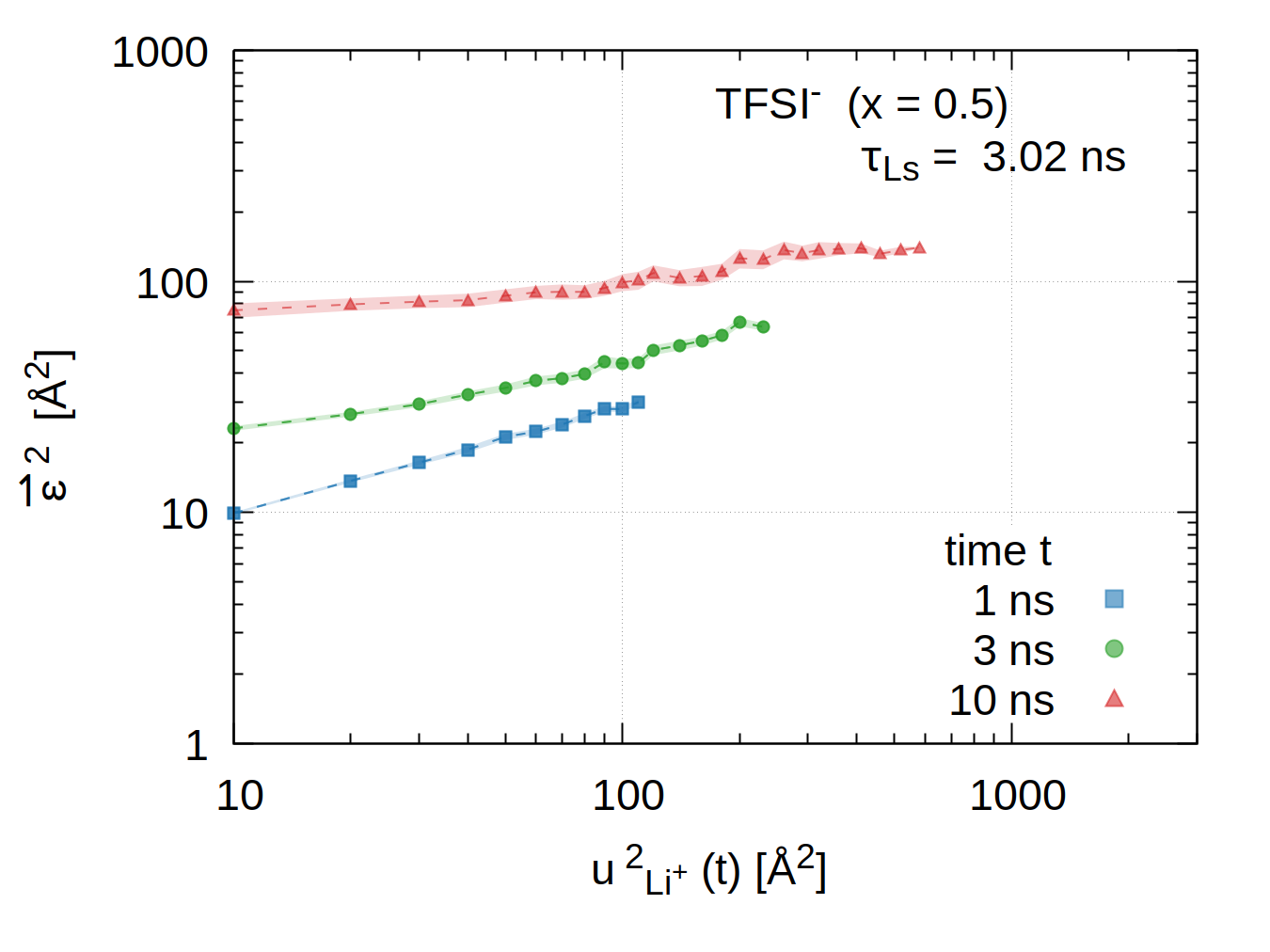}}
\end{figure}

\begin{figure}[H]
  \centering
  \subfloat{\includegraphics[width=0.5\textwidth]{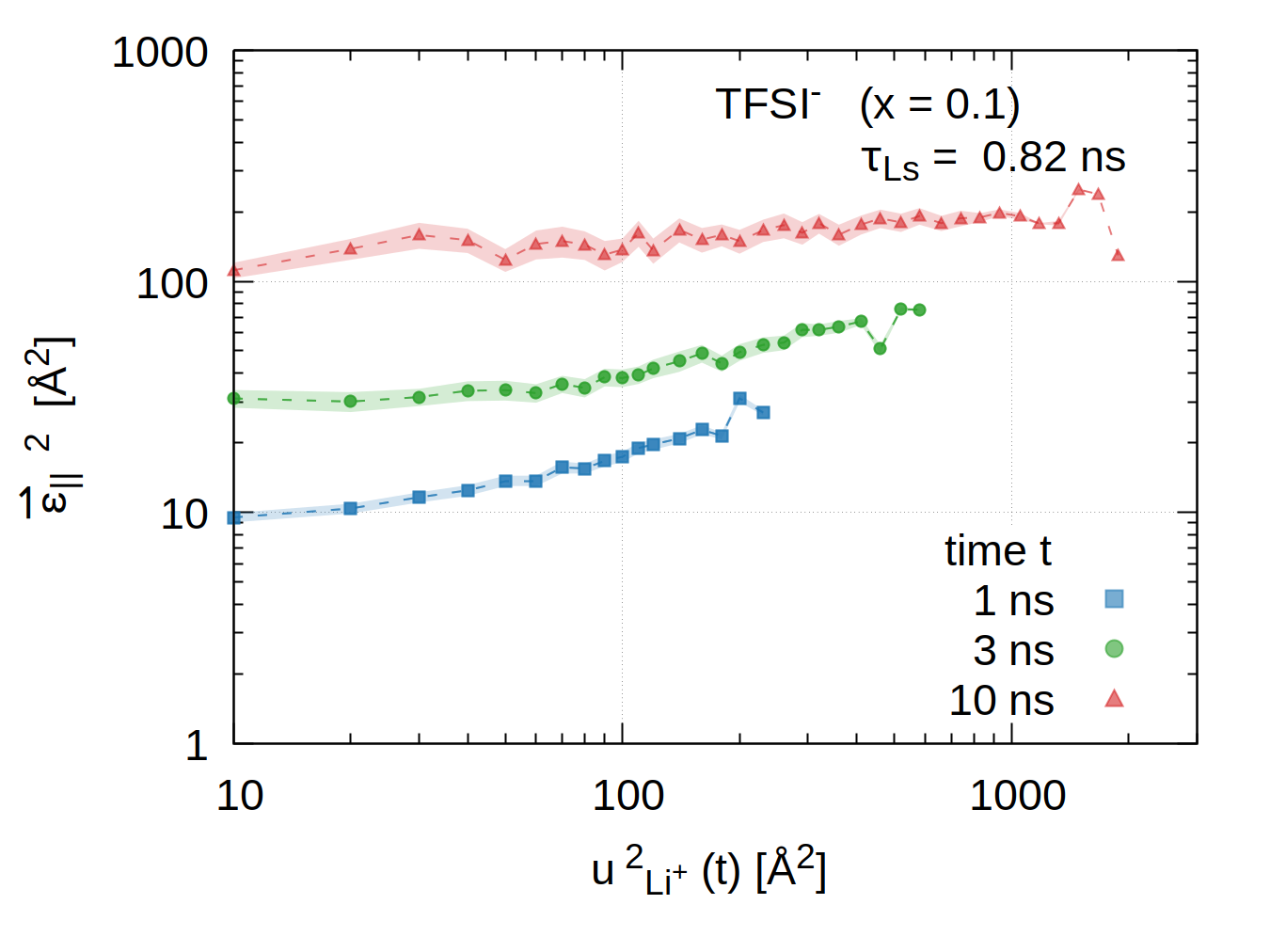}}
  \hfill
  \subfloat{\includegraphics[width=0.5\textwidth]{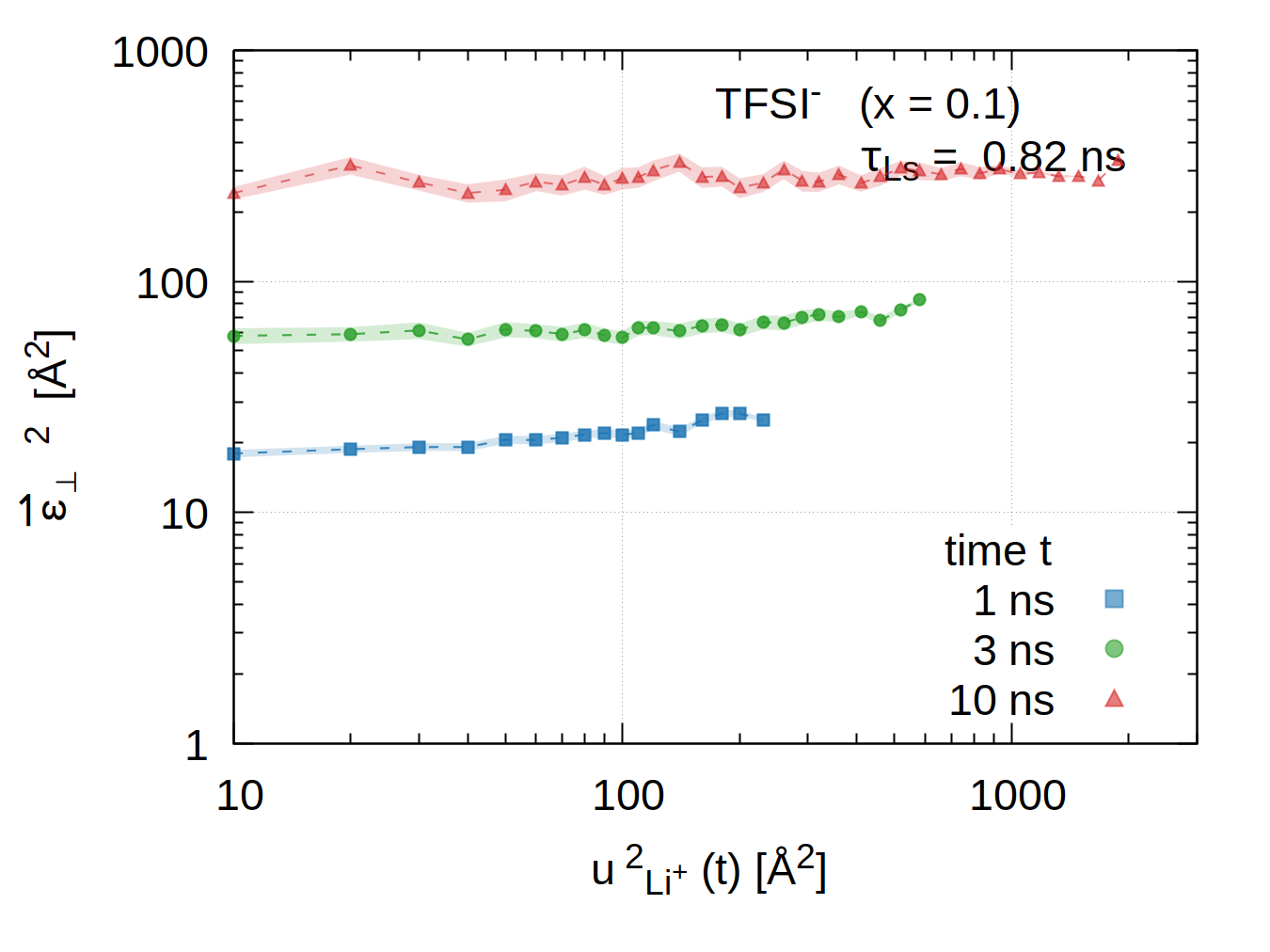}}
\end{figure}

\begin{figure}[H]
  \centering
  \subfloat{\includegraphics[width=0.5\textwidth]{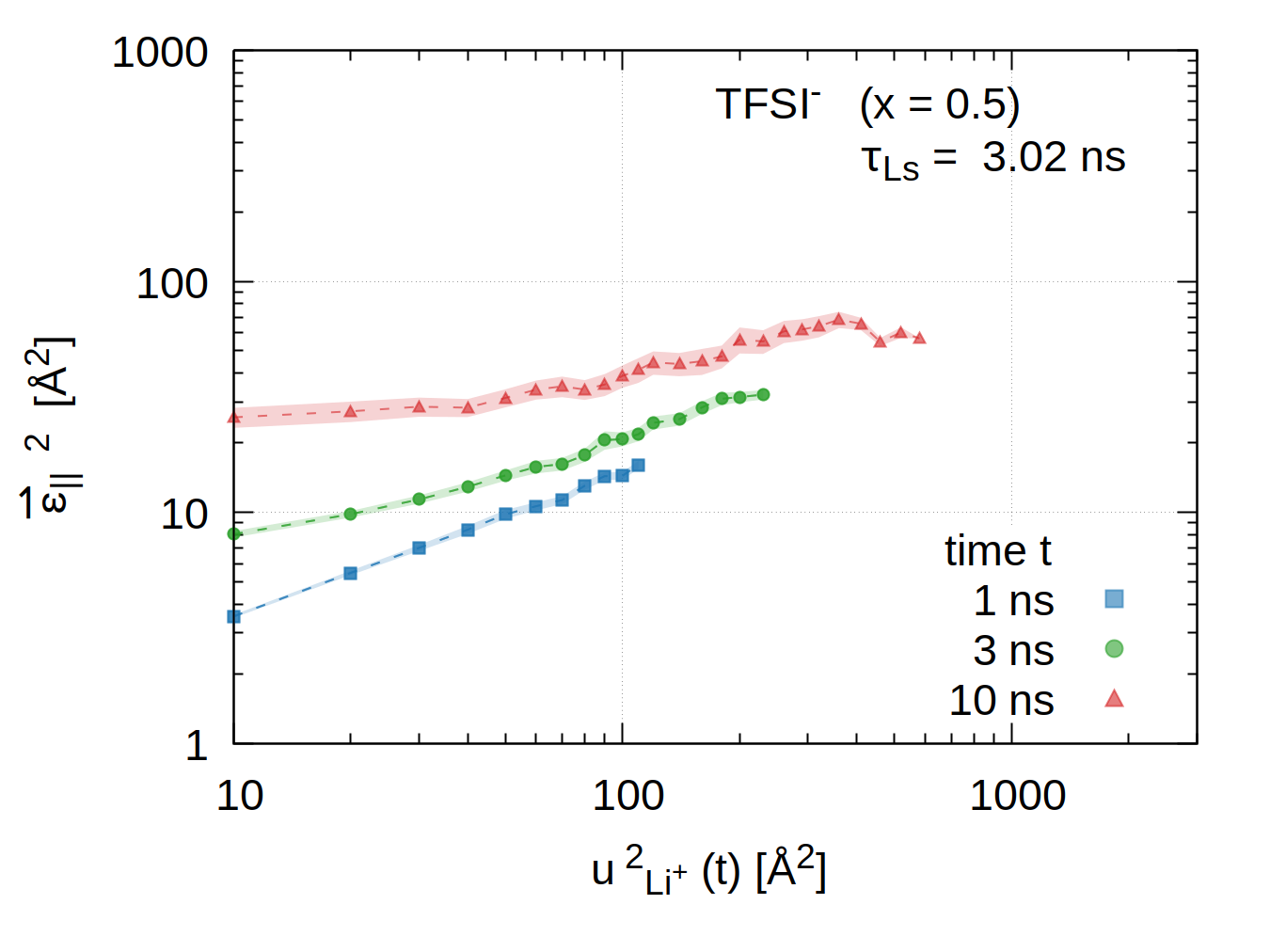}}
  \hfill
  \subfloat{\includegraphics[width=0.5\textwidth]{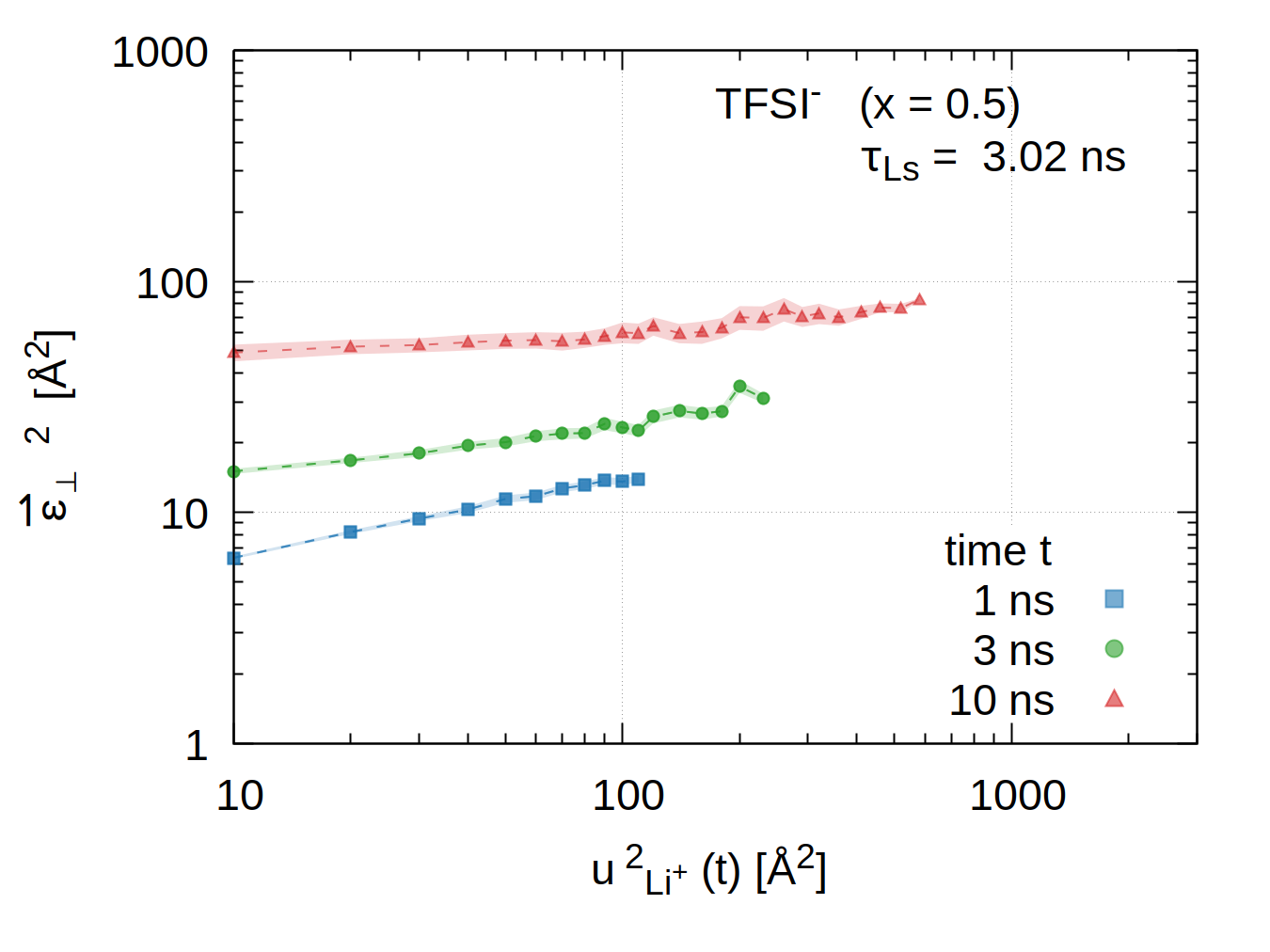}}
 
 \caption{Variances $\vec{\epsilon}^2$, $\vec{\epsilon}_{\parallel}^2$ and $\vec{\epsilon}_{\perp}^2$   as a function of $u^2(\text{t})_{\text{Li}^+}$ for $\text{TFSI}^-$ exemplary for salt contents x\,=\,0.1 and x\,=\,0.5.}
 \label{fig:eps_para_ortho_x2_tfsi}
\end{figure}

\begin{figure}[H]
  \centering
  \subfloat{\includegraphics[width=0.5\textwidth]{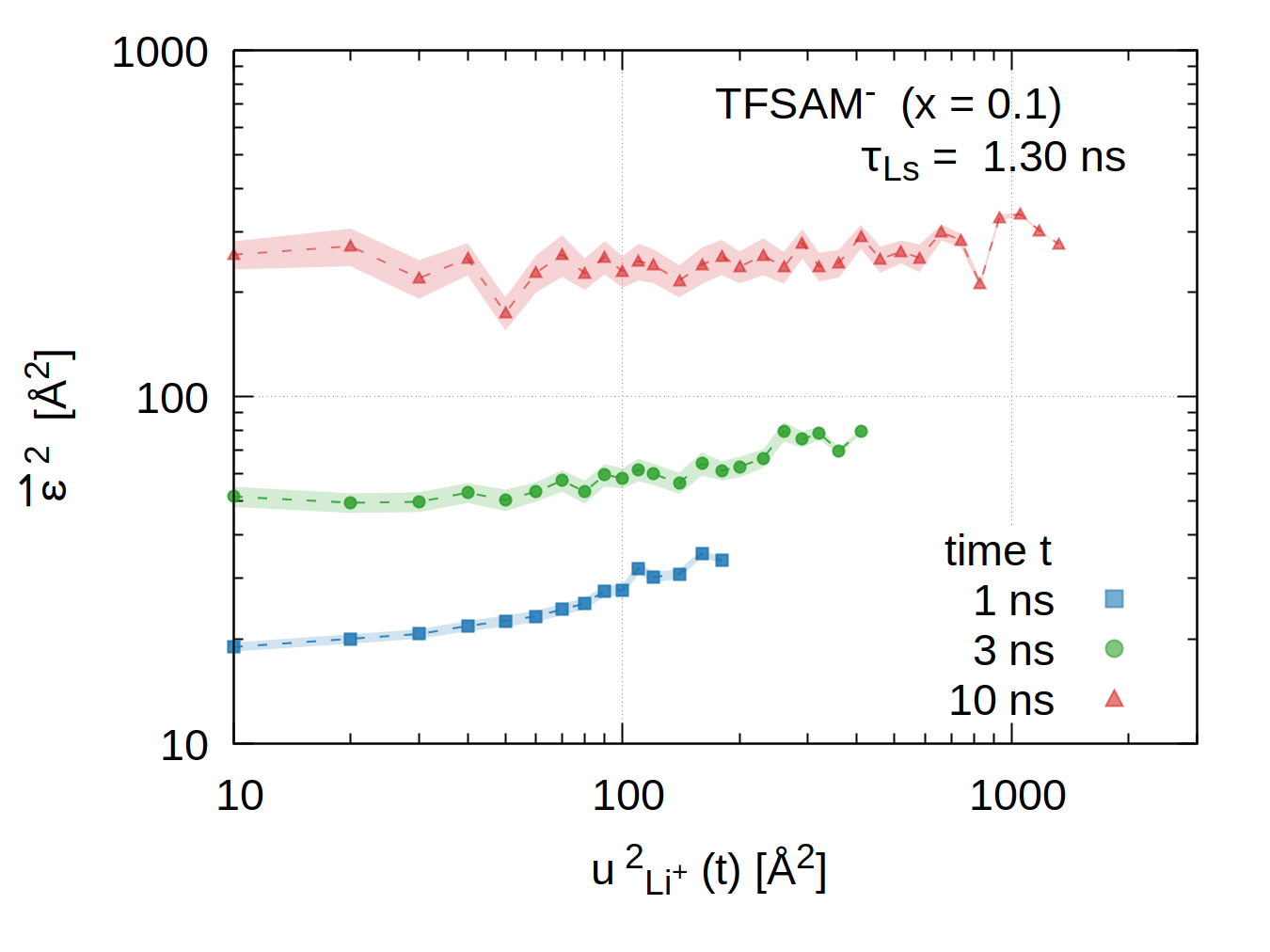}}
  \hfill
  \subfloat{\includegraphics[width=0.5\textwidth]{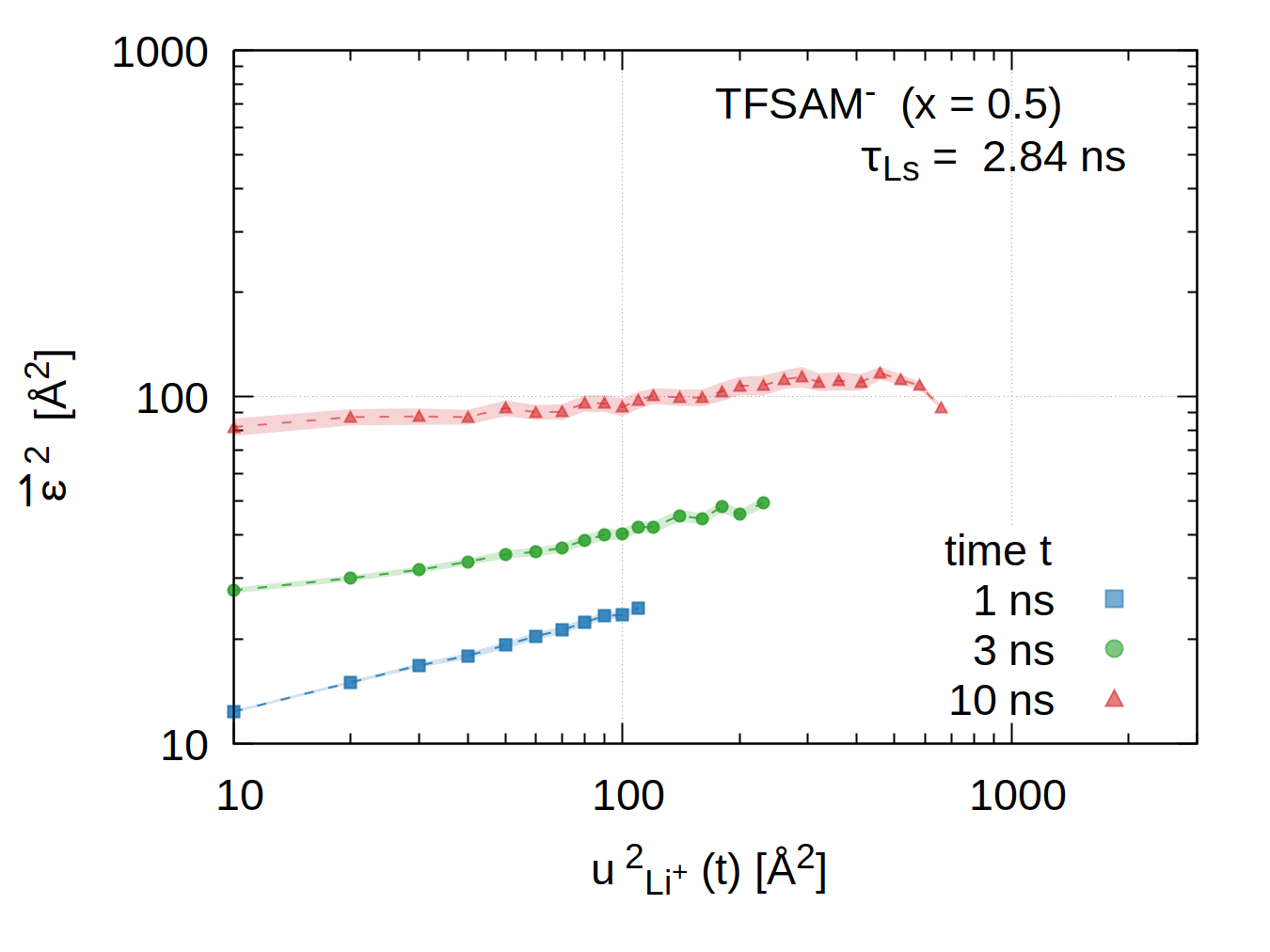}}
\end{figure}

\begin{figure}[H]
  \centering
  \subfloat{\includegraphics[width=0.5\textwidth]{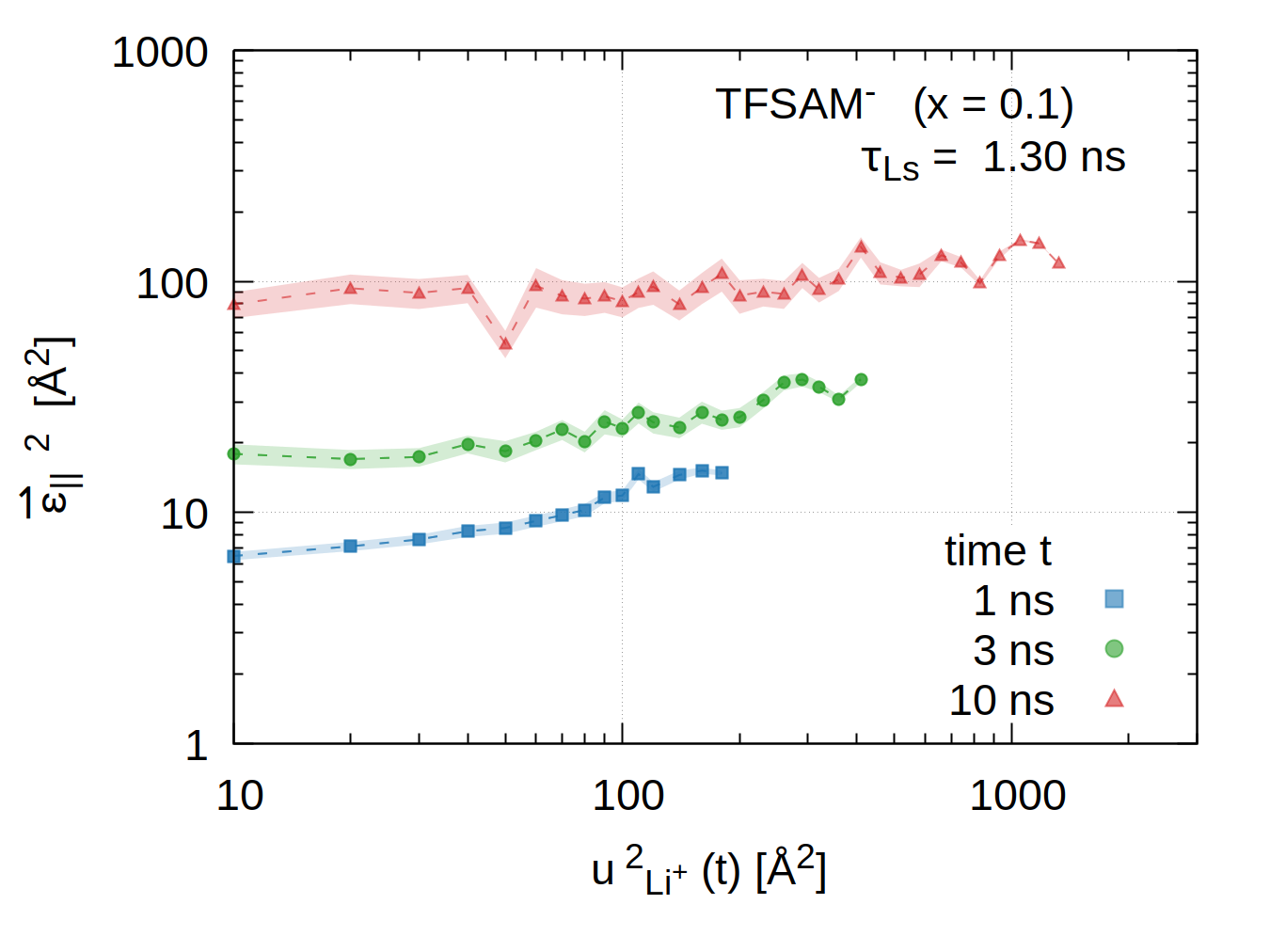}}
  \hfill
  \subfloat{\includegraphics[width=0.5\textwidth]{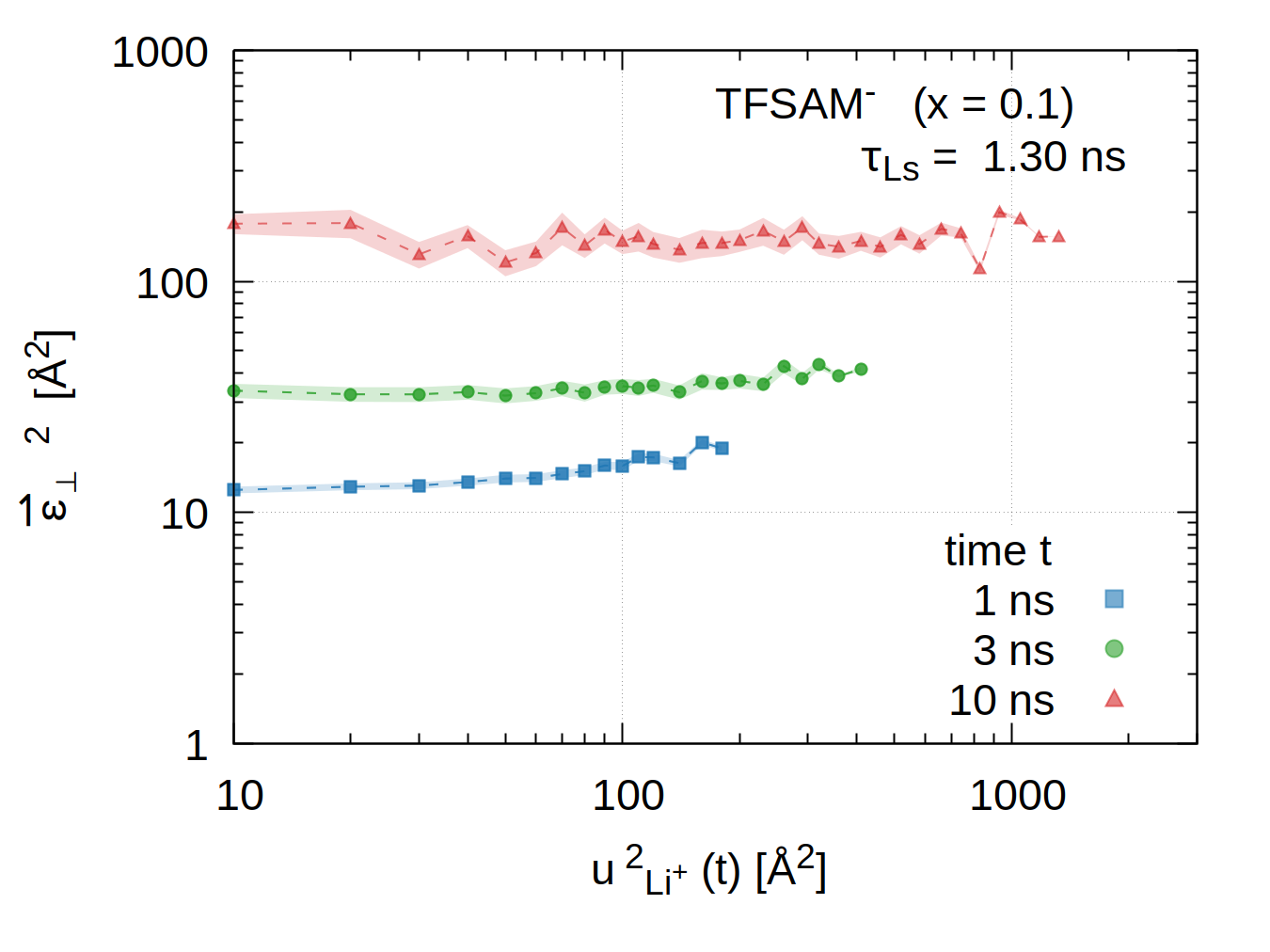}}
\end{figure}

\begin{figure}[H]
  \centering
  \subfloat{\includegraphics[width=0.5\textwidth]{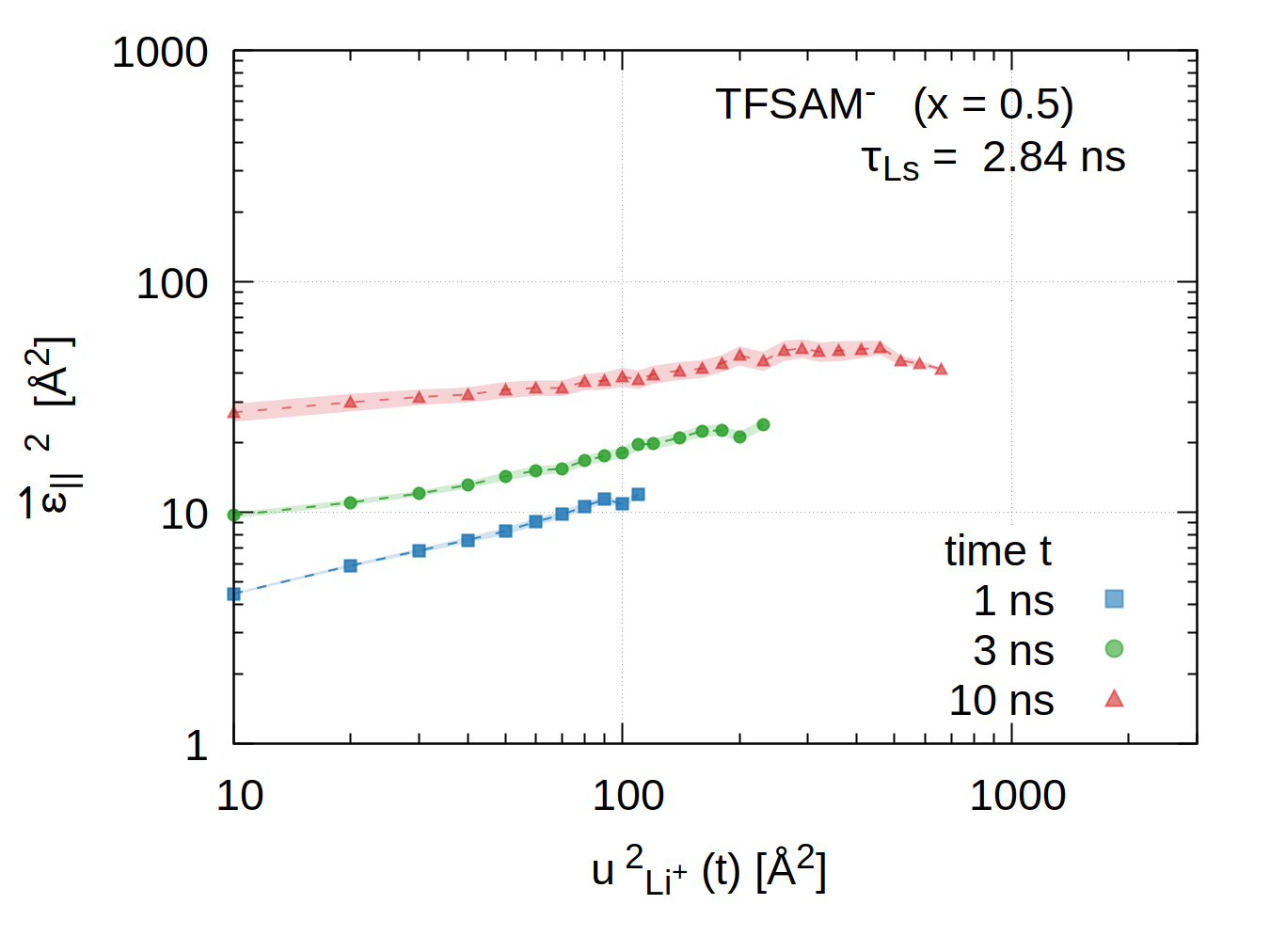}}
  \hfill
  \subfloat{\includegraphics[width=0.5\textwidth]{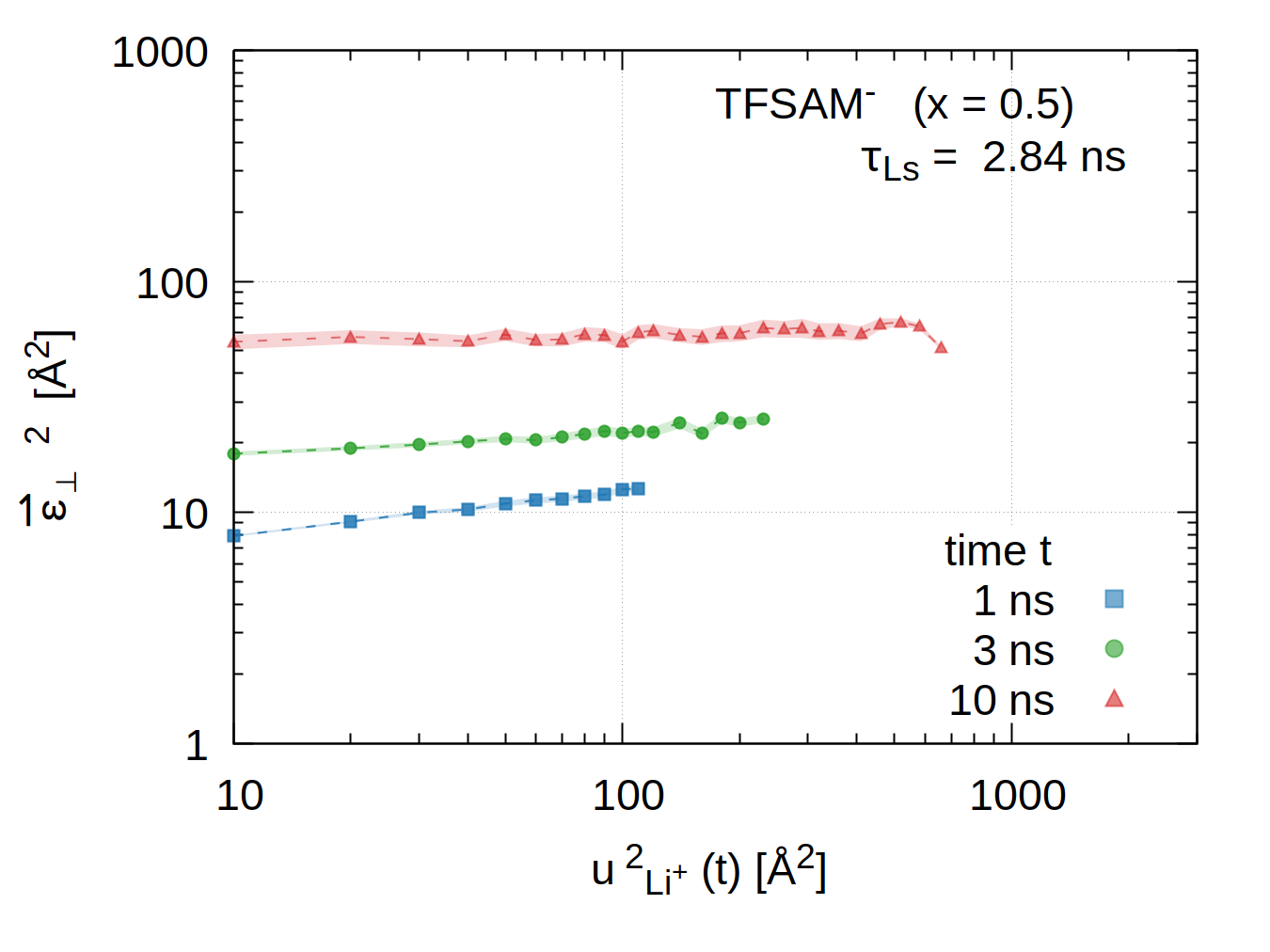}}
 
  \caption{Variances $\vec{\epsilon}^2$, $\vec{\epsilon}_{\parallel}^2$ and $\vec{\epsilon}_{\perp}^2$  as a function of $u^2(\text{t})_{\text{Li}^+}$ for $\text{TFSAM}^-$ exemplary for salt contents x\,=\,0.1 and x\,=\,0.5.}
 
\label{fig:eps_para_ortho_x2_tfsam}

\end{figure}

%%%%%%%%%%%%%%%%%%%%%%%%%%%%%%%%%%%%%%%%%%%%%%%%%%%%%%%%%%%%%%%%%%%%%%%%%%%%%%%%%%
%%%%%%%%%%%%%%%%%%%%%%%%%%%%%%%%%%%%%%%%%%%%%%%%%%%%%%%%%%%%%%%%%%%%%%%%%%%%%%%%%%
%%%%%%%%%%%%%%%%%%%%%%%%%%%%%%%%%%%%%%%%%%%%%%%%%%%%%%%%%%%%%%%%%%%%%%%%%%%%%%%%%%
%%%%%%%%%%%%%%%%%%%%%%%%%%%%%%%%%%%%%%%%%%%%%%%%%%%%%%%%%%%%%%%%%%%%%%%%%%%%%%%%%%
%%%%%%%%%%%%%%%%%%%%%%%%%%%%%%%%%%%%%%%%%%%%%%%%%%%%%%%%%%%%%%%%%%%%%%%%%%%%%%%%%%
%%%%%%%%%%%%%%%%%%%%%%%%%%%%%%%%%%%%%%%%%%%%%%%%%%%%%%%%%%%%%%%%%%%%%%%%%%%%%%%%%%
%%%%%%%%%%%%%%%%%%%%%%%%%%%%%%%%%%%%%%%%%%%%%%%%%%%%%%%%%%%%%%%%%%%%%%%%%%%%%%%%%%
%%%%%%%%%%%%%%%%%%%%%%%%%%%%%%%%%%%%%%%%%%%%%%%%%%%%%%%%%%%%%%%%%%%%%%%%%%%%%%%%%%
%%%%%%%%%%%%%%%%%%%%%%%%%%%%%%%%%%%%%%%%%%%%%%%%%%%%%%%%%%%%%%%%%%%%%%%%%%%%%%%%%%
%%%%%%%%%%%%%%%%%%%%%%%%%%%%%%%%%%%%%%%%%%%%%%%%%%%%%%%%%%%%%%%%%%%%%%%%%%%%%%%%%%
%%%%%%%%%%%%%%%%%%%%%%%%%%%%%%%%%%%%%%%%%%%%%%%%%%%%%%%%%%%%%%%%%%%%%%%%%%%%%%%%%%
%%%%%%%%%%%%%%%%%%%%%%%%%%%%%%%%%%%%%%%%%%%%%%%%%%%%%%%%%%%%%%%%%%%%%%%%%%%%%%%%%%

\newpage

\textbf{K: Correlation of the random motion of next-neighbor anions in a solvation shell}

To investigate the interaction between two initially adjacent anions $\text{anion}_1$ and $\text{anion}_2$ in a lithium solvation shell, we compute the correlation $\left( \vec{\epsilon}_1\cdot\vec{\epsilon}_2\right)/(\vec{\epsilon}^2)$
as well as quantify the contributions parallel $\left( \vec{\epsilon}_{1,\parallel}\cdot\vec{\epsilon}_{,\parallel}\right)/(\vec{\epsilon}_{\parallel}^2)$ and orthogonal $\left( \vec{\epsilon}_{1,\perp}\cdot\vec{\epsilon}_{,\perp}\right)/(\vec{\epsilon}_{\perp}^2)$ to the lithium pathway.

\begin{figure}[H]
  \centering
  \subfloat{\includegraphics[width=0.5\textwidth]{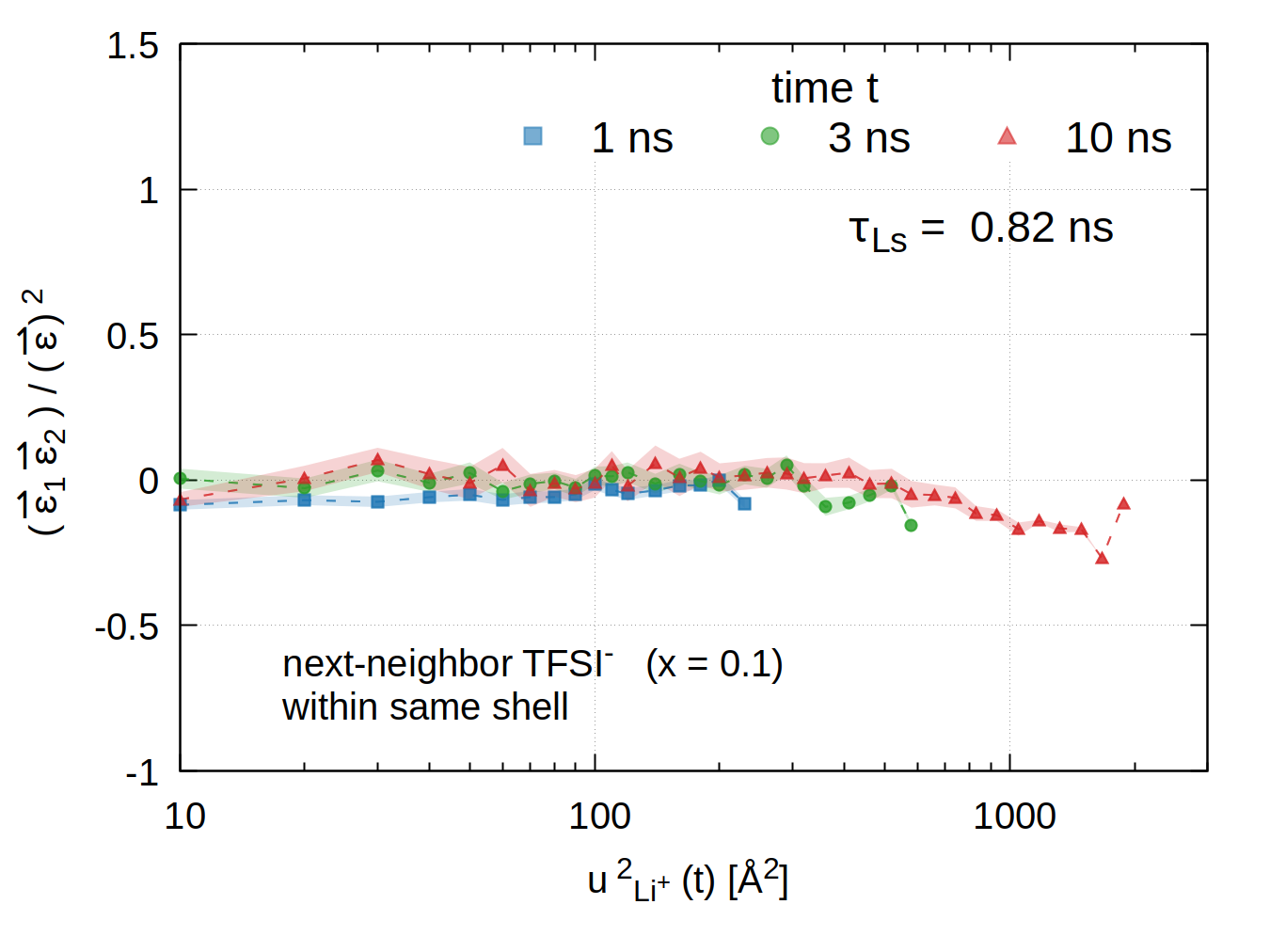}}
  \hfill
  \subfloat{\includegraphics[width=0.5\textwidth]{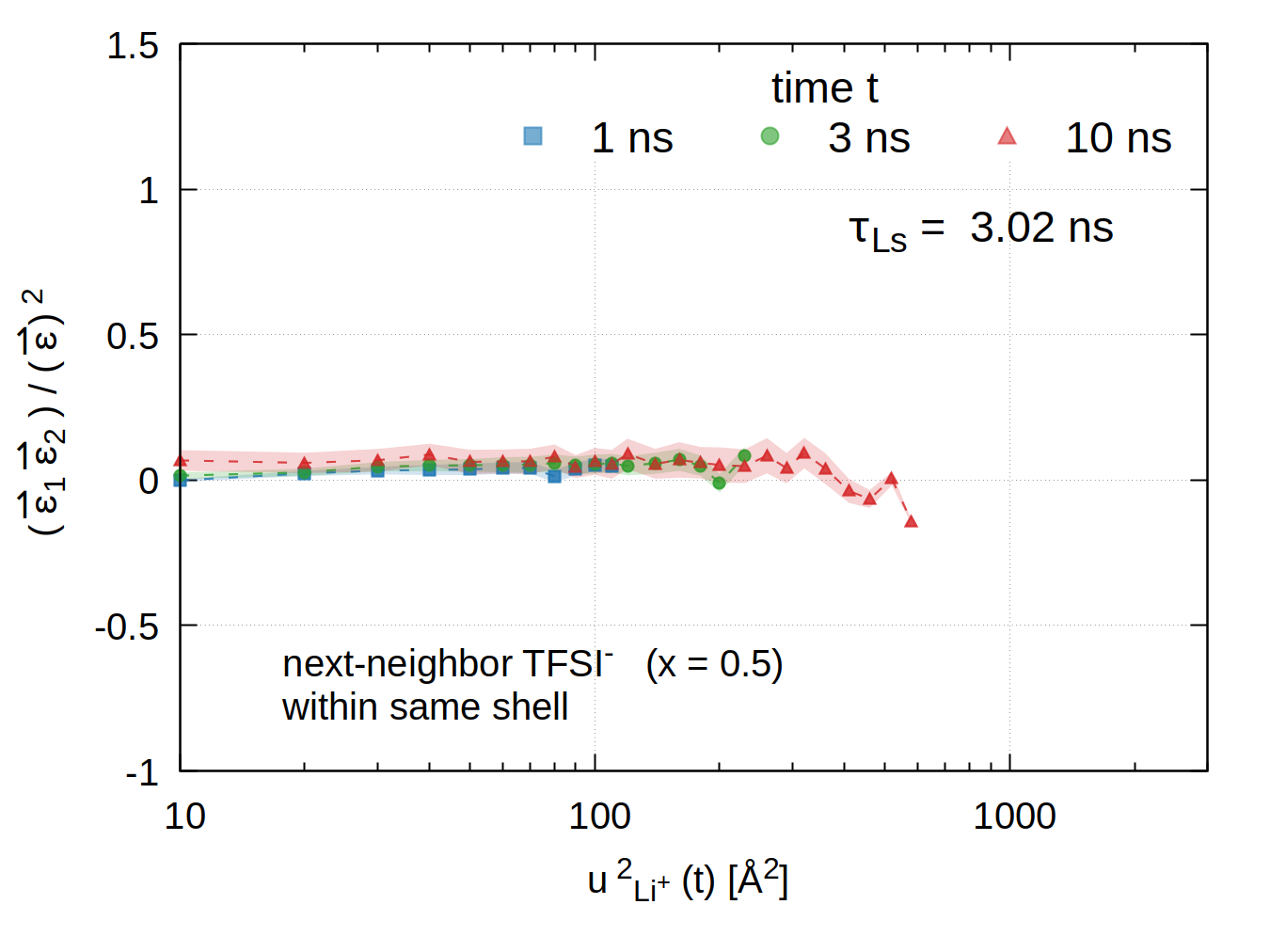}}
\end{figure}

\begin{figure}[H]
  \centering
  \subfloat{\includegraphics[width=0.5\textwidth]{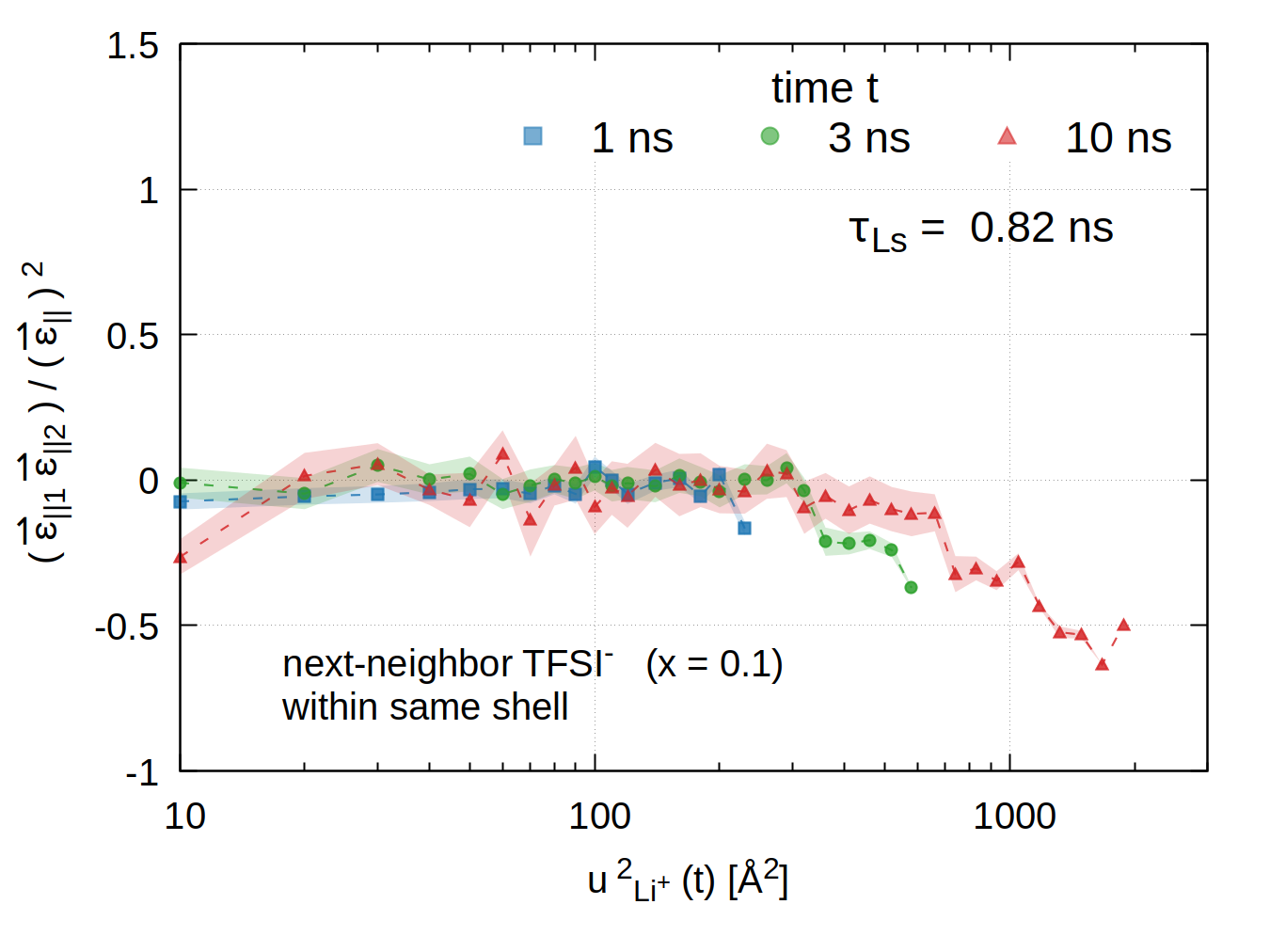}}
  \hfill
  \subfloat{\includegraphics[width=0.5\textwidth]{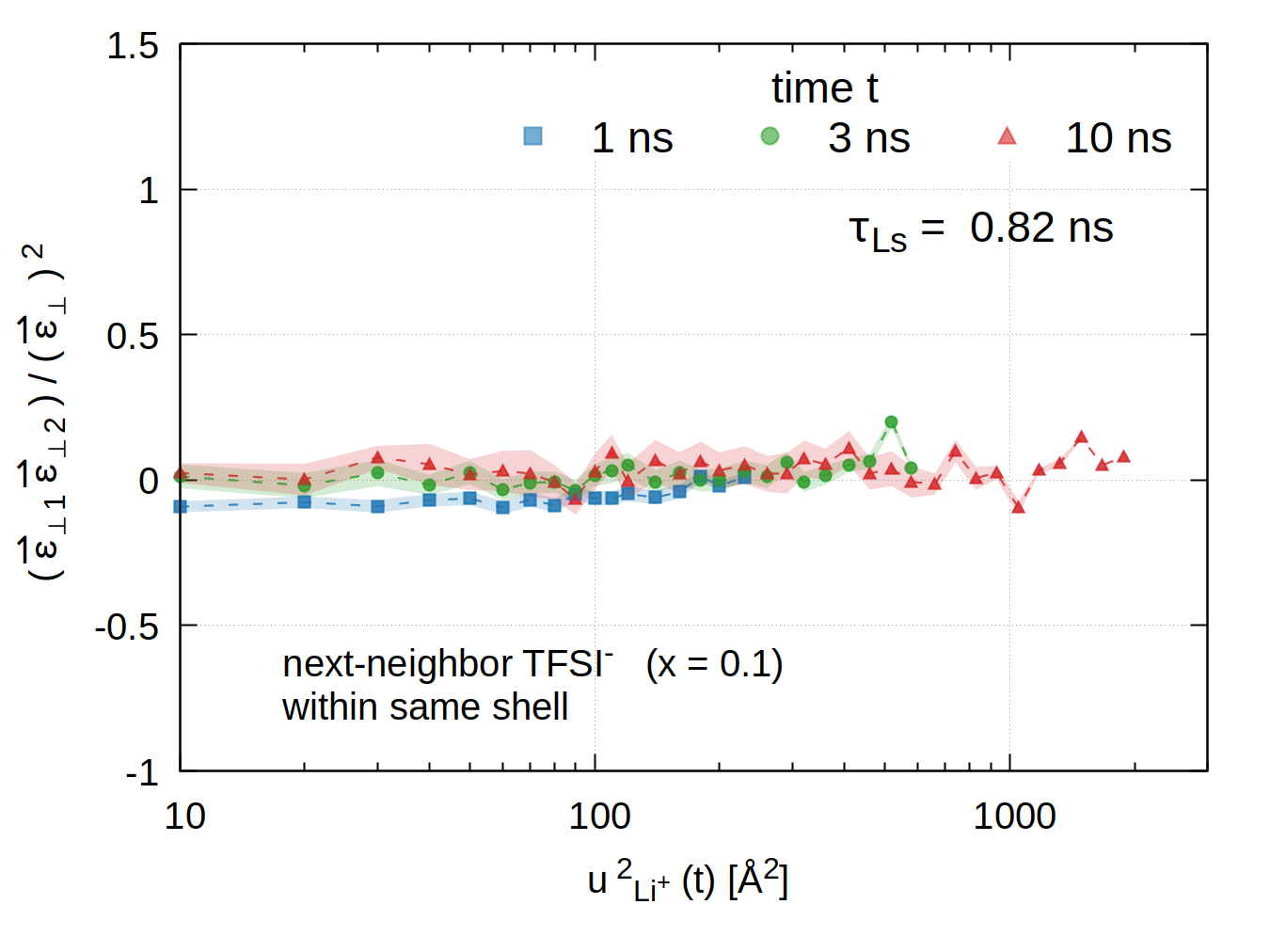}}
\end{figure}

\begin{figure}[H]
  \centering
  \subfloat{\includegraphics[width=0.5\textwidth]{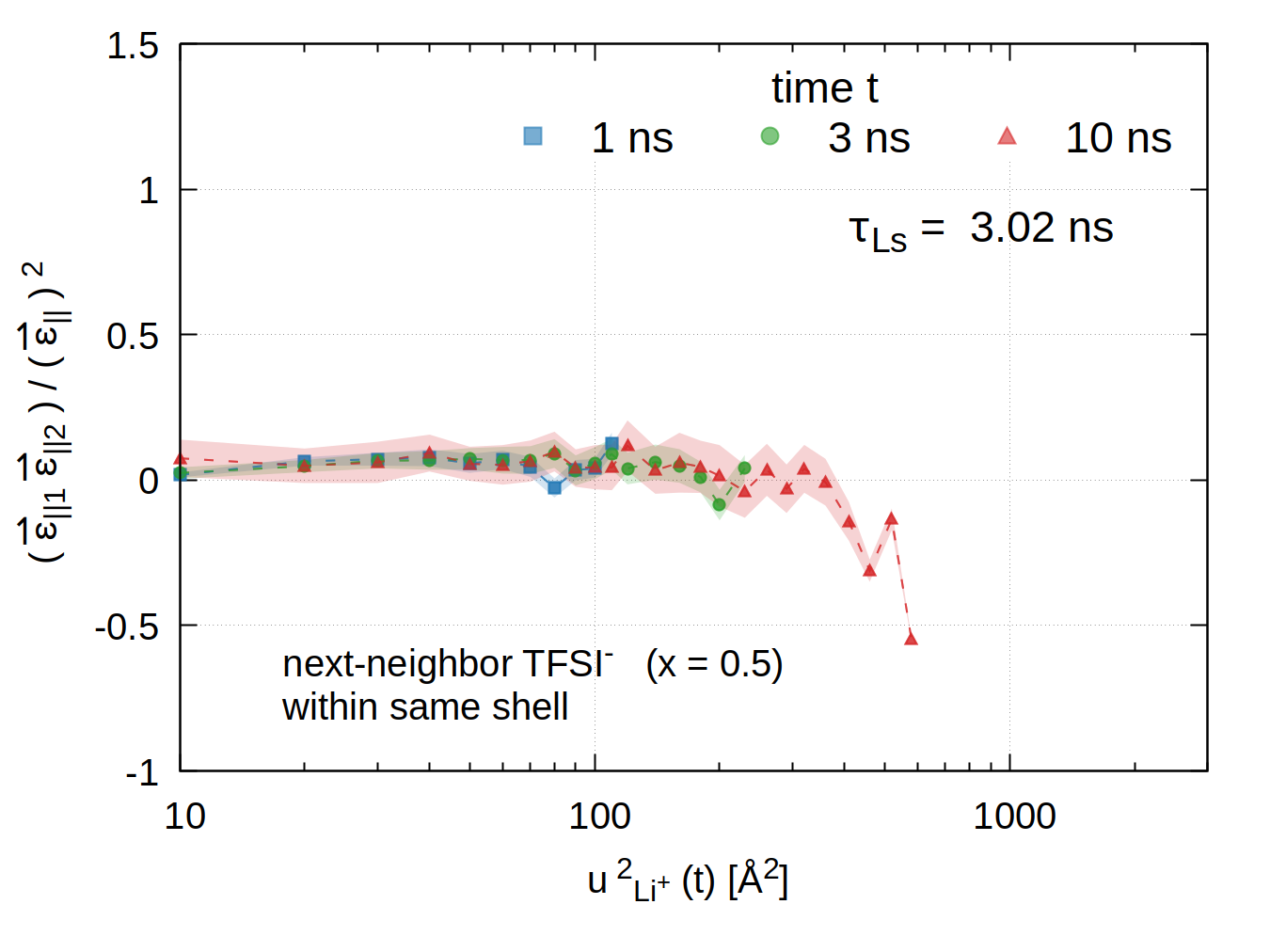}}
  \hfill
  \subfloat{\includegraphics[width=0.5\textwidth]{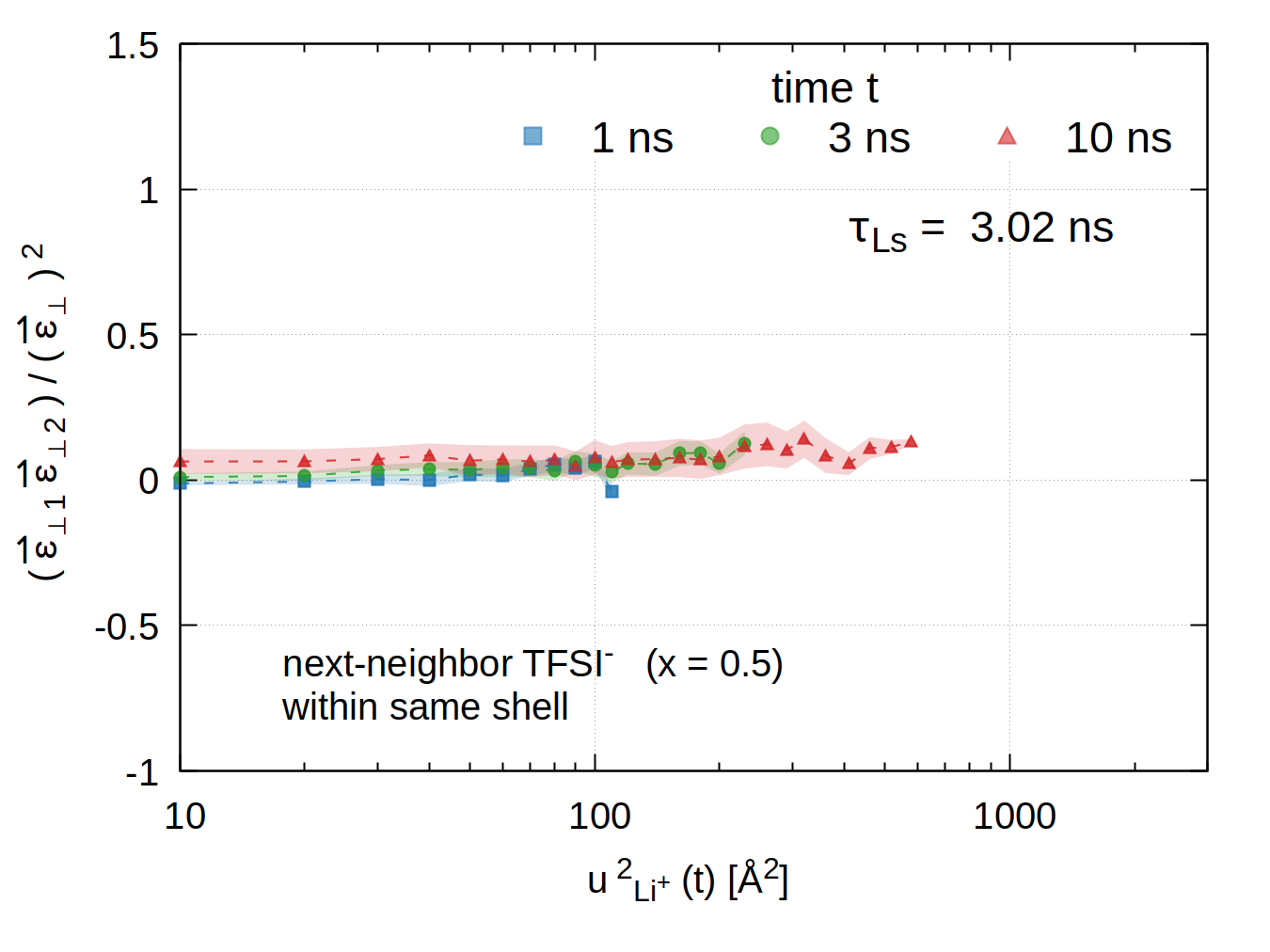}}
 
 \caption{Correlations of initial next-neighbor shell anions $\text{TFSI}^-_1$ and $\text{TFSI}^-_2$ as a function of $u^2(\text{t})_{\text{Li}^+}$ exemplary for salt contents x\,=\,0.1 and x\,=\,0.5.}
\end{figure}

\begin{figure}[H]
  \centering
  \subfloat{\includegraphics[width=0.5\textwidth]{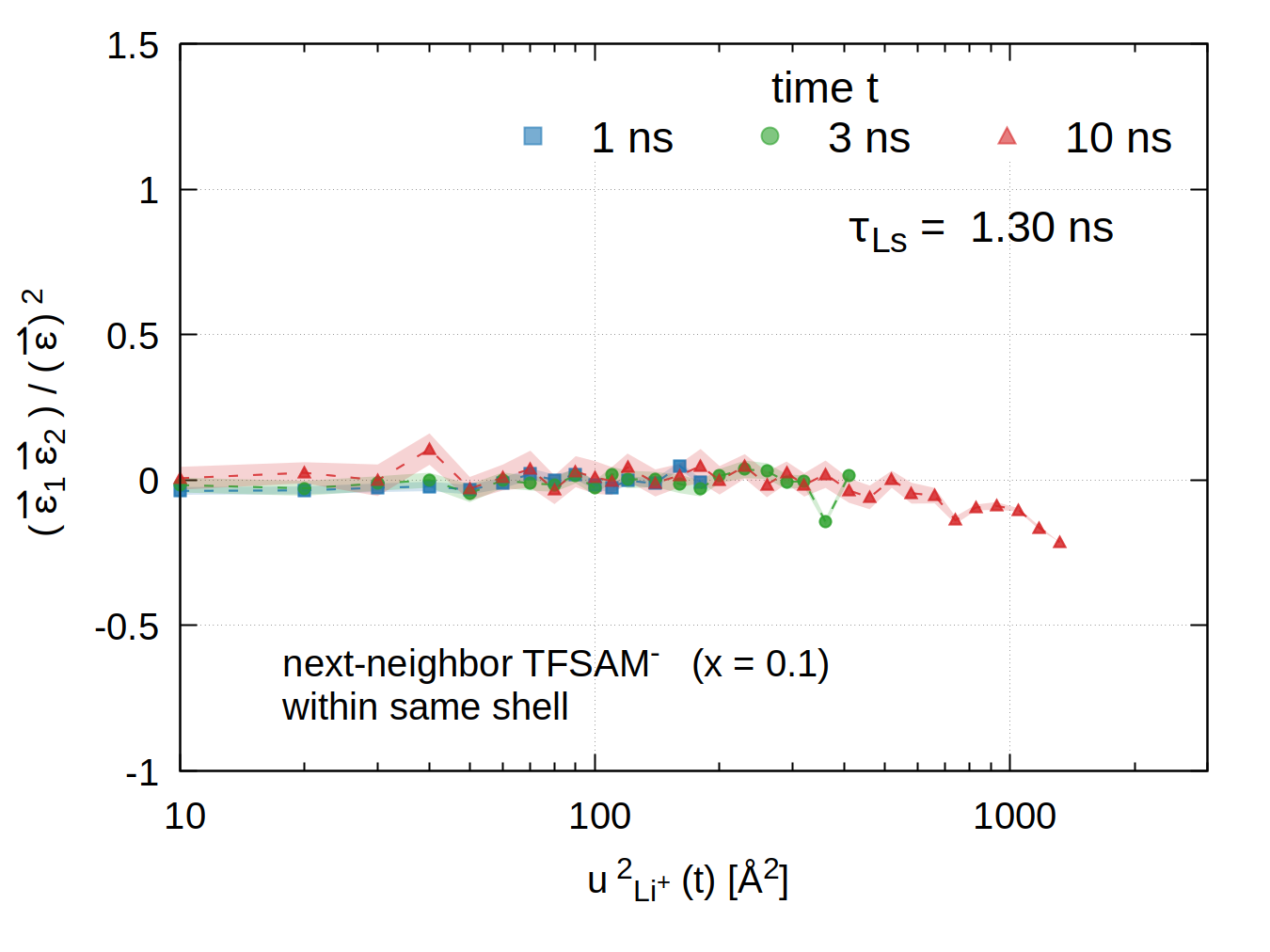}}
  \hfill
  \subfloat{\includegraphics[width=0.5\textwidth]{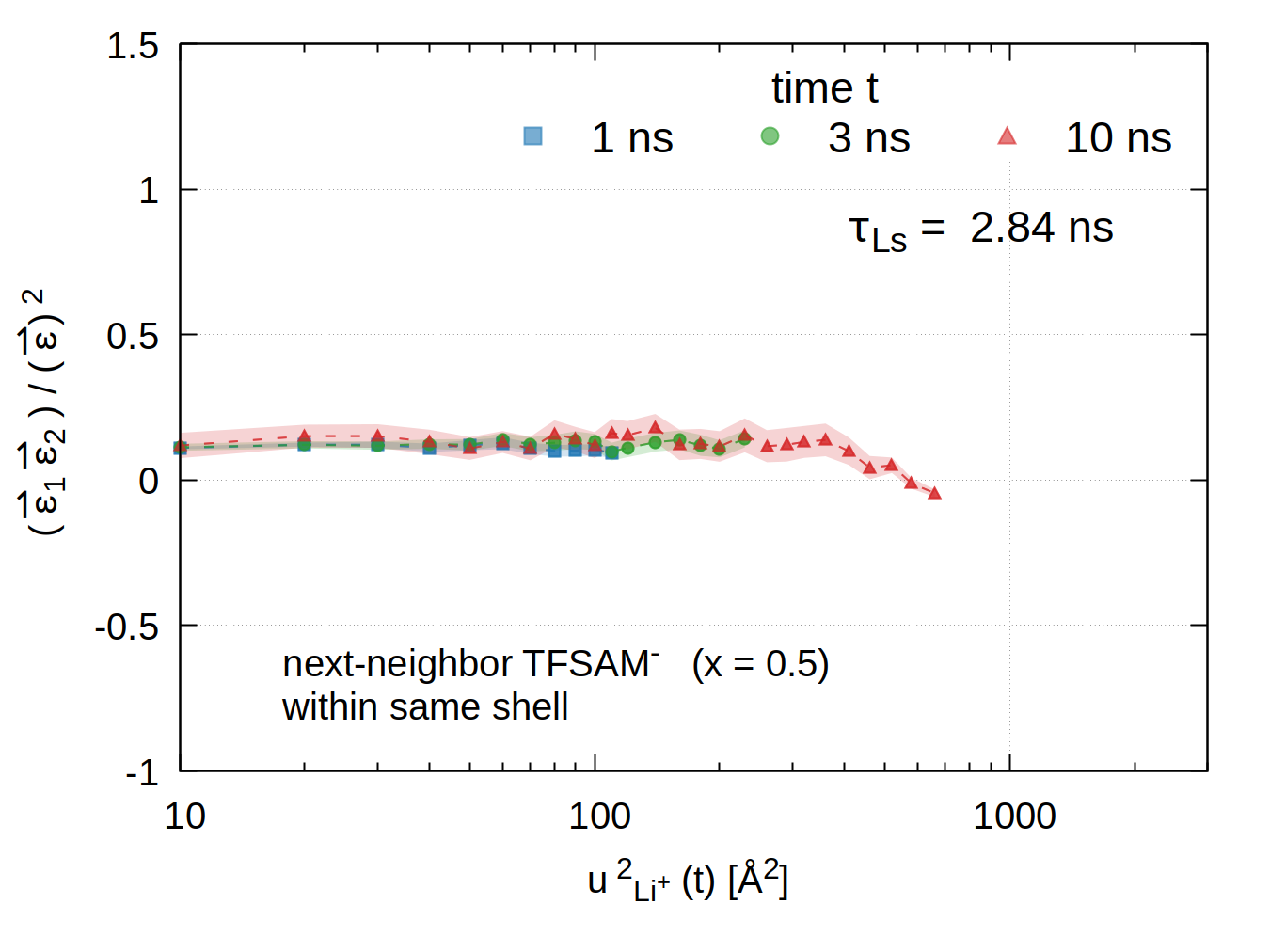}}
\end{figure}

\begin{figure}[H]
  \centering
  \subfloat{\includegraphics[width=0.5\textwidth]{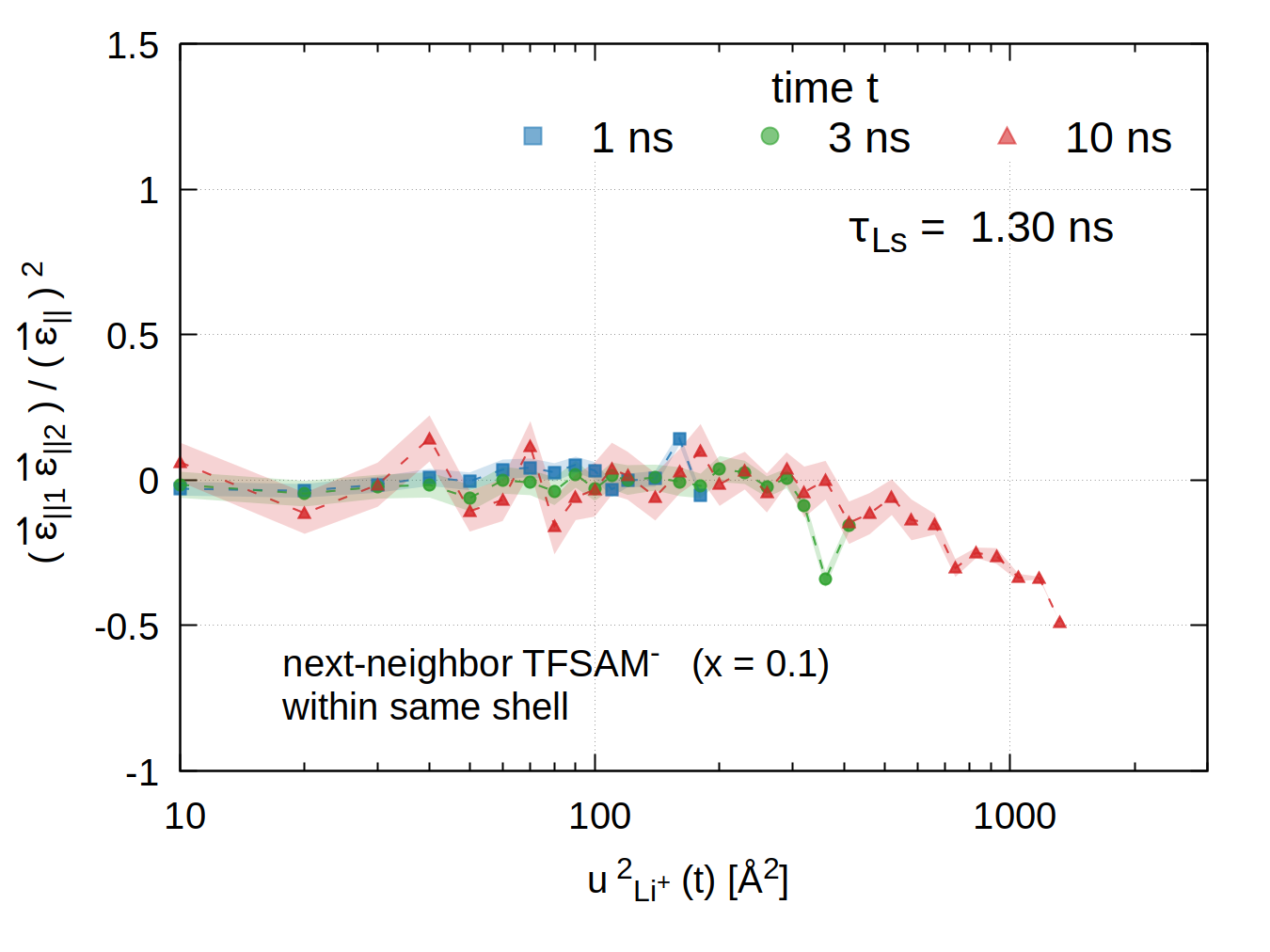}}
  \hfill
  \subfloat{\includegraphics[width=0.5\textwidth]{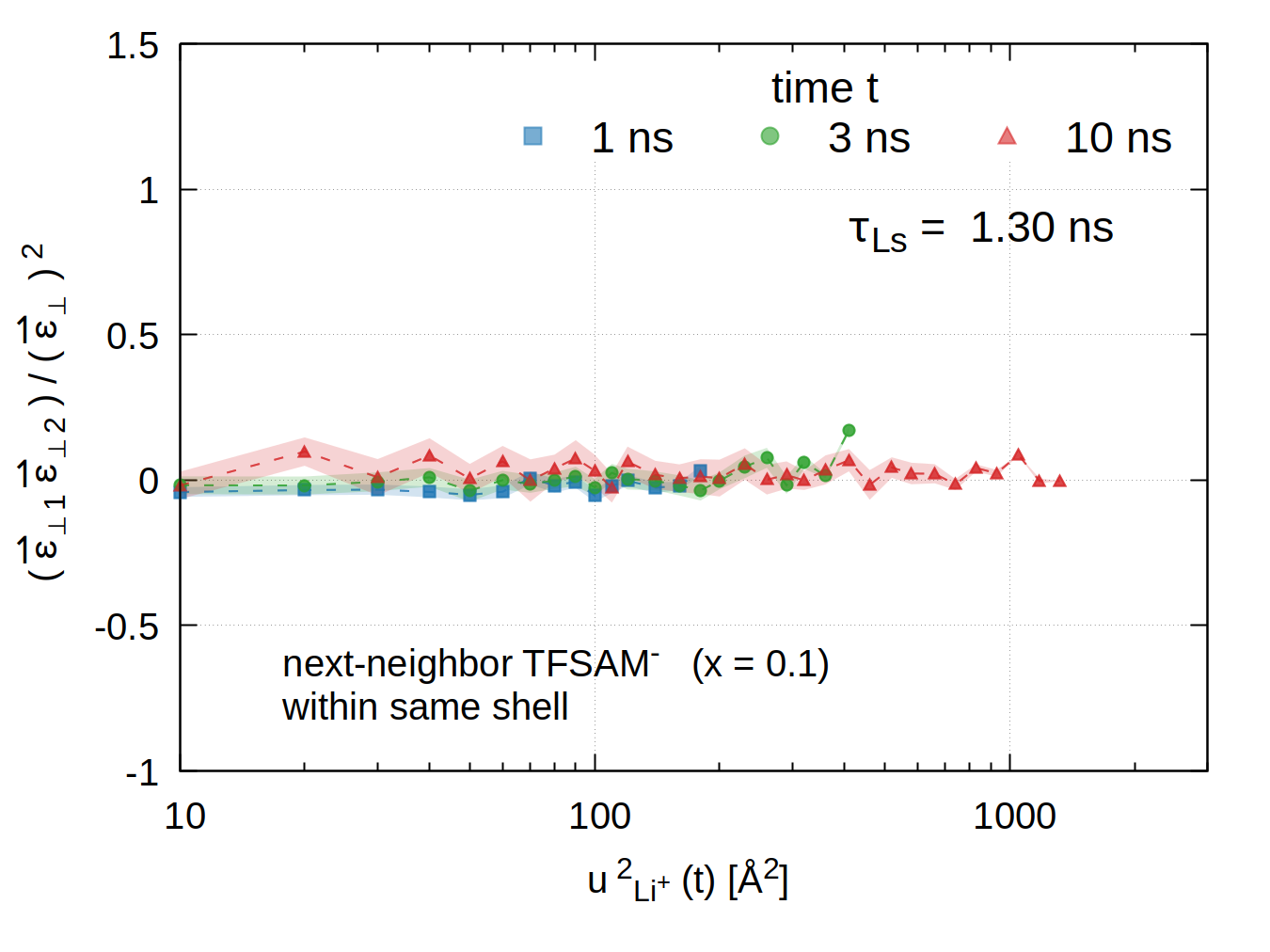}}
\end{figure}

\begin{figure}[H]
  \centering
  \subfloat{\includegraphics[width=0.5\textwidth]{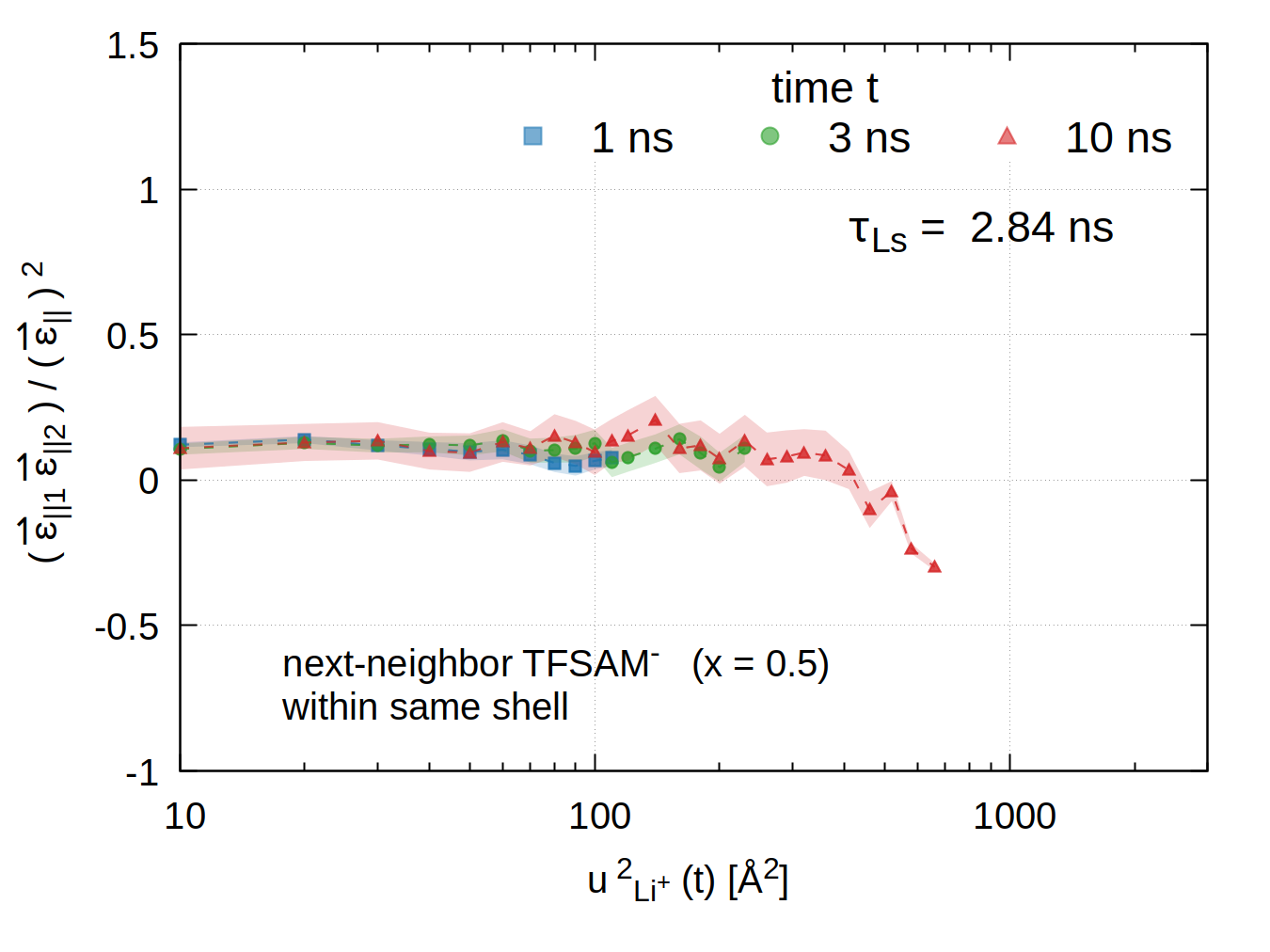}}
  \hfill
  \subfloat{\includegraphics[width=0.5\textwidth]{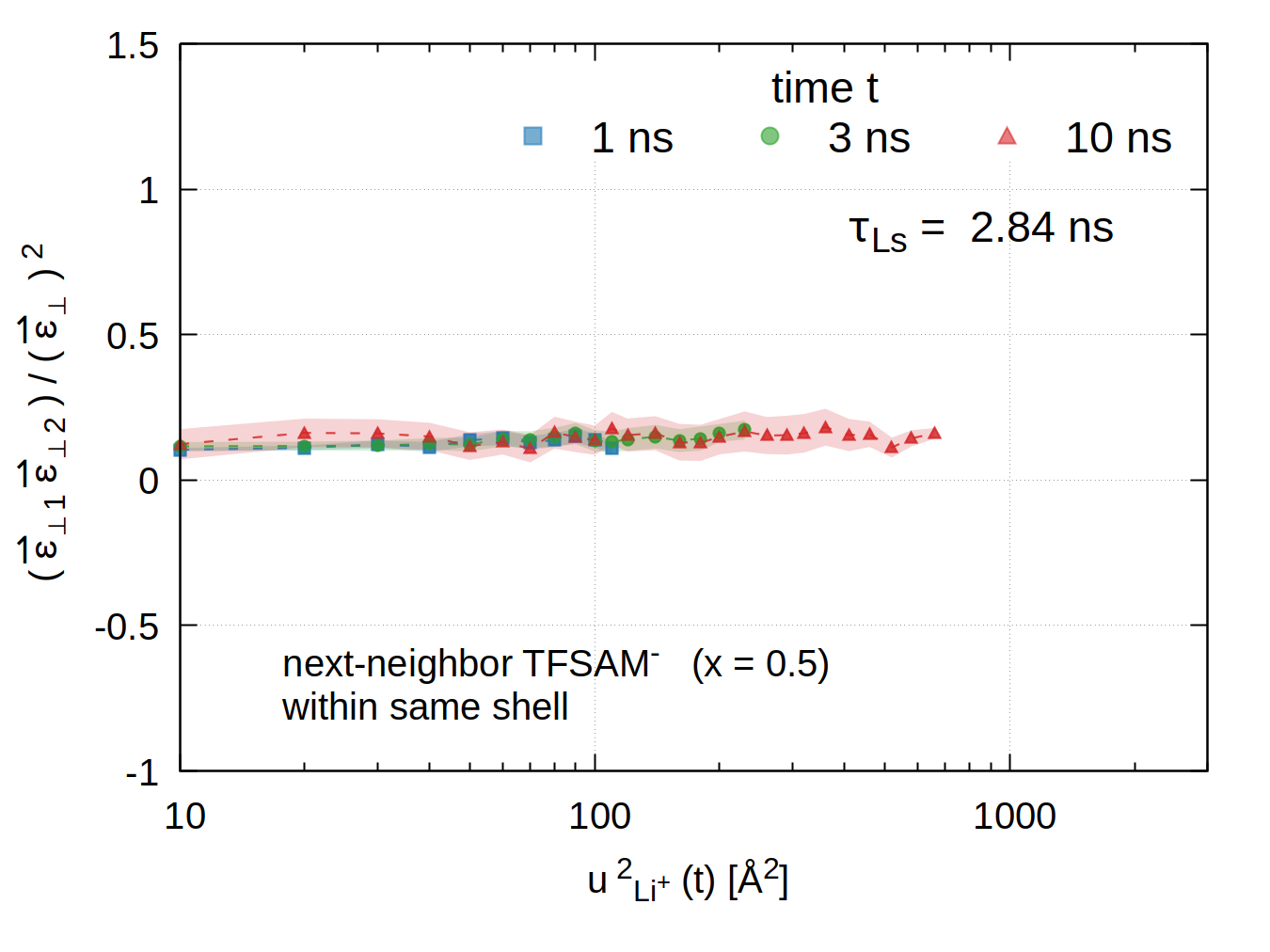}}
 
 \caption{Correlations of initial next-neighbor shell anions $\text{TFSAM}^-_1$ and $\text{TFSAM}^-_2$ as a function of $u^2(\text{t})_{\text{Li}^+}$ exemplary for salt contents x\,=\,0.1 and x\,=\,0.5.}

\end{figure}

\newpage
\textbf{L: LCF $\lambda$ as a function of time $t$}

%%% Li- TFSAM - NH,NC,OP

\begin{figure}[H]
  \centering
    \subfloat{\includegraphics[width=0.5\textwidth]{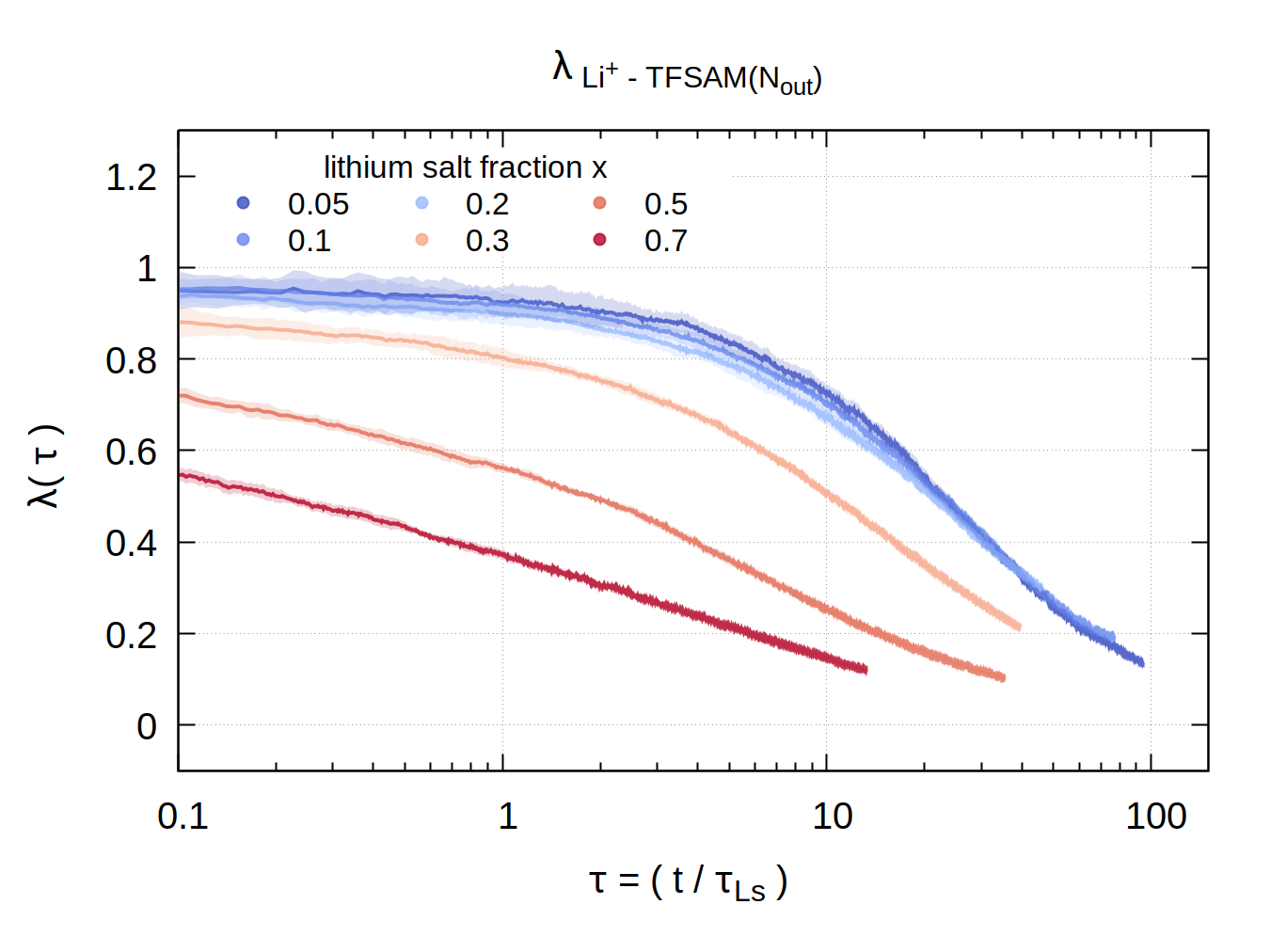}}
  \hfill
    \subfloat{\includegraphics[width=0.5\textwidth]{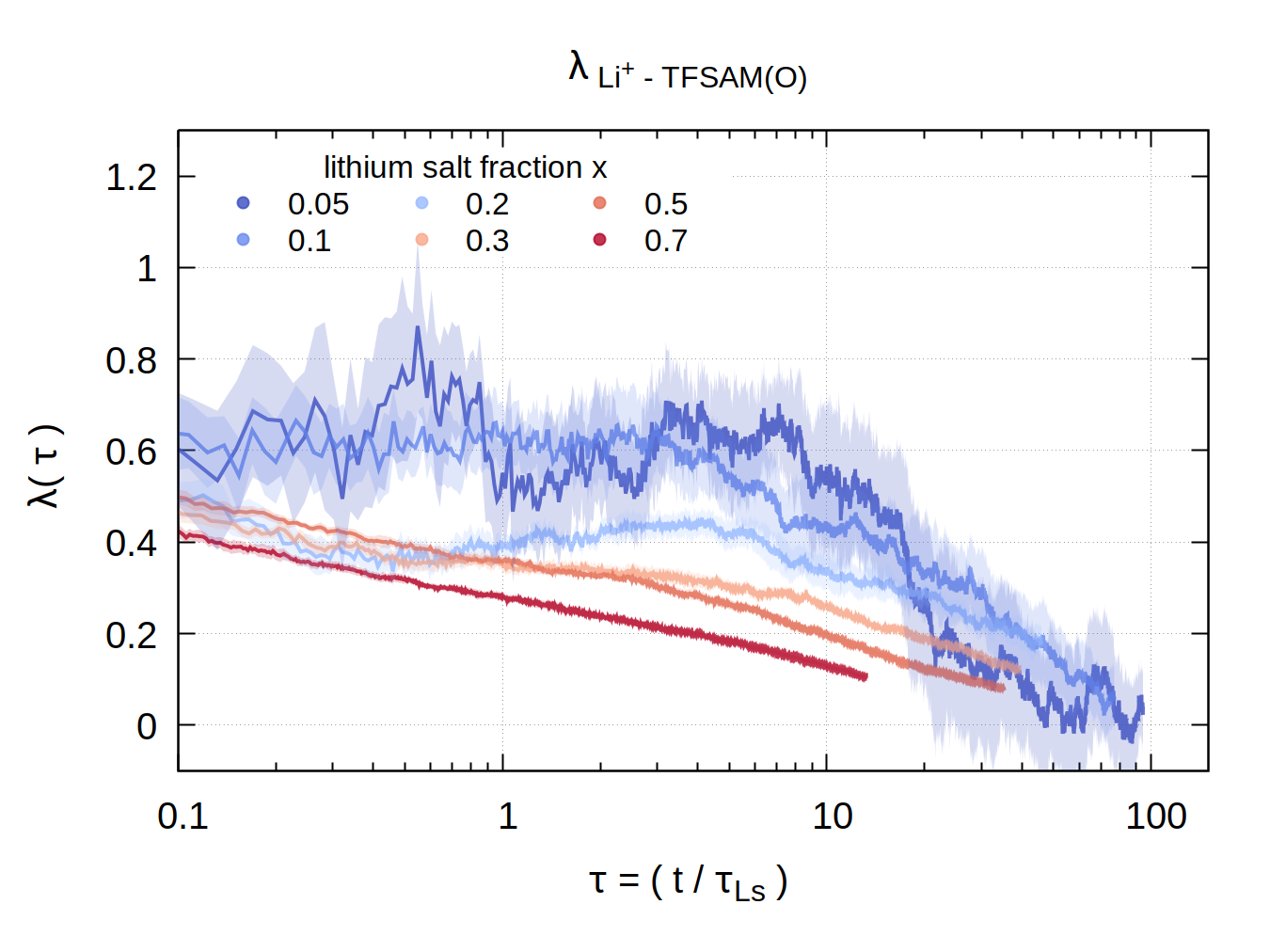}}
  \hfill
  \centering
    \subfloat{\includegraphics[width=0.5\textwidth]{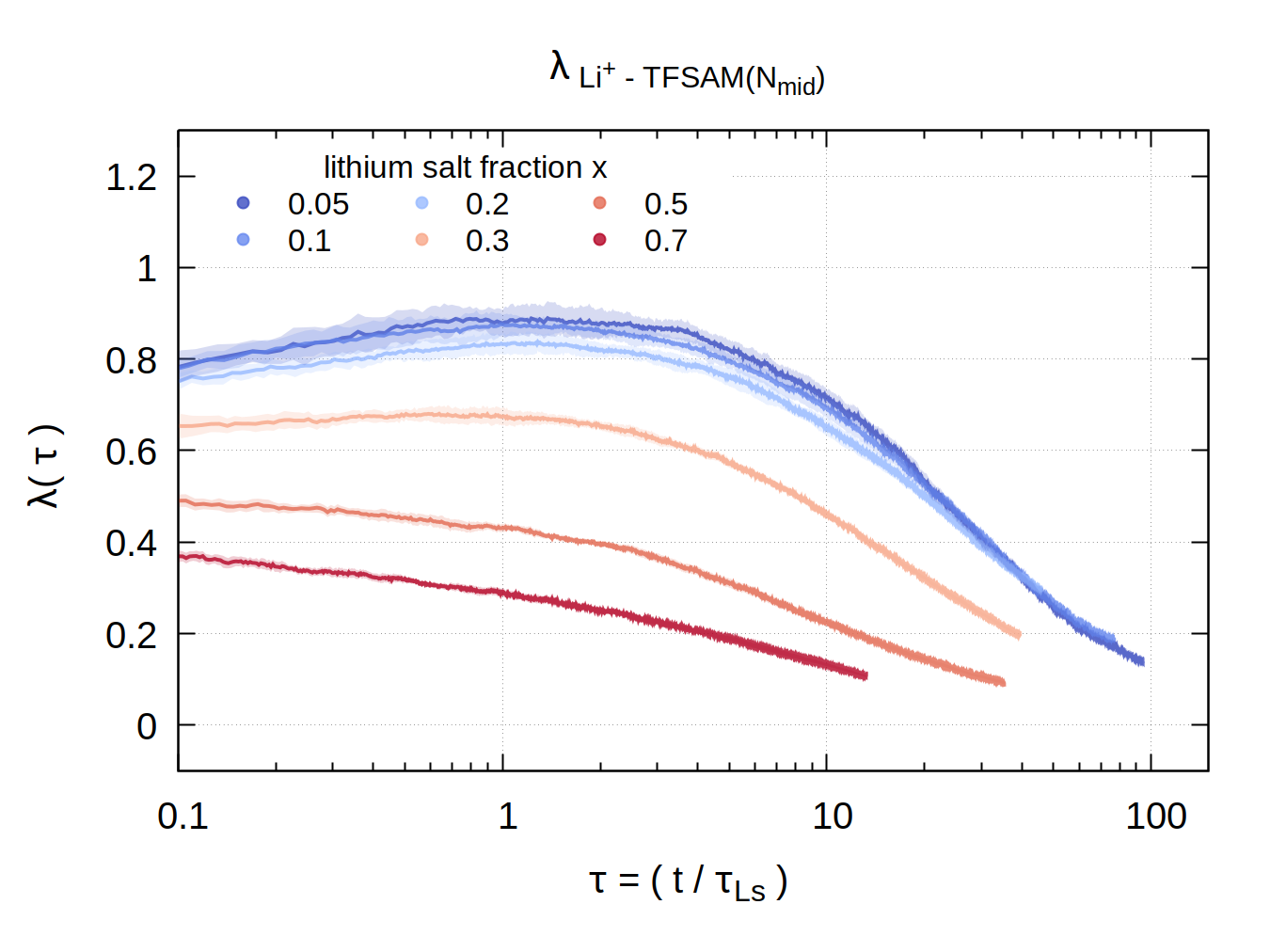}}
 \caption{Time dependence of LCF $\lambda$ for $\text{Li}^+$ to different binding sites provided by $\text{TFSAM}^-$, \textit{i.e.}, the outer nitrogen atoms $\text{N}_{\text{out}}$ (top left) or the oxygen atoms (top right). Using the second minimum position of $g_{\text{Li}^+-\text{N}_{\text{mid}}}$ as a cutoff distance to determine $\text{Li}^+-\text{TFSAM}^-$-binding as discussed in the manuscript, contains all possible coordination geometries. For a structurally equivalent comparison with $\text{TFSI}^-$, we analyse the LCF of $\text{Li}^+$ and the middle nitrogen atoms $\text{N}_{\text{mid}}$ in the initial $\text{TFSAM}^-$ solvation cage. To compare the time dependence of $\lambda$ for different salt contents x, $t$ is scaled by the characteristic self diffusion time $\tau_{\text{Ls}}$ (see Figure 5B) of each electrolyte composition. }
  \label{fig:lambda_OV_li_tfsam_nh_nc_op}

\end{figure}

%%% Li- TFSI - NI,OS

\begin{figure}[H]
  \centering
    \subfloat{\includegraphics[width=0.5\textwidth]{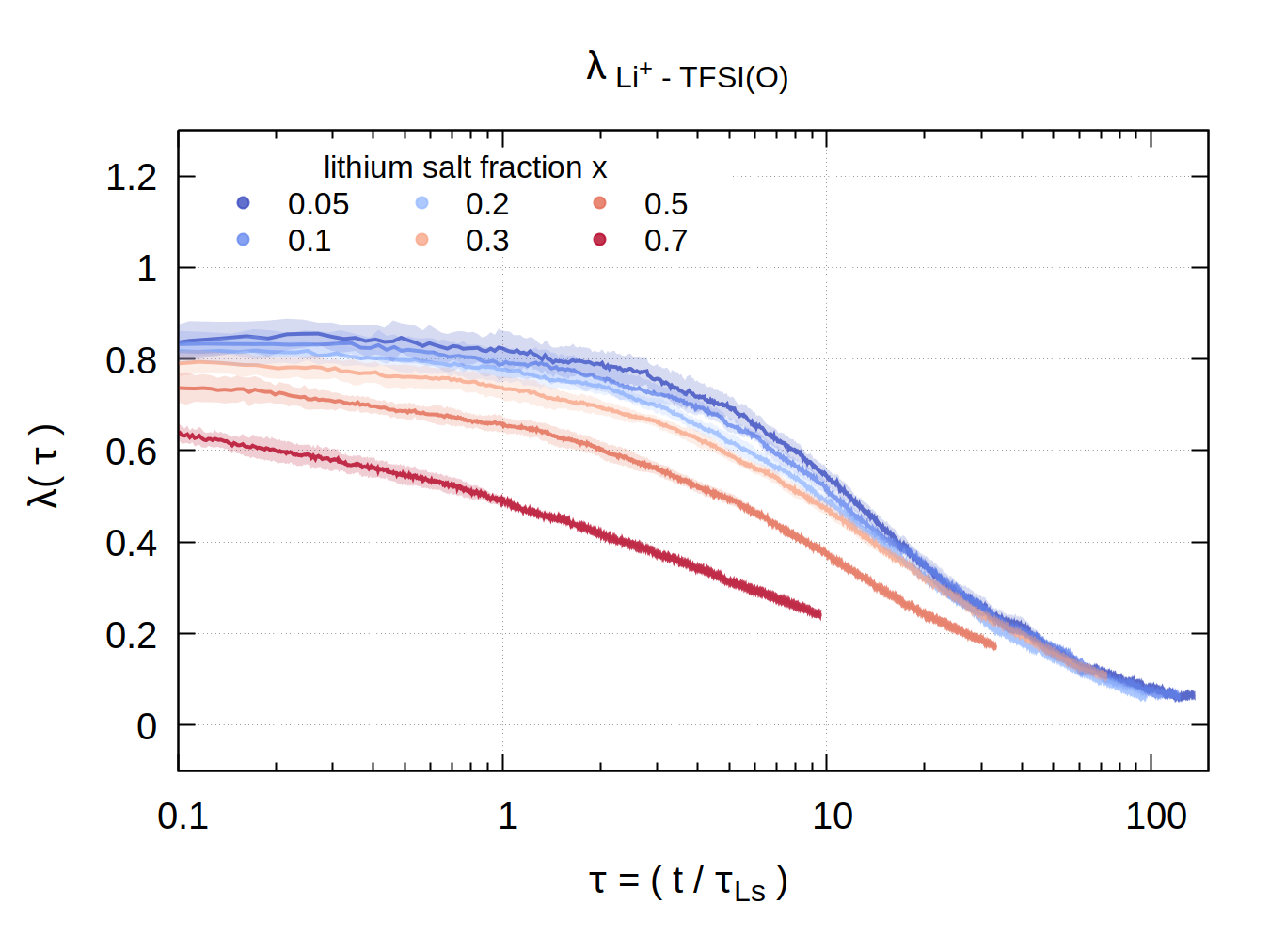}}
  \hfill
    \subfloat{\includegraphics[width=0.5\textwidth]{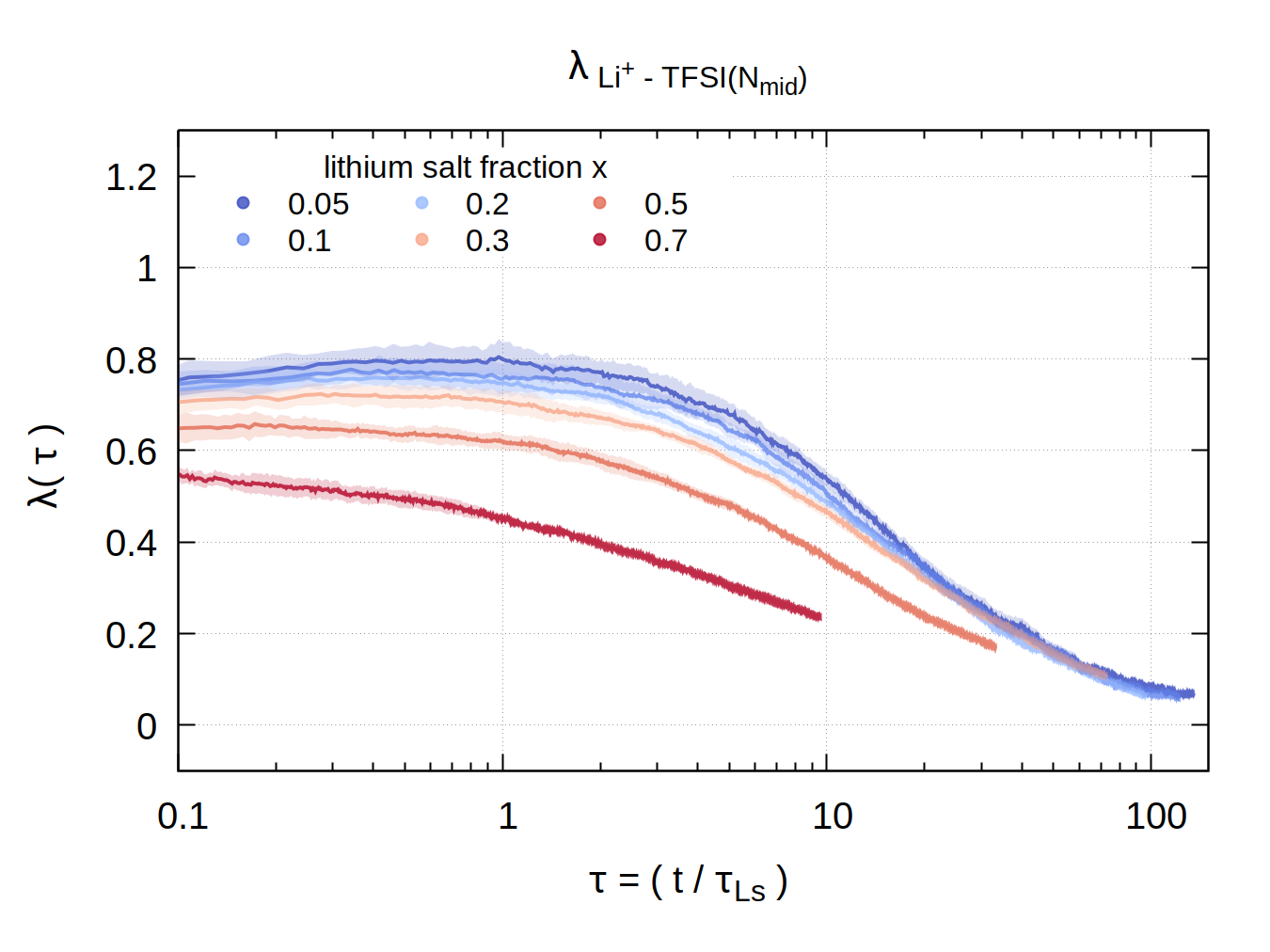}}
    
 \caption{Time dependence of LCF $\lambda$ for $\text{Li}^+$ to the oxygen atoms (top left) of $\text{TFSI}^-$. Employing the second minimum position of $g_{\text{Li}^+-\text{N}_{\text{mid}}}$ as a cutoff distance to determine $\text{Li}^+-\text{TFSI}^-$-binding as discussed in the main manuscript allows for a structurally equivalent comparison with $\text{TFSAM}^-$. Thus, we analyse the LCF of $\text{Li}^+$ and the middle nitrogen atoms $\text{N}_{\text{mid}}$ in the initial $\text{TFSI}^-$ solvation cage. To compare the time dependence of $\lambda$ for different salt contents x, $t$ is scaled by the characteristic self diffusion time $\tau_{\text{Ls}}$ (see Figure 5A) of each electrolyte composition. }
  \label{fig:lambda_OV_li_tfsi_ni_os}

\end{figure}

\newpage
\textbf{M: Comparison of $\lambda, \lambda_1, \lambda_2$ and $\Lambda_2$ as a function of salt content x for characteristic times of $3\cdot\tau_{\text{Ls}}$ and $5\cdot\tau_{\text{Ls}}$}

\begin{figure}[H]
  \centering
  \subfloat{\includegraphics[width=0.5\textwidth]{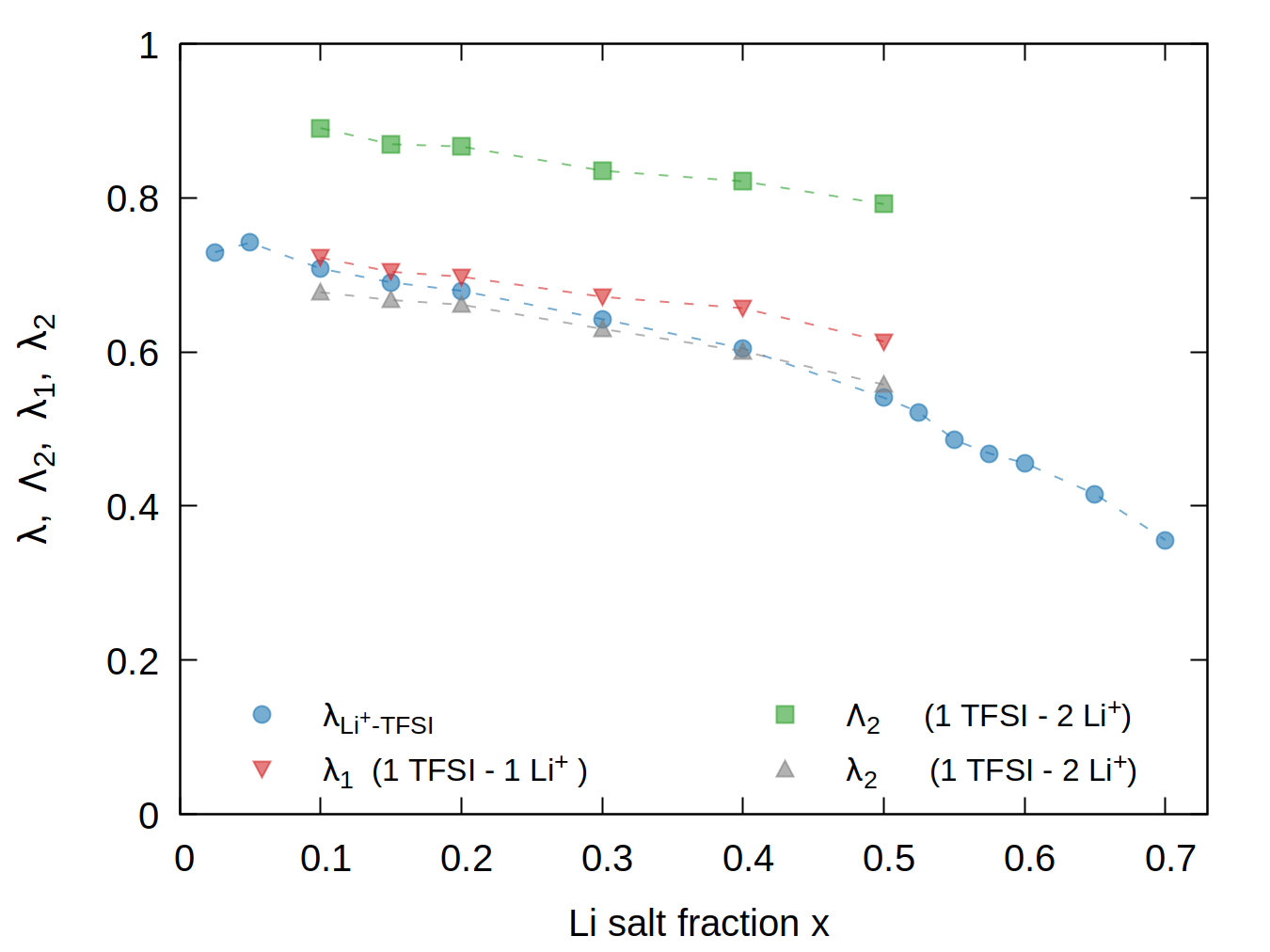}}
  \hfill
  \subfloat{\includegraphics[width=0.5\textwidth]{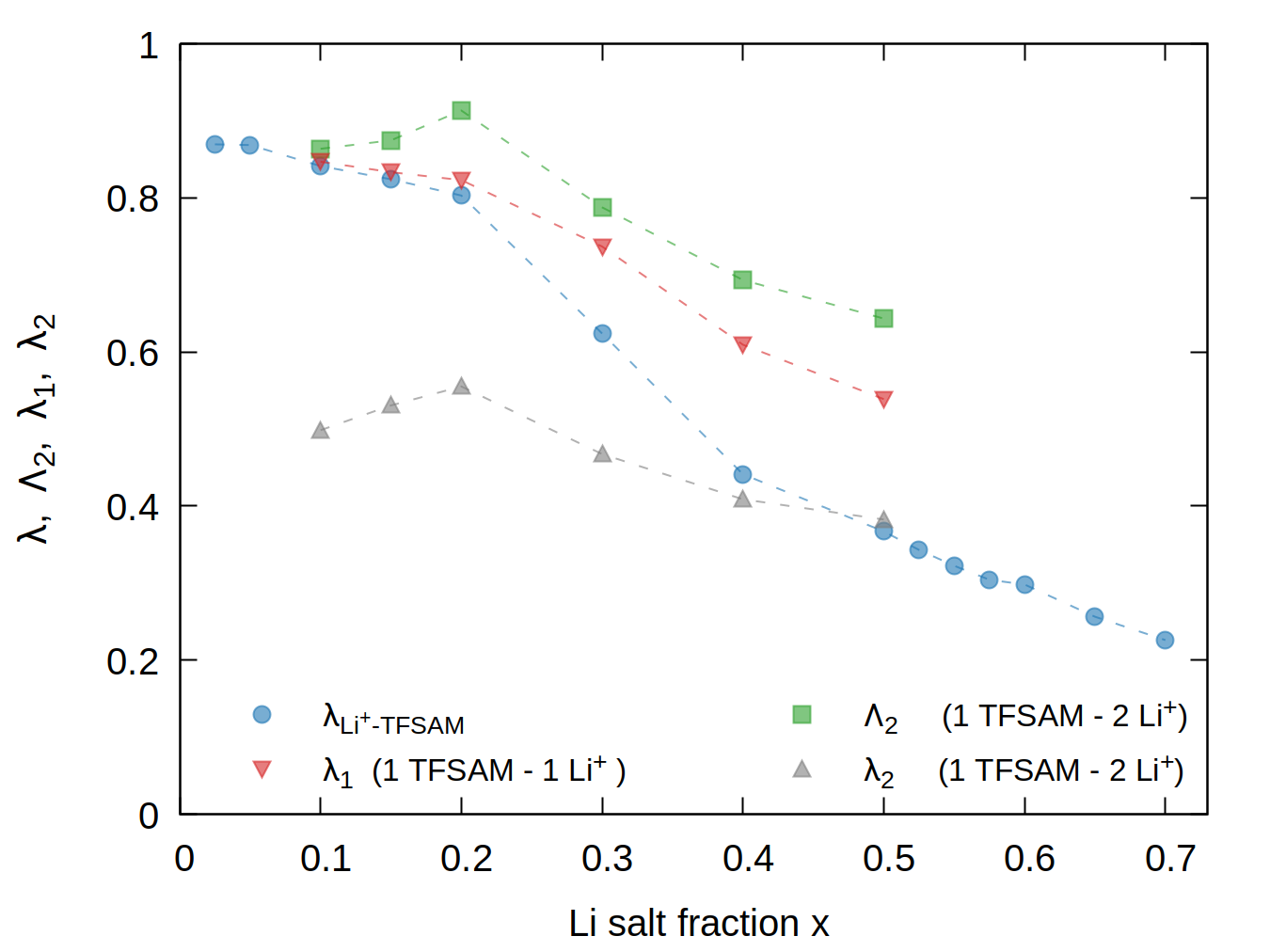}}

\end{figure}

\begin{figure}[H]
  \centering

  \subfloat{\includegraphics[width=0.5\textwidth]{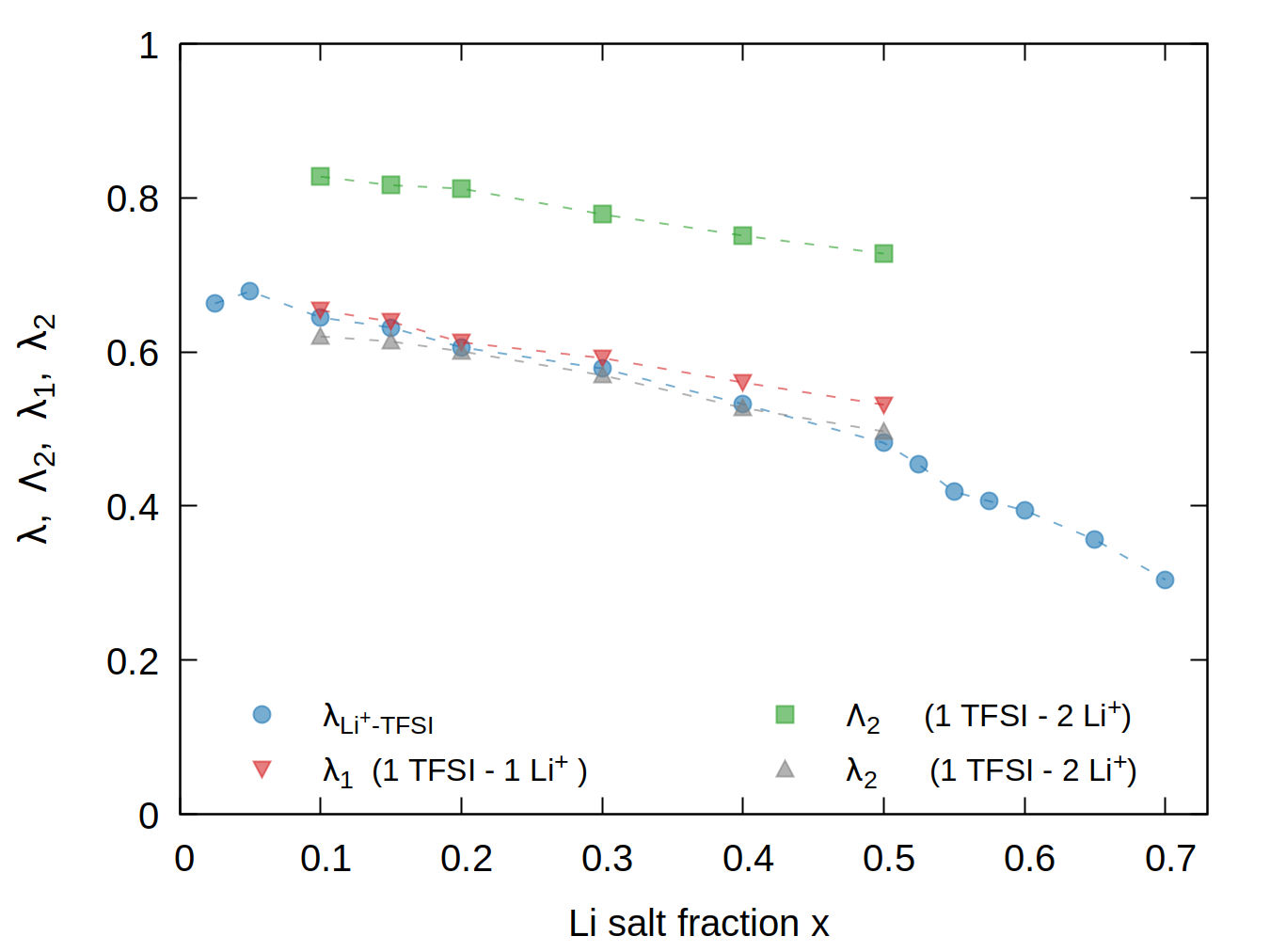}}
  \hfill
  \subfloat{\includegraphics[width=0.5\textwidth]{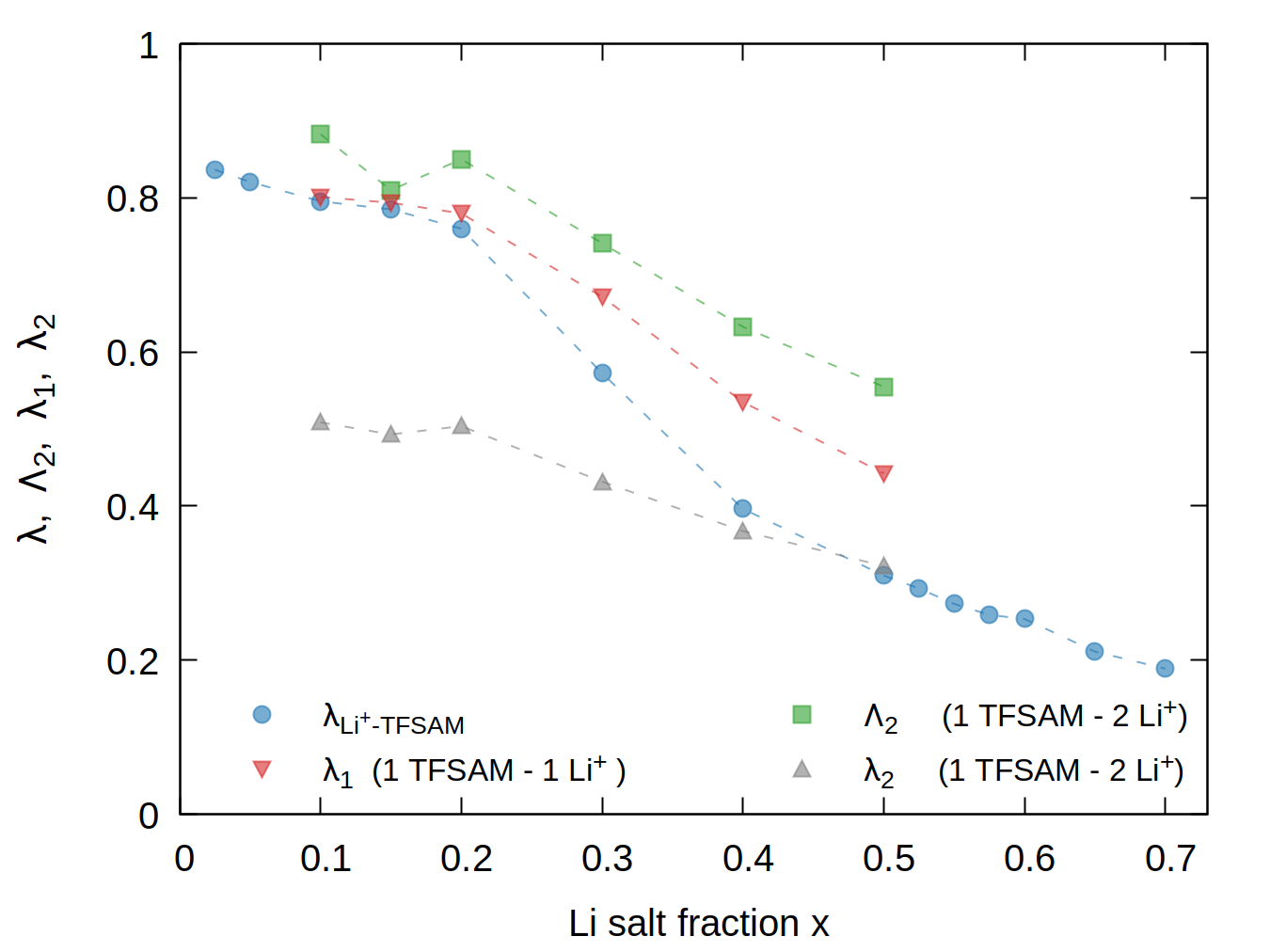}}

  \caption{Comparison of lithium coupling factors ($\lambda, \lambda_1, \lambda_2, \Lambda_2$) as a function of salt content for both $\text{TFSI}^-$ (left) and $\text{TFSAM}^-$ (right) - based mixtures for characteristic times of $3\cdot\tau_{\text{Ls}}$ (top) and $5\cdot\tau_{\text{Ls}}$ (bottom).}
\label{fig:lambda_subensembles_function_salt_content}

\end{figure}

\begin{figure}[H]
  \centering
    \subfloat{\includegraphics[width=0.5\textwidth]{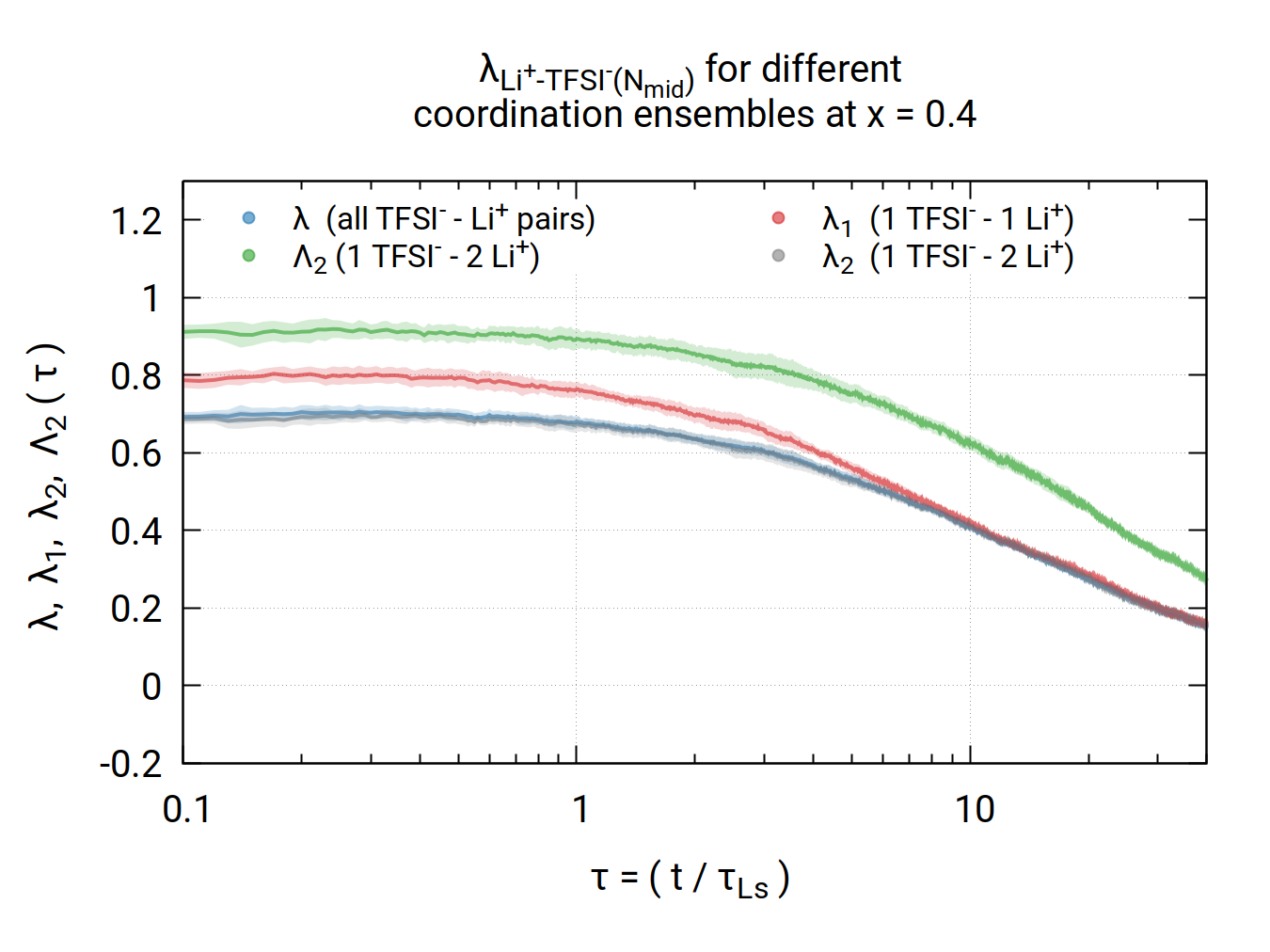}}
  \hfill
    \subfloat{\includegraphics[width=0.5\textwidth]{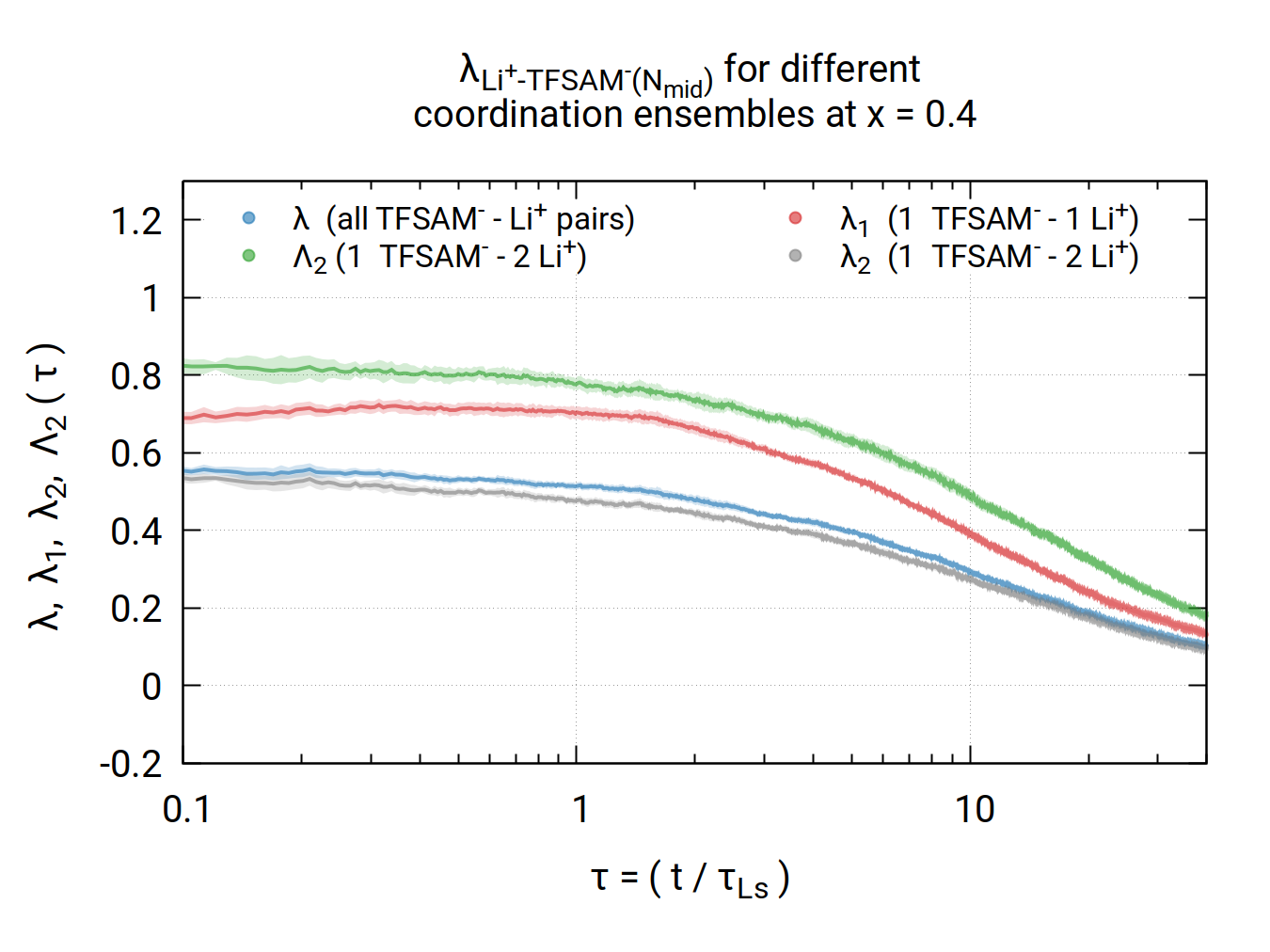}}
    
 \caption{Time dependence of LCF $\lambda, \lambda_1, \lambda_2$ and $\Lambda_2$ exemplary shown for the x=0.4 $\text{TFSI}^-$- (left) and $\text{TFSAM}^-$-based (right) electrolyte compositions.}
  \label{fig:lambda_subensembles_time_dependence}

\end{figure}

%%%%%%%%%%%%%%%%%%%%%%%%%%%%%%%%%%%%%%%%%%%%%%%%%%%%%%%%%%%%%%%%%%%%%%%%%%%%%%%%%%
%%%%%%%%%%%%%%%%%%%%%%%%%%%%%%%%%%%%%%%%%%%%%%%%%%%%%%%%%%%%%%%%%%%%%%%%%%%%%%%%%%
%%%%%%%%%%%%%%%%%%%%%%%%%%%%%%%%%%%%%%%%%%%%%%%%%%%%%%%%%%%%%%%%%%%%%%%%%%%%%%%%%%
%%%%%%%%%%%%%%%%%%%%%%%%%%%%%%%%%%%%%%%%%%%%%%%%%%%%%%%%%%%%%%%%%%%%%%%%%%%%%%%%%%
%%%%%%%%%%%%%%%%%%%%%%%%%%%%%%%%%%%%%%%%%%%%%%%%%%%%%%%%%%%%%%%%%%%%%%%%%%%%%%%%%%
%%%%%%%%%%%%%%%%%%%%%%%%%%%%%%%%%%%%%%%%%%%%%%%%%%%%%%%%%%%%%%%%%%%%%%%%%%%%%%%%%%
%%%%%%%%%%%%%%%%%%%%%%%%%%%%%%%%%%%%%%%%%%%%%%%%%%%%%%%%%%%%%%%%%%%%%%%%%%%%%%%%%%
%%%%%%%%%%%%%%%%%%%%%%%%%%%%%%%%%%%%%%%%%%%%%%%%%%%%%%%%%%%%%%%%%%%%%%%%%%%%%%%%%%
%%%%%%%%%%%%%%%%%%%%%%%%%%%%%%%%%%%%%%%%%%%%%%%%%%%%%%%%%%%%%%%%%%%%%%%%%%%%%%%%%%
%%%%%%%%%%%%%%%%%%%%%%%%%%%%%%%%%%%%%%%%%%%%%%%%%%%%%%%%%%%%%%%%%%%%%%%%%%%%%%%%%%
%%%%%%%%%%%%%%%%%%%%%%%%%%%%%%%%%%%%%%%%%%%%%%%%%%%%%%%%%%%%%%%%%%%%%%%%%%%%%%%%%%
%%%%%%%%%%%%%%%%%%%%%%%%%%%%%%%%%%%%%%%%%%%%%%%%%%%%%%%%%%%%%%%%%%%%%%%%%%%%%%%%%%

\newpage
\textbf{N: Additional information for discussing ${\text{D}_{\text{anion}}}/{\text{D}_{\text{Li}^+}}$ \newline as a function of salt content x } \newline
It can be easily shown that the line of argumentation for a decreasing ratio $\langle \vec{v}\rangle/\langle \vec{u}\rangle$ as a consequence of double $\text{Li}^+$-anion coordination is applicable to higher $\text{Li}^+$ over-coordination of the anion.
Assume the anion $j$ is bound to $n$ $\text{Li}^+$ and tries to couple with the strength $\Lambda_n$ to the average $\text{Li}^+$ displacement $\vec{U}_i^j = \dfrac{1}{n}\cdot( \vec{u}_1 + ... + \vec{u}_n):$
\begin{equation}
    \vec{v}_j = \Lambda_n \cdot \vec{U}_i^j + \vec{E}_j.
\end{equation}
Squaring and rearranging yields for the ratio $\langle \vec{v}\rangle/\langle \vec{u}\rangle$:
\begin{equation}
\begin{aligned}
      \dfrac{\langle\vec{v}^2\rangle}{\langle \vec{u}\rangle} &= \Lambda_n^2\,\cdot \dfrac{1}{n^2} \cdot \bigg( n  + 2  {{n}\choose{2}}\dfrac{\langle \vec{u}_1\vec{u}_2\rangle }{\langle \vec{u}\rangle} \bigg) +   \dfrac{\vec{E}^2}{\langle \vec{u}\rangle} \\
      &= \Lambda_n^2\,\cdot \underbrace{ \dfrac{1}{n} \cdot \bigg( 1  + (n-1)\underbrace{\dfrac{\langle \vec{u}_1\vec{u}_2\rangle }{\langle \vec{u}\rangle}}_{<1} \bigg)}_{\rightarrow \dfrac{\langle \vec{u}_1\vec{u}_2\rangle }{\langle \vec{u}\rangle} }  + \dfrac{\vec{E}^2}{\langle \vec{u}\rangle} 
\end{aligned}
\end{equation}
If the remaining terms and factors do not change considerably, one can easily see that the ratio drops further with $n$-fold coordination.

\begin{figure}[H]
  \centering

  \subfloat{\includegraphics[width=0.5\textwidth]{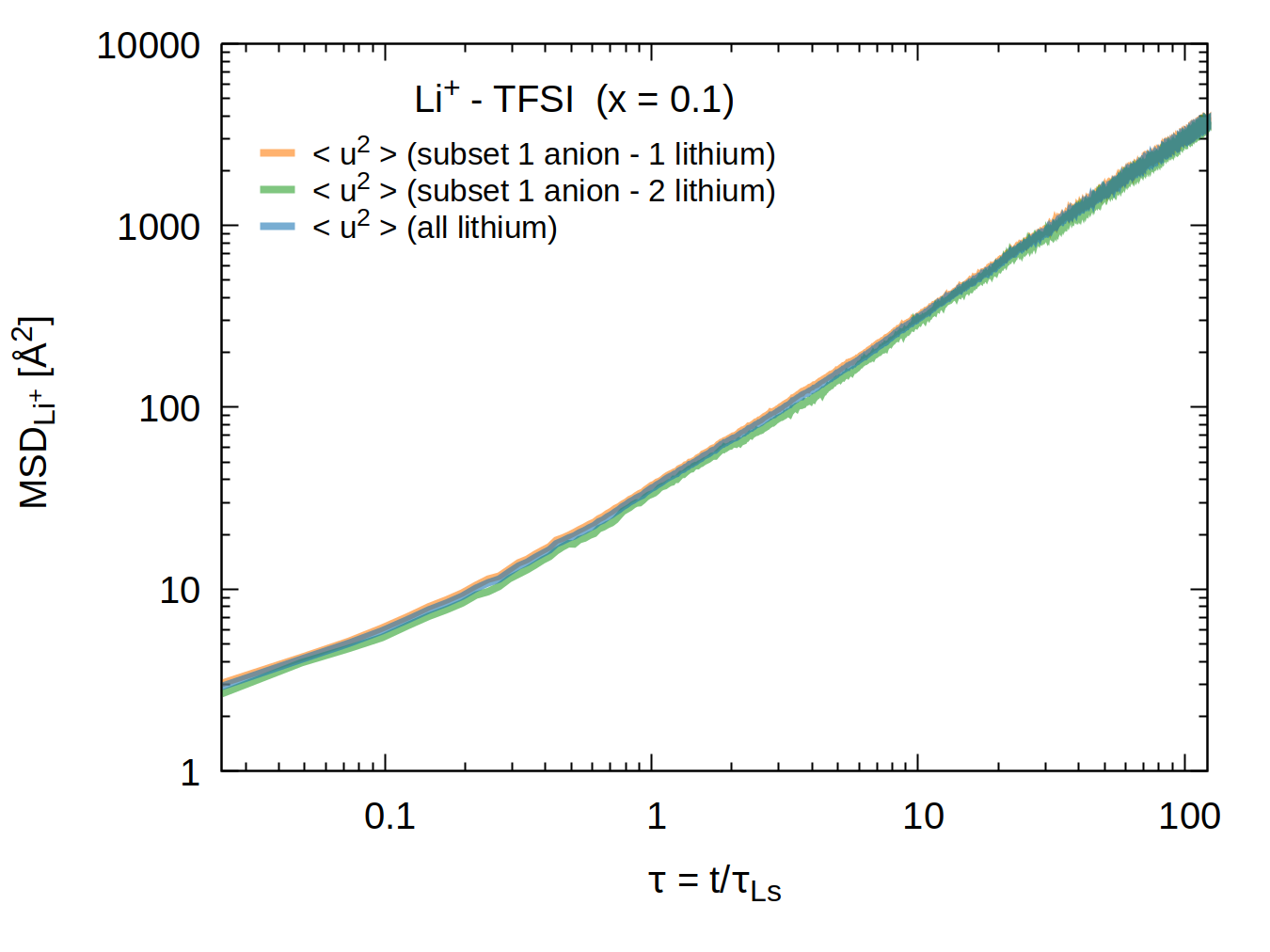}}
  \hfill
  \subfloat{\includegraphics[width=0.5\textwidth]{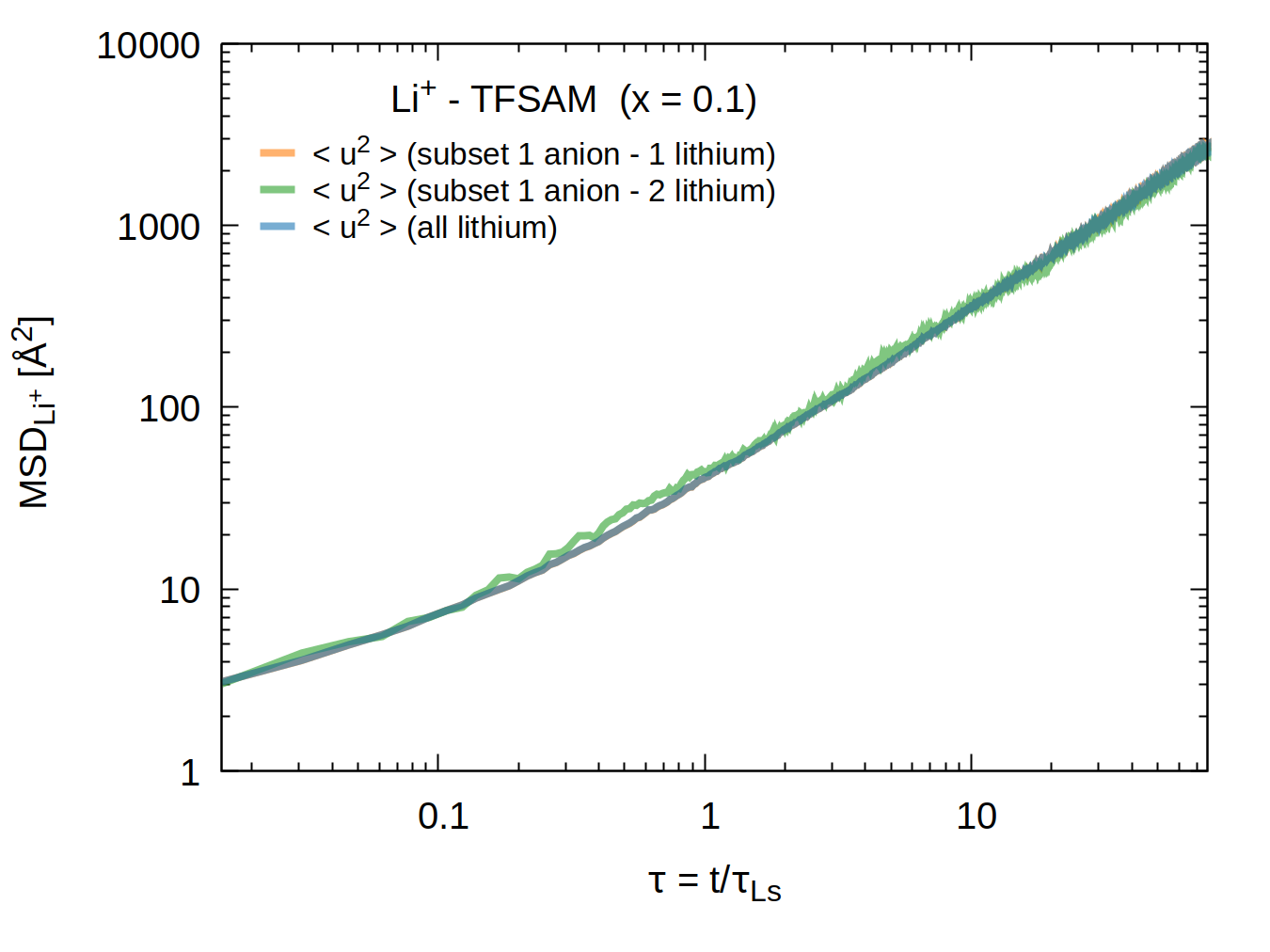}}

\end{figure}

\begin{figure}[H]
  \centering

  \subfloat{\includegraphics[width=0.5\textwidth]{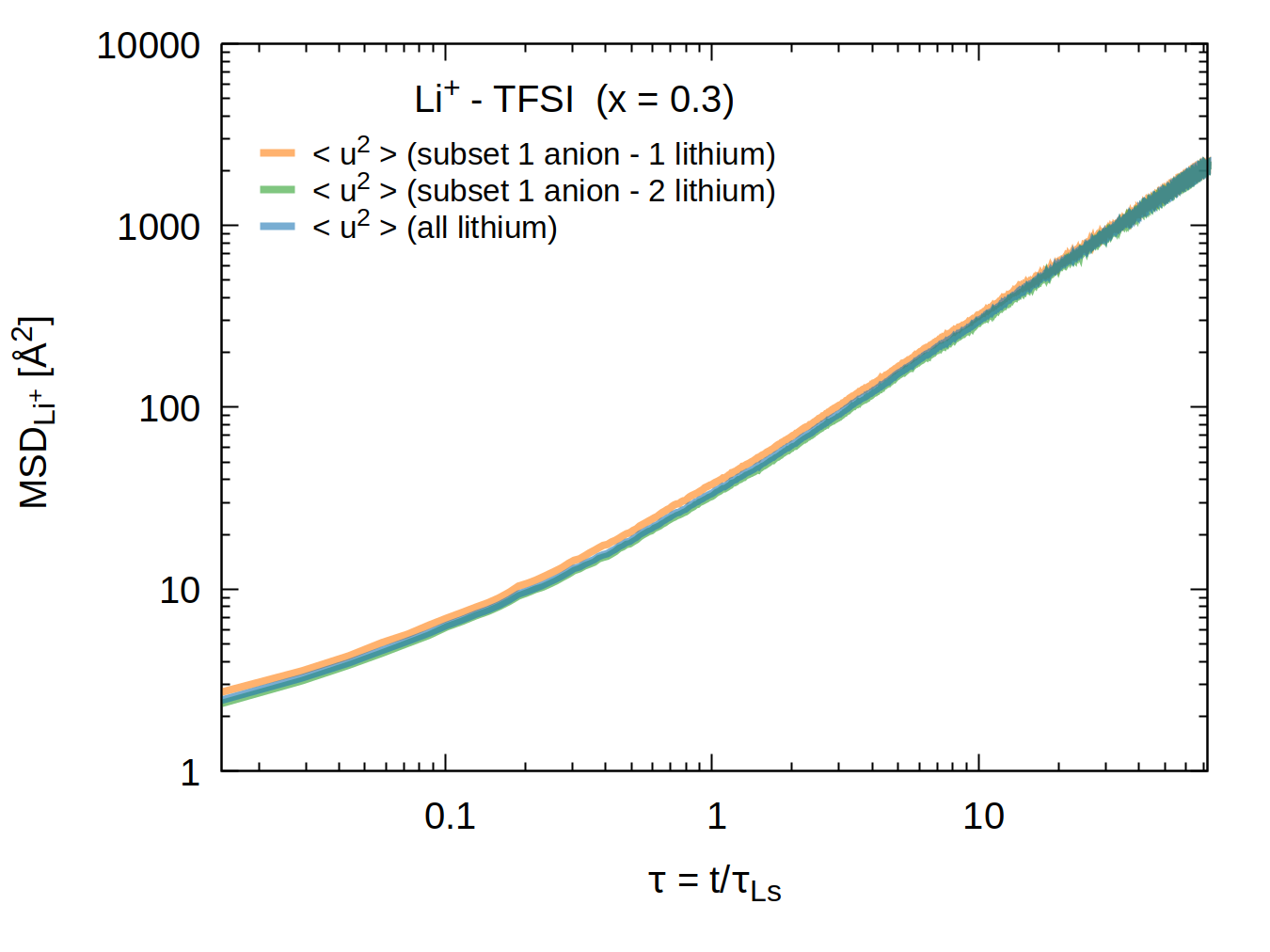}}
  \hfill
  \subfloat{\includegraphics[width=0.5\textwidth]{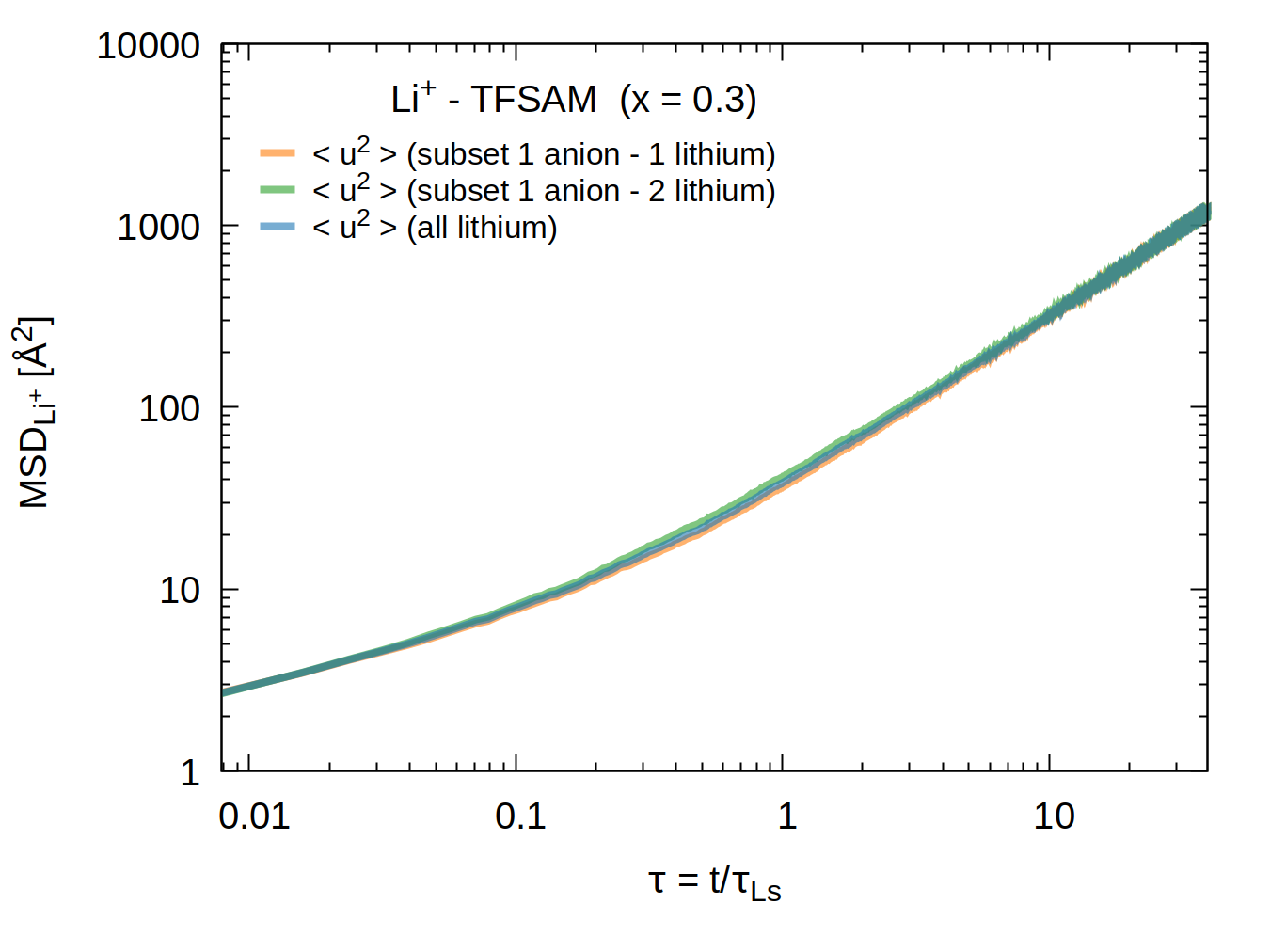}}

\end{figure}

\begin{figure}[H]
  \centering

  \subfloat{\includegraphics[width=0.5\textwidth]{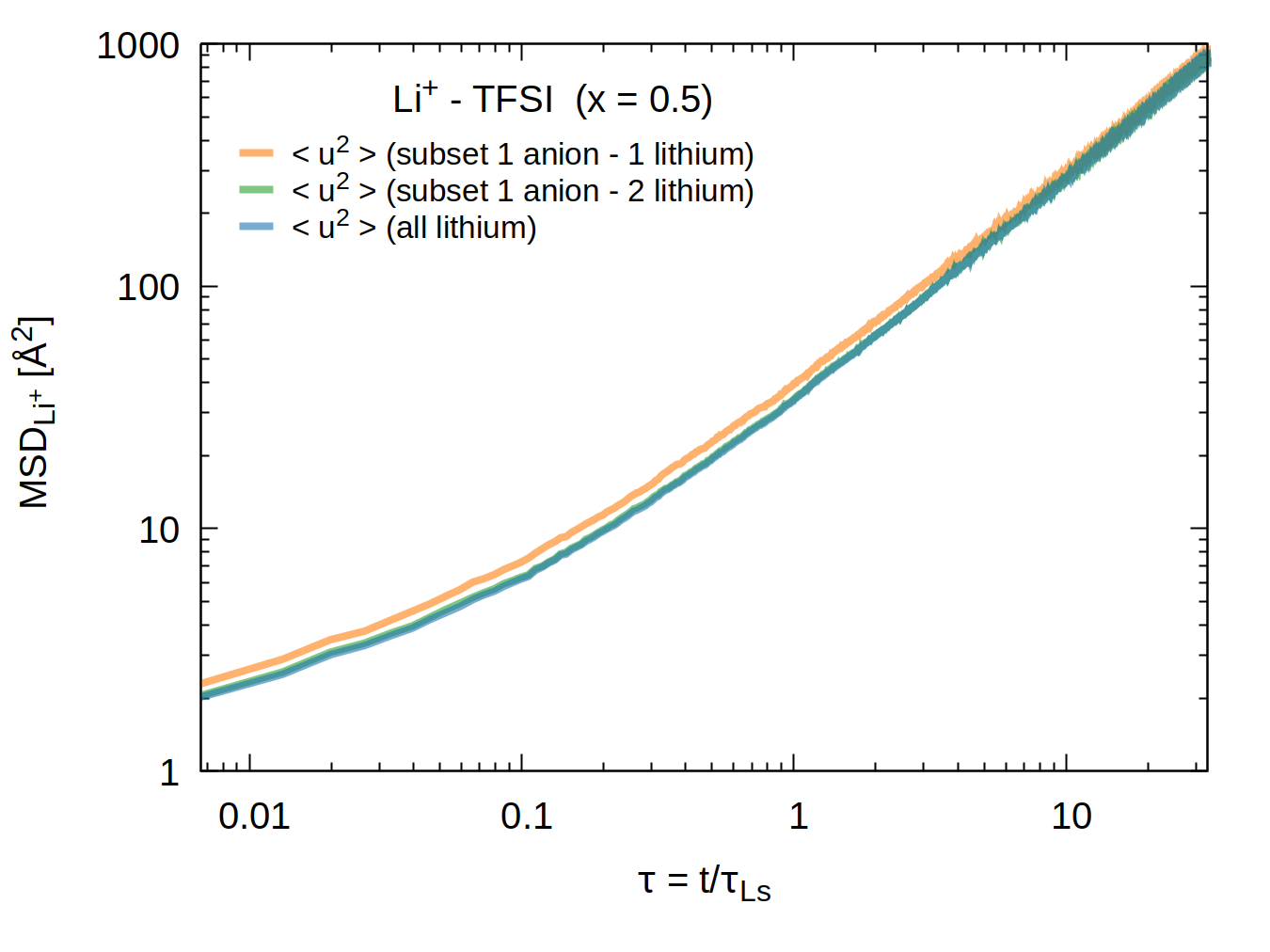}}
  \hfill
  \subfloat{\includegraphics[width=0.5\textwidth]{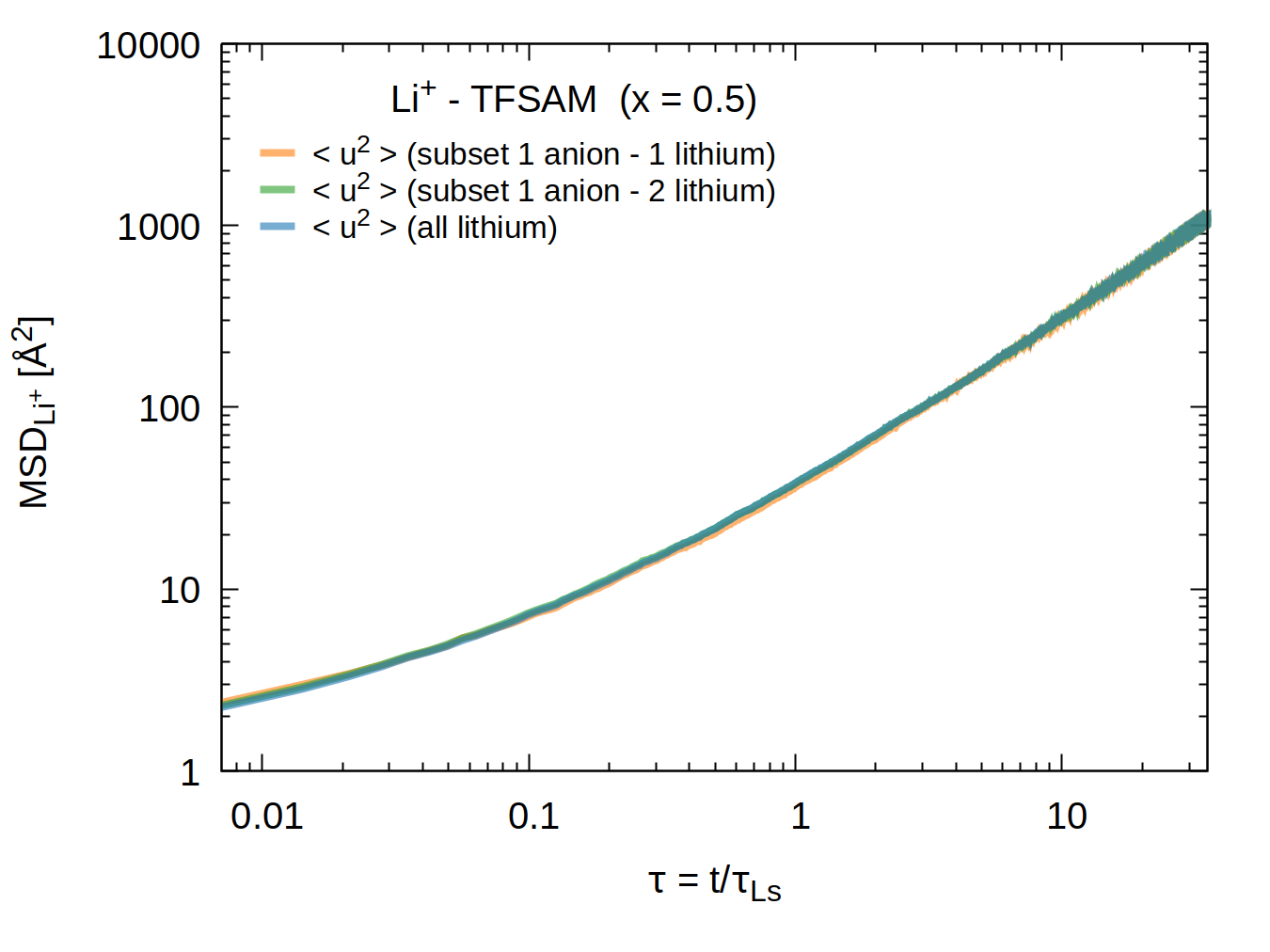}}

  \caption{Comparison of lithium MSD to $\langle u^2 \rangle$ in the subensembles of $\lambda_1$ and $\lambda_2/\Lambda_2$.}
\label{fig:lithium_MSDs_ensembles}

\end{figure}

\begin{figure}[H]
  \centering

  \subfloat{\includegraphics[width=0.5\textwidth]{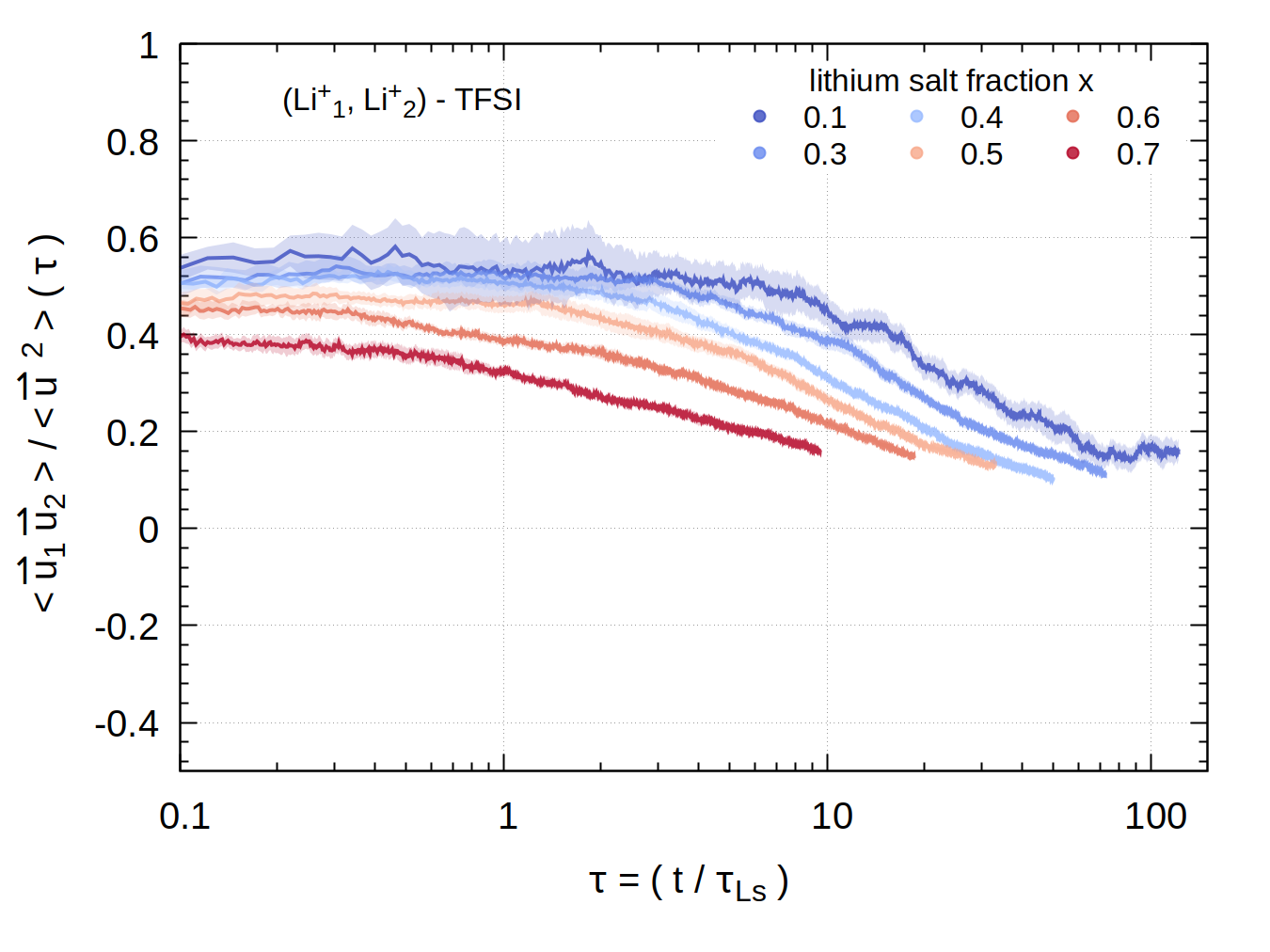}}
  \hfill
  \subfloat{\includegraphics[width=0.5\textwidth]{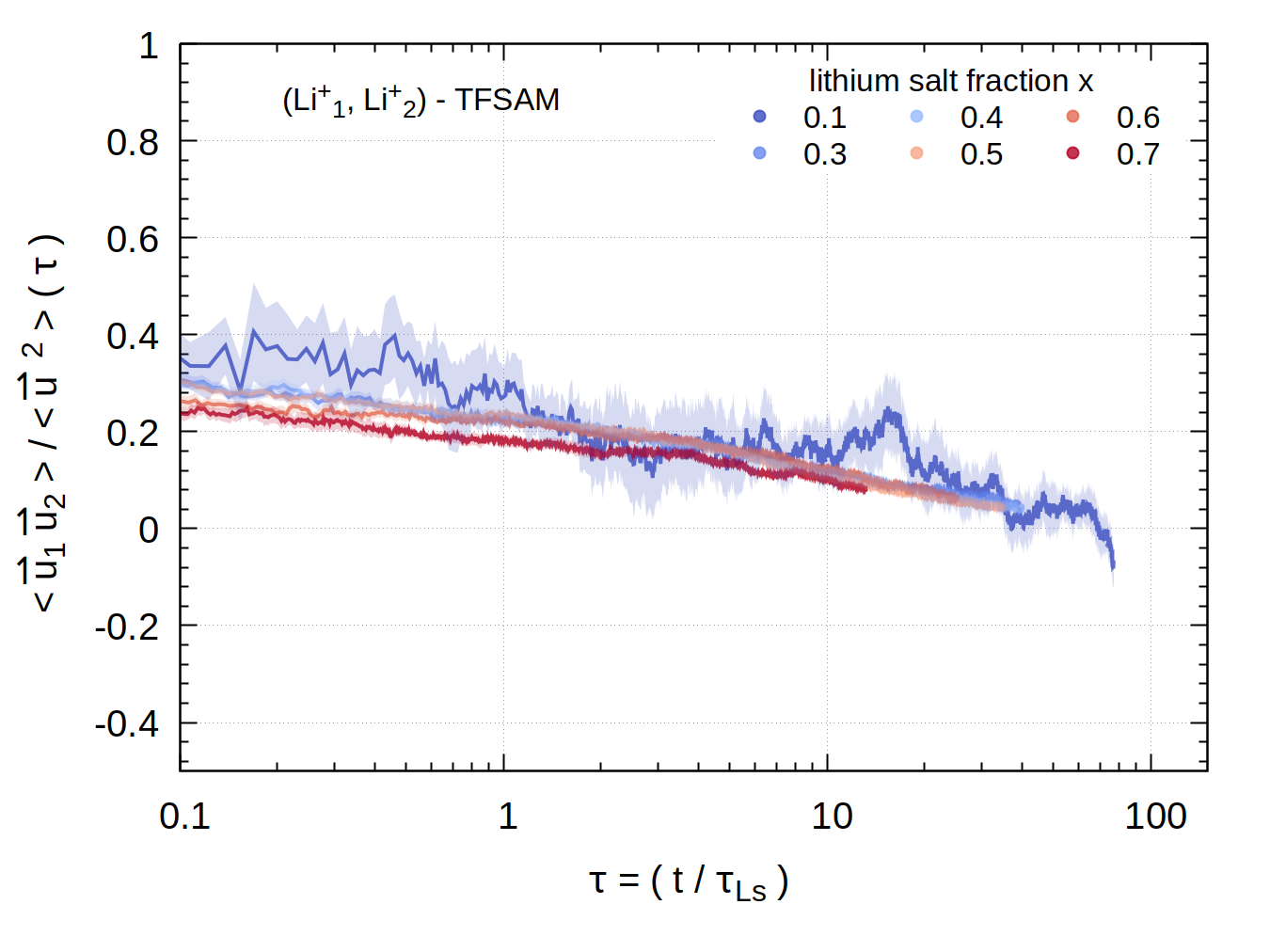}}

  \caption{$\text{Li}^+ - \text{Li}^+$ - correlation ${\langle \vec{u}_1\cdot\vec{u}_2\rangle}/{\langle \vec{u}^{\,2}\rangle}$ for lithium ions that are bound to the same anion at time $\tau = 0$.}
\label{fig:lithium_lithium_correlation_ensemble_Lambda2}

\end{figure}

\begin{figure}[H]
  \centering

  \subfloat{\includegraphics[width=0.5\textwidth]{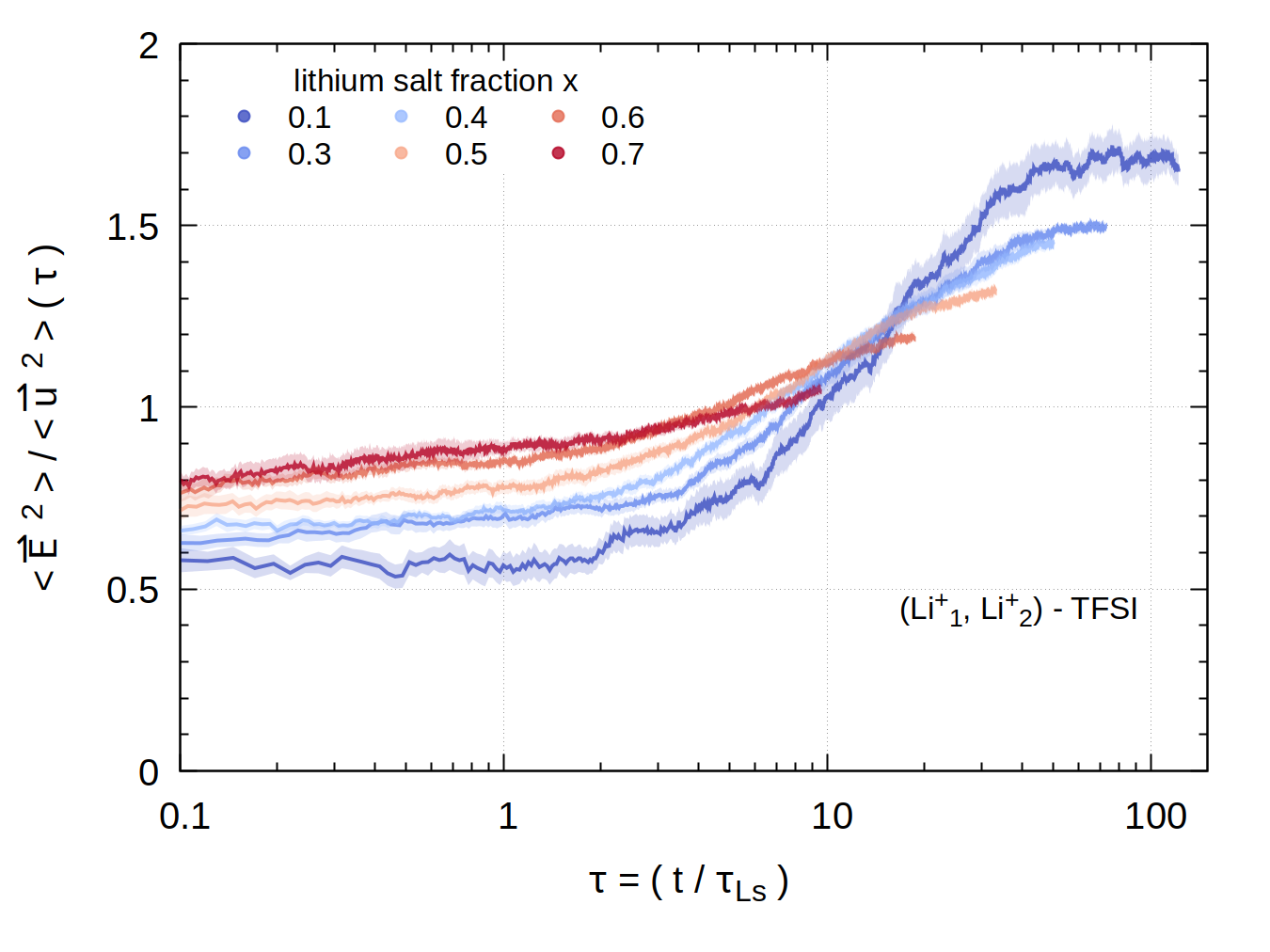}}
  \hfill
  \subfloat{\includegraphics[width=0.5\textwidth]{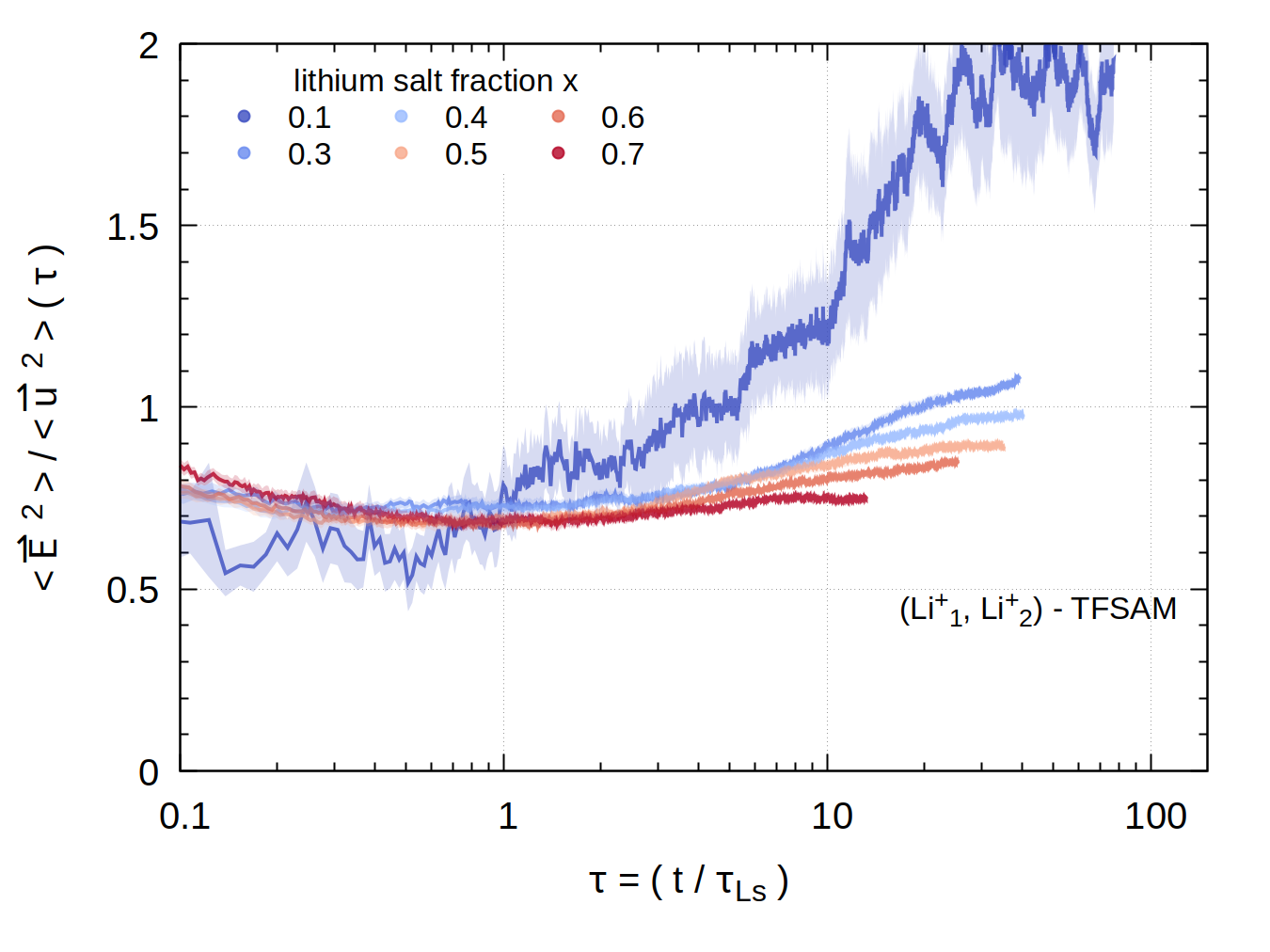}}

  \caption{Random motion $\langle \vec{ \mathcal{E}}^2 \rangle$ of anions, which have two $\text{Li}^+$ neighbours at time $\tau = 0$, scaled by the average squared lithium displacement $\langle \vec{u}^{\,2}\rangle$. $\langle .. \rangle$ denotes the ensemble average over the $\text{Li}^+$ that are involved in the double coordination of the anion at $\tau = 0$.}
\label{fig:variance_eps_lithium_path_Lambda2}

\end{figure}

\begin{figure}[H]
  \centering

  \subfloat{\includegraphics[width=0.5\textwidth]{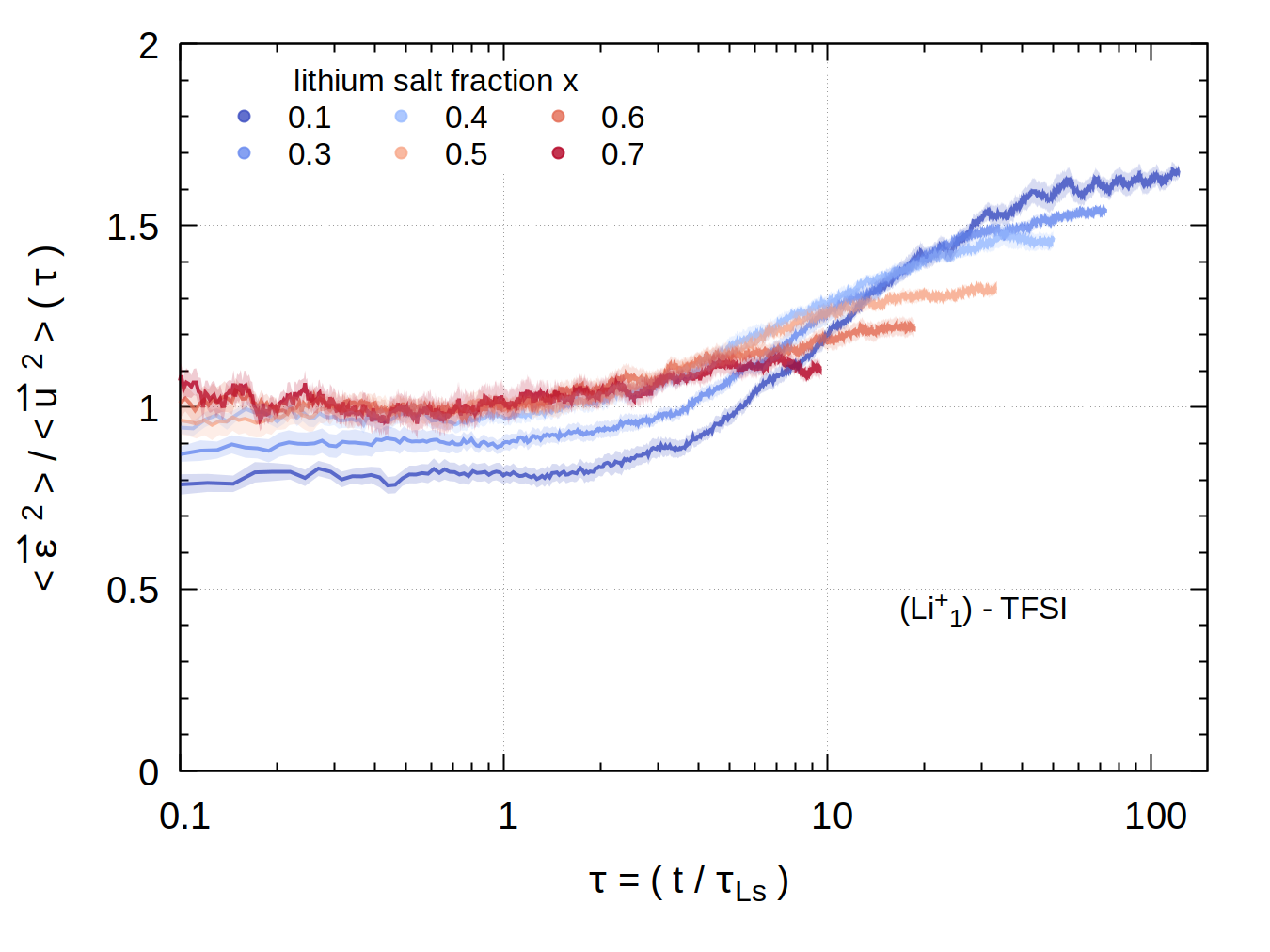}}
  \hfill
  \subfloat{\includegraphics[width=0.5\textwidth]{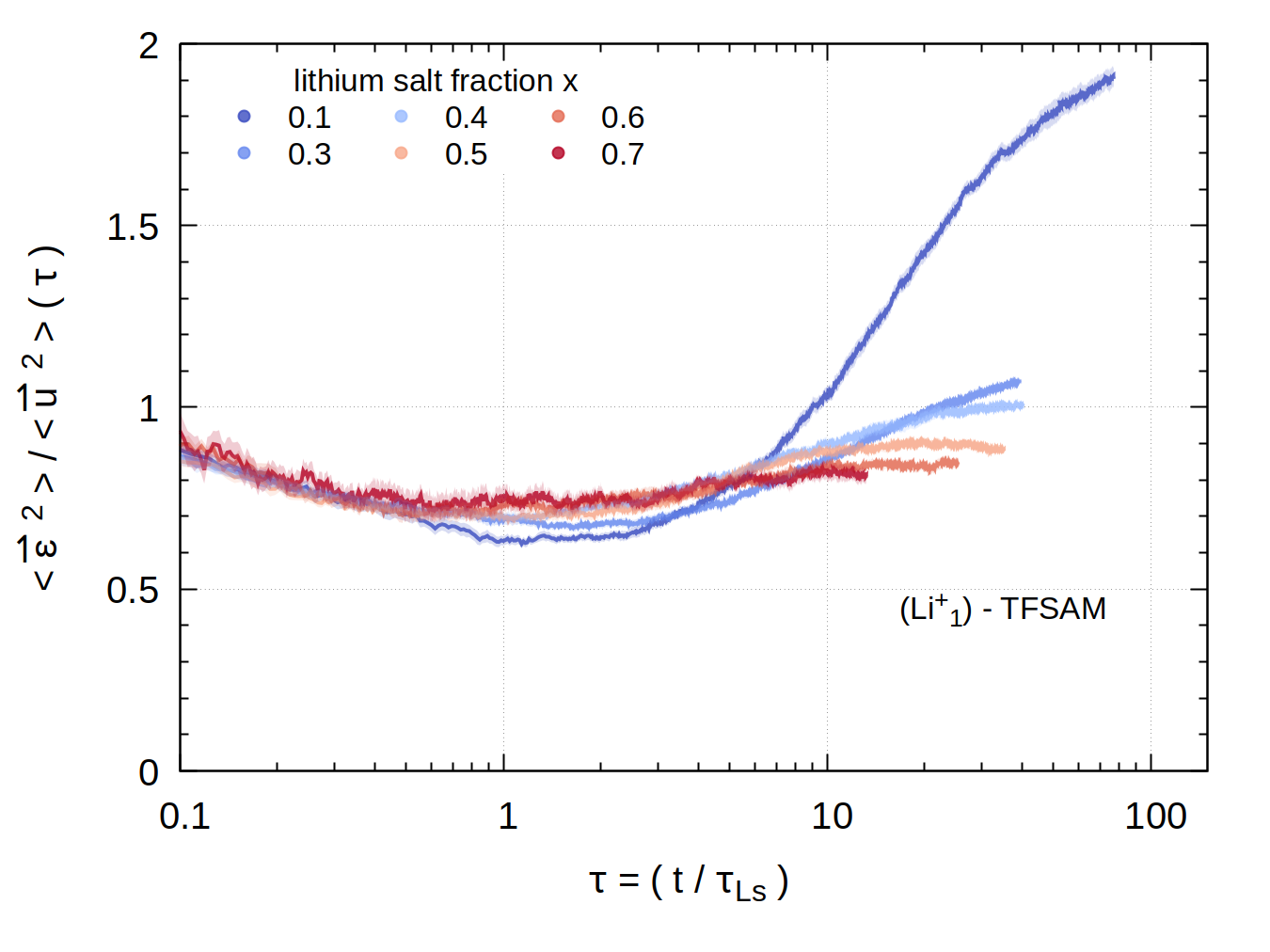}}

  \caption{Random motion $\langle \vec{\epsilon}^2 \rangle$ of anions, which have only a single $\text{Li}^+$ neighbour at time $\tau = 0$, scaled by the average squared lithium displacement $\langle \vec{u}^{\,2}\rangle$. $\langle .. \rangle$ denotes the ensemble average over the $\text{Li}^+$ that are involved in the single coordination of the anion at $\tau = 0$.} 
\label{fig:variance_eps_lithium_path_lambda1}

\end{figure}

%%%%%%%%%%%%%%%%%%%%%%%%%%%%%%%%%%%%%%%%%%%%%%%%%%%%%%%%%%%%%%%%%%%%%%%%%%%%%%%%%%
%%%%%%%%%%%%%%%%%%%%%%%%%%%%%%%%%%%%%%%%%%%%%%%%%%%%%%%%%%%%%%%%%%%%%%%%%%%%%%%%%%
%%%%%%%%%%%%%%%%%%%%%%%%%%%%%%%%%%%%%%%%%%%%%%%%%%%%%%%%%%%%%%%%%%%%%%%%%%%%%%%%%%
%%%%%%%%%%%%%%%%%%%%%%%%%%%%%%%%%%%%%%%%%%%%%%%%%%%%%%%%%%%%%%%%%%%%%%%%%%%%%%%%%%
%%%%%%%%%%%%%%%%%%%%%%%%%%%%%%%%%%%%%%%%%%%%%%%%%%%%%%%%%%%%%%%%%%%%%%%%%%%%%%%%%%
%%%%%%%%%%%%%%%%%%%%%%%%%%%%%%%%%%%%%%%%%%%%%%%%%%%%%%%%%%%%%%%%%%%%%%%%%%%%%%%%%%
%%%%%%%%%%%%%%%%%%%%%%%%%%%%%%%%%%%%%%%%%%%%%%%%%%%%%%%%%%%%%%%%%%%%%%%%%%%%%%%%%%
%%%%%%%%%%%%%%%%%%%%%%%%%%%%%%%%%%%%%%%%%%%%%%%%%%%%%%%%%%%%%%%%%%%%%%%%%%%%%%%%%%
%%%%%%%%%%%%%%%%%%%%%%%%%%%%%%%%%%%%%%%%%%%%%%%%%%%%%%%%%%%%%%%%%%%%%%%%%%%%%%%%%%
%%%%%%%%%%%%%%%%%%%%%%%%%%%%%%%%%%%%%%%%%%%%%%%%%%%%%%%%%%%%%%%%%%%%%%%%%%%%%%%%%%
%%%%%%%%%%%%%%%%%%%%%%%%%%%%%%%%%%%%%%%%%%%%%%%%%%%%%%%%%%%%%%%%%%%%%%%%%%%%%%%%%%
%%%%%%%%%%%%%%%%%%%%%%%%%%%%%%%%%%%%%%%%%%%%%%%%%%%%%%%%%%%%%%%%%%%%%%%%%%%%%%%%%%

\newpage
\textbf{O: Self van Hove function $\text{G}_{\text{s}}(r,\Delta t) $}

The self van Hove function $\text{G}_{\text{s}}(r,\Delta t)$ describes the probability distribution that a particle has moved a distance r within a time lag $\Delta$t away from its original position:

\begin{equation}
\text{G}_{\text{s}}(r,\Delta t) = \dfrac{1}{N} \langle \sum_{i=1}^{N} \delta\left(r - |\vec{r}_i(t+\Delta t) -\vec{r}_i(t) | \right) \rangle
\end{equation}

\begin{figure}[H]
  \centering
  \subfloat{\includegraphics[width=0.5\textwidth]{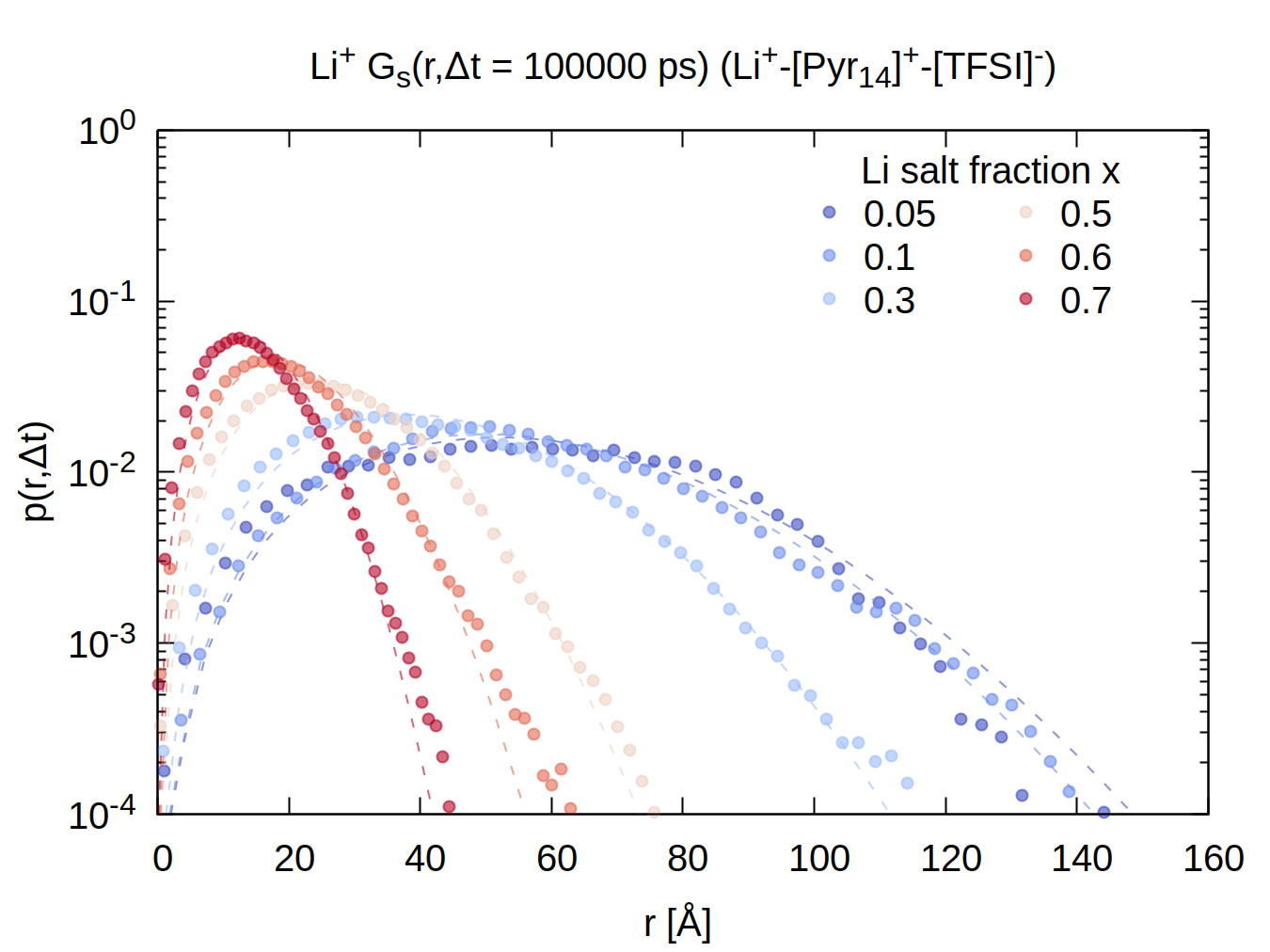}}
  \hfill
  \subfloat{\includegraphics[width=0.5\textwidth]{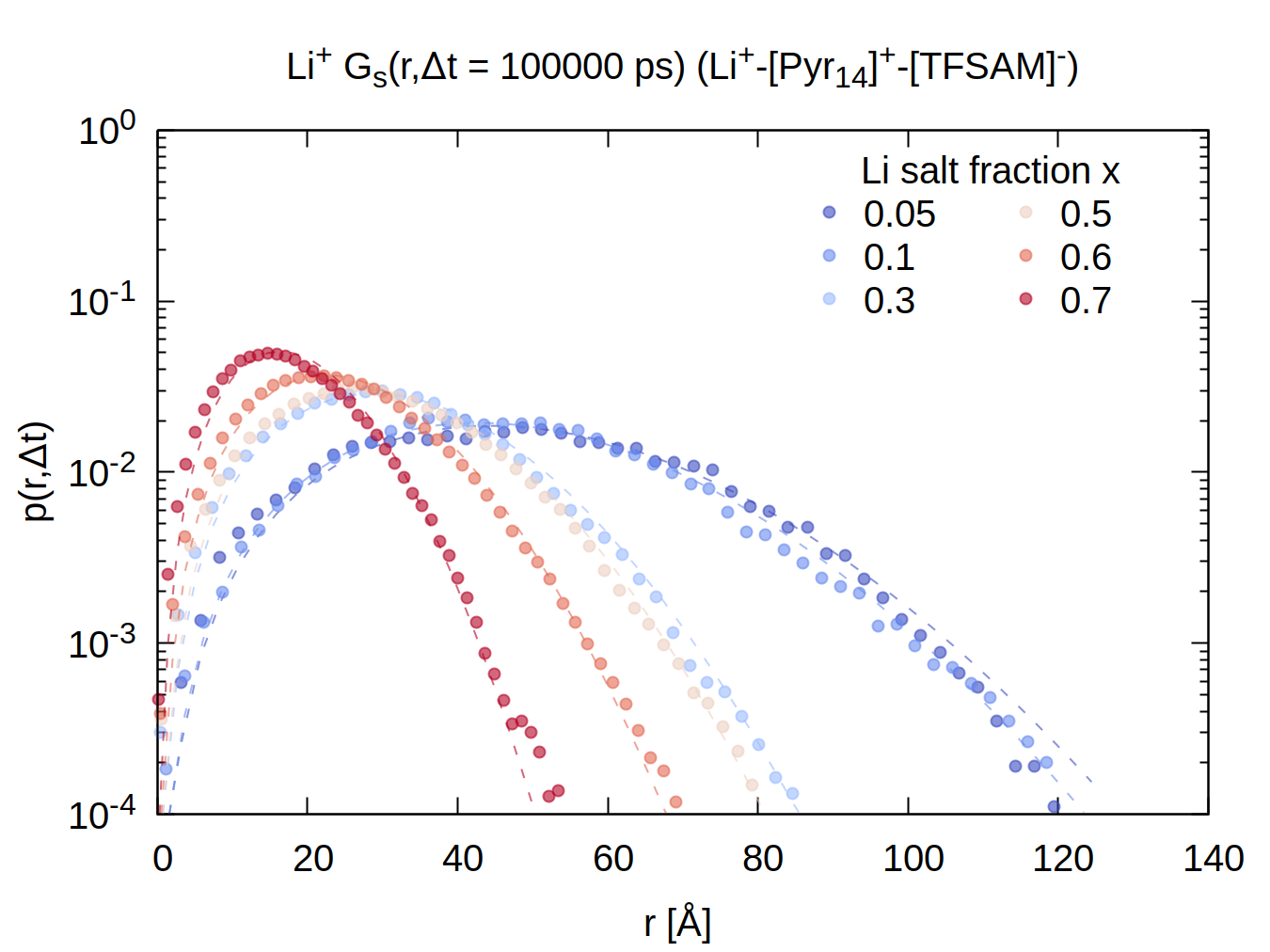}}

  \caption{The self van Hove functions $\text{G}_{\text{s}}(r,\Delta t)$ of the lithium ions in $\text{TFSI}^-$ (left) and  $\text{TFSAM}^-$ (right) containing electrolytes for the spectrum of investigated lithium salt concentrations x. $\text{G}_{\text{s}}(r,\Delta t)$ is compared to the corresponding ideal Gaussian distribution $\text{G}_{\text{0,s}}(r,\Delta t)$, which is depicted by the dashed lines.}
  \label{fig:Gsrt_lithium_TFSI_TFSAM}
\end{figure}

Figure S\ref{fig:Gsrt_lithium_TFSI_TFSAM} gives an overview of the displacement distributions $\text{G}_{\text{s}}(r,\Delta t)$ of the lithium ions at time lags $\Delta t$\,=\,10 ps, 100 ps, 1 ns, 10 ns and 100 ns for the broad spectrum of lithium salt contents. The dashed lines represent the displacement distribution expected for an ideal diffusive motion that exhibits a Gaussian behavior \newline $\text{G}_{\text{0,s}}(r,\Delta t) = \left(\dfrac{3}{2\pi \cdot \langle r^2(\Delta t)\rangle}\right)^{3/2}\exp\left(-\dfrac{3}{2} \dfrac{r^2}{\langle r^2(\Delta t)\rangle}\right)  $.
Comparison of the probed displacement distribution $\text{G}_{\text{s}}(r,\Delta t)$ to $\text{G}_{\text{0,s}}(r,\Delta t)$ shows that the non-Gaussian characteristics of the lithium dynamics increase in both electrolyte series with increasing lithium salt content. The tails of the distributions at high displacement distances $r$ are indicative of a fraction of lithium ions that display a higher mobility.\newline
We note that although no secondary peaks emerge in the distributions at elevated salt content, which would indicate that lithium transport is achieved through discrete "hopping" events, the pronounced tails of $\text{G}_{\text{s}}(r,\Delta t)$ might stem from lithium jumps.

\begin{figure}[H]
  \centering
  \subfloat{\includegraphics[width=0.5\textwidth]{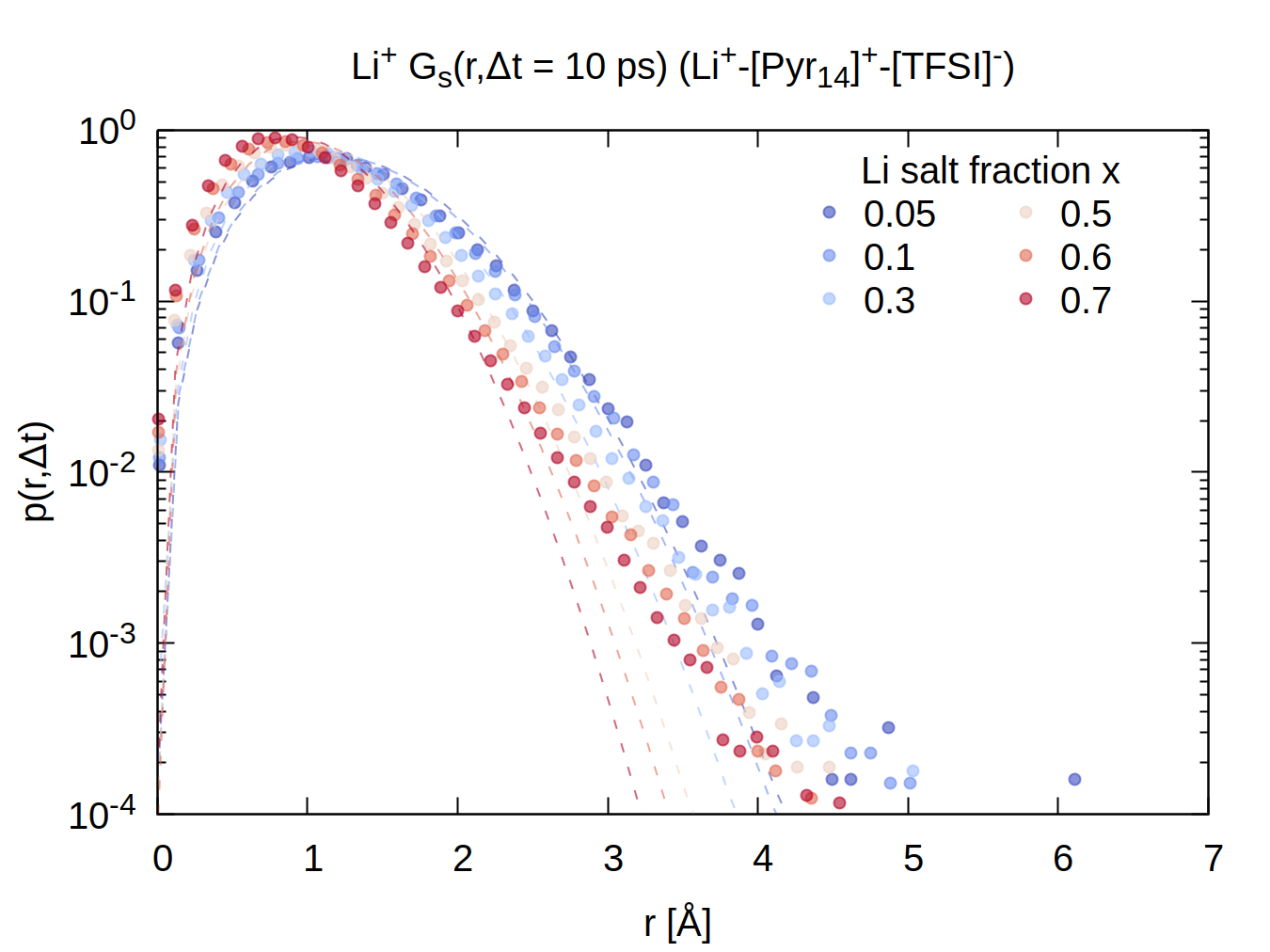}}
  \hfill
  \subfloat{\includegraphics[width=0.5\textwidth]{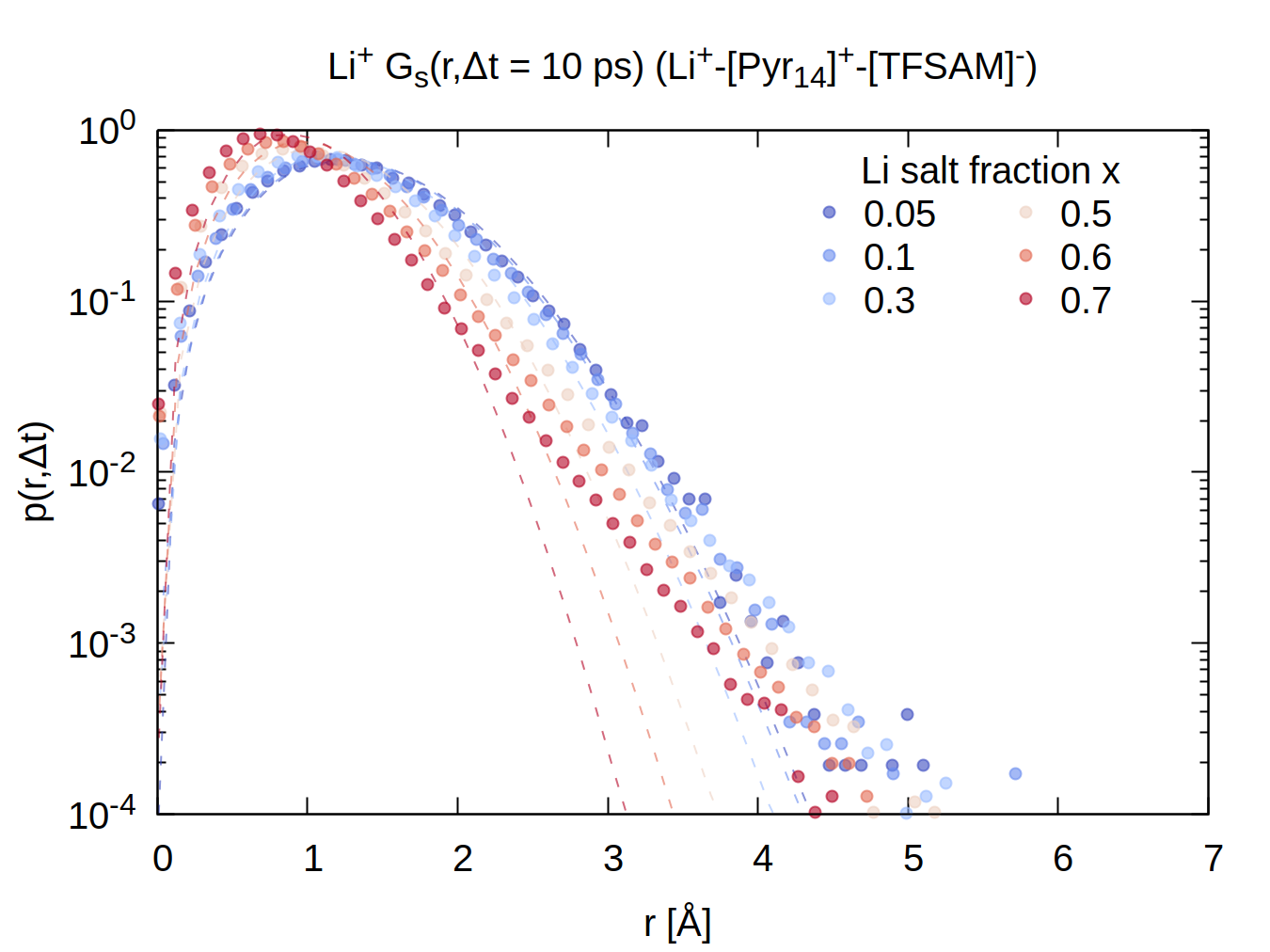}}
  
\end{figure}

\begin{figure}[H]
  \centering
  \subfloat{\includegraphics[width=0.5\textwidth]{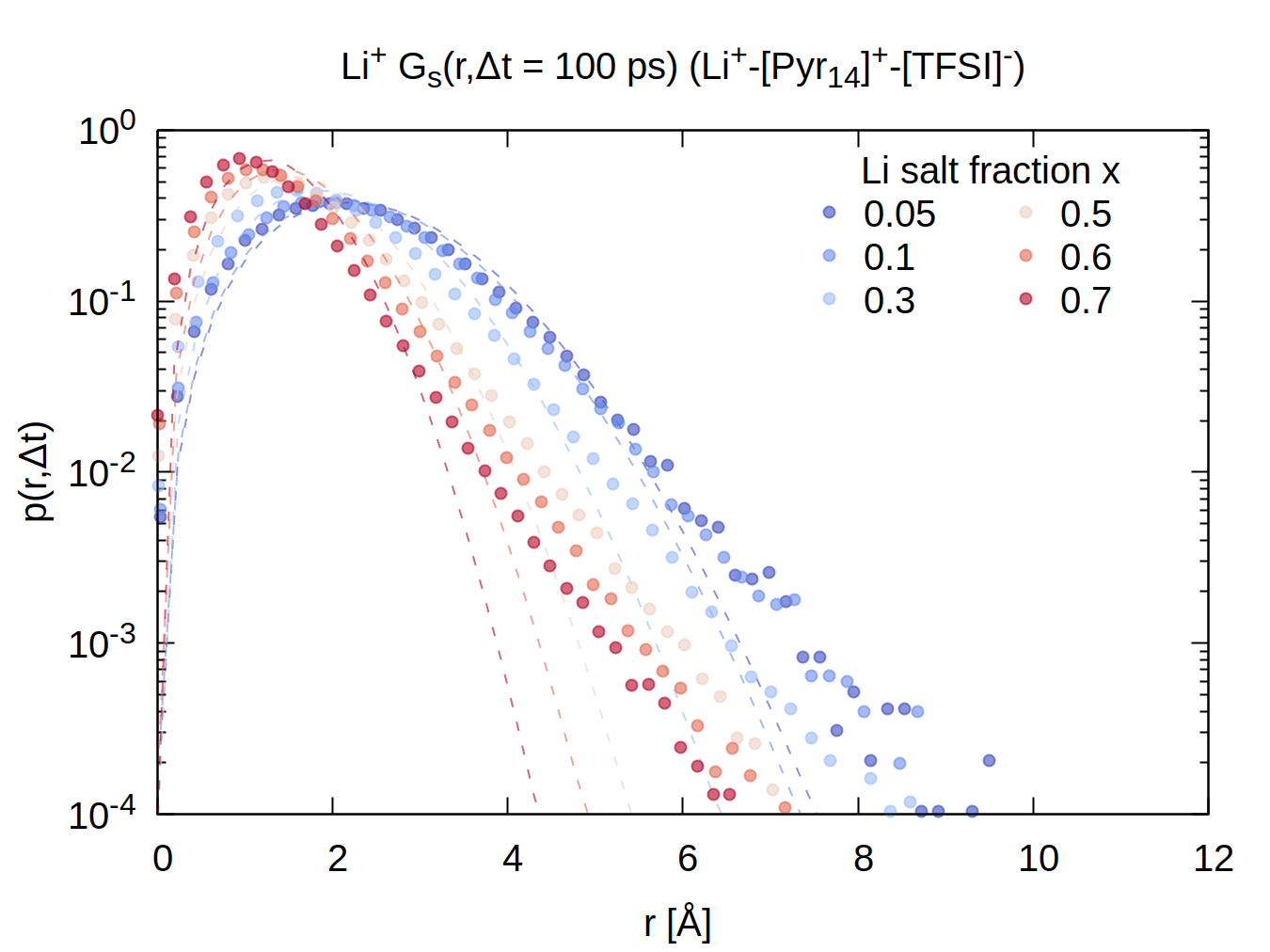}}
  \hfill
  \subfloat{\includegraphics[width=0.5\textwidth]{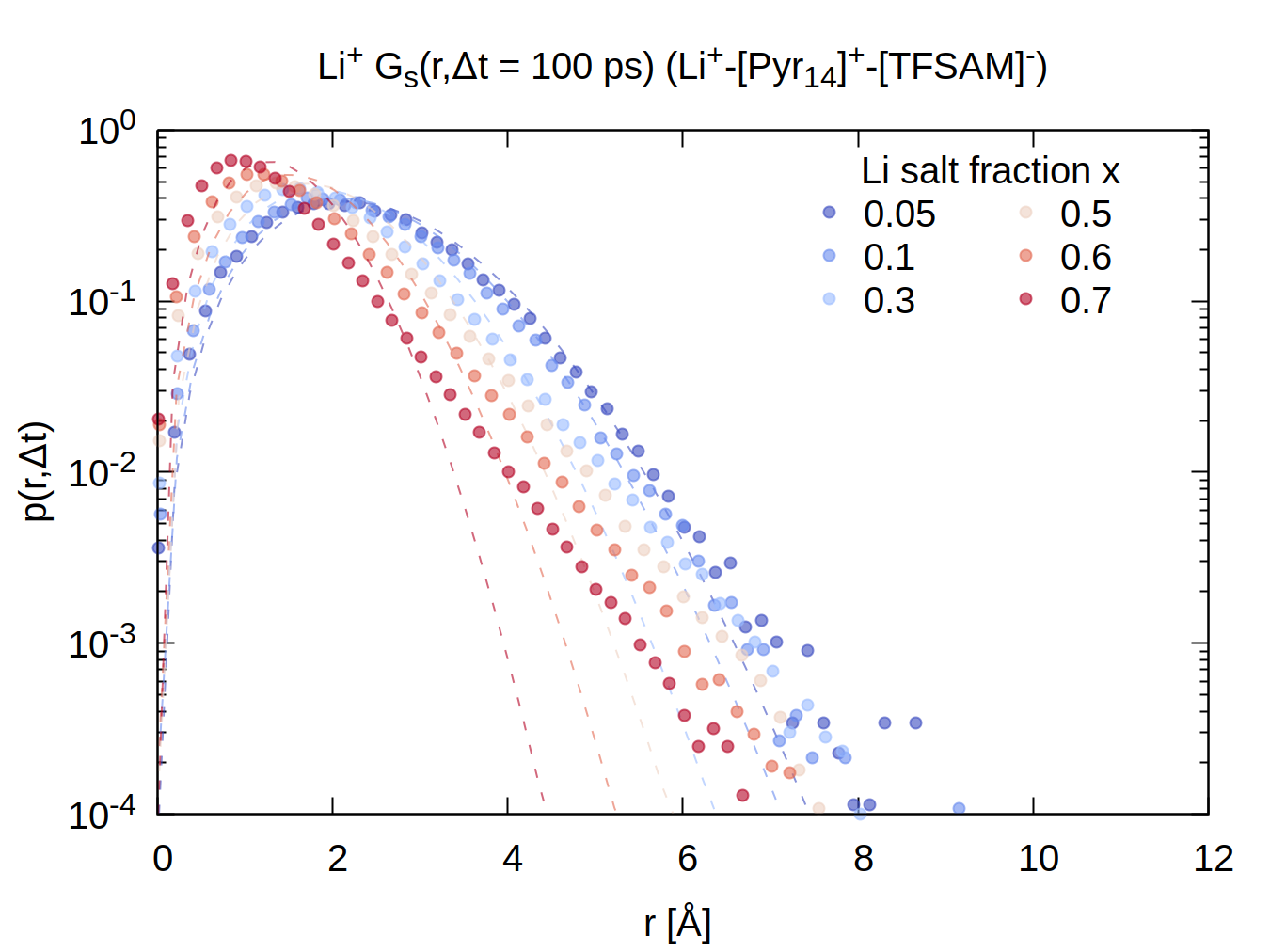}}
  
\end{figure}

\begin{figure}[H]
  \centering
  \subfloat{\includegraphics[width=0.5\textwidth]{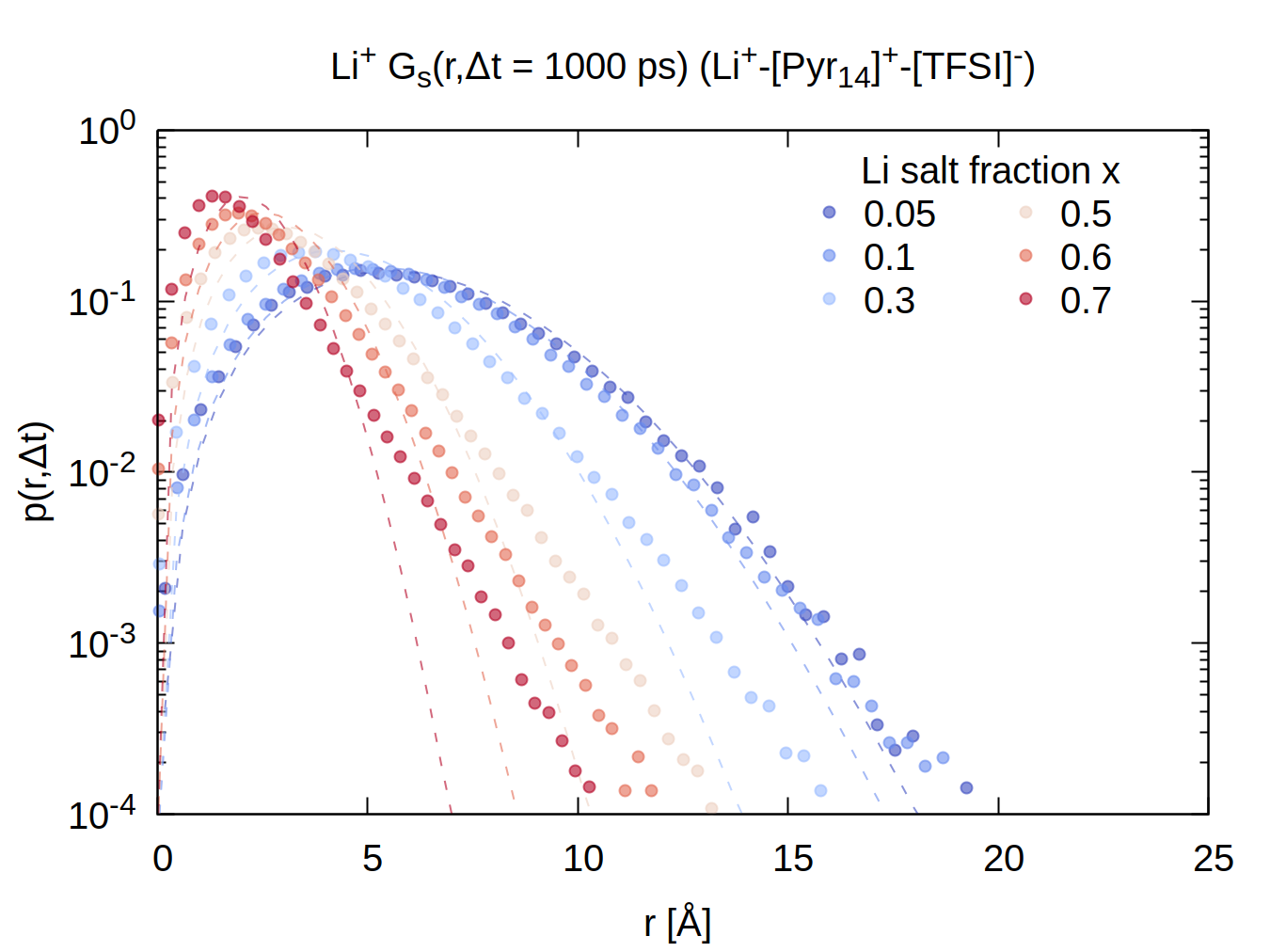}}
  \hfill
  \subfloat{\includegraphics[width=0.5\textwidth]{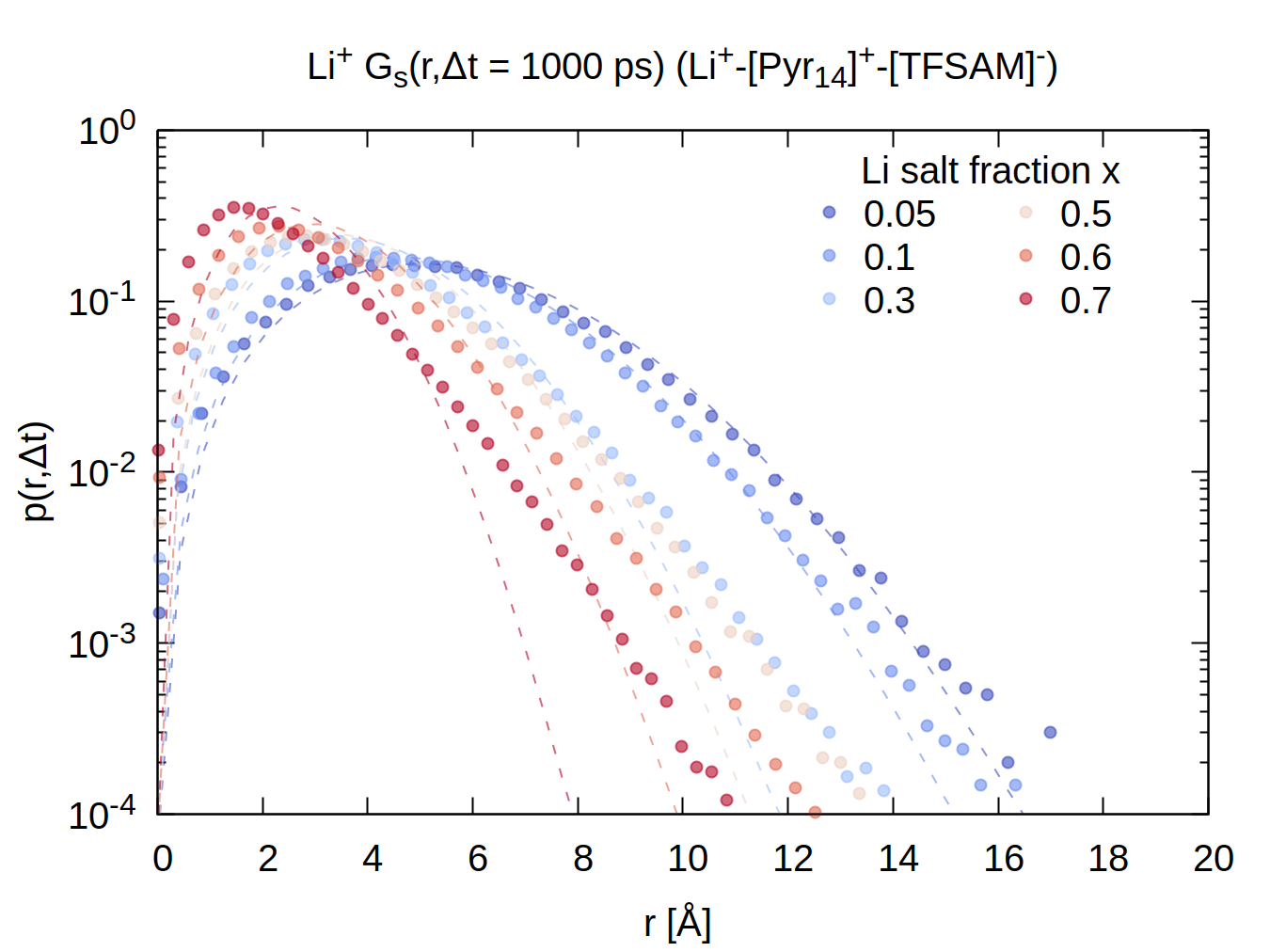}}

\end{figure}

\begin{figure}[H]
  \centering
  \subfloat{\includegraphics[width=0.5\textwidth]{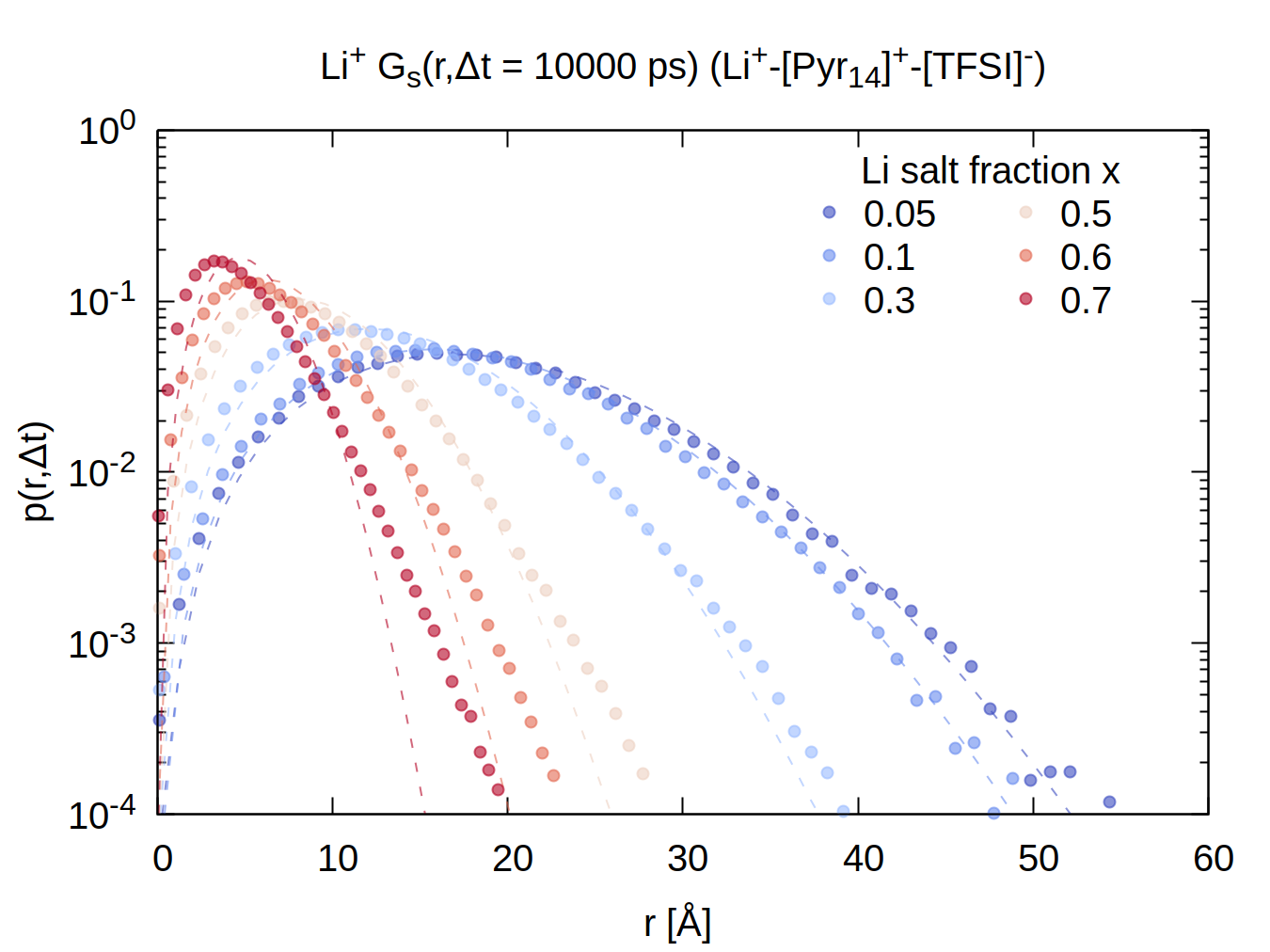}}
  \hfill
  \subfloat{\includegraphics[width=0.5\textwidth]{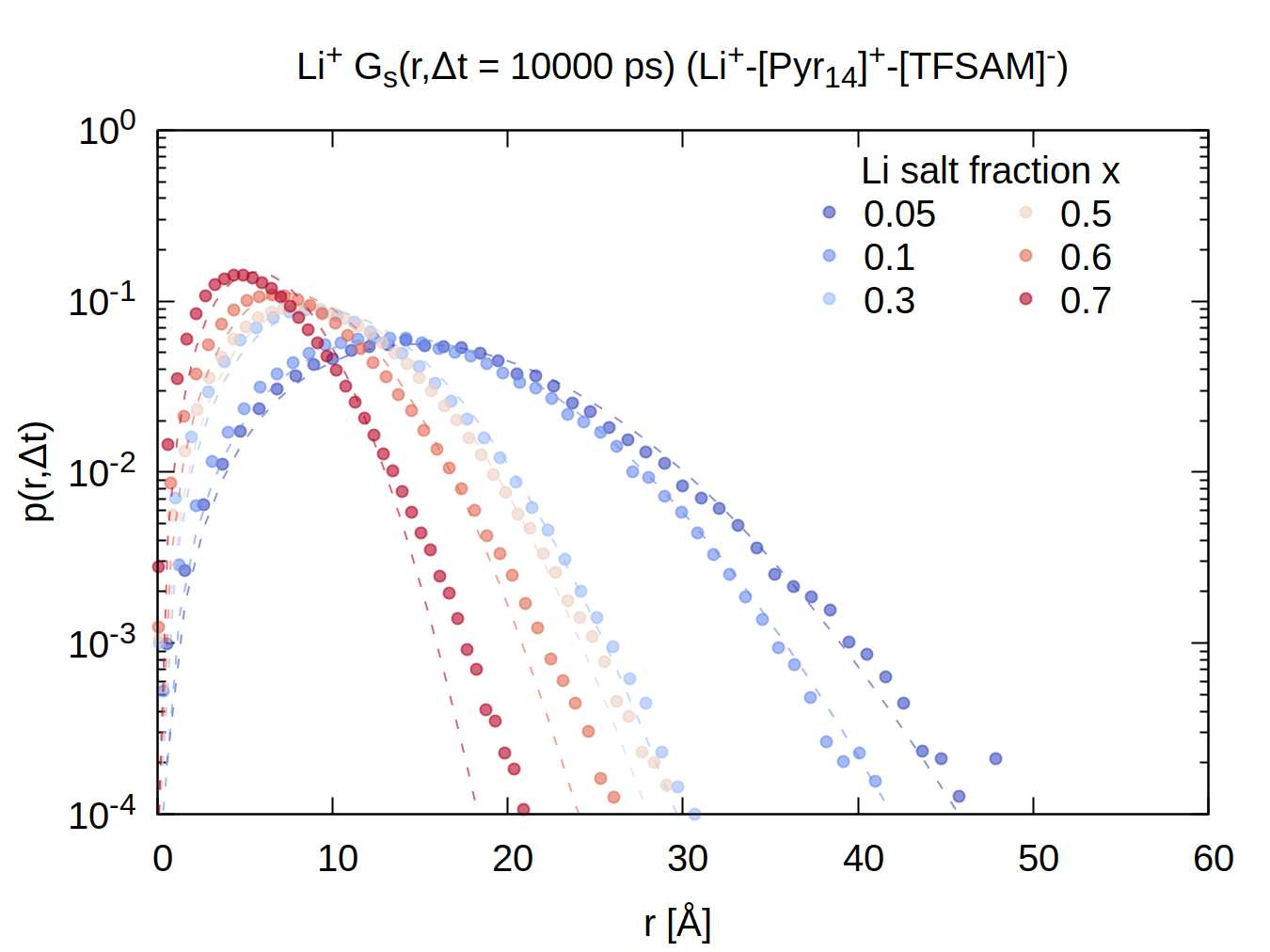}}
  
\end{figure}

\newpage
\textbf{P: Non-Gaussian parameter $\alpha_2$}

The non-Gaussian parameter (NGP) $\alpha_2$ which probes the deviation from truly Gaussian dynamics, is extracted from the second and fourth moment of the particle displacements:

\begin{equation}
\alpha_2(t) = \dfrac{3}{5}\cdot\dfrac{\langle \Delta \vec{r}^{\,\,4}(t) \rangle}{\langle \Delta \vec{r}^{\,\,2}(t) \rangle^2} - 1,
\end{equation}
where $\Delta \vec{r}(t)\,=\,(\vec{r}(t)-\vec{r}(0))$ denotes the particle displacement within the time $t$ and the brackets $<..>$ indicate the ensemble average over all particles for the given time lag.

\begin{figure}[H]
  \centering
  \subfloat{\includegraphics[width=0.5\textwidth]{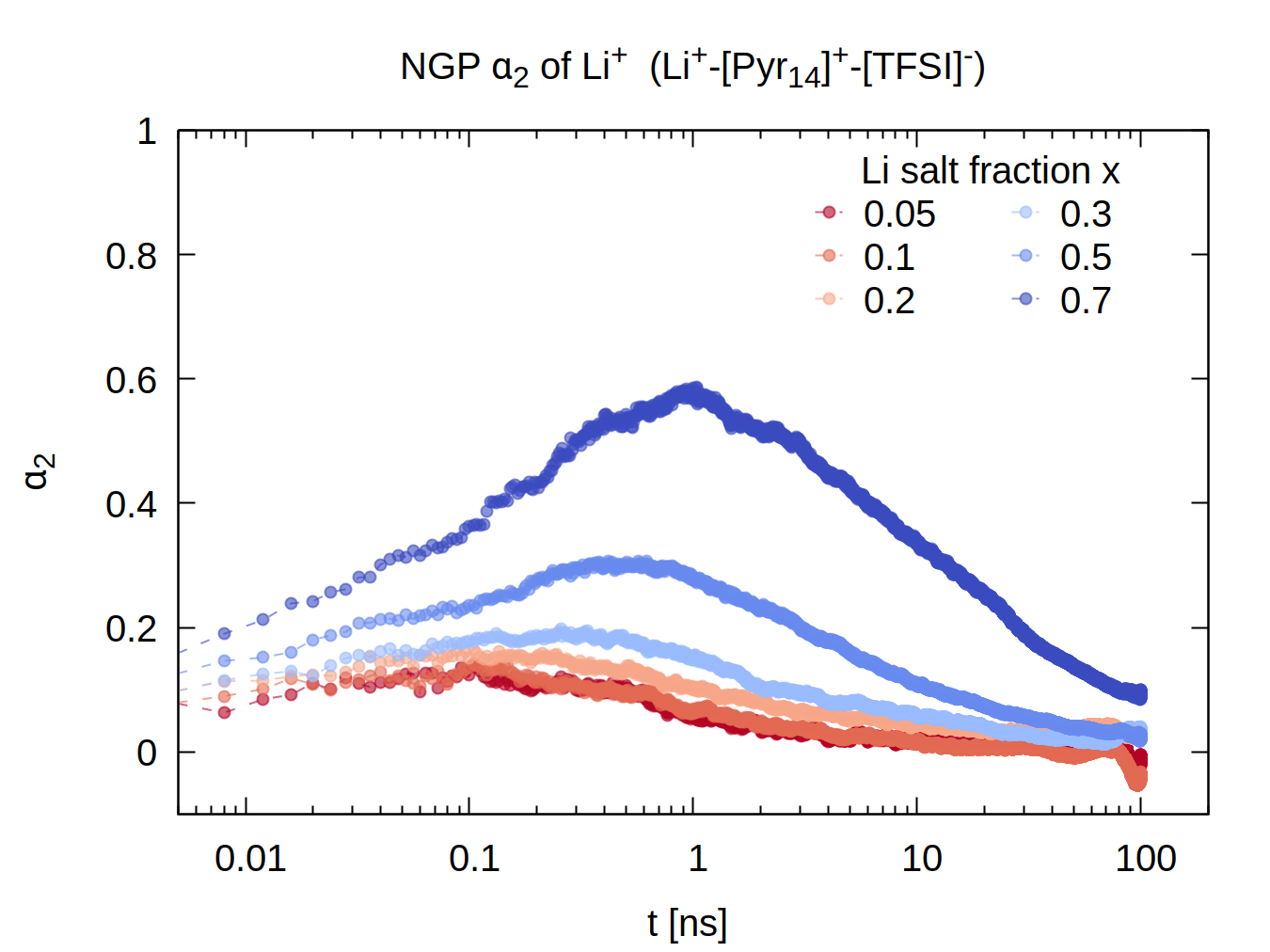}\label{fig:alpha2_lithium_in_tfsi}}
  \hfill
  \subfloat{\includegraphics[width=0.5\textwidth]{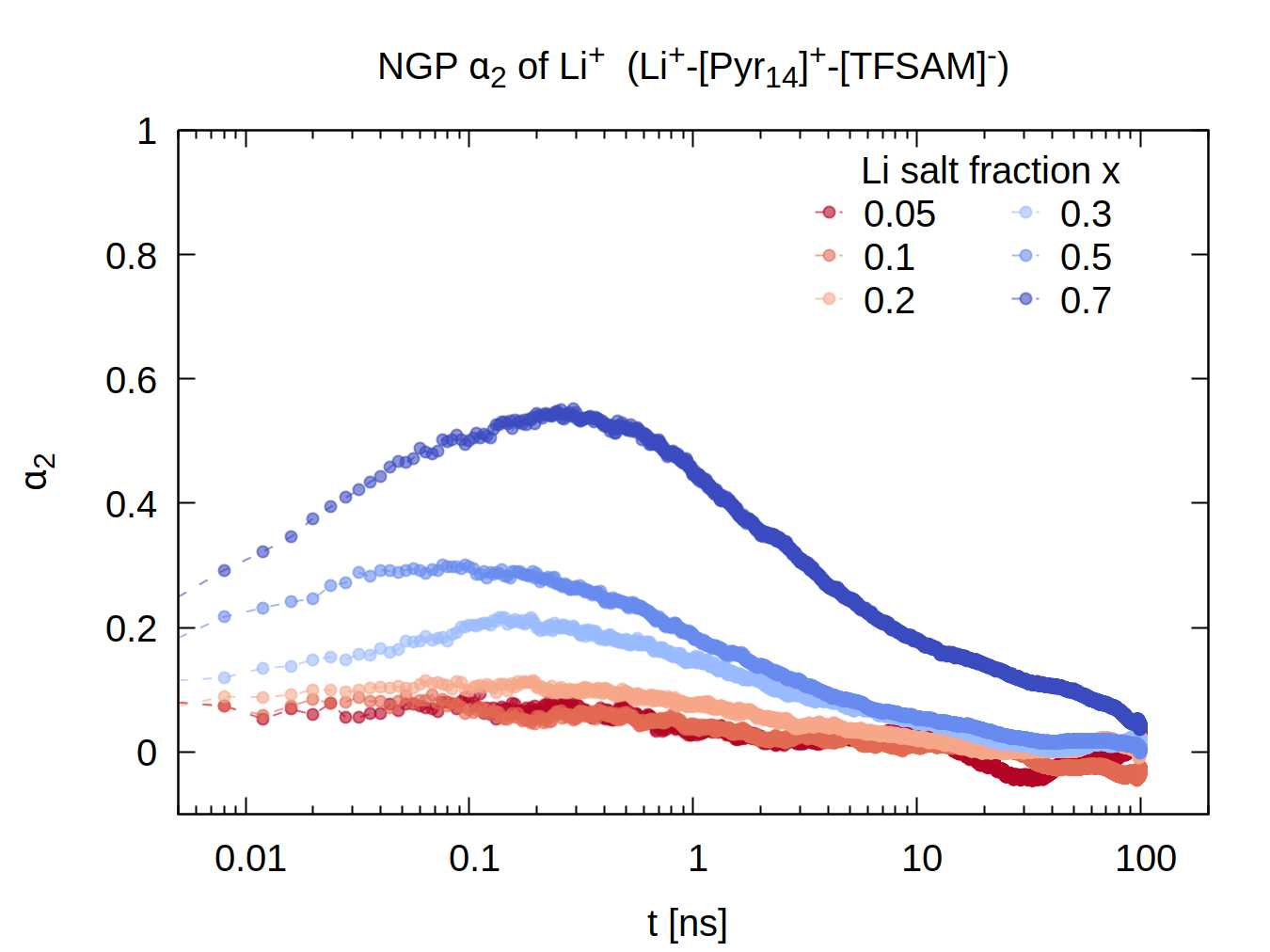}\label{fig:alpha2_lithium_in_tfsam}}
\hfill
  \centering
  \subfloat{\includegraphics[width=0.5\textwidth]{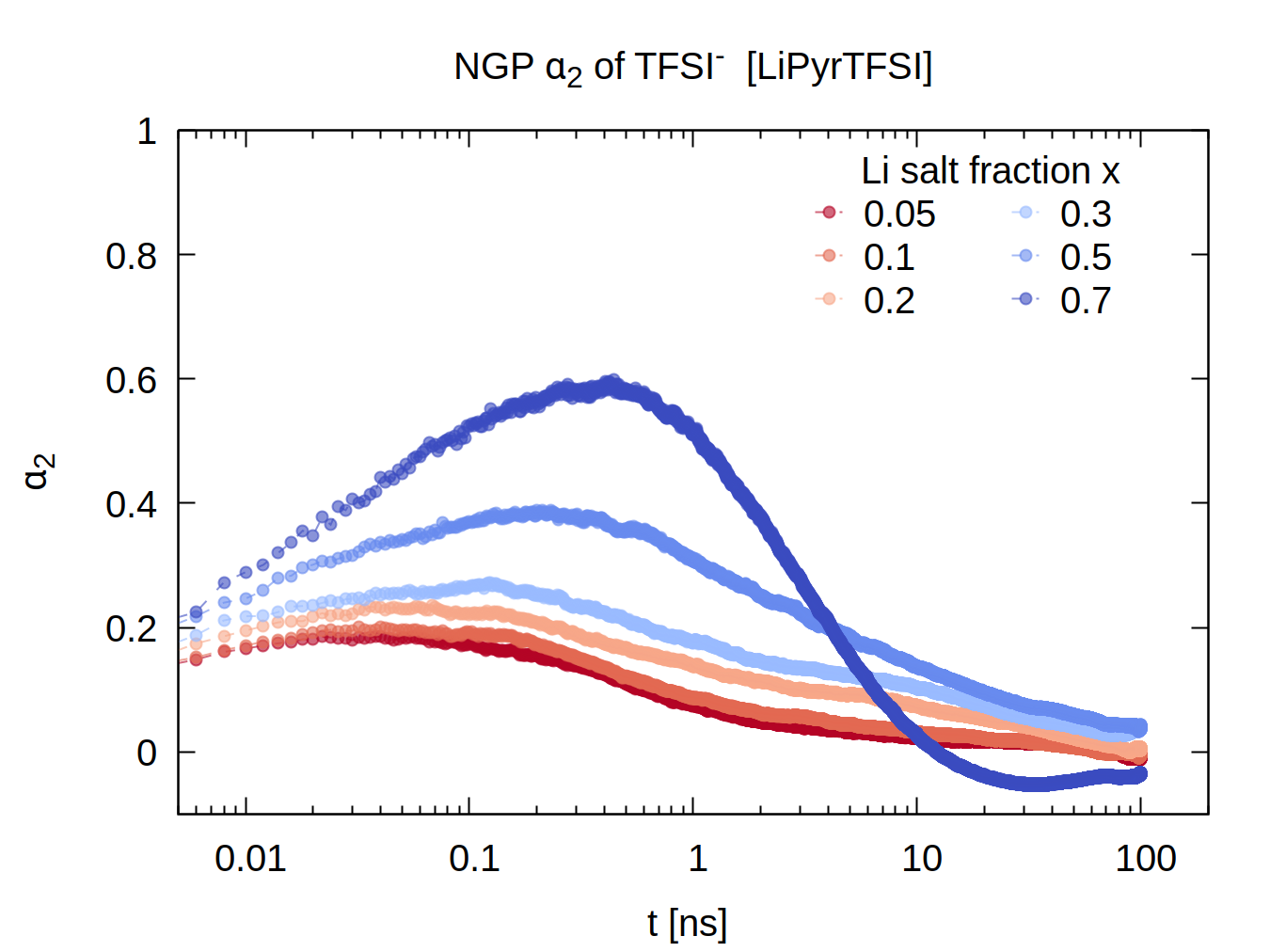}\label{fig:alpha2_tfsi}}
  \hfill
  \subfloat{\includegraphics[width=0.5\textwidth]{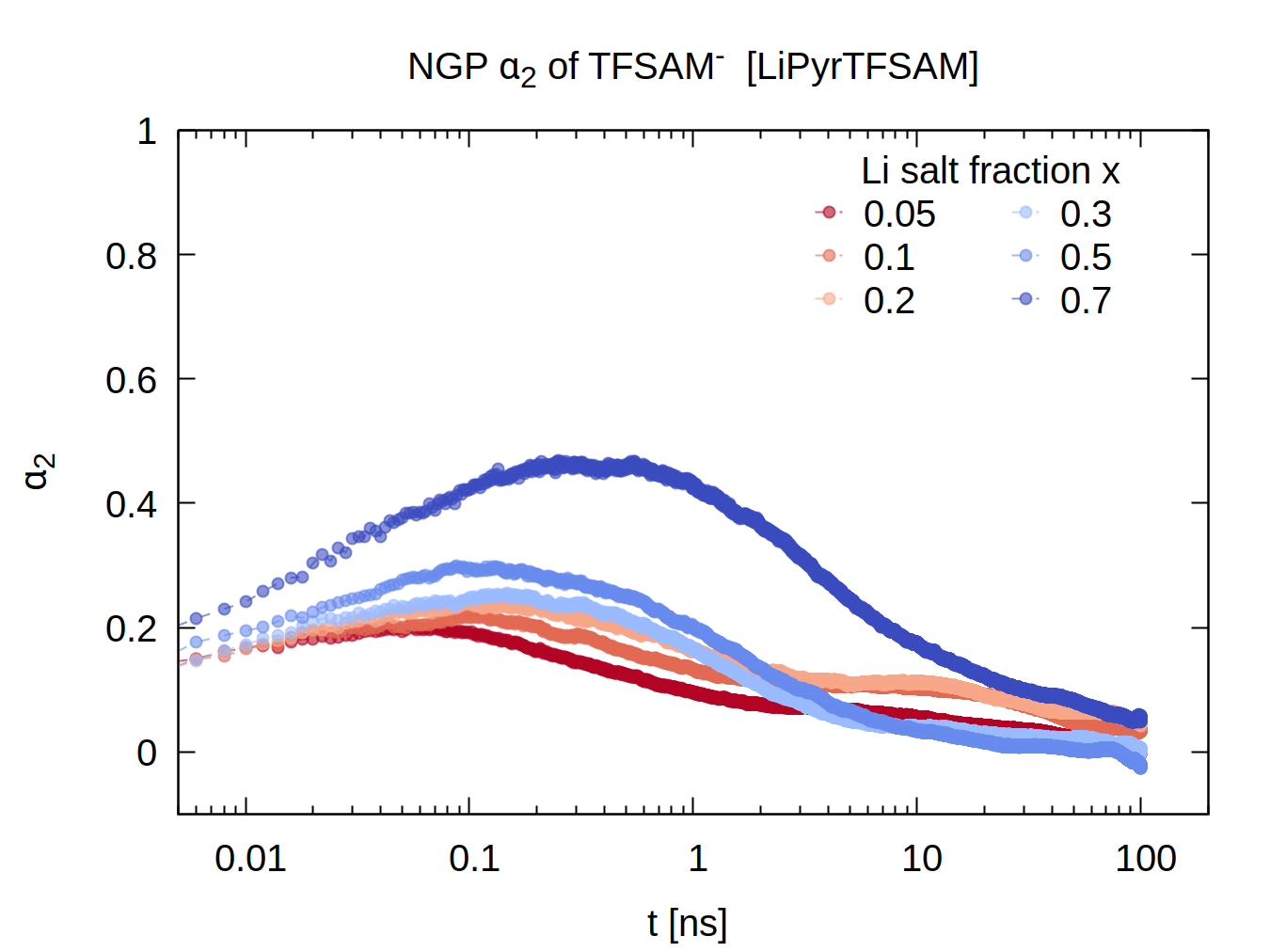}\label{fig:alpha2_tfsam}}
\end{figure}
\begin{figure}[H]

  \centering
  \subfloat{\includegraphics[width=0.5\textwidth]{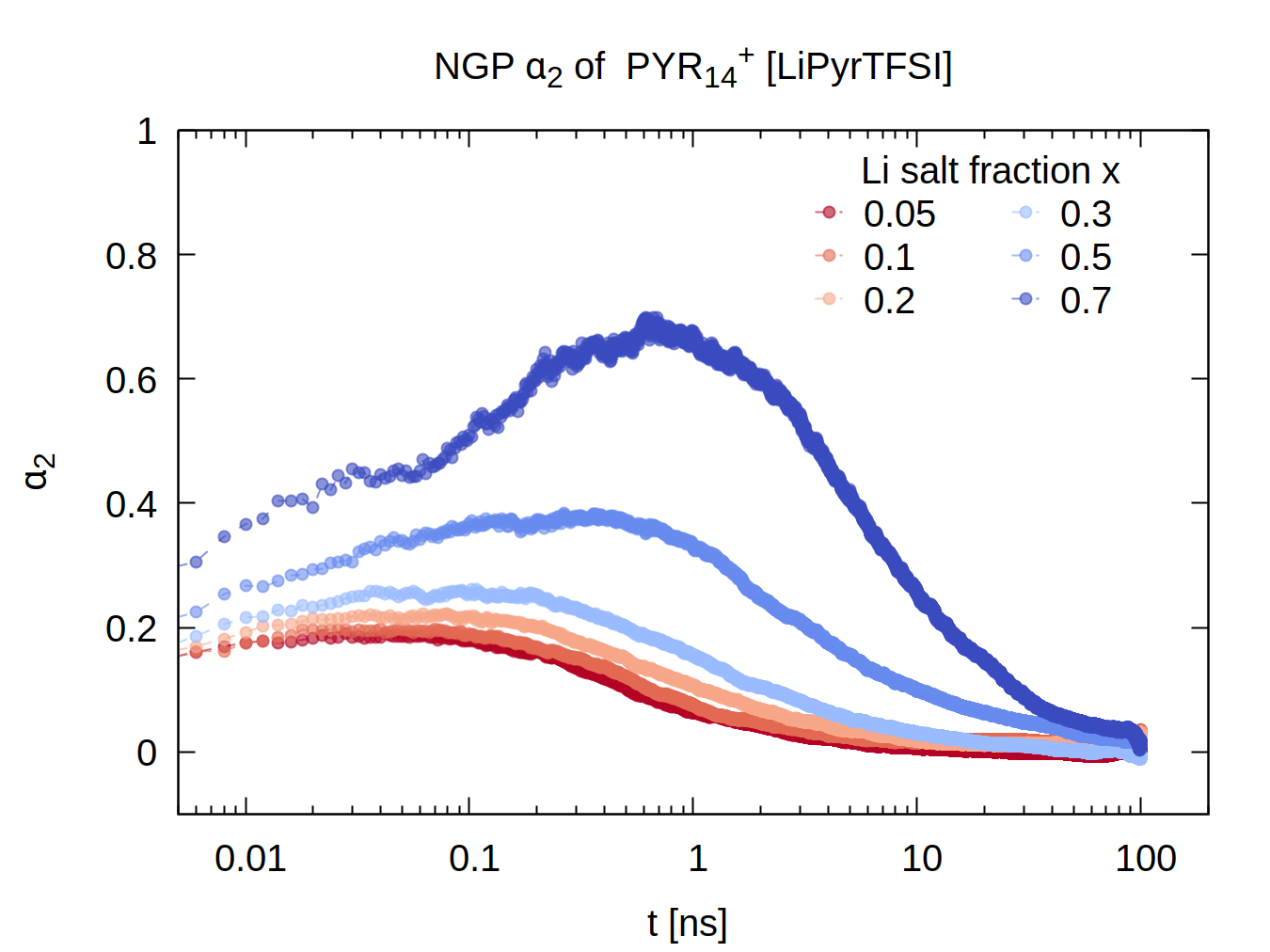}\label{fig:alpha2_pyr_in_tfsi}}
  \hfill
  \subfloat{\includegraphics[width=0.5\textwidth]{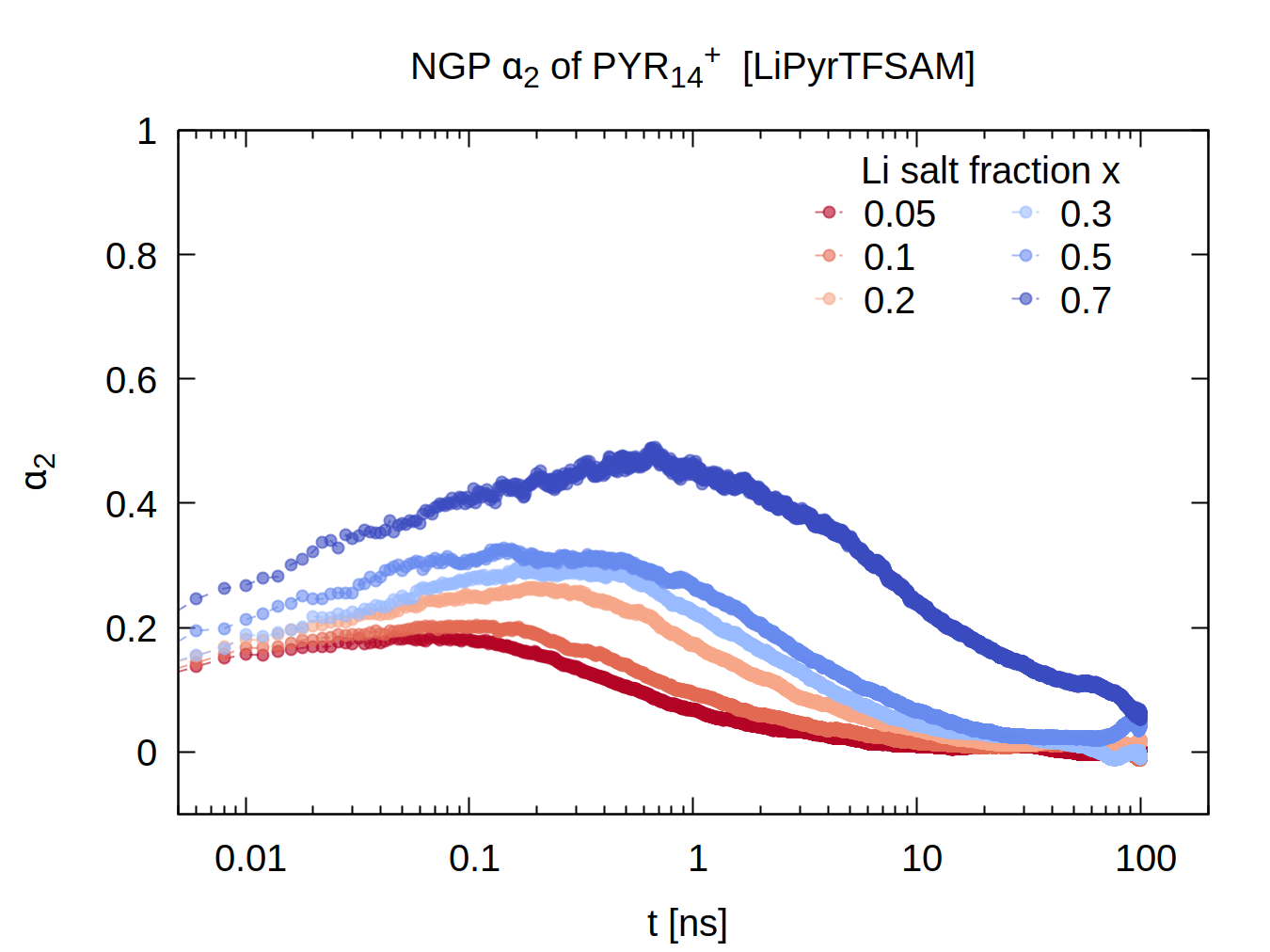}\label{fig:alpha2_pyr_in_tfsam}}
  \caption{Non-gaussian parameters $\text{Li}^+$ (top), anions (middle) and $\text{Pyr}_{14}^+$ (bottom) dynamics in $\text{TFSI}^-$ (left) and $\text{TFSAM}^-$ (right) containing electrolytes for various lithium salt fractions x and averaging over 4 blocks of 100\,ns duration each.}
  \label{fig:NGP_lithium}
\end{figure}

\bibliography{literature}
%%%%%%%%%%%%%%%%%%%%%%%%%%%%%%%%%%%%%%%%%%%%%%%%
%%%%%%%%%%%%%%%%%%%%%%%%%%%%%%%%%%%%%%%%%%%%%%%%
%%%%%%%%%%%%%%%%%%%%%% TOC graphic
%%%%%%%%%%%%%%%%%%%%%%%%%%%%%%%%%%%%%%%%%%%%%%%%
\newpage
\textbf{TOC graphic:}

\begin{figure}[H]
  \centering
  \includegraphics[width=0.7\textwidth]{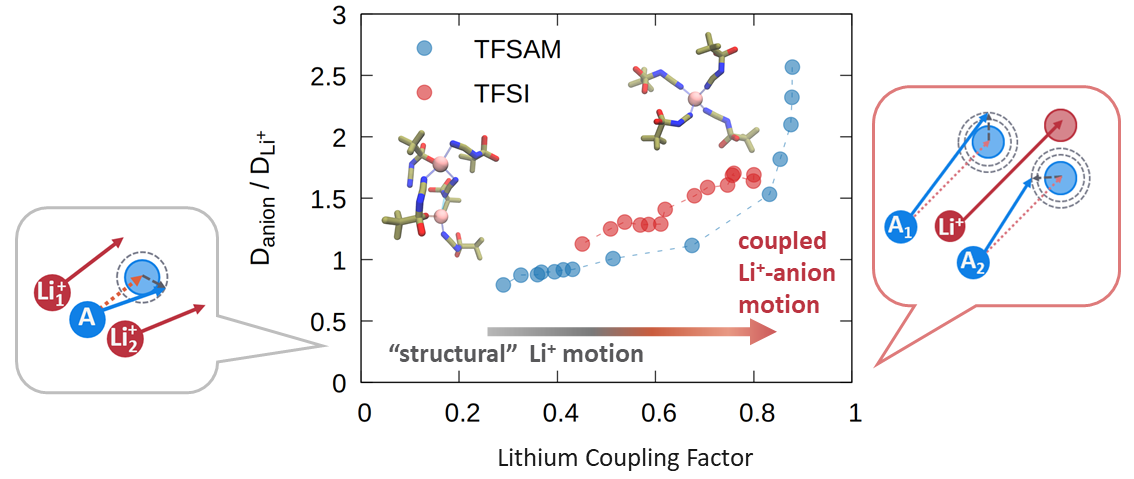}
  \caption*{Increased information content of the LCF $\lambda$ compared to $\text{D}_{\text{anion}}/\text{D}_{\text{Li}^+}$ and subsequently derived $\text{Li}^+$ transport principles in ionic liquid electrolytes.}
  
\end{figure}